\documentclass[fleqn,usenatbib]{mnras}

\usepackage{newtxtext,newtxmath}

\usepackage[T1]{fontenc}
\usepackage{array}
\usepackage{tabularx}
\usepackage{rotating}
\usepackage{longtable}
\usepackage{pdflscape}
\usepackage{tablefootnote}
\usepackage{subfig}  
\usepackage{graphicx}

\newcommand{\orcid}[1]{\href{https://orcid.org/#1}{\textsuperscript{\includegraphics[width=8pt]{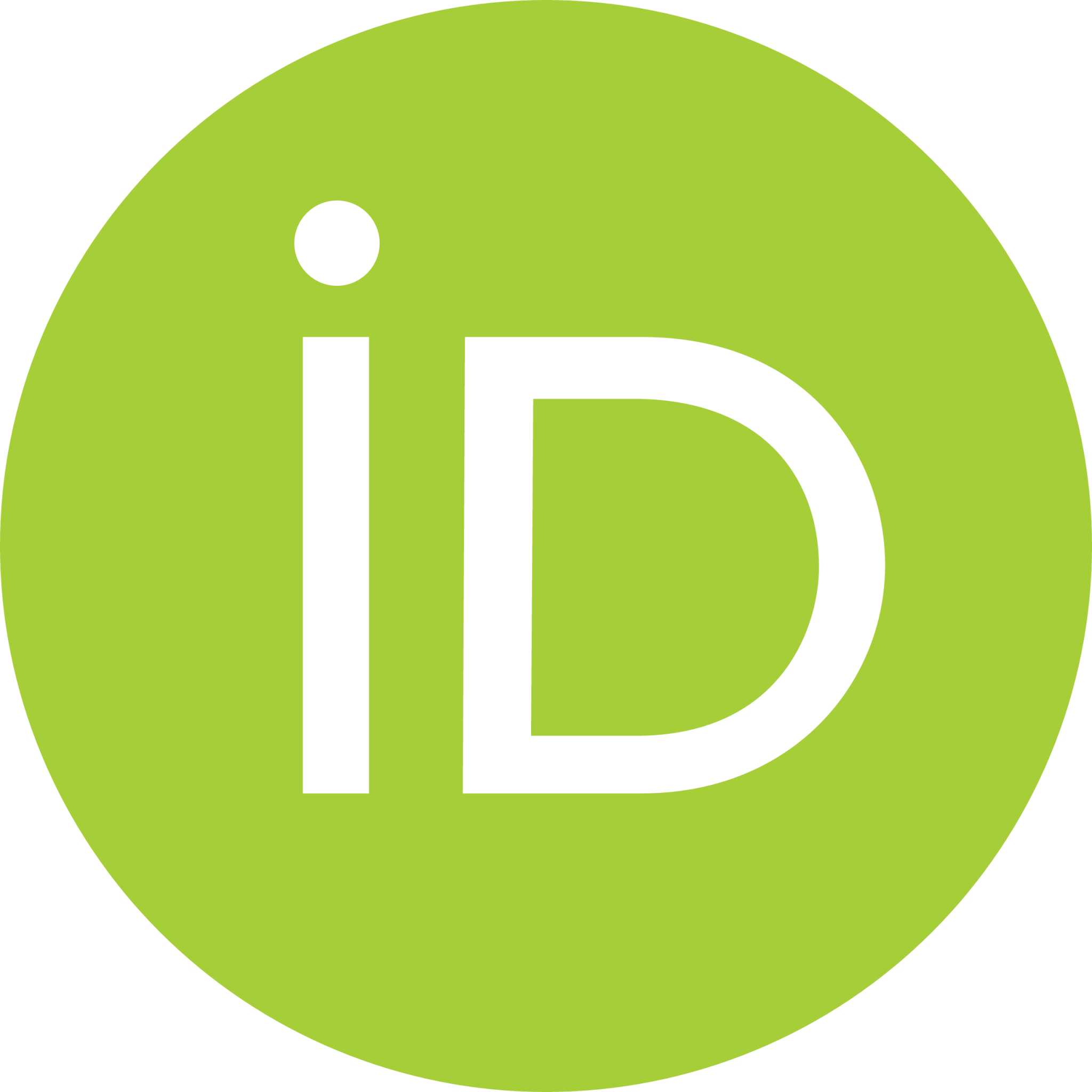}}}}

\DeclareRobustCommand{\VAN}[3]{#2}
\let\VANthebibliography\thebibliography
\def\thebibliography{\DeclareRobustCommand{\VAN}[3]{##3}\VANthebibliography}

\usepackage{graphicx}	
\usepackage{amsmath}	

\title[ALMA-detected galaxies in JWST PRIMER COSMOS]{JWST PRIMER: A deep JWST study of all ALMA-detected galaxies in PRIMER COSMOS -- dust-obscured star-formation history back to z $\simeq$ 7}

\author[F.-Y. Liu et al.]{Feng-Yuan Liu\orcid{0000-0003-1386-3676},$^{1}$\thanks{E-mail: Fengyuan.Liu@ed.ac.uk}
James\,S.\,Dunlop\orcid{0000-0002-1404-5950},$^{1}$
Ross\,J.\,McLure,$^{1}$
Derek\,J.\,McLeod\orcid{0000-0003-4368-3326},$^{1}$
Laia Barrufet\orcid{0000-0003-1641-6185},$^{1}$ \and
Adam\,C.\,Carnall\orcid{0000-0002-1482-5818},$^{1}$
Ryan Begley\orcid{0000-0003-0629-8074},$^{1}$
Pablo\,G.\,P\'erez-Gonz\'alez\orcid{0000-0003-4528-5639},$^{2}$
Callum\,T.\,Donnan\orcid{0000-0002-7622-0208},$^{1}$ 
Richard\,S. Ellis\orcid{0000-0001-7782-7071},$^{3}$ \and
Norman\,A.\,Grogin\orcid{0000-0001-9440-8872},$^{4}$ 
Dan Magee\orcid{0000-0002-6668-2011},$^4$ 
Garth\,D.\,Illingworth\orcid{0000-0002-8096-2837},$^5$
Fergus Cullen\orcid{0000-0002-3736-476X},$^{1}$
Struan\,D.\,Stevenson\orcid{0000-0001-5642-752X},$^{1}$ \and
Anton\,M.\,Koekemoer\orcid{0000-0002-6610-2048},$^4$
Adriano Fontana\orcid{0000-0003-3820-2823},$^{6}$
Rebecca\,A.\,A.\,Bowler\orcid{0000-0003-3917-1678}$^{7}$\\
\\
$^{1}$Institute for Astronomy, University of Edinburgh, Royal Observatory, Edinburgh, EH9 3HJ, UK\\
$^{2}$Centro de Astrobiolog\'ia (CAB), CSIC-INTA, Ctra. de Ajalvir km 4, Torrej\'on de Ardoz, E-28850, Madrid, Spain\\
$^{3}$Department of Physics \& Astronomy, University College London. Gower St., London WC1E 6BT, UK\\
$^{4}$Space Telescope Science Institute, 3700 San Martin Drive, Baltimore, MD 21218, USA\\
$^{5}$Department of Astronomy and Astrophysics, UCO/Lick Observatory, University of California, Santa Cruz, CA 95064, USA\\
$^{6}$INAF - Osservatorio Astronomico di Roma, via di Frascati 33, 00078 Monte Porzio Catone, Italy\\
$^{7}$Jodrell Bank Centre for Astrophysics, Department of Physics \& Astronomy, School of Natural Sciences, The University of Manchester, Manchester, M13 9PL, UK
}

\date{Accepted XXX. Received YYY; in original form ZZZ}

\pubyear{2025}

\begin{document}
\label{firstpage}
\pagerange{\pageref{firstpage}--\pageref{lastpage}}
\maketitle

\begin{abstract}
We use deep NIRCam and MIRI imaging from the {\it JWST} PRIMER survey to study the properties of ALMA detected (sub)mm sources
in the COSMOS field, with the aim of defining the
cosmic history of dust-enshrouded star formation. The wealth of ALMA
data in this field enabled us to isolate a robust sample of 128
(sub)mm sources within the 175\,arcmin$^2$ PRIMER COSMOS survey
footprint, spanning two decades in (sub)mm flux density.
The {\it JWST} imaging is deep and red enough to reveal secure galaxy
counterparts for {\it all} of these sources.
This 100\% identification
completeness is accompanied by a high level of redshift
completeness: 52\% of the sources have spectroscopic redshifts, and
this has enabled us to refine the photometric redshifts for the
remaining galaxies.
Armed with robust redshift information, we
calculate the star-formation rates (SFR) and stellar masses ($M_*$) of
all 128 ALMA-detected galaxies, and place them in the context of other
galaxies in the field.
We find that the vast majority of star
formation is dust-enshrouded in all of the ALMA-detected galaxies,
with SFR ranging from $\simeq 1000\,{\rm M_{\odot}\, yr^{-1}}$ down to
$\simeq 20\,{\rm M_{\odot}\, yr^{-1}}$. We also find that virtually
all (126/128) have high stellar masses, $M_* > 10^{10}\, {\rm
M_{\odot}}$, independent of redshift. The unusually high quality of
our sample enables us to make a robust estimate of the contribution of
the ALMA-detected galaxies to cosmic star-formation rate density,
$\rho_{\rm SFR}$. The existing ALMA imaging only covers $< 20$\% of the PRIMER COSMOS area, but based on our knowledge of all other massive galaxies in the field, we
produce a completeness-corrected estimate of dust-enshrouded $\rho_{\rm SFR}$. This confirms that UV-visible
star formation dominates $\rho_{\rm SFR}$ at $z > 4$, but also
indicates that dust-enshrouded star formation still makes a contribution of $\simeq 20$\%
at $z \simeq 8$, and $\simeq 5$\% at $z \simeq 10$.

\end{abstract}


\section{Introduction}
A complete understanding of the prevalence, history and physical drivers of dust-obscured star-formation activity is of vital importance for advancing our knowledge of galaxy formation and evolution. It is now well established that the majority of the star formation in galaxies during the era of peak activity (`cosmic noon’, corresponding to $z \simeq 2$) is enshrouded in dust, with most of the rest-frame ultraviolet (UV) light emitted by newly born massive stars being absorbed and re-emitted by the warmed dust grains in the rest-frame far-infrared \citep{2014ARA&A..52..415M, 2017MNRAS.466..861D, 2020ApJ...902..112B, 2021ApJ...909..165Z}. 

However, the picture at earlier times is much less clear, in large part due to the limitations of existing ground-based and space-based observing facilities in the near-mid infrared prior to the advent of the {\it James Webb Space Telescope} ({\it JWST}) .
Specifically, the limited wavelength coverage offered by the (relatively warm) {\it Hubble Space Telescope} ({\it HST}) ($\lambda < 1.6\,{\rm \mu m}$) has hampered its ability to uncover  dust-reddened galaxies beyond $z \simeq 3$ \citep{2023MNRAS.522..449B, 2023A&A...672A..18X, 2023ApJ...946L..16P, 2025ApJ...980...12S}, while the poor angular resolution of the {\it Spitzer} Space Telescope ultimately limited its ability to deliver near/mid-infrared imaging of sufficient depth and resolution to detect all but the most extreme objects at high redshifts. As a result, previous optical-IR follow-up of sources uncovered in (sub)mm surveys has generally been incomplete \citep[e.g.,][]{2017MNRAS.469..492M, 2020MNRAS.494.3828D}, leading to considerable uncertainty over the importance/evolution of dust-enshrouded star-formation activity during the first $\simeq 2$\,Gyr of cosmic history \citep[e.g.,][]{2016MNRAS.461.1100R, 2017MNRAS.471.4155K, 2019A&A...624A..98W, 2019ApJ...884..154W, 2021A&A...646A..76L, 2020A&A...643A...8G, 2021ApJ...923..215C}.

At very early cosmic times most/all star-formation activity is of course expected to be visible in the rest-frame UV, due to the delayed production of dust from the first generation of stars. At some point star formation in the Universe must therefore transition from mainly unobscured to primarily dust-obscured, and mapping out this transition is a key challenge. Interestingly, recent studies of very early galaxies with {\it JWST} indicate that most discovered to date at $z > 10$ are largely dust-free \citep[e.g.,][]{2024MNRAS.531..997C, 2025arXiv250111099C, 2025A&A...694A.215F}, in stark contrast to the aforementioned later dominance of dust-enshrouded activity by $z \simeq 1-3$. However, the comparison is complicated by the fact that the dust-obscured star formation activity at cosmic noon is dominated by the contribution of massive galaxies, with stellar masses $M_* > 10^{10}\,{\rm M_{\odot}}$ \citep{2018MNRAS.476.3991M}, and none of the extreme-redshift galaxies uncovered so far lie above this threshold \citep{2024MNRAS.533.3222D, 2025arXiv250103217D}.  Indeed, the cosmic volume covered by the deepest contiguous ALMA survey completed to date -- in the Hubble Ultra-Deep Field \citep[HUDF:][]{2017MNRAS.466..861D, 2020ApJ...902..112B} -- proved too small to contain any such massive galaxies at $z > 3$.

Now, however, the situation is being transformed due to the unique power of {\it JWST} in the near-mid infrared, and by a gradual growth in the availability of deep ALMA (sub)mm imaging over significant areas of sky. Indeed, the data provided by the major {\it JWST} Cycle-1 Public Release IMaging for Extragalactic Research \citep[PRIMER,][]{2021jwst.prop.1837D} survey
has already proved its worth in this arena, with \citet{2024A&A...691A.299G} using the PRIMER imaging of the COSMOS and UDS survey fields to successfully identify and study the submm-bright galaxies discovered through ALMA follow-up of the 850-${\rm \mu m}$ sources originally uncovered with SCUBA-2 on the James Clerk Maxwell Telescope (JCMT) \citep{2017MNRAS.465.1789G, 2019ApJ...880...43S, 2020MNRAS.495.3409S, 2020MNRAS.494.3828D}. However, such extreme (sub)mm galaxies, forming stars at a rate $\simeq 500\,{\rm M_{\odot}\,yr^{-1}}$ are extremely rare at high redshift, and only scratch the surface of the full dust-enshrouded star-forming galaxy population. What is required to complete our inventory is deep {\it JWST} imaging of a large and representative sample of (sub)mm galaxies, deep enough to reach down to the luminosities of the faintest sources uncovered by ALMA in the HUDF, and large enough to contain significant numbers of very high-redshift dusty galaxies (should they exist).

In this new investigation we attempt to bridge this gap by using the data from the aforementioned {\it JWST} PRIMER survey to identify and study the properties of all known ALMA sources in the central region of the COSMOS field \citep[corresponding to the PRIMER COSMOS survey footprint, which in turn is based on the {\it HST} COSMOS footprint in CANDELS;][]{2011ApJS..197...35G, 2011ApJS..197...36K}. While the ALMA imaging of this $\simeq 175$\,arcmin$^{2}$ field is a long way from complete, it transpires that a sufficiently large number of deep ALMA observing programmes have been conducted in this region for us to construct an archival sample of 128 (sub)mm sources \citep[only 17 of which are bright enough to feature in the AS2COSMOS survey;][]{2020MNRAS.495.3409S}. The outcome of our detailed {\it JWST}+{\it HST}+UltraVISTA optical-infrared investigation of this newly assembled ALMA sample is the key focus of the work presented here.

The rest of this paper is structured as follows. 
In Section \ref{sec:data} we present and summarize the key {\it JWST}, UltraVISTA and ALMA data (along with the important supporting archival {\it HST}, {\it Spitzer}, CFHT, and Subaru imaging) used in this new study of the PRIMER COSMOS field, detailing the reduction of the {\it JWST} PRIMER data, and how the ALMA catalogue in the PRIMER COSMOS survey footprint  was assembled from the A$^3$COSMOS database. 
In Section \ref{sec:construction} we then describe how the multi-frequency catalogue was produced for the ALMA sources, discussing first the accurate positional cross-matching of the (sub)mm and optical-infrared catalogues, and then explaining the process by which complete redshift information was assembled for the 128 ALMA-detected galaxies in our final sample. In Section \ref{sec:analysis} we use this unusually robust redshift information along with the full multi-frequency photometry to derive the basic physical properties of these galaxies, in particular their star-formation rates (SFR) (as estimated through different techniques) and stellar masses ($M_*$). We conclude this section by making a new estimate of the contribution of dust-enshrouded star formation to the evolution of total cosmic star-formation density ($\rho_{\rm SFR}$) out to $z \simeq 7$, and comparing our findings to other recent results in the literature. Finally, our main conclusions are summarized in Section \ref{sec:sum}.

Throughout the paper we assume a flat cold dark matter cosmology with $H_0 = 67.4\,{\rm km\,s^{-1}\,Mpc^{-1}}$, ${\rm \Omega_m} = 0.315$ and ${\rm \Omega_{\Lambda}} = 0.685$
\citep{2020A&A...641A...6P}.
All quoted magnitudes are in the AB system \citep{1974ApJS...27...21O, 1983ApJ...266..713O} and all derived star-formation rates (SFR) and stellar masses ($M_*$) assume a Chabrier IMF \citep{2003PASP..115..763C}.

\section{Data}
\label{sec:data}
In this study, we use the best available space-based and ground-based optical-infrared data to attempt to gain a complete understanding of the (sub)mm-detected galaxies which have been uncovered by a range of studies with ALMA in the centre of the COSMOS survey field over recent years.

Because the primary aim of this work is to explore what new information about the known sub-mm/mm sources in the COSMOS field can be revealed by the new high-quality {\it JWST} NIRCam and MIRI imaging, we first describe the {\it JWST} PRIMER data in the COSMOS field, as this defines the chosen survey area (and hence the cosmological volumes probed). We then stay at near-infrared wavelengths, summarizing the ground-based UltraVISTA data that was (by necessity) used in those areas of the COSMOS PRIMER survey footprint which only received coverage with MIRI. Lastly, we summarize the sample of (sub)mm-selected sources within the PRIMER COSMOS footprint which we assembled from the ALMA A$^{3}$COSMOS database. This sample of 128 sources, detected via their rest-frame far-infrared/sub-mm dust emission, is the primary focus of this study. However, to attempt to correct for the inevitable incompleteness of the ALMA `survey', we also exploit the highly-complete infrared-selected catalogues of massive galaxies produced from the {\it JWST} PRIMER and UltraVISTA survey data.

\begin{figure}

	\includegraphics[width=1.0\columnwidth, angle=0.3]{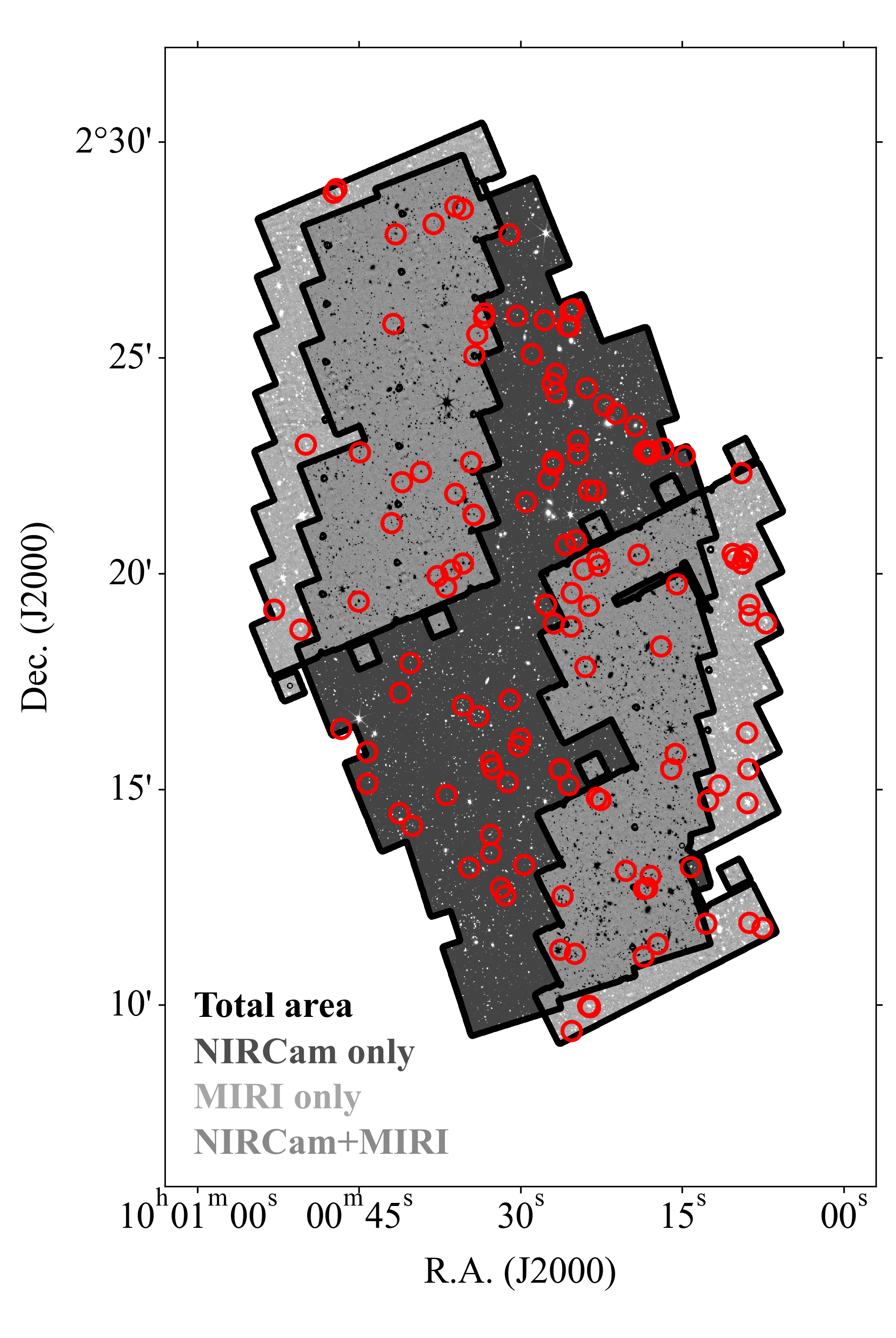}
    
    \caption{The complete 175.3 arcmin$^2$ coverage provided by the {\it JWST} PRIMER imaging within the COSMOS field. The NIRCam-only coverage (dark grey) is 66.0 arcmin$^2$, and the MIRI-only coverage (light grey) is 33.1 arcmin$^2$, with 76.2 arcmin$^2$ of overlapping imaging provided by both instruments (grey). The positions of the 128 ALMA sources in our final sample within this footprint (as specified in Section \ref{sec:construction}) are indicated by the red circles. The near/mid-infrared counterparts of all of these ALMA sources are securely detected by the NIRCam and/or MIRI imaging provided by PRIMER, enabling a complete study of their redshift distribution and physical properties.}
    \label{fig:coverage}

\end{figure}

\begin{table}
\centering
\caption{The median global 5$\sigma$ limiting magnitudes for each of the photometric filters used in this study. For all \textit{JWST} NIRCam and \textit{HST} imaging, these are measured in $0.5^{\prime\prime}-$diameter apertures and corrected to total assuming a point-source correction. For MIRI F770W, this is measured using $0.7^{\prime\prime}-$diameter apertures. For all ground-based imaging, these are measured in $2^{\prime\prime}-$diameter apertures and corrected to total assuming a point-source correction. For IRAC imaging, the median global depths are calculated using the uncertainties as output by the deconfusion software \textsc{TPHOT}, including a correction to total.}
\label{tab:depths}
    \begin{tabular}{lc}
    \hline
    Filter & 5$\sigma$ limit (mag) \\
        \hline
    CFHT u$^{*}$ & 27.2 \\
    CFHT g & 27.0 \\
    CFHT r & 26.5 \\
    CFHT i & 26.2 \\
    CFHT z & 25.2 \\
    Subaru SSC z$^{\prime}$ & 26.2 \\
    VISTA Y & 25.5 \\
    VISTA J & 25.3 \\
    VISTA H & 25.1 \\
    VISTA K$_{s}$ & 24.9 \\
    Spitzer IRAC 3.6$\mu$m & 26.0\\
    Spitzer IRAC 4.5$\mu$m & 26.2\\
    HST ACS F435W & 27.4\\
    HST ACS F606W & 27.5\\
    HST ACS F814W & 27.3\\
    JWST NIRCam F090W & 27.4\\
    JWST NIRCam F115W & 27.6\\
    JWST NIRCam F150W & 27.8\\
    JWST NIRCam F200W & 27.9\\
    JWST NIRCam F277W & 28.1\\
    JWST NIRCam F356W & 28.2\\
    JWST NIRCam F410M & 27.5\\
    JWST NIRCam F444W & 27.9\\
    JWST MIRI F770W & 25.8\\
    \hline
    \end{tabular}
\end{table}

\subsection{PRIMER COSMOS}
\label{sec:primer_cat}
The Public Release IMaging for Extragalactic Research \citep[PRIMER,][]{2021jwst.prop.1837D} survey is the largest public GO `Galaxies' programme undertaken with {\it JWST} in Cycle 1. Now complete, PRIMER is a deep extragalactic imaging survey which exploits the broad near-infrared wavelength coverage of NIRCam {\it and} the mid-infrared capabilities of MIRI to provide 10-band imaging across the same central regions of the two equatorial extragalactic survey fields that were previously imaged at UV/optical wavelengths with {\it HST} as part of the Cosmic Assembly NIR Deep Extragalactic Legacy Survey \citep[CANDELS:][]{2011ApJS..197...35G, 2011ApJS..197...36K} Treasury programme. Specifically, the two PRIMER fields each provide $\simeq$ 200\,arcmin$^2$ of high-quality space-based imaging within each of the degree-scale Cosmic Evolution Survey \citep[COSMOS,][]{2007ApJS..172....1S} and UKIDSS Ultra-Deep Survey \citep[UDS,][]{2007MNRAS.379.1599L} imaging programmes.

The PRIMER NIRCam imaging provides contiguous coverage of the near infrared out to $\lambda \simeq 5$\,$\mu$m through 8 filters: F090W, F115W, F150W, F200W, F277W, F356W, F410M, and F444W. To the extent allowed by parallel imaging and the layout of the {\it JWST} focal plane, the same field was also imaged in the mid-infrared with MIRI through 2 filters: F770W and F1800W.

As noted above, due to the richness of the ALMA data in the COSMOS field, in this (sub)mm-focused study we confine our attention to the COSMOS component of the (now complete) PRIMER survey, i.e., the PRIMER COSMOS field. Here, PRIMER provides contiguous NIRCam coverage of 142.2 arcmin$^2$ and contiguous MIRI coverage of 109.3 arcmin$^2$, which overlap by 76.2 arcmin$^2$. The combined {\it JWST} PRIMER imaging of the field provided by NIRCam and/or MIRI thus provides a total sky coverage of 175.3 arcmin$^2$ (see Fig. \ref{fig:coverage}), which defines the chosen survey area for this study.

The PRIMER NIRCam data were reduced using the PRIMER Enhanced NIRCam Image Processing Library ({\sc pencil}; Magee et al., in preparation, Dunlop et al. in preparation) software. The astrometry of all the reduced images was aligned to {\it Gaia} Data Release 3 \citep{2023A&A...674A...1G} and stacked to the same pixel scale of 0.03\,arcsec. The PRIMER MIRI data were reduced with a bespoke version of the official {\it JWST}  pipeline described in \citet{2024ApJ...968....4P} and \citet{2024arXiv241119686O} with calibrations based on {\it JWST} pipeline version 1.13.4, context 1263.pmap. Apart from the regular 3 stages of the pipeline, the calibration of the PRIMER MIRI observations included two additional offline steps. First, a background homogenization step was applied after the flat-fielding, based on the construction of a median super-background frame for each exposure (combining the rest of the data taken in the same filter and after sources are masked). This super-background procedure is able to remove striping (mainly in the vertical direction, but also horizontal) and global gradients which especially affect the lower region of the MIRI detector. For the PRIMER MIRI observations we found the structure of the background to be time-dependent, and therefore the super-background frame was based only on data taken in the same epoch (i.e., within 2 weeks). 
The second bespoke procedure was developed because the WCS calibration algorithm, tweakreg, in the official pipeline failed to provide a good astrometric calibration for our large $20\arcmin\times12\arcmin$ mosaics. Therefore, we used the tweakreg package version developed by the CEERS team \citep{2023ApJ...946L..12B} and switched off the official pipeline version in stage 3, obtaining a significantly better astrometrically calibrated mosaic across the whole area covered by our survey.

The NIRCam source selection utilised here is based on the F356W imaging. At F356W, the NIRCam detection has a sensitivity of $\sigma=$ 0.0039 $\mu$Jy (point-source total, based on 0.5-arcsec apertures) and an angular resolution (PSF FWHM) of $0.12$\,arcsec.
The MIRI catalogue utilised here is constructed based on the F770W data (Donnan et al., in preparation). This reaches a sensitivity of $\sigma=$ 0.035 $\mu$Jy (point-source total, based on 0.7-arcsec apertures) with an angular resolution (PSF FWHM) of $0.27$\,arcsec.

Crucially, by design, the new {\it JWST} near/mid-infrared imaging is complemented by pre-existing imaging of comparable angular resolution at shorter (optical) wavelengths, obtained with {\it HST} through the CANDELS Treasury Program. Specifically, imaging with the Advanced Camera for Surveys/Wide Field Channel (ACS/WFC) is available across the full PRIMER COSMOS survey footprint in the F814W and F606W filters and for most of the area in the F435W filters. The 5-$\sigma$ point-source corrected depths of the {\it JWST} and {\it HST} photometry used to create the PRIMER COSMOS source catalogue utilised here are given in Table \ref{tab:depths}.

The description of the catalogue construction and subsequent photometric redshift determination is described in detail in \citet{2025MNRAS.537.3245B}, to which we refer interested readers. Briefly, following PSF homogenisation to the angular resolution of the F444W imaging, we construct a multi-frequency catalogue by running \textsc{Source Extractor} in dual-image mode, with F356W as the detection image and aperture photometry measured using 0.5-arcsec diameter apertures. 

To account for light outside of the aperture for extended sources, we require a further correction to total beyond point-source corrections. Following \citet{mcleod2024}, we therefore scale the fluxes and hence SED-derived properties (e.g., stellar masses, see Section \ref{sec:SFR}) to the FLUX\_AUTO value \citep{Kron1980} measured by \textsc{Source Extractor}, and add a further 10\% to account for light outside of the Kron aperture.

The photometry provided by the 8 {\it JWST} NIRCam bands and the 3 {\it HST} ACS bands was then combined to derive photometric redshifts for all of the PRIMER-detected sources. Here, the median redshift of seven different spectral energy distribution (SED) fitting runs was taken as the final photometric redshift for each {\it JWST} source (although in practice spectroscopic redshifts proved to be available for 52\% of the sources - see Section \ref{sec:specz_sed}).

Specifically, we perform two \textsc{LePhare} \citep{2011ascl.soft08009A} SED-fitting runs, using the BC03 \citep{BC03} and \textsc{Pegase} \citep{Fioc1999} template libraries. We also perform five SED-fitting runs using \textsc{EAZY-PY} \citep{2008ApJ...686.1503B}. The first run included the standard template set with additional high equivalent-width emission lines. The second run utilised the \textsc{Pegase} template library \citep{Fioc1997,Fioc1999,Fioc2019}. We also use the $AGN\_BLUE\_SFHZ\_13$ template set, and finally, the two SED template sets as described in \citet{Larson2023} and \citet{Hainline2024}.

\subsection{UltraVISTA}
\label{sec:uvista_cat}
A significant portion (19\%) of the PRIMER COSMOS field is only covered by MIRI, which makes it impossible to infer galaxy properties solely on the basis of the new {\it JWST} data.
In order to provide the crucial near-infrared photometry in these regions we used the ground-based imaging data from the UltraVISTA survey \citep{2012A&A...544A.156M}.
This deep, near-infrared survey of the COSMOS field was conducted over the last $\sim$\, 14 years using ESO's Visible and Infrared Survey Telescope for Astronomy (VISTA).  
The survey spans a sky area of 1.9 deg$^2$ and provides imaging in 4 bands: $Y$, $J$, $H$, and $K_s$.

The ground-based data that we use corresponds to the central $\simeq1$ degree overlapping region between UltraVISTA Data Release 4 (DR4) and the optical $u^{*}griz$ imaging from the CFHT Legacy Survey D2 T0007 data release \citep{2012yCat.2317....0H}. We supplement this optical imaging with deep Subaru Suprime-Cam $z'_{\rm{new}}$ imaging \citep[][see \citealp{2014MNRAS.440.2810B} for further details]{2016ApJ...822...46F}.

The UltraVISTA DR4 imaging consists of “deep” and “ultra-deep” stripes, with the area of each depth tier of the survey being about $\sim50\%$ of the total area. However, the PRIMER footprint lies within one of the ultradeep stripes and, hence, our ground-based photometry is confined to regions of uniform depth in a given photometric band.

A detailed description of the UltraVISTA $K_S$-band catalogue construction is provided in \citet{2021MNRAS.503.4413M}, to which we refer interested readers. Briefly, source detection was performed on the UltraVISTA $K_S$-band imaging, by running \textsc{Source Extractor} in dual-image mode and aperture photometry measured in 2-arcsec diameter apertures. To maintain consistency across the multi-frequency dataset, all UV, optical, and NIR images in the UltraVISTA field were PSF homogenized to a Moffat profile with a full width at half maximum (FWHM) of $1.0$\,arcsec. For the lower resolution {\it Spitzer} IRAC imaging, photometry was extracted using the deconfusion software {\sc TPHOT} \citep{merlin2015}. The 5-$\sigma$ point-source corrected depths of all the VISTA, CFHT, Subaru and {\it Spitzer} IRAC photometry used to create the UltraVISTA source catalogue utilised here are summarized in Table \ref{tab:depths}.

Photometric redshifts for the UltraVISTA sources were derived from the combined VISTA, CFHT, Subaru, and {\it Spitzer} IRAC photometry. 
The methodology is described in detail in \citet{2021MNRAS.503.4413M}, and is similar to that adopted for the {\it JWST}+{\it HST} photometric redshifts as outlined above.

As well as providing the crucial near-infrared photometry in those regions of PRIMER COSMOS only covered by MIRI, where NIRCam imaging {\it does} exist the inferred galaxy properties from UltraVISTA can still often provide a useful independent check on those derived from the NIRCam+ACS data. However, where available, we decided to always adopt the final galaxy properties based on the latter, because of the improved depth and extended wavelength coverage provided by NIRCam in particular.

Finally, we note that, in this study, for a few sources with ambiguous photometric redshifts, the optical-infrared photometric redshifts were further refined for the final catalogue by exploiting the (sub)mm data (see Section \ref{sec:mm_fit}).

\subsection{ALMA sources in the PRIMER COSMOS field}
\label{sec:alma_cat}
The COSMOS field is particularly rich in archival ALMA data. It has been targeted in a systematic way, for example through the AS2COSMOS survey \citep{2020MNRAS.495.3409S} which 
uncovered 260 ALMA sources via band-7 continuum follow-up observations of a highly-complete flux-limited ($S_{850} > 6.2$\,mJy) sample of the brightest submillimeter sources 
detected by the JCMT SCUBA-2 S2COSMOS single-dish survey over the full 1.6 deg$^2$ of the COSMOS field \citep{2019ApJ...880...43S}. This survey has also stimulated further ALMA follow-up, such as the band-3 ALMA spectroscopic survey, AS2COSPEC \citep{2022ApJ...929..159C, 2024ApJ...961..226L}. However, the COSMOS field, and in particular the central area of most relevance for the present study, has also been the subject of numerous pointed ALMA follow-up observations of objects selected at other wavelengths, as exemplified by the REBELS survey which undertook targeted ALMA follow-up of the brightest rest-frame-UV selected $z \simeq 6-7$ galaxies in the field \citep[e.g.,][]{2022ApJ...931..160B, 2022MNRAS.515.3126I, 2023MNRAS.522..449B, 2024MNRAS.527.5808B}.

The most comprehensive source of archival ALMA data in the COSMOS field is provided by the  latest version of the A$^3$COSMOS survey \citep[\texttt{version 20220606},][]{2024A&A...685A...1A}. The A$^3$COSMOS database currently contains all publicly available ALMA observations taken before 6 June 2022 in the COSMOS field, calibrated, reduced, and analysed in a consistent manner. To search for ALMA-detected sources in the PRIMER COSMOS footprint, we used the `blind' photometric catalogue provided by \cite{2024A&A...685A...1A},
in which no prior information at other wavelengths was used to aid source extraction.
This catalogue contains all ALMA detections (at any frequency) with a Gaussian-fitted peak flux density signal-to-noise ratio (S/N$_{\rm peak}$) of 5.40 or higher.
Across the full COSMOS field, there are 2,204 entries in total in this latest blind catalogue,
which is almost double the number of detections (1,134) reported five years earlier in the \texttt{version 20180102} of A$^3$COSMOS \citep{2019ApJS..244...40L}.
Basic source properties, such as positional coordinates and flux densities are based on multiple-component Gaussian fitting performed by PYBDSF \citep[for details see][]{2019ApJS..244...40L}.

Restricting attention to the relatively small PRIMER COSMOS footprint as defined by our {\it JWST} observations still leaves a perhaps surprisingly large number of 347 A$^3$COSMOS detections. To ensure the quality of the A$^3$COSMOS detections, we required the Gaussian-fitted signal-to-noise ratio of the total observed flux density (S/N$_{\rm{total}}$) to also be higher than 3.1 at all frequencies. Since both S/N$_{\rm peak}$ and S/N$_{\rm{total}}$ are from the same Gaussian fitting, a significant mismatch between them is strongly suggestive of a spurious detection from a bad fit on noise artefact. This cut let to the rejection of one apparent ALMA detection (R.A. = 150.11128 deg, Dec. = 2.41047 deg), which was confirmed to be spurious due to the lack of any other co-spatial ALMA detection or any {\it JWST} counterpart within a search radius of 5\,arcsec.
Most of the remaining 346 entries in the database represent repeat observations of the same ALMA sources at different frequencies or resolutions (or both).
We therefore visually grouped the sources to avoid duplication, and produced a sample of 129 unique ALMA sources in the PRIMER COSMOS footprint. To this 
sample we were also able to add one extra source from the new large-area ALMA blank-field Extended Mapping Obscuration to Reionization with ALMA survey \citep[Ex-MORA,][]{2024arXiv240814546L}. This is a 2-mm imaging survey of a 577 arcmin$^2$ area of COSMOS which includes the PRIMER COSMOS field at its centre. However, because it is relatively shallow, almost all of the sources uncovered by Ex-MORA are sources which were already known from pre-existing bright (sub)mm surveys, and have been reported by A$^3$COSMOS, utilising either its progenitor MORA \citep{2021ApJ...923..215C, 2021ApJ...909..165Z} data or other ALMA archival observations. Nonetheless, to ensure our ALMA source sample is as up-to-date as possible, we added this single new Ex-MORA source (eMORA.29), yielding a final ALMA source list of 130 objects within the PRIMER COSMOS field. 

\begin{figure}
    \centering
    \includegraphics[width=\columnwidth]{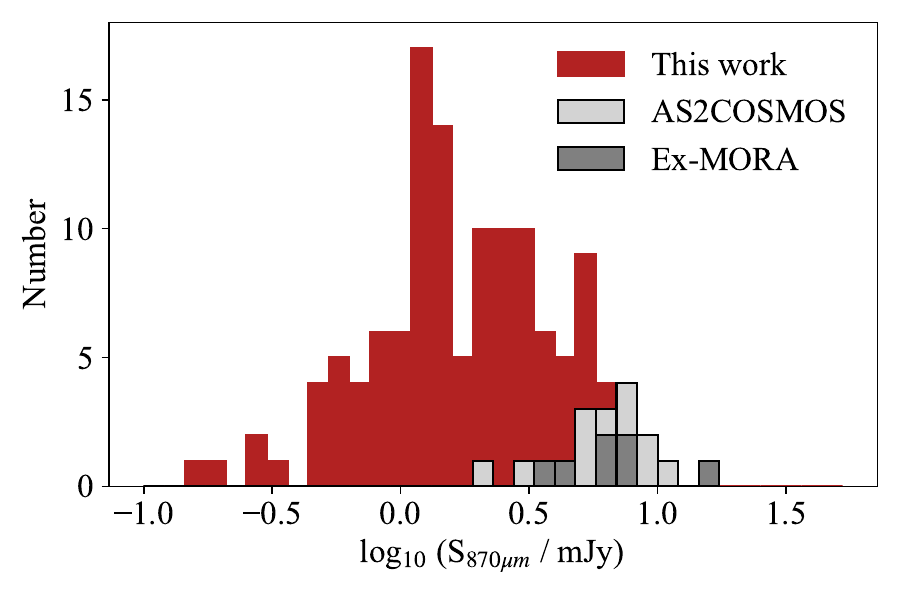}
    \caption{The distribution of 870$\rm{\mu}$m flux density for the 128 ALMA sources in our final sample (red). For sources where the only (or highest S/N) detection in the A$^3$COSMOS catalogue was obtained at a wavelength other than 870${\rm \mu}$m, the flux density was converted to a common observed wavelength of 870$\rm{\mu}$m assuming that the (sub)mm spectral energy distribution has a form $\propto \nu^3$ (see text). The 17 sources detected by AS2COSMOS (light grey) and the 7 sources detected at 2\,mm by Ex-MORA (dark grey; 6 in common with A2SCOSMOS) included in our sample are over-plotted on the histogram. Not only is our ALMA sample $\sim 10$ times larger than previous (sub)mm samples studied within the same footprint, but it also bridges the luminosity gap between the brightest sources (e.g., those uncovered with SCUBA-2) and the fainter (sub)mm detections resulting from targeted ALMA follow-up of distant optical/infrared selected galaxies \citep[e.g., as in the REBELS programme;][]{2022MNRAS.515.3126I}.} 
    \label{fig:flux_dist}
\end{figure}
One complication of utilising the A$3$COSMOS database is that different ALMA detections of a given source, even when observed at similar wavelengths, can sometimes report quite different flux densities (either because the observations are made at different angular resolutions or because of variation in the S/Ns achieved). For clarity and convenience we therefore cleaned the sample by only retaining the highest S/N ALMA detections in every 10 $\rm{\mu}$m wavelength bin for each ALMA source.
This left 302 potentially useful independent ALMA detections for the 130 ALMA sources at different wavelengths. Because most of these detections are at (or close to) an observing wavelength of 870\,$\mu$m, we decided to construct a pseudo 870\,$\mu$m selected sample, by 
converting the highest S/N detections of the 130 ALMA sources to 870\,$\mu$m using the relation $f_{870\rm{\mu m}}/f_{\rm{obs}} = (\nu_{870\rm{\mu m}}/\nu_{\rm{obs}})^3$.

Another consideration is that the ‘blind’ A$^3$COSMOS catalogue does not explicitly correct the reported continuum flux densities for potential contamination from emission lines. At the (sub)mm wavelengths and redshift range relevant to this work ($z \lesssim 7$), only two types of emission lines are expected to contribute significantly: the [C\,\textsc{ii}] 158\,$\mu$m fine-structure line and various rotational transitions of CO.
Reassuringly, both theoretical predictions, based on the [C\,\textsc{ii}] luminosity–SFR relation \citep[][]{2018A&A...609A.130L}, and actual detected line strengths (see Section~\ref{sec:specz_sed} for (sub)mm line searches and Table~\ref{tab:SED1} for references) show that any potential line contamination is unlikely to be significant. Specifically, even if present, line emission contributes only $\simeq 10 - 15$\% or lower to the total continuum flux density measured within the 8\,GHz ALMA bandwidth. This level of (potential) contamination is generally comparable to or smaller than our typical continuum flux-density uncertainties and can thus be safely neglected in the subsequent analysis.

Two sources in this list were later excluded from the final catalogue used in the present study (for reasons explained below in Section \ref{sec:construction}). Our final sample for the full multi-frequency analysis undertaken here thus consists of 128 unique ALMA sources.
The 870\,$\mu$m flux-density distribution of this sample of 128 ALMA sources in the PRIMER COSMOS footprint is shown in Fig. \ref{fig:flux_dist}. It can be seen that our sample spans two orders-of-magnitude in flux density, extending from the brightest (sub)mm sources uncovered by the wide-area blank-field surveys (our sample contains 17 sources from AS2COSMOS and 7 from Ex-MORA, albeit 6 of the Ex-MORA sources are already included in the 17-source AS2COSMOS list) down to the lowest luminosity sources typically uncovered by deep ALMA pointed follow-up observations (approaching flux densities of $S_{870} \simeq 0.1\,{\rm mJy}$). We note here that \citet{2024A&A...691A.299G} used {\it JWST} data to study 17 AS2COSMOS sources in COSMOS. Our sample only contains 13 of these because the other 4 lie outside the high-quality PRIMER COSMOS footprint defined here \citep[i.e., they lie in a region only covered by {\it JWST} imaging from the COSMOS-Web survey;][]{2023ApJ...954...31C}. However, we recovered another 3 sources (namely AS2COS0017.1, AS2COS0032.1, and AS2COS0035.1) within the MIRI-only footprint that were reported in the original AS2COSMOS catalogue \citep{2020MNRAS.495.3409S}. We also include the lensed source (AS2COS0005.1), as its properties have been obtained by detailed modeling \citep{2024A&A...690L..16J}. Therefore, the total number of AS2COSMOS sources in our catalogue is still 17.

\section{Construction of the full multi-frequency ALMA source catalogue}
\label{sec:construction}

In this section we describe how we established the full multi-frequency SEDs, and hence ultimately the redshifts and other physical properties of the 128 ALMA-detected galaxies in the PRIMER COSMOS field.

First, we cross-matched the ALMA positions with the positions of the galaxies in the {\it JWST} and UltraVISTA imaging. As demonstrated and discussed below in Section \ref{sec:x_match}, as a result of the high positional accuracy of the ALMA detections {\it and} the infrared imaging, this proved to be a relatively straightforward task, with only a few objects requiring additional in-depth analysis to reveal the true infrared-detected galaxy counterpart. Crucially, as a result of the depth of the {\it JWST} imaging, we have been able to detect the infrared galaxy counterpart of {\it every} ALMA source in the sample, something rarely before achieved in the optical/infrared follow-up of large (sub)mm-selected samples. This 100\% identification completeness, not possible before {\it JWST}, is obviously of enormous value in subsequent analysis.

With successful and unambiguous identification of the optical/infrared galaxy counterpart comes photometric redshift information from the {\it JWST} and UltraVISTA photometric catalogues (as described in Sections \ref{sec:primer_cat} and \ref{sec:uvista_cat}). However, for a few objects, especially those where the photometry in the infrared catalogues may have been contaminated by flux from nearly sources, careful manual photometry was required to yield reliable flux densities for the determination of robust photometric redshifts. This process is described in Section \ref{sec:bespoke_photo}.

Even after assembling the best available optical-infrared photometry, there remain some objects for which the photometric redshift remains unclear or ambiguous, often in the sense that there are two possible alternative solutions at very different redshifts. To improve this situation, we used the multi-frequency ALMA data for each source (where available) to check the plausibility of a given photometric redshift given the expected form of a modified black body for a reasonable assumed dust temperature. This process, which generally helped to exclude some highly uncertain extreme redshift solutions, is summarised in Section \ref{sec:mm_fit}. 

Finally, notwithstanding all the above efforts to produce reliable photometric redshifts, we undertook a comprehensive search of the literature and all available spectroscopic databases at optical, infrared and (sub)mm wavelengths to check for the existence of a reliable spectroscopic redshift for each source in our catalogue. Pleasingly, the result of this extensive search is that 66 sources (i.e., 52\% of our sample) have reliable spectroscopic redshifts. As described in Section \ref{sec:specz_sed}, this relatively high spectroscopic completeness has enabled us to validate the accuracy of our photometric redshifts, and hence to deliver a robust redshift distribution for our full sample. 

\begin{figure}
	\includegraphics[width=\columnwidth]{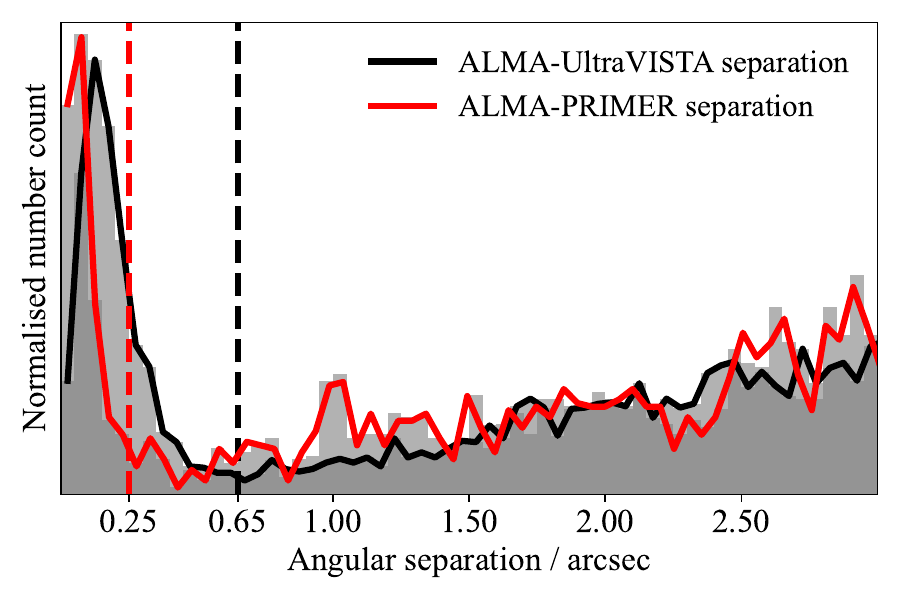}
    \caption{The distribution of the angular separation between the position of each ALMA source in the PRIMER COSMOS footprint and its nearest near-infrared neighbour in the {\it JWST} (red) or UltraVISTA (black) imaging. Both distributions peak at an angular separation $< 0.2$ arcsec, reflecting the high positional accuracy of all catalogues. The first local minimum of the ALMA-NIRCam cross-matching distribution occurs at 0.25\,arcsec (red vertical dashed line) while that for the ALMA-UltraVISTA cross-matching distribution appears at a somewhat larger radius of 0.65\,arcsec (black vertical dashed line), as expected due to the limitations of ground-based seeing. These minima were used to define the search radius for near-infrared counterparts to the ALMA sources in the {\it JWST} and UltraVISTA catalogues respectively.}
    \label{fig:separation}
\end{figure}

\subsection{Infrared source identification/cross-matching}
\label{sec:x_match}
To determine the counterparts of the ALMA sources at near-infrared wavelengths, we first cross-matched the ALMA detection coordinates to the coordinates of the sources in the PRIMER COSMOS and UltraVISTA catalogues discussed in Section \ref{sec:data}.
Fig.~\ref{fig:separation} shows the distribution of the angular separation between each ALMA detection and the nearest {\it JWST} or UltraVISTA source. Both distributions peak at very small angles ($< 0.15$ arcsec), as expected for secure identifications given the high-positional accuracy delivered by ALMA, {\it JWST} and UltraVISTA (albeit the distribution for UltraVISTA peaks at a slightly larger angle due to the limitations of ground-based seeing). This implies that a relatively small search radius can be used to secure the near-infrared galaxy counterparts of all ALMA sources, and indeed the distributions both show a clear minimum at reassuringly small angular separation beyond which the number of counterparts starts to rise systematically with increasing angular separation as expected for spurious identifications with galaxies in the field. 

We found that an unambiguous NIRCam counterpart for each ALMA source within the NIRCam footprint could be secured using a search radius of only 0.25\,arcsec. As can be seen from 
the vertical red dashed line in Fig.~\ref{fig:separation} this corresponds to a first minimum in the ALMA-PRIMER source separation distribution. The use of a larger search radius delivered no additional identifications, but rather just added a few spurious companion objects. This is also indicated by the minor rise in the distribution after the first minimum and before the global minimum (at $\simeq 0.4$\,arcsec), which reveals the presence of close optical/IR companions to the (sub)mm sources. For sources not covered by NIRCam, we searched for UltraVISTA-detected counterparts within a radius of 0.65\,arcsec, as indicated by the vertical black dashed line in Fig.~\ref{fig:separation}. This was necessitated by the poorer angular resolution of the data, and resulted in a few multiple possible identifications, but these were able to be refined (and the true galaxy counterpart unambiguously identified) with the use of the MIRI imaging.

The adoption of these relatively small search radii reduces the probability of a spurious identification for each source to $< 1$\%. We note that this situation is very different to the statistically more complex problem of identifying the true galaxy counterparts of (sub)mm sources detected in single dish surveys, where the large (sub)mm beam size (typically FWHM $>10$\,arcsec) necessitates the use of much larger search radii, vastly increasing the probability of random associations in deep optical/infrared imaging \citep[see, for example,][]{2007MNRAS.380..199I}.

\begin{figure*}
    \centering
    \includegraphics[width=2.0\columnwidth]{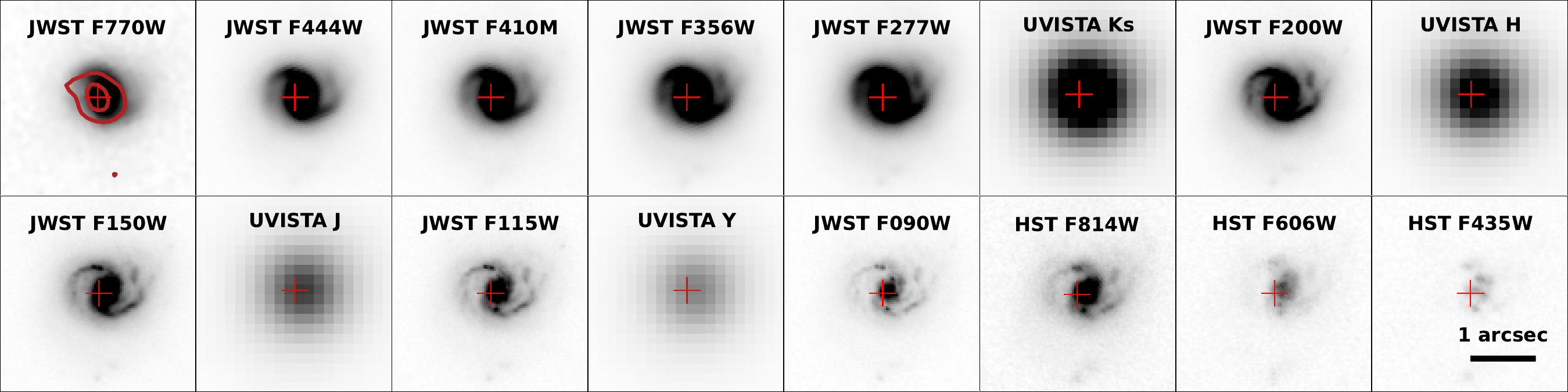}\\
    \,\\
    \,\\
    \includegraphics[width=2.0\columnwidth]{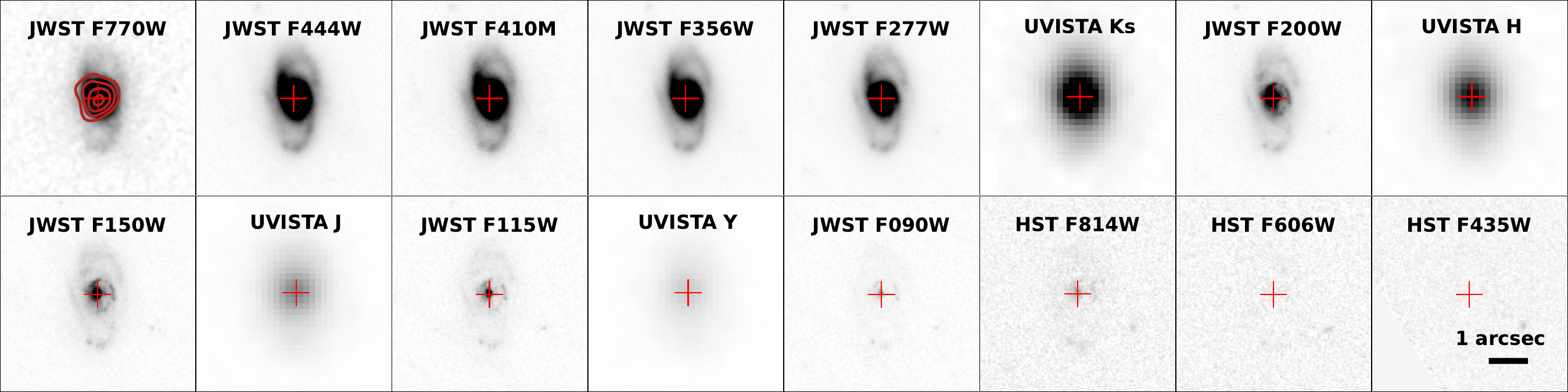}\\
    \,\\
    \,\\
    \includegraphics[width=2.0\columnwidth]{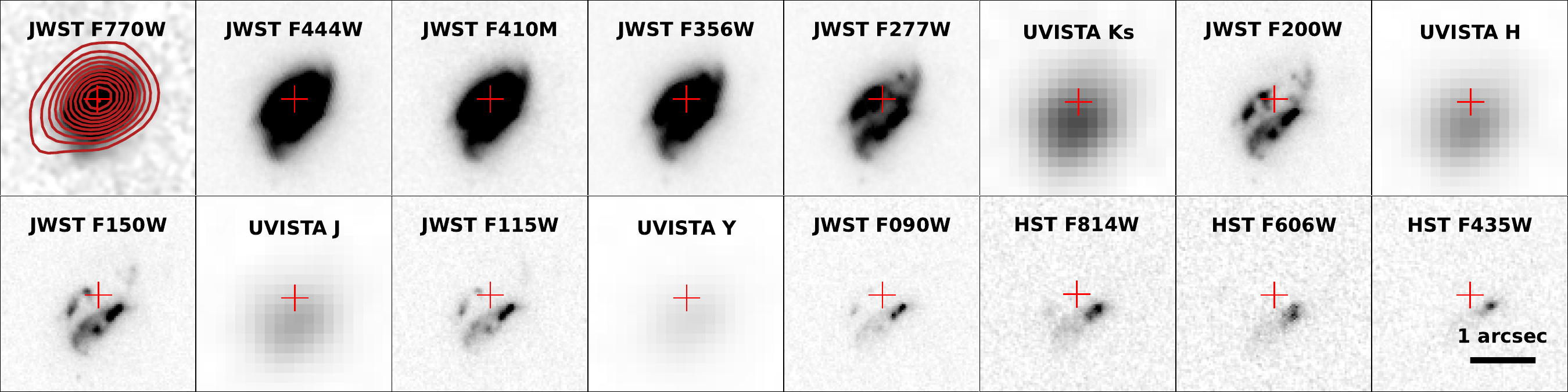}\\
    \,\\
    \,\\
    \includegraphics[width=2.0\columnwidth]{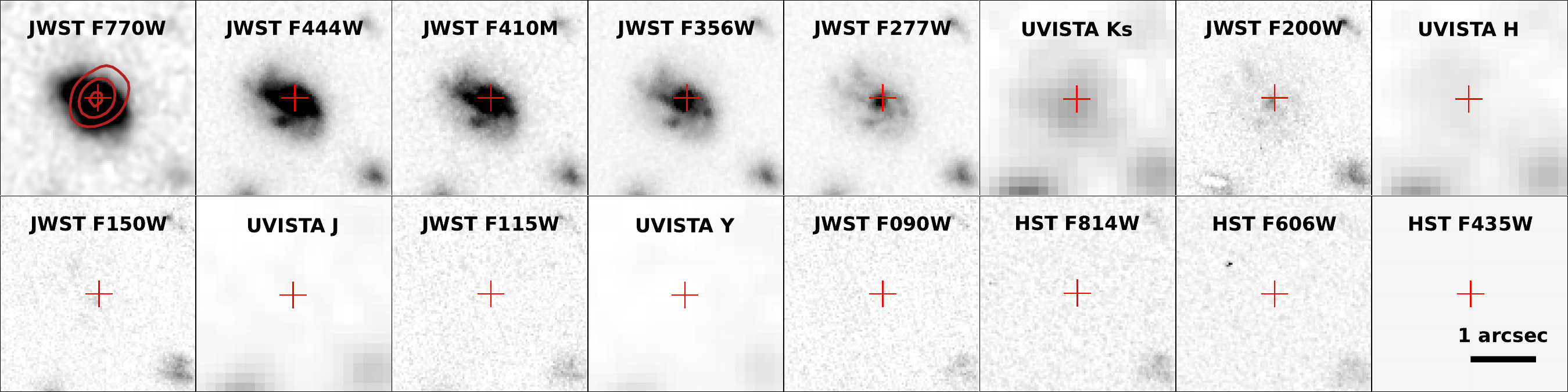}\\
    \,\\
    
    \caption{Image stamps in all the available {\it JWST}, UltraVISTA and {\it HST} bands for 4 of the sources in our sample (ID 095, ID 053, ID 007, and ID 034). For each of the 4 sources, 16 stamps are shown, ranked in order of decreasing wavelength with filter names marked at the top of each panel, and with the position of the highest-S/N ALMA detection marked by the red cross. ALMA contours are illustrated in the longest wavelength JWST image, F770W, starting from 3.5 $\sigma$ with 1.5-$\sigma$ spacing for ID 095 and 3-$\sigma$ spacing for the rest. These illustrate different but also common features of the typical galaxies in the sample. ID 095, which lies at a redshift of only $z=0.726$ is detected with {\it HST} at optical wavelengths (albeit only just at F435W). It is also securely detected in all UltraVISTA near-infrared bands, but the vastly improved angular resolution provided by NIRCam is key to revealing its spiral disk morphology. ID 053, with $z=1.552$ is dark on HST F435W to HST 814W images, but has a consistent disk-like morphology across all infrared bands. ID 007, at $z=2.086$ appears to have an {\it HST} detection but the longer-wavelength imaging reveals that this does not correspond to the centre of the galaxy; it is either a blue clump or a separate object, and the true mass-dominant galaxy is not revealed until wavelengths longward of 2\,$\mu$m. Finally, ID 034 is a higher-redshift `{\it HST}-dark' galaxy, with $z_{\rm phot} \simeq 3.55$, and is basically not detected at all until 2\,$\mu$m. Despite these differences caused by a mix of redshift and the degree of dust attenuation, all 4 of the example objects shown here, and in fact all of the galaxies in our sample are detected (and resolved) with relative ease in the longest wavelength NIRCam and/or MIRI 7.7$\mu$m imaging.}
    \label{fig:stamp}
\end{figure*}

The search of the {\it JWST} and UltraVISTA catalogues within these adopted search radii yielded a near-infrared counterpart for all but one ALMA source
(ALMA R.A. = 150.07486 deg, Dec. = 2.3806 deg). However, this ALMA source is based purely on a highly marginal 3-mm `detection' with S/N$_{\rm{total}} = 3.26$ in the A$^3$COSMOS catalogue, and is not reported elsewhere in the literature. Increasing the search radius for this object (even by a factor of several) does not reveal a counterpart in either the {\it JWST} or UltraVISTA imaging. The 2$\sigma$ upper limit on flux density in the F444W PRIMER imaging is an order-of-magnitude fainter than the measured F444W flux densities of even the faintest confirmed sources in our sample. Hence any non-detection is significantly disconnected from the faint tail of the F444W flux density distribution of all the other ALMA sources (see Table \ref{tab:CAT1}). Such a situation is clearly highly unlikely even for the most dusty high-$z$ systems. We thus decided to exclude this source as most likely a spurious ALMA detection.

This leaves a sample of 129 ALMA sources with a 100\% near/mid-infrared identification rate. All are detected with {\it JWST}, with the coverage of the PRIMER COSMOS field resulting in 104 being detected with NIRCam, 112 being detected by MIRI/UltraVISTA, and 87 being detected by both. 
The reliability of this cross-matching was carefully checked by visually inspecting the near/mid-infrared images at the positions of every ALMA source. Image stamps in all the available {\it JWST}, UltraVISTA and {\it HST} bands are shown for four example sources in Fig. \ref{fig:stamp}. These stamps illustrate different aspects of the power of the multi-frequency imaging, including the vastly improved angular resolution provided by NIRCam over VISTA, and the relative ease with which all of these massive dust-attenuated galaxies are detected at the longest near-mid infrared wavelengths now made possible by {\it JWST} NIRCam and MIRI imaging. 

It is clear that the depth and extended wavelength coverage provided by the {\it JWST} PRIMER imaging is vital for uncovering the secure near-infrared counterpart of {\it every} ALMA source in our sample. While the depth of the $K_S$ imaging provided by UltraVISTA is impressive, without the addition of the {\it JWST} imaging, 11 of the sources would have remained unidentified given only the previously available ground-based and {\it HST} imaging. Secondly, in a very small number of cases, the high astrometric accuracy delivered by {\it JWST} has allowed us to distinguish the true identification from a nearby companion galaxy that lies within the (by necessity) larger UltraVISTA counterpart search radius, as exemplified by our successful identification of the true massive dusty ALMA-detected galaxy Hyde \citep[][ID 064 in our final sample, see Table \ref{tab:CAT1}]{2018A&A...611A..22S}, which lies only 0.43\,arcsec away from a brighter quiescent galaxy, ZF-COSMOS-20115.

\subsection{Bespoke photometry and the final 128-source catalogue}
\label{sec:bespoke_photo}
We produced three-colour (RGB) infrared images of the NIRCam-detected galaxy counterparts of the ALMA sources using the NIRCam F444W (red), F277W (green), and F115W (blue) imaging (where available) after convolving the imaging through the latter two filters to the same resolution as the F444W imaging. Where NIRCam imaging was not available we used the UltraVISTA $J$-band, the UltraVISTA $K_S$-band, and MIRI $7.7\,\mu$m imaging to produce the RGB images (albeit these inevitably have poorer resolution than the NIRCam examples). An example (NIRCam) RGB image is shown in Fig. \ref{fig:rgb}, with the three-colour images for all of the ALMA sources presented in Appendix B.

\begin{figure}
    \centering
    \includegraphics[width=\columnwidth]{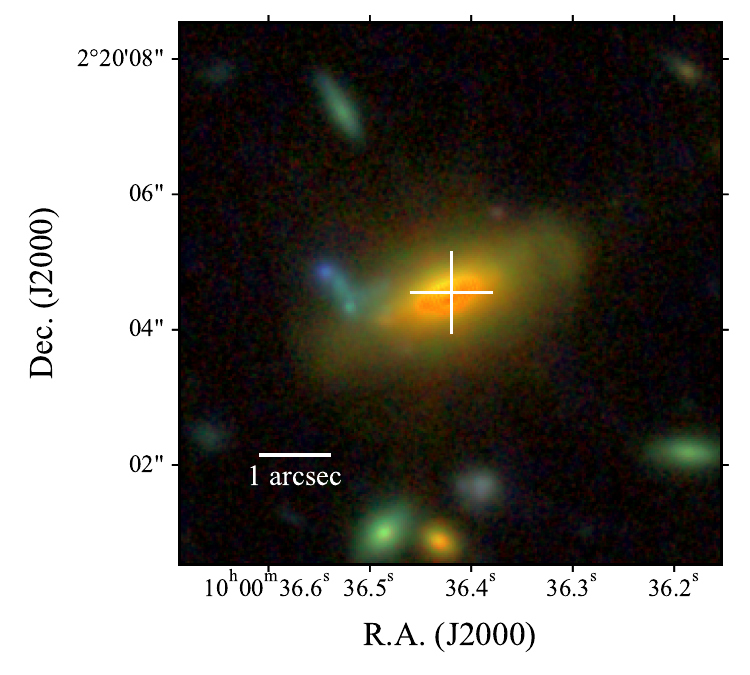}
    \caption{An example RGB image of one of the ALMA-detected galaxies in our sample (ID 100), created from the {\it JWST} PSF-homogenised NIRCam imaging: F444W (red), F277W (green), and F115W (blue). The position of the ALMA detection is marked by the white cross. While there is no ambiguity over the correct near-infrared counterpart, a nearby very blue object/component
    can be seen $\simeq 1$\,arsec to the east. Such a source is too distant to contaminate the NIRCam photometry, but could potentially contaminate ground-based photometry, especially at optical (i.e., rest-frame UV) wavelengths. A total of 26 sources were revisited and subjected to manual (rather than automated) photometry to ensure reliable colours and total flux densities. RGB images of all sources in the sample are presented in Appendix \ref{fig:stamp1}.}
    \label{fig:rgb}
\end{figure}

Fig. \ref{fig:rgb} illustrates an important point, in particular the dangers of simply assuming that the automated catalogue photometry is correct. In this case, while there is no issue over the correct (red) near-infrared counterpart to the ALMA source, it can be seen that a blue companion object is present (likely in projection) which, while lying well outside the search radius for even an UltraVISTA counterpart, could well contaminate the photometry of this source in ground-based imaging, especially at optical (rest-frame UV) wavelengths. While potential contamination of the NIRCam photometry due to nearby sources is obviously less likely (due to higher resolution and hence the use of smaller apertures) we nonetheless carefully inspected all of the images of the near-infrared counterparts to search for potential complications due to merging, lensing, or image contamination issues such as diffraction spikes from nearby stars. 

Through this process we identified 26 sources for which we judged that the automated catalogue photometry was potentially not to be trusted. To be safe, for these objects we manually re-measured the photometry at all bands. This involved careful masking of companions where necessary, but in the main enabled a better representation of the SExtractor isophotal footprint of each object, and hence a better sense of what SExtractor deemed to be the appropriate Kron aperture. This analysis improved the overall centroiding and deblending and delivered a more robust measurement of total flux density which, as well as potentially leading to improved photometric redshifts, ultimately yields more robust physical properties such a stellar mass. This is a painstaking process but one which nonetheless proved invaluable. For example, with the re-measured photometry, the photometric redshift of the (possibly merging) galaxy (ID 075 in the final sample, see Table \ref{tab:SED1}) changes from $z_{\rm{photo}}=1.65$ to $z_{\rm{photo}}=2.82$, which is reassuringly much closer to its spectroscopic redshift $z_{\rm{spec}}=2.989$ \citep[][from the DAWN JWST Archive, see Section \ref{sec:specz_sed}]{2024jwst.prop.6585C}.

The near-infrared counterpart of one source (ALMA RA = 150.10968 deg, Dec = 2.18797 deg) lies on top of a strong diffraction spike from a nearby bright source in the NIRCam and MIRI images and is blended into that same nearby bright source in the UltraVISTA imaging.
Since it is impossible to derive reliable information from the contaminated photometry of this source, we decided to exclude it from the final catalogue. We note that since it is clearly detected in the PRIMER imaging, the removal of this source is unrelated to its actual observed or physical properties, so its rejection on the basis of quality control should not bias our  sample.

The final catalogue used for subsequent analysis thus contains a total of 128 ALMA sources, all with robust near-infrared identifications. The coordinates and flux densities of these sources are presented in Table~\ref{tab:CAT1}, ranked (and labelled) by descending ALMA flux density (converted to $\lambda_{\rm obs} = 870\,\mu$m, as discussed in Section \ref{sec:alma_cat}).

\begin{figure}
    \centering
    \includegraphics[width=\columnwidth]{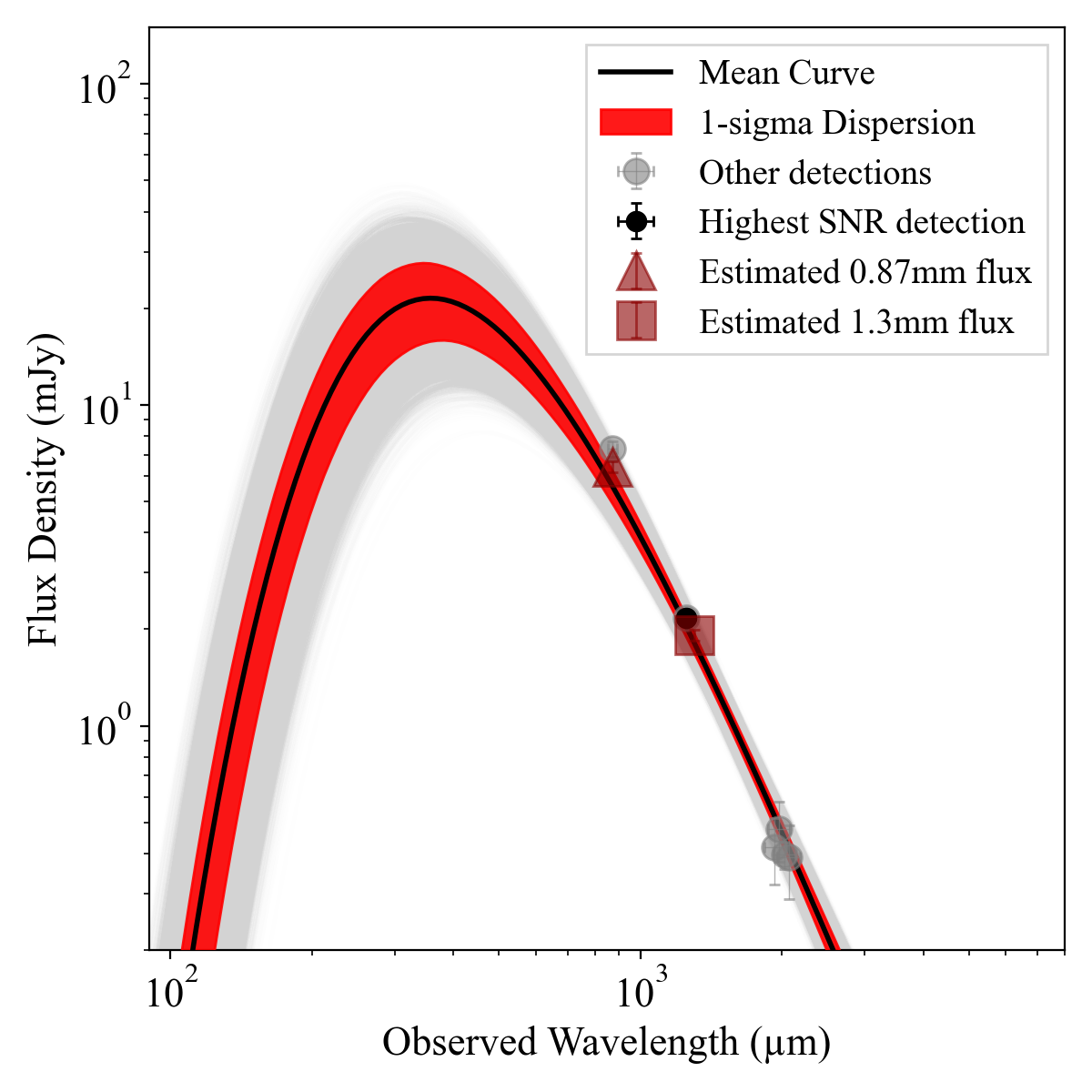}
    \caption{An example of fitting an isothermal modified blackbody to the ALMA photometry for one of our sources, which in this case has a spectroscopic redshift (ID 010, $z_{\rm{spec}}=2.510$). The mean curve is plotted in black with the $1\sigma$ dispersion shown in red. The 10,000 MC runs (light grey curves) are plotted in the background. The highest-S/N ALMA detection is marked in black and the rest are shown in grey, with error bars indicating both the flux-density and frequency (i.e., bandwidth) errors. The ALMA detections are consistent with the fitted curve at the adopted redshift of $z = 2.51$.}
    \label{fig:greybody}
\end{figure}

\subsection{Redshift constraints from ALMA photometry}
\label{sec:mm_fit}
As mentioned in Section \ref{sec:alma_cat}, 46 (36\%) of the ALMA sources in our catalogue have detections in the A$^3$COSMOS database at more than one (sub)mm wavelength. This means that, at least for this subset of sources, it makes sense to perform a sanity check on the optical-infrared photometric redshifts by exploring whether the ALMA photometry (essentially the (sub)mm colour) is consistent with that expected from thermal dust emission of reasonable temperature at the adopted redshift. To do this we fitted a simple isothermal modified blackbody to the available ALMA photometry, as given by the expression:

\vspace*{-0.04in}
\begin{equation}
\label{eq:greybody}
S_\nu = \Omega \frac{2h\nu^3}{c^2} \frac{1}{\text{exp}(h\nu/kT-1)} \left\{ 1-\text{exp} \left[ -(\nu/\nu_0)^\beta \right] \right\}
\end{equation}

\vspace*{0.07in}

\noindent
where $\nu$ is the rest frequency (given the adopted redshift $z$), $F_\nu$ is the flux density observed at a frequency equal to $\nu/(1+z)$, $\Omega$ is the solid angle subtended by the source, $\beta$ is the emissivity index of the dust, $\nu_0=3.0$~THz is the rest-frame frequency at which the dust emission is assumed to become optically thick (and at all higher frequencies, because the optical depth, $\tau$, at frequency $\nu$ is $(\nu/\nu_0)^\beta$), and $T$ is the intrinsic temperature of the dominant dust mass.

During fitting, as well as testing alternative photometric redshifts where appropriate, we allowed the dust emissivity index $\beta$ to vary in the range $1.5\leq\beta\leq2.0$, and adopted a redshift-dependent dust temperature $T = 32.9 + 4.60 \times (z_{\rm{photo}} - 2.0)\, \text{K}$ \citep{2018A&A...611A..22S} consistent with $T=35$\,K at the median redshift of the final ALMA sample ($z_{\rm med} \simeq2.5$: see Sec. \ref{sec:specz_sed}).
The ALMA fluxes have Gaussian errors and we assumed that the frequency follows a uniform distribution within an ALMA bandwidth of 8 GHz (equivalent to an uncertainty of 4 GHz).
We then ran a Monte Carlo (MC) simulation with 10,000 runs for each source to derive means and standard deviations, with each run weighted by the S/N$_{\rm{total}}$ of the detections.
An example of one of the resulting fits, in this case for a source with a robust spectroscopic redshift, is presented in Fig. \ref{fig:greybody}.

For the vast majority of sources, the modified blackbody fits proved to be consistent with the ALMA detections at different frequencies. However, in 5 cases, the fitted curves deviated by more than $5\sigma$ from at least one of the ALMA detections (given the assumed optical-infrared photometric redshift). It transpired that these sources all have rather high photometric redshifts ($z_{\rm{photo}}>5$) but in fact the optical-infrared photometric redshifts also proved to be highly uncertain, with different code/template combinations delivering very different results and at least some code/template combinations favouring more modest redshifts. In these cases we therefore decided to also allow the redshift to vary while re-performing the modified blackbody fitting, and adopted the optical-infrared photometric solution which lay closest to that implied by the best-fitting (sub)mm photometric redshift. For these 5 sources, this changed the final adopted photometric redshift by $\Delta z\simeq2-4$.

\begin{figure}
    \centering
    \includegraphics[width=\columnwidth]{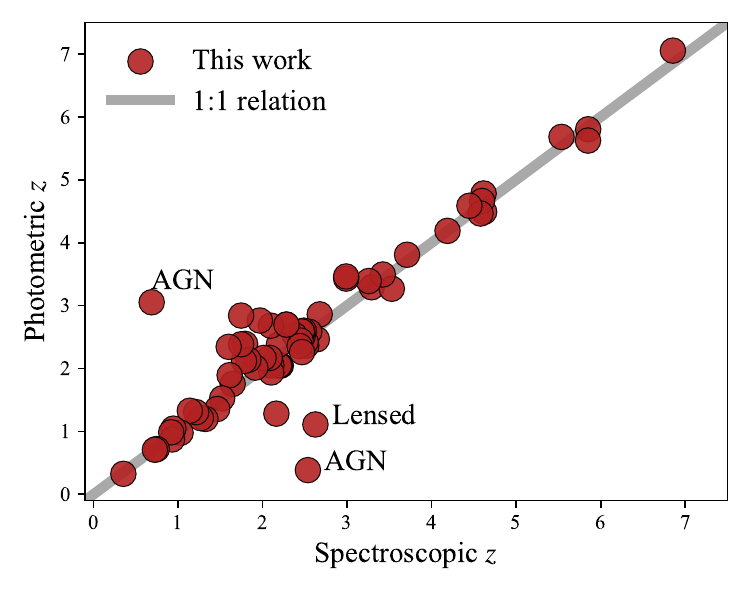}
    \caption{A comparison between the spectroscopic redshifts, $z_{\rm{spec}}$, and our final improved photometric redshifts, $z_{\rm{photo}}$, for the 66 sources with known $z_{\rm{spec}}$ in our sample. Apart from the one known lensed galaxy, two bright AGN, and some modest focusing at $2<z<3$, $z_{\rm{photo}}$ and $z_{\rm{spec}}$ are in excellent agreement, following the 1:1 relation (indicated by the grey line) right out to the highest redshifts.}
    \label{fig:pho_spec_z}
\end{figure}

\subsection{Spectroscopic redshifts}
\label{sec:specz_sed}

Finally, having established the most probable photometric redshifts for all 128 sources in the sample, we undertook an extensive search of the literature to establish which sources already possessed robust optical/infrared or (sub)mm spectroscopic redshifts.

We found that 53 sources (41\% of the entire sample) already have secure optical-infrared spectroscopic redshifts. Unsurprisingly, the vast majority of these come from major spectroscopic survey programmes (with overlaps), such as 3D HST \citep[23 sources,][]{2012ApJS..200...13B, 2015ApJS..219...29M}, DEIMOS \citep[6 sources,][]{2018ApJ...858...77H}, zCOSMOS \citep[13 sources,][]{2007ApJS..172...70L, 2009ApJS..184..218L}, MOSDEF \citep[7 sources,][]{2015ApJS..218...15K}, LEGA-C \citep[3 sources,][]{2021ApJS..256...44V}, and those in the DAWN JWST Archive \citep[DJA, 6 sources,][]{2022zndo...7299500B, 2024Sci...384..890H, 2024arXiv240905948D}.

To search for reliable (sub)mm spectroscopic redshifts, we matched the ALMA coordinates with ALMA archival targets using ALminer \citep{2023ApJ...951..111C}, an automated ALMA archive mining tool, which enabled us to find published spectroscopic redshifts based on (sub)mm line observations taken up to 2023. We found that 17 sources (13\% of the whole sample) had reliable (sub)mm spectroscopic redshifts based on archival ALMA observations. Of these, 11 are new in the sense that these sources do not have secure published optical-infrared spectroscopic redshifts. Among the 6 sources with both optical-infrared and (sub)mm $z_{\rm{spec}}$, the results for 4 sources are highly consistent within the (small) measurement uncertainties. However, for the remaining 2 sources, the optical-infrared $z_{\rm{spec}}$ from 3D HST is somewhat uncertain, and so for these we adopted the more precise (sub)mm $z_{\rm{spec}}$ resulting from the ALMA line detections. As a final check, we matched our sample with the COSMOS Spectroscopic Redshift Compilation \citep{khostovan2025cosmosspectroscopicredshiftcompilation} to verify the quality of the known redshifts and find possible missed spectroscopic information. Three new spectroscopic redshifts with quality flags 4 (`very reliable redshift') and 3 (`reliable redshift') were obtained, which are originally from surveys C3R2 \citep{2017ApJ...841..111M, 2019ApJ...877...81M}, VUDS \citep{2015A&A...576A..79L, 2017A&A...600A.110T}, and zFIRE \citep{2016ApJ...828...21N}.

\begin{figure}
    \centering
    \includegraphics[width=\columnwidth]{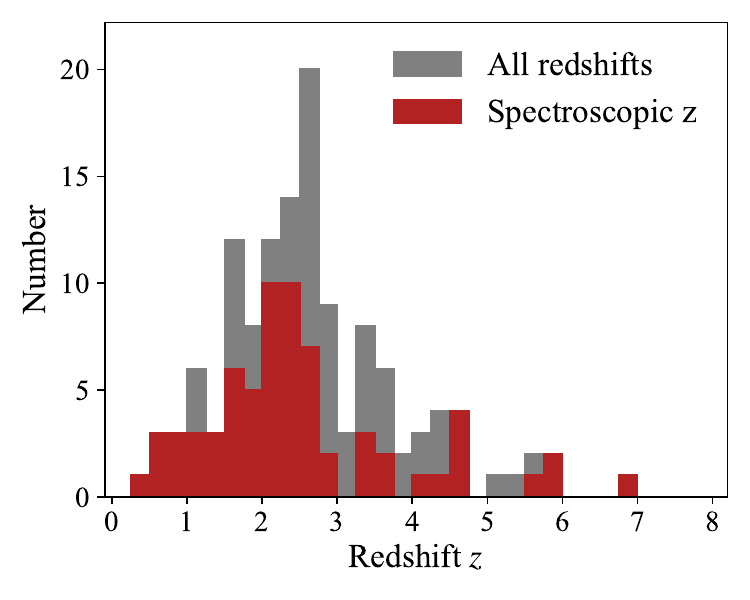}
    \caption{The final redshift distribution of our 128-source sample. The 66 sources for which a spectroscopic redshift has been secured are indicated by the red histogram. It can be seen that the spectroscopic redshift distribution is highly comparable to the overall distribution of the entire sample (grey histogram), with extra coverage in the high-redshift tail.}
    \label{fig:z_dist}
\end{figure}

In total, therefore, we have established secure spectroscopic redshifts for 66/128 sources, equivalent to 52\% of the sample (see Table \ref{tab:SED1}). This is a pleasingly high fraction, especially for a (sub)mm selected sample in which the rest-frame optical-UV is often highly obscured. This, plus the fact that the spectroscopic redshifts span the wide redshift range $0 < z < 7$ allows us to test the validity of our photometric redshifts. The photometric and spectroscopic redshifts for these 66 sources are compared in Fig. \ref{fig:pho_spec_z}.
It can be seen that, apart from 3 obvious outliers, the photometric and spectrosopic redshifts are in excellent agreement. The only serious outliers are two AGN and one lensed galaxy \citep[ID 002,][]{2024A&A...690L..16J}.
One of the two outlying AGN, ID 092, exhibits spiral galaxy morphology at shorter wavelengths, but then becomes a saturated point source at wavelengths longer than F200W, resulting in a substantially over-estimated photometric redshift. The other AGN, ID 039, is a saturated point source in all available images and thus unsurprisingly does not have a meaningful constraint on its photometric redshift. Fortunately, these two AGN obviously have secure spectroscopic redshifts, and so their poor photometric redshift estimates are not of importance in this study.
No other lensed galaxies are evident in the rest of the sample, and the (very few) AGN are not of serious concern because optical-infrared spectroscopic redshifts are usually readily available for these sources (as in the 2 unusual cases discussed above).
In summary, Fig. \ref{fig:pho_spec_z} confirms that the processes we have used to estimate and refine the photometric redshifts of the sources in our ALMA-selected sample have delivered reliable results.

\begin{table}
    \centering
    \caption{Parameter values adopted as input for {\sc bagpipes} SED fitting, where \texttt{eta} is the birth-cloud scaling factor, \texttt{B} is the amplitude of the 2175\,\AA\ bump, \texttt{alpha} is the power‐law slope of the rising part of the star-formation history (SFH), \texttt{beta} is the slope of the declining part of the SFH, and \texttt{tau} is the age of universe at the SFH turnover \citep{2018MNRAS.480.4379C}. Any parameters not explicitly listed here were set to their default values.}
    \label{tab:bagpipes}
    \begin{tabular}{lll}
        \hline
         Component & Model parameter & Initial values \\
         \hline
         Dust & type & Salim \\
        & \texttt{eta} & 2.0 \\
        & \texttt{Av} & (0.0, 8.0) \\
        & \texttt{delta} & (-0.3, 0.3) \\
        & \texttt{delta\_prior} & Gaussian \\
        & \texttt{delta\_prior\_mu} & 0.0 \\
        & \texttt{delta\_prior\_sigma} & 0.1 \\
        & \texttt{B} & (0.0, 5.0) \\
        Nebular & \texttt{log\_U} & -3.0 \\
        Stellar & \texttt{massformed} & (0.0, 13.0)\\
        (dblplaw) & \texttt{metallicity} & (0.2/z\_factor, 3.5/z\_factor) \\
        & & z\_factor = (0.02/0.014) \\
        & \texttt{metallicity\_prior} & "log\_10" \\
        & \texttt{alpha} & (0.1, 1000.0) \\
        & \texttt{alpha\_prior} & "log\_10" \\
        & \texttt{beta} & (0.1, 1000.0) \\
        & \texttt{beta\_prior} & "log\_10" \\
        & \texttt{tau} & (0.01, 15.0) \\
        & \texttt{tau\_prior} & "log\_10" \\
        \hline
    \end{tabular}
\end{table}

The resulting final, complete redshift distribution of our sample is shown in Fig. \ref{fig:z_dist}, with the spectroscopic measurements indicated in red. Here, it can again be seen that the 
spectroscopic redshifts are representative of the overall redshift distribution, and in particular provide excellent coverage of the key high-redshift tail. The majority (59\%) of the final sample lies at so-called cosmic noon, in the redshift range $1.5<z<3$.
The mean redshift of the sample is $\langle z\rangle=2.68$ and the median is $z_{\rm med}=2.52$. 

Compared to previous blind ALMA surveys in the same field, our sample has more sources, both in number and fraction, at higher redshifts. Specifically, almost 30\% (37/128) of the ALMA sources in our sample lie at $z>3$. However, none lie beyond $z = 7$, with the redshift of the most distant source in our sample being spectroscopically confirmed at $z = 6.854$ \citep{2018Natur.553..178S}. It is worth noting that the spectroscopic redshift coverage at $z>4$ (10/18, 56\%) is even higher than that at lower redshifts (54/110, 49\%).
This provides additional confidence that our multi-frequency sample can be used to undertake a robust estimate of dust-enshrouded cosmic star-formation activity in the young Universe.

\section{Analysis \& Discussion}
\label{sec:analysis}
Armed with the redshifts and full multi-frequency photometry for all 128 ALMA-detected sources in the PRIMER COSMOS field, we next proceeded to derive best estimates of the basic physical properties of these dusty galaxies. As discussed further in the following subsections, we used the the SED-fitting code {\sc bagpipes} \citep{2018MNRAS.480.4379C} to determine stellar masses and to provide one estimate of star-formation rate (SFR). We also derived an alternative measure of SFR for each galaxy directly from the raw rest-frame UV and (sub)mm photometry. Below, we present and discuss the resulting derived properties of the galaxies before going on to explore the implications of our results for cosmic star-formation history.

\begin{figure}
	\includegraphics[width=\columnwidth]{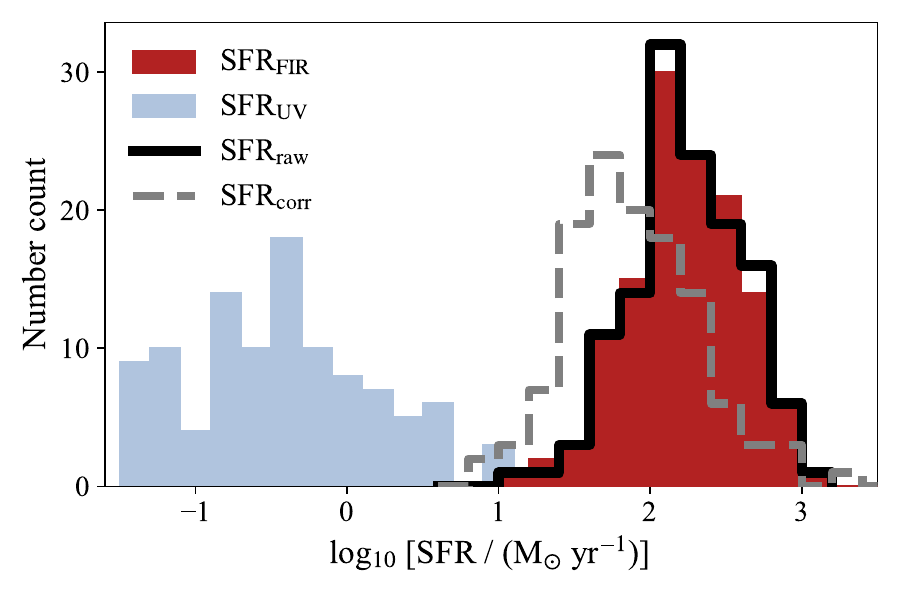}
    \caption{The distribution of different measures of star-formation rate (SFR) for the 128 ALMA-detected galaxies in the PRIMER COSMOS sample. Total `raw' SFR (SFR$_{\rm raw}$, black solid profile) is the sum of the dust-enshrouded SFR (SFR$_{\rm FIR}$, red; derived directly from the (sub)mm photometry) and the UV-visible SFR (SFR$_{\rm UV}$, blue; derived from directly from the optical-infrared photometry {\it without} correction for dust attenuation). Unsurprisingly, the vast majority of the star-formation activity for all of these ALMA-detected sources is reflected in the (sub)mm emission. The SFR derived from fitting to the optical-infrared SEDs allowing for dust attenuation typically gives a somewhat lower estimate of star-formation rate (SFR$_{\rm corr}$, dashed grey profile) than the total `raw' SFR, likely due to the failure of this method to account for star formation in the most highly-obscured regions of the galaxies.}
    \label{fig:SFR_dist}
\end{figure}

\subsection{Star-formation rates}
\label{sec:SFR}

We derived estimates of the SFR for each ALMA-detected galaxy using two different approaches, with the results of both calculations presented in Table \ref{tab:SED1}.

The first calculation of SFR is based on fitting to the observed optical-infrared 
SEDs of the galaxies (as derived from the {\it JWST} or UltraVISTA photometry, with aperture size effect corrected) using the {\sc bagpipes} code. In effect this assumes that the true SFR can be recovered by allowing the attenuation, $A_V$, to float while seeking the best-fitting stellar population model which can account for the rest-frame UV-optical emission from the galaxy. This estimate of SFR is denoted here as SFR$_{\rm corr}$ (because it involves an obscuration `correction'). For most types of galaxy, with moderate dust obscuration, this approach is known to be as reliable as any other \citep[e.g.,][]{2017MNRAS.466..861D}. Where available, we always used the {\it JWST} photometry for the SED fitting, and only used the UltraVISTA data for those objects that lack NIRCam imaging. Consistent with the fact that all of these galaxies have been detected in the (sub)mm, we allowed a large range of dust attenuation ($0<A_V<8$) during the fitting process.
The initial parameter inputs to the {\sc bagpipes} code are listed in Table \ref{tab:bagpipes}. We note that this {\sc bagpipes} SED fitting also yielded the stellar masses of the galaxies. These stellar mass estimates are also tabulated in Table \ref{tab:SED1}, and are discussed further in the next subsection.

The second method we used to determine SFR may seem somewhat cruder, but it is less model dependent, and allows for the possibility that at least some of the star-formation activity in these ALMA-detected sources is actually {\it completely} hidden from view in the rest-frame UV-optical. This involves the calculation of what we term total `raw' star-formation rate, SFR$_{\rm raw}$ by adding the dust-obscured SFR as determined from the (sub)mm photometry (SFR$_{\rm FIR}$) to the 
UV-visible SFR (SFR$_{\rm UV}$) as determined from the rest-frame UV emission {\it without} any correction for dust attenuation. For the reasons demonstrated and explained below, we ultimately adopted SFR$_{\rm raw}$ as the preferred most robust measure of SFR for the ALMA-detected galaxies under study here.

\begin{figure}
	
	\includegraphics[width=\columnwidth]{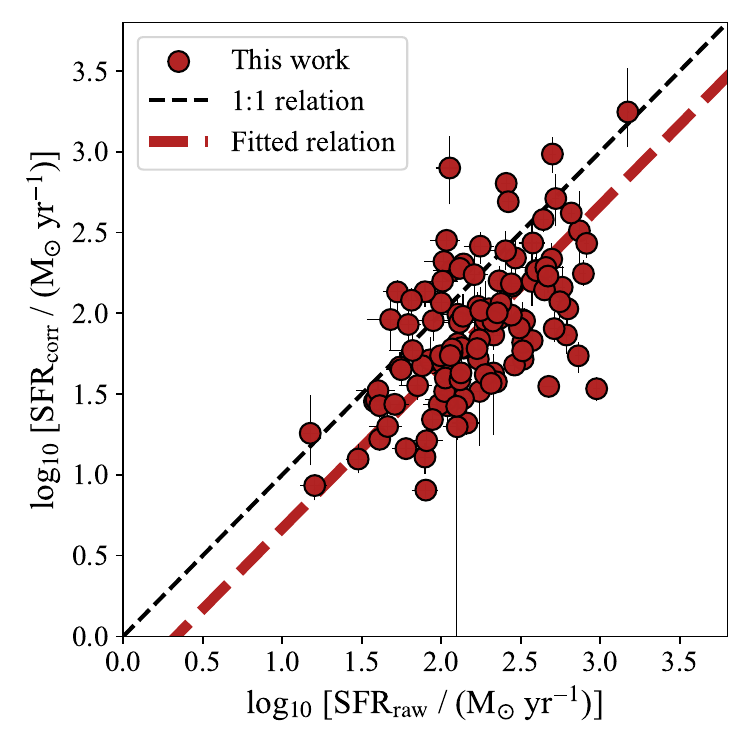}
    \caption{The values of SFR$_{\rm corr}$ (based on SED fitting with {\sc bagpipes}) compared with the corresponding values of SFR$_{\rm raw}$ (created by adding the raw sub(mm) and raw UV-derived values of SFR).
    The fitted dashed red line (with a fixed slope of unity in log space) shows that, on average, SFR$_{\rm raw}$ is 0.33 dex larger than SFR$_{\rm corr}$. Given this, and the obvious explanation that some of the star formation in these ALMA-detected galaxies is too obscured to be visible at all in optical-infrared regime used for the SED fitting, we adopt 
    SFR$_{\rm raw}$ as the best estimate of SFR in all subsequent analysis.}
    \label{fig:sfr_comparison}
\end{figure}

\begin{figure}
	\includegraphics[width=\columnwidth]{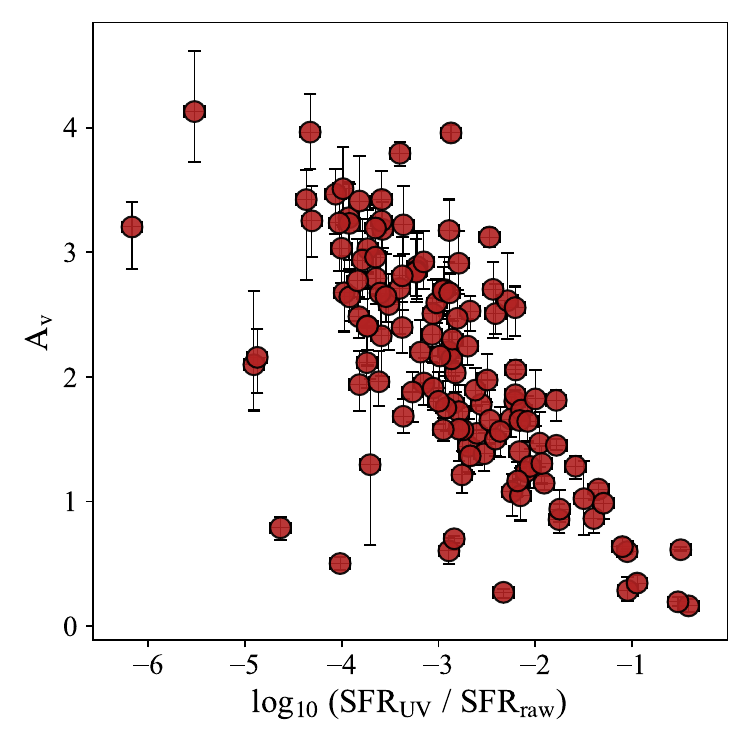}
    \caption{The values of dust attenuation, $A_{\rm V}$ derived from the optical-infrared SED fitting with {\sc bagpipes}, plotted against the log of the ratio of raw UV and raw total SFR. Given that these are two different and independent measures of inferred onscuration, the strong correlation is reassuring. It can also be seen that the correlation is tighter for more modestly obscured sources where the SED fitting can provide a a reasonable estimate of the dust attenuation.}
    \label{fig:av_comparison}
\end{figure}

\begin{figure}
	\includegraphics[width=\columnwidth]{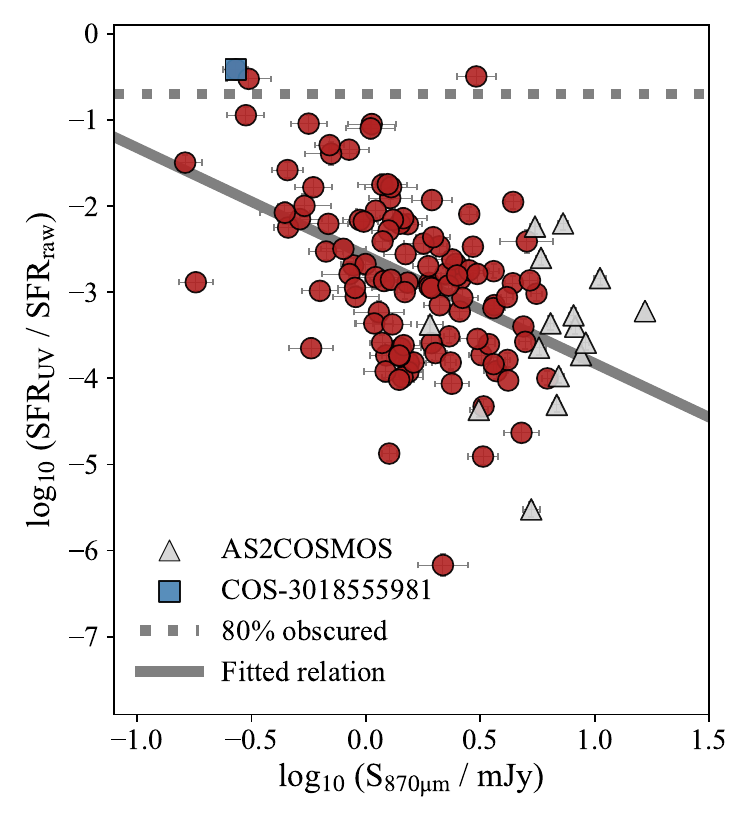}
    \caption{The level of dust attenuation, 
    as quantified by the ratio of UV-visible to total SFR (SFR$_{\rm UV}$/SFR$_{\rm raw}$) plotted against the observed sub-mm flux density for each galaxy in our 128-source sample. All except 3 of the galaxies are more than 80\% obscured. Moreover, while an anti-correlation is expected (see text), this plot confirms that the brightest (sub)mm sources are typically also the most heavily dust obscured as judged by this SFR ratio.
    We highlight the brightest and generally most heavily obscured sources in our sample which were also detected by AS2COSMOS \citep[grey triangles,][]{2020MNRAS.495.3409S}, as well as the one (relatively faint) (sub)mm source which also features in the REBELS sample \citep[blue rectangle, namely COS-3018555981 in][]{2018Natur.553..178S}, and which was in practice first selected at rest-frame UV wavelengths (see text).}
    \label{fig:ratio_flux}
\end{figure}

SFR$_{\rm FIR}$ is estimated from the observed-frame 1.3-mm continuum flux, using the conversion from \citet{2017MNRAS.466..861D}, which was based on the typical mid-far infrared SED displayed by the ALMA-detected galaxies in the Hubble Ultra Deep Field (HUDF), assuming a Chabrier IMF. Specifically we use the simple relation:

\vspace*{-0.04in}

\begin{equation}
\label{eq:sfr_convert}
    \text{SFR$_{\rm FIR}$ (M$_{\odot}$ yr$^{-1}$) = 0.30 $\times$ S$_{\rm 1.3mm}$ (${\rm \mu}$Jy)}
\end{equation}

\vspace*{0.07in}
\noindent
where, for each galaxy, the adopted 1.3-mm flux density $S_{\rm 1.3mm}$ and associated uncertainty was converted from the highest-S/N ALMA detection assuming the form of the (sub)mm SED follows $S_{\nu}\propto\nu^3$. We checked that the converted fluxes are all consistent with the greybody fit and other existing ALMA observations around the same frequency (see Fig. \ref{fig:greybody} for an example). We also verified that a calculation of total FIR luminosity, $L_{\rm 8-1000 \mu m}$, derived from integration of the fitted greybody curves, when converted to SFR (assuming, again, a Chabrier IMF) using the relation \citep{2018MNRAS.476.3991M}:

\vspace*{-0.04in}

\begin{equation}
\label{eq:sfr_integ}
    \mathrm{log_{10} (SFR_{FIR, integ}~/~M_{\odot}\,yr^{-1})} = \mathrm{log_{10} (L_{\mathrm 8-1000\mu m}~/~\rm{W})} - 36.44
\end{equation}

\noindent 
yields, on average, identical results to the simple mono-chromatic calculation of SFR$_{\rm FIR}$ described above.

The `raw' SFR$_{\rm UV}$ is derived from the UV luminosity of each source at rest-frame $\lambda = 1500$\,\AA\ ($L_{\rm 1500\AA}$, as determined from the {\sc bagpipes} SED fitting) using the relation:

\vspace*{-0.04in}

\begin{equation}
    \text{SFR$_{\rm UV}$ (M$_{\odot}$ yr$^{-1}$) = $\mathscr{K}_{\rm UV}$ $\times$ $L_{\rm 1500\text{\AA}}$ (erg s$^{-1}$ Hz$^{-1}$)}
\end{equation}

\vspace*{0.07in}
\noindent
where $\mathscr{K}_{\rm UV} = 0.72\times10^{-28}$, corrected from a Salpeter IMF \citep{1998ARA&A..36..189K} to a Chabrier IMF \citep{2014ARA&A..52..415M}.

Finally, we note that one of our sources (ID 002, ALMA R.A. = 150.09991 deg, Dec. = 2.29722 deg) is known to be a lensed galaxy \citep{2024A&A...690L..16J}. Because derivation of its intrinsic physical properties requires much more complicated modeling than is required for the other sources in our sample, for this galaxy we adopted the values of SFR, $M_*$ and $A_V$ deduced by \citet{2024A&A...690L..16J}, correcting the derived values by the magnification factor given there.

\begin{figure*}
	\includegraphics[width=2\columnwidth]{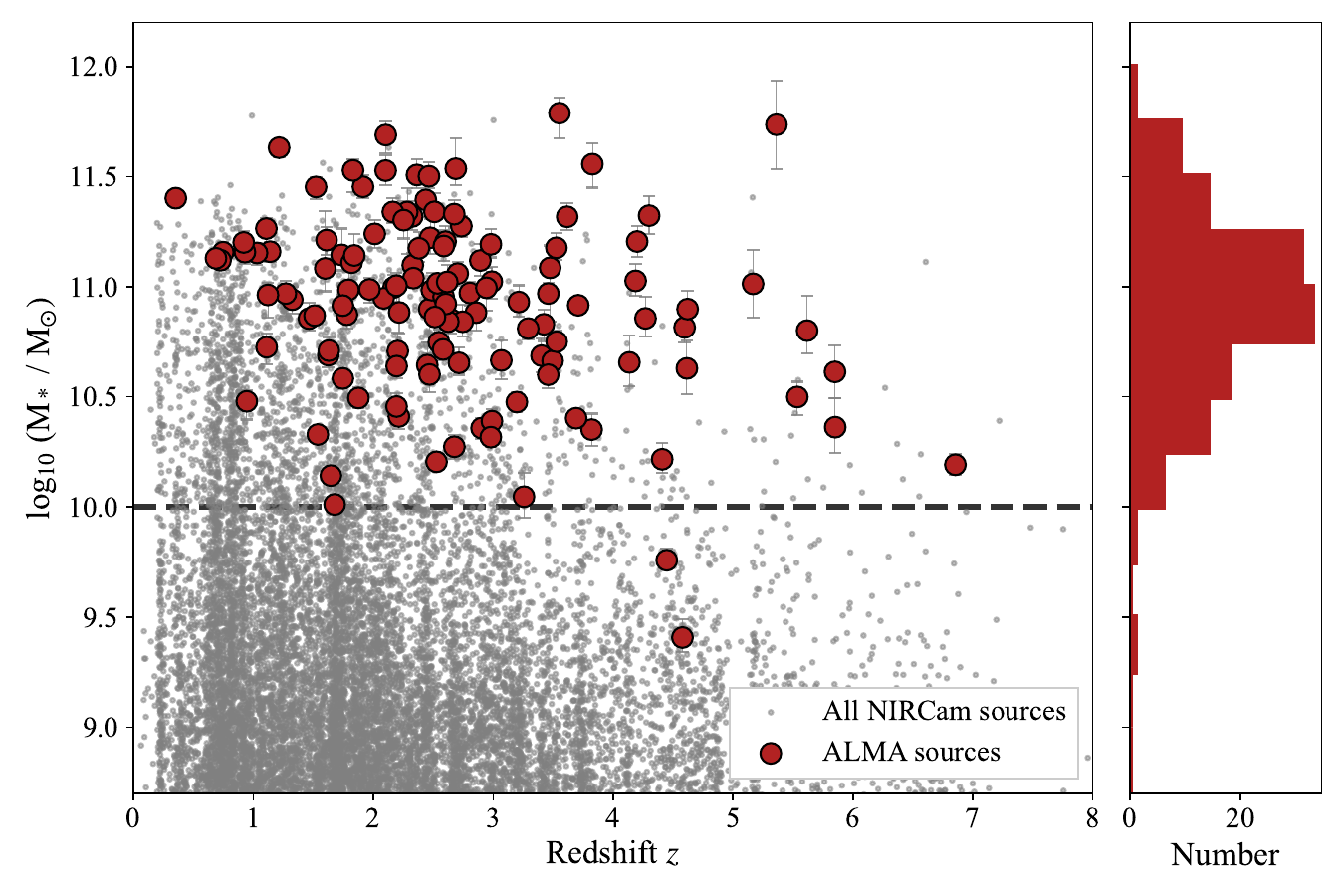}
    \caption{The stellar masses of all sources in the complete NIRCam PRIMER COSMOS galaxy catalogue, plotted as a function of redshift (small grey dots). The 128 sources in our ALMA-detected  sample are highlighted by the red filled circles and can be seen to be confined to the high-mass end of the galaxy mass distribution at all epochs. Specifically, almost all (126/128) having a stellar mass $M_* > 10^{10}$ M$_{\odot}$ (a threshold indicated by the dashed horizontal line). It can also be seen that that the ALMA detected subsample becomes an increasingly dominant subset of the general high-mass galaxy population with increasing redshift (at $z < 1$ the majority of such high-mass galaxies are known to be quiescent). The distribution of $M_*$ for the ALMA-detected galaxies is shown in the right-hand panel. We remind the reader that because complete ALMA coverage of this field has not yet been achieved, several/many of the high-redshift high-mass NIRCam sources plotted here in grey could yet transpire to be detectable in the (sub)mm given new ALMA observations.}
    
    \label{fig:M*_z}
\end{figure*}

Fig. \ref{fig:SFR_dist} shows the distribution of SFR for the 128 sources in our sample, as derived from these different methods. It can be seen that SFR$_{\rm FIR}$ dominates SFR$_{\rm UV}$ by on average a factor $\simeq 100$. Quantifying this source-by-source via the ratio SFR$_{\rm UV}$/SFR$_{\rm raw}$, we find that the vast majority of the sources in the sample (125/128) are highly obscured with more than 80\% of SFR$_{\rm raw}$ composed of SFR$_{\rm FIR}$.

It can also be seen that the (FIR dominated) total SFR$_{\rm raw}$ for the sample is generally higher than SFR$_{\rm corr}$, confirming that, for this class of source, optical-infrared SED fitting runs the risk of under-estimating the true  SFR when at least some of the SFR in these galaxies is highly obscured. For the sample as a whole the offset between SFR$_{\rm raw}$ and SFR$_{\rm corr}$ is shown and quantified in Fig. \ref{fig:sfr_comparison}, which reveals that while the two values are clearly well correlated, there exists an offset of $\simeq 0.33$\,dex.

As a sanity check that our results are reasonable, in Fig.\,\ref{fig:av_comparison} we show that, reassuringly, the fitted value of dust attenuation, $A_{\rm V}$, from the {\sc bagpipes} SED fitting correlates well with SFR$_{\rm UV}$/SFR$_{\rm raw}$. Moreover, as expected \citep[and discussed by][]{2017MNRAS.466..861D} the correlation is tightest for the less obscured galaxies in our sample, where less of the star formation is deeply obscured (and hence the SED fitting is more likely to yield unbiased results).  

Finally, for completeness, in Fig. \ref{fig:ratio_flux} we plot raw obscuration (SFR$_{\rm UV}$/SFR$_{\rm raw}$) versus the sub-mm flux density for all of the galaxies in our sample. An anti-correlation is arguably expected here (because SFR$_{\rm raw}$ is dominated by SFR$_{\rm FIR}$ which is proportional to $S_{\rm 870\mu m}$ according to Equation (\ref{eq:sfr_convert})) provided SFR$_{\rm UV}$ is not strongly related (correlated or anti-correlated) to SFR$_{\rm FIR}$. Indeed, this is what is seen, with the best-fitting relation given by:

\vspace*{-0.04in}

\begin{equation}
    \text{log}_{10}\frac{\rm SFR_{UV}}{\rm SFR_{raw}} = (-1.23\pm0.04) \times {\rm log_{10}}\,S_{\rm 870\mu m}\,{\rm(mJy)} - (2.57\pm0.01)
\end{equation}

\noindent
In effect this plot, and the fitted relation, confirm that the galaxies which are brightest in the (sub)mm are indeed the most heavily obscured. We also use Fig. \ref{fig:ratio_flux} to show how our sample connects the most luminous (and most dust-obscured) (sub)mm galaxies originally detected by SCUBA-2 in the field (i.e., the 17 AS2COSMOS sources) down to the much less luminous (sub)mm sources, which are in turn much less obscured as exemplified by the UV-selected REBELS source (ID 126) (for which SFR$_{\rm UV}$ is comparable to SFR$_{\rm FIR}$).

\subsection{Galaxy stellar masses}
\label{sec:Mass}

The stellar masses of all 128 ALMA-detected galaxies in our sample were determined by SED fitting with {\sc bagpipes} as described in Section \ref{sec:SFR}. The resulting values, which assume a Chabrier IMF,  are tabulated for all the sources in Table \ref{tab:SED1} and plotted in Fig.\,\ref{fig:M*_z}. To place these results in context, in Fig.\,\ref{fig:M*_z} we have also plotted the stellar masses (versus their redshifts) of all the galaxies in the NIRCam PRIMER COSMOS catalogue (also derived with {\sc bagpipes}) and highlight in red the sources that feature in our ALMA-detected sample (the stellar mass histogram for the ALMA-detected sources is projected into the right-hand panel). 

\begin{figure}
	\includegraphics[width=\columnwidth]{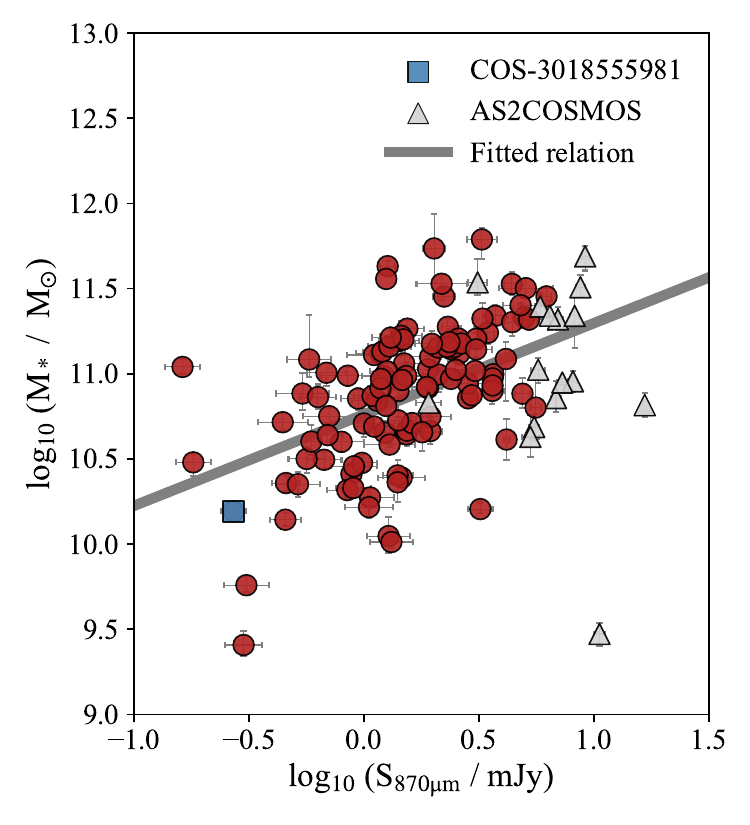}
    \caption{The relation between stellar mass and (sub)mm flux-density displayed by the 128 ALMA-detected galaxies  in our sample. Again we highlight the 17 bright sources within our sample which were detected by the AS2COSMOS follow-up of the SCUBA-2 survey \citep{2020MNRAS.495.3409S, 2024A&A...691A.299G} and the one (UV-selected) REBELS source \citep[namely COS-3018555981 in ][]{2018Natur.553..178S} in our sample.}
    \label{fig:M*_flux}
\end{figure}

From Fig.\,\ref{fig:M*_z} it can be seen that virtually all (126/128) of the ALMA-detected galaxies have stellar masses, $M_*$, greater than the mass threshold $M_* = 10^{10}\,{\rm M_{\odot}}$, at all redshifts. Moreover, they extend upwards in mass to include the most massive galaxies in the field.
This result echoes the results from the early blank-field ALMA surveys, such as those conducted in the Hubble Ultra Deep Field \citep[HUDF;][]{2016ApJ...833...72B, 2017MNRAS.466..861D}, but the HUDF was too small a field to contain galaxies with $M_* > 10^{10}\,{\rm M_{\odot}}$ at $z > 3$. Now, with the larger cosmological volumes sampled here by the PRIMER COSMOS field, it can be seen that this galaxy stellar mass threshold for bright (sub)mm emission applies back towards the epoch of reionisation. 

\begin{figure}
	\includegraphics[width=\columnwidth]{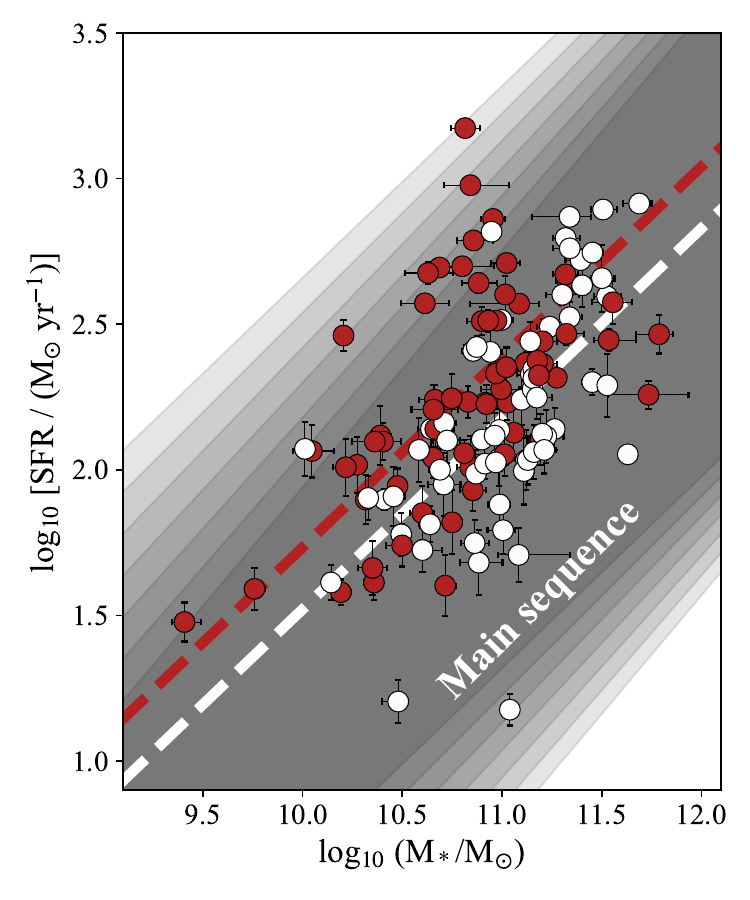}
    \caption{Star-formation rate (SFR$_{\rm raw}$ plotted versus stellar mass ($M_*$) for the 128 ALMA-detected galaxies. Shown in grey is the region occupied by `normal' star-forming galaxies - the so-called main sequence \citep[MS,][]{2014ApJS..214...15S}. The redshift evolution of the MS is illustrated by the offset of the translucent grey bands (each has the same shade) from lower right ($z=1$) to upper left ($z=7$), i.e., the darkest region is the shared MS across the whole redshift range of this study. Despite the (arguably realistic) scatter it is clear that the vast majority of the ALMA-detected galaxies lie on the MS, suggesting that these galaxies are representative of star-forming galaxies over most of cosmic time. As shown here the sample is divided in half into a low-redshift group (white filled circles) and a high-redshift group (red filled circles) around the median redshift of the sample ($z_{\rm med}=2.52$). The two groups exhibit a $\simeq0.2$ dex difference in normalisation (i.e., sSFR) when fitted with a linear relation with the same slope (fitted from the low-redshift group) under the logarithmic scale, showing that our sample is powerful enough to reveal the redshift evolution of the MS.}
    \label{fig:MS}
\end{figure}

It can also been seen that the ALMA-detected sample becomes an increasingly dominant subset of the general high-mass galaxy population with increasing redshift. Here we remind the reader that the ALMA 'survey' of the PRIMER COSMOS field utilised here is highly incomplete (being based simply on $\simeq 100$ pointings) and so the fraction of high-mass galaxies detected by ALMA (highlighted in red) is undoubtedly a lower limit. It is entirely possible, indeed likely, that many of the other high-mass galaxies in the field indicated by the grey dots in Fig.\,\ref{fig:M*_z} will also yield (sub)mm detections if observed with ALMA. At the highest masses, $M_* > 10^{11}\,{\rm M_{\odot}}$, the majority of the galaxies in the field at $z > 2$ are ALMA-detected. This reflects the fact that, within the ALMA-detected sample of galaxies under study here, it transpires that (sub)mm flux density, $S_{\rm 870\mu m}$, is actually a rather strong function of stellar mass, as shown in Fig.\,\ref{fig:M*_flux}. This is perhaps not unexpected, given the well-established strong stellar-mass dependence of dust attenuation in galaxies at $z \simeq 2$ \citep[e.g.,][]{2018MNRAS.476.3991M}.

In summary, the clear implication of this still incomplete ALMA `survey' is that virtually all massive galaxies ($M_* > 2 \times 10^{10}\,{\rm M_{\odot}}$) at high redshift are dust-enshrouded star-forming galaxies, with the most massive being detectable in the (sub)mm with bright (but also reasonably complete) surveys as have been conducted with SCUBA-2.
At lower redshifts, it can be seen that the ALMA-detected fraction of massive galaxies declines fairly rapidly at $z < 2$, with only a handful remaining at $z < 1$. This is as expected given that an increasing majority of the most massive galaxies at $ z < 1.5$ are known to be quiescent, with most star-formation activity in such objects having been quenched by this epoch.

Finally, in Fig.\,\ref{fig:MS}, we plot the raw total value of SFR (derived and discussed in Section \ref{sec:SFR}) for each galaxy versus its stellar mass. This allows us to explore where the ALMA-detected galaxies in our sample lie relative to the so-called `main sequence' (MS) of star-forming galaxies. There is quite a lot of scatter in this plot, but this is probably realistically due to the fact that SFR$_{\rm raw}$ and $M_*$ have been derived in a completely independent manner (as compared to studies where SFR and $M_*$ are {\it both} derived from SED fitting). Nonetheless, there is enough dynamic range in our sample for it to be clear that these galaxies, with a few possible exceptions, lie on the MS \citep{2014ApJS..214...15S}. Moreover, dividing our sample into a low-redshift half and a high-redshift half (around the median redshift, $z_{\rm med} = 2.52$) there is enough signal to reveal the redshift evolution of the MS, and hence infer the redshift dependence of the typical specific SFR (sSFR) of the ALMA-detected galaxies. This information becomes important in the final analysis section below, where we attempt to correct for the incompleteness in our ALMA `survey' in order to properly estimate the redshift evolution of dust-enshrouded star-formation rate density.

\begin{figure*}
	\includegraphics[width=2.1\columnwidth]{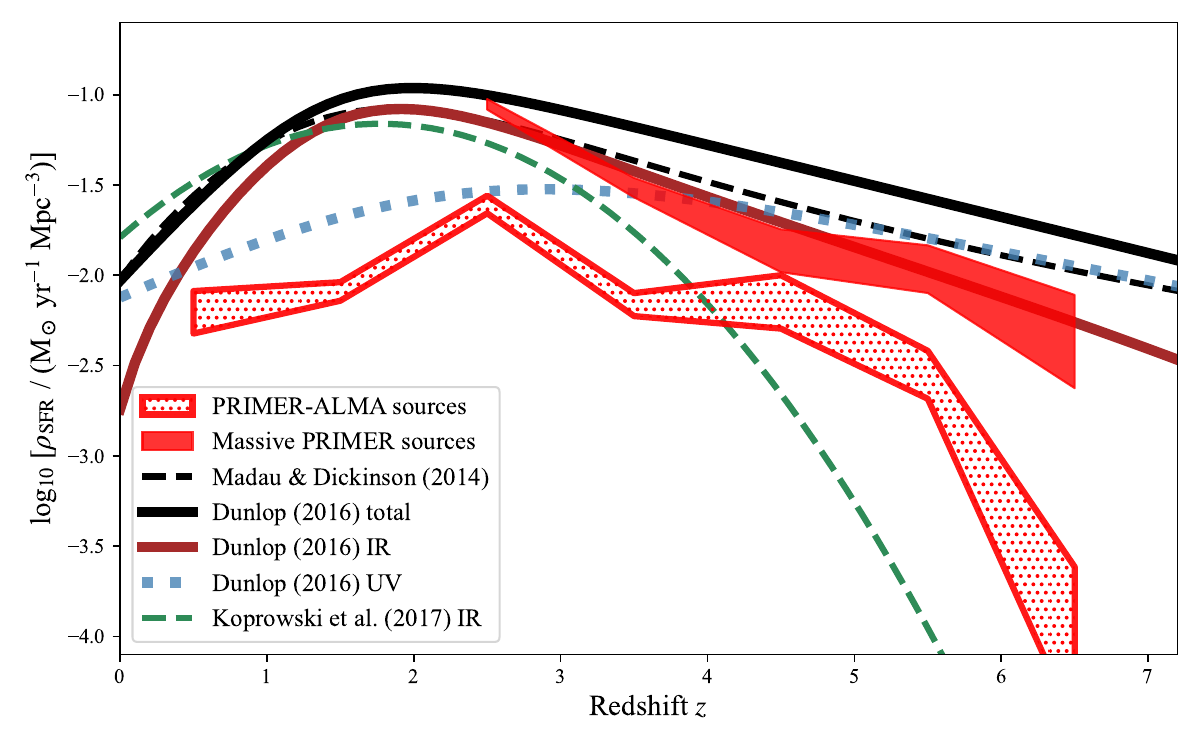}
    \vspace*{-0.3in}
    \caption{Co-moving cosmic star-formation rate density, $\rho_{\rm SFR}$. The contribution made by dust-enshrouded star formation ($\rho_{\rm SFR\,IR}$) as calculated directly from the 128 ALMA-detected galaxies in our sample is shown (in integer redshift bins) by the dot-hatched red region. In effect, given the high incompleteness of the ALMA coverage of the COSMOS PRIMER field, this represents a hard lower limit to the true contribution of dust-enshrouded galaxies. Our corrected version of $\rho_{\rm SFR\,IR}$, derived by adding in the anticipated contributions of other massive ($M_* > 10^{10}\,{\rm M_{\odot}}$) galaxies in the field is shown by the filled bright red region
  (see Section \ref{sec:CSFH}). Five  parametric fits to previous data are plotted for comparison: the total  $\rho_{\rm SFR}$ from \citet{2014ARA&A..52..415M} (corrected to a Chabrier IMF), the total, obscured and UV-visible values of $\rho_{\rm SFR}$ from \citet{2016Msngr.166...48D}, and the extrapolated contribution of obscured star formation  inferred by \citet{2017MNRAS.471.4155K}. The very rapid drop in dust-obscured star-formation activity suggested by \citet{2017MNRAS.471.4155K} can be excluded, even by our raw ALMA-only results, while our final incompleteness-corrected results are in excellent agreement with the relation for $\rho_{\rm SFR\,IR}$ given by \citet{2016Msngr.166...48D}.}
    \label{fig:CSFH}
\end{figure*}

\begin{figure}
	\includegraphics[width=\columnwidth]{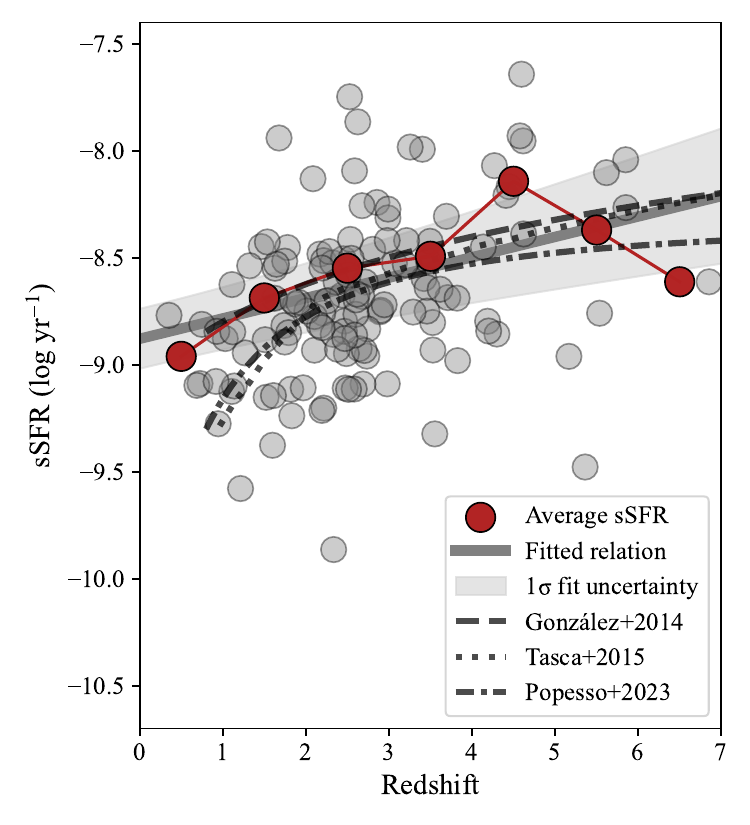}
    \caption{Specific star-formation rate (sSFR) as a function of redshift for the 128 ALMA sources (grey filled circles) in our final sample. The fitted relation (grey line) is consistent with the average sSFR of the sample in virtually all redshift bins from $z=1$ to $z=7$ (red filled circles) within the 1-$\sigma$ uncertainty (light-grey region). The growing uncertainty towards higher redshifts simply reflects the reduction in source numbers, yielding a less well constrained average sSFR at early cosmic epochs. Our relation is consistent with previous studies, specifically \citet{2014ApJ...781...34G}, \citet{2015A&A...581A..54T}, and \citet{2023MNRAS.519.1526P} (where we adopt $M_*=10^{10.5}\,{\rm M_{\odot}}$ as representative of the stellar masses in our high-$z$ ALMA-detected sample) at $2<z<7$, which is the redshift range over which we use this relation to estimate the contribution of massive galaxies which have not yet been observed with ALMA (see text).}
    \label{fig:ssfrz}
\end{figure}

\begin{figure*}
	\includegraphics[width=2.1\columnwidth]{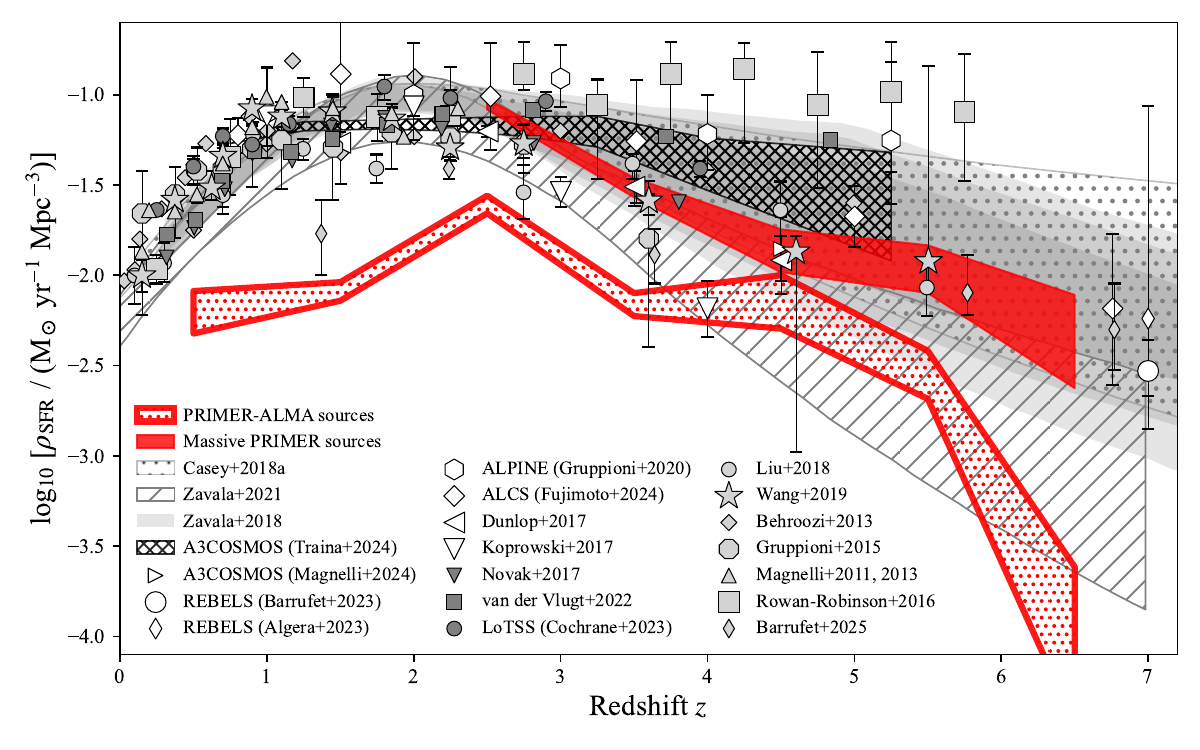}
    \vspace*{-0.3in}
    \caption{Our raw and incompleteness-corrected results for dust-obscured $\rho_{\rm SFR\,IR}$ as shown above in Fig.\,\ref{fig:CSFH}, but this time compared with a compilation of  results for obscured star-formation rate density from the literature. White data-points are derived from  (sub)mm observations, light-grey points are from mid-to-far-infrared observations, and dark-grey data-points are from data at other wavelengths (e.g., radio observations). Data include \citet{2011A&A...528A..35M, 2013A&A...553A.132M}, a collection of IR observations in \citet{2013ApJ...770...57B}, \citet{2015MNRAS.451.3419G}, \citet{2016MNRAS.461.1100R}, \citet{2017MNRAS.471.4155K}, \citet{2017A&A...602A...5N}, \citet{2018ApJ...853..172L}, \citet{2018ApJ...869...71Z} \citep[fitted from models in][colours lightened from 68\% to 99.7\% confidence interval]{2018ApJ...862...77C, 2018ApJ...862...78C}, \citet{2018ApJ...862...77C} (dust-rich and dust-poor models as upper and lower borders), \citet{2019A&A...624A..98W}, ALPINE \citep{2020A&A...643A...8G}, \citet{2021ApJ...909..165Z}, \citet{2022ApJ...941...10V}, REBELS \citep{2023MNRAS.518.6142A, 2023MNRAS.522.3926B}, LoTSS \citep{2023MNRAS.523.6082C}, ALCS \citep{2024ApJS..275...36F}, A$^3$COSMOS \citep{2024A&A...688A..55M, 2024A&A...690A..84T}, PRIMER COSMOS \citep[][red galaxies]{2025arXiv250805740B}.}
    \label{fig:CSFH2}
\end{figure*}

\subsection{Cosmic star formation history}
\label{sec:CSFH}
Finally, we used our results to make a new estimate of the contribution of dust-enshrouded star formation to cosmic star-formation history, reaching back to $z \simeq 7$. Here we made two calculations. 

\subsubsection{ Star-formation rate density from the ALMA sources}

First, we simply calculated the cosmic history of dust-enshrouded star-formation rate density, $\rho_{\rm SFR\,IR}$, on the basis of the 128 ALMA-detected galaxies studied here. Because this ALMA `survey' of the PRIMER COSMOS field simply consists of a series of individual pointings, it is inevitably highly incomplete (except at the very bright end where the AS2COSMOS ALMA sources were originally uncovered by the contiguous SCUBA-2 CLS survey). This means that the result of summing the SFR contributions of the ALMA sources in our sample can only yield a lower limit to $\rho_{\rm SFR\,IR}$ as a function of redshift/time. However, this lower limit is robust, and significantly higher (and hence more useful) than the results achieved by simply summing the contributions of the very bright A2SCOSMOS sources (of which there are only 17 in our sample). 

The results of this calculation, in integer reshift bins ($z=0-1$, $z=1-2$ etc), are tabulated in the second row of Table \ref{tab:sfrd}. The error in each redshift bin was estimated by bootstrapping, where we randomly re-sampled the galaxies 10,000 times in each redshift bin and calculated the standard deviation of all 10,000 runs. We also calculated the errors in each bin resulting from the propagation of the individual errors in SFR$_{\rm raw}$ for each galaxy and from the overall sampling uncertainty, but found that these errors were relatively small compared to those calculated from the bootstrap analysis. 

\begin{table*}
\centering
\setlength{\tabcolsep}{9pt} 
\caption{Calculated estimates of co-moving cosmic star-formation rate density, $\rho_{\rm SFR}$, as a function of redshift. The first row indicates the redshift bins, with the second row giving the contribution to $\rho_{\rm SFR}$ made by the 128 ALMA-detected galaxies in our sample ($\rho_{\rm SFR\,mm}$). The third row shows the result of our efforts to correct for the incompleteness of the ALMA `survey' used here, by assigning dust-obscured SFR estimates to all galaxies in the {\it JWST} PRIMER COSMOS sample with $M_* > 10^{10}\,{\rm M_{\odot}}$, assuming a sSFR as derived from the ALMA-detected galaxies at comparable redshifts. This latter calculation is confined to $z > 2$ because, at lower redshifts, it is unreasonable to assume that the vast majority of massive galaxies are star-forming. This final row thus gives our best incompleteness-corrected estimate of the evolution of dust-obscured star-formation rate density, $\rho_{\rm SFR\, IR}$, from $z \simeq 2$ to $z \simeq 7$. Errors are calculated from bootstrapping and propagated to the logarithmic scale. The bootstrapping for $\rho_{\rm SFR\,IR}$ used a perturbation method to reflect the uncertainty in the measurement of SFR (see Section \ref{sec:CSFH}).}
\label{tab:sfrd}
    \begin{tabularx}{\textwidth}{cccccccc}
    \hline
    Redshift range & $0 < z \leq 1$ & $1 < z \leq 2$ & $2 < z \leq 3$ & $3 < z \leq 4$ & $4 < z \leq 5$ & $5 < z \leq 6$ & $6 < z \leq 7$ \\
        \hline
    log$_{10}\rho_{\rm SFR\,mm}$ & $-2.21\pm0.12$ & $-2.09\pm0.05$ & $-1.61\pm0.05$ & $-2.16\pm0.06$ & $-2.15\pm0.15$ & $-2.55\pm0.13$ & $-4.05\pm0.43$ \\
    log$_{10}\rho_{\rm SFR\,IR}$ & - & - &  $-1.00\pm0.03$ & $-1.48\pm0.06$ & $-1.85\pm0.12$ & $-1.97\pm0.16$ & $-2.42\pm0.32$ \\
    \hline
    \end{tabularx}
        \vspace*{0.1in}
        
\end{table*}

These values, and the associated errors are plotted as the {\it dot-hatched} red band in 
Fig.\,\ref{fig:CSFH}. The resulting `curve' rises by two orders-of-magnitude from $z \simeq 7$ to a peak at cosmic noon ($z \simeq 2.5$) before declining by $\simeq 0.5$\,dex by $z \simeq 0.5$. Unsurprisingly, given that this curve essentially represents a lower limit on
$\rho_{\rm SFR\,IR}$, this curve lies below previous estimates of total $\rho_{\rm SFR}$ 
\citep[e.g.,][]{2014ARA&A..52..415M, 2016Msngr.166...48D}, but interestingly it already lies above some estimates of the high-redshift evolution of $\rho_{\rm SFR\,IR}$ based on certain functional extrapolations from lower redshift (sub)mm detections \citep[e.g.,][]{2017MNRAS.471.4155K}. It is also notable that, at $z \simeq 2.5$, this incomplete determination of $\rho_{\rm SFR\,IR}$ essentially matches the value of $\rho_{\rm SFR\,UV}$ as indicated by the dotted blue curve in Fig.\,\ref{fig:CSFH}, which is given by the equation

\vspace*{-0.04in}

\begin{equation}
\label{eq:rho_uv}
\rho_{\rm SFR\,UV} = \frac{0.055}{10^{-0.38(z -2.2)}+10^{0.16(z-2.2)}}
\end{equation}

\vspace*{0.07in}

\noindent
as derived by \citet{2016Msngr.166...48D}. However, this is perhaps not surprising, given that a full calculation of $\rho_{\rm SFR\,IR}$ from complete ALMA surveys such as those conducted in the HUDF, shows that $\rho_{\rm SFR\,IR}$ dominates $\rho_{\rm SFR\,UV}$ at cosmic noon 
\citep{2017MNRAS.466..861D, 2020ApJ...902..112B, 2021ApJ...909..165Z}.

\subsubsection{Final incompleteness-corrected star-formation rate density}
Second, armed with the {\it full} galaxy sample in the PRIMER COSMOS NIRCam imaging  resulting from the deep {\it JWST}+{\it HST} observations (including derived photometric/spectroscopic redshifts and stellar masses), we have attempted to correct for the incompleteness in our ALMA survey. This calculation was primarily motivated by Fig. \ref{fig:M*_z}, which confirms that essentially all the ALMA-detected galaxies are massive with $M_* > 10^{10}\,{\rm M_{\odot}}$, but that there remain many comparably massive galaxies in the full PRIMER COSMOS catalogue that have yet to be observed with ALMA. We therefore proceeded to estimate the additional contribution to $\rho_{\rm SFR\,IR}$ that would be made by these other massive galaxies in the field, on the assumption that their properties are similar to those of the ALMA-detected objects at comparable redshift and of comparable stellar mass. 

In detail, we proceeded as follows.
Firstly, we calculated the sSFR of each ALMA source, and fitted the sSFR as a function of redshift for the 128 ALMA sources. This gives:

\vspace*{-0.04in}
\begin{equation}
\label{eq:ssfr}
    \rm{log}_{10} \rm{sSFR} ~(\rm{yr}^{-1}) = (0.10\pm0.04)\times z -(8.88 \pm 0.14)
\end{equation}

\vspace*{0.07in}
\noindent
fitted over the redshift range $0<z<7$, which is illustrated in Fig. \ref{fig:ssfrz}. This fitted relation, although cast in a different functional form, is fully consistent with the sSFR-$z$ relations found in several previous studies of massive star-forming galaxies at $2<z<7$ (within the uncertainties, e.g., \citet{2014ApJ...781...34G}, \citet{2015A&A...581A..54T}, and \citet{2023MNRAS.519.1526P}). We then required the NIRCam sources in the PRIMER COSMOS catalogue to have a SED-fitted $M_*>10^{10}\,{\rm M_{\odot}}$, to also have $\rm{S/N}>10$ in the F356W imaging (already true for almost all high-mass galaxies in the sample), and for the SED fitting to achieve a reduced $\chi_{\nu}^2<5$.
We also carefully checked the photometric redshift solutions, especially in the highest redshift bins, to exclude any sources with unreliable redshift estimates.
The fitted sSFR from the ALMA sample as given by Equation (\ref{eq:ssfr}) is then applied to the SED-fitted $M_*$ values of all the galaxies in this cleaned high-mass galaxy sample, which contained 2,225 sources in total.
The deduced SFR, SFR$_{\rm sSFR}$, has an error propagated from both the uncertainty in the individual values of $M_*$ and the uncertainty in the fitted $z$-$\rm{sSFR}$ relation.
We then calculated the contribution to $\rho_{\rm SFR}$ made by this much larger galaxy sample as a function of redshift, calculating the appropriate cosmological volumes corresponding to the 142.2\,arcmin$^2$ of NIRCam coverage. During the error derivation, we perturbed the SFR$_{\rm sSFR}$ of each galaxy with its error, and ran bootstrap on the perturbed sample in each redshift bin 10,000 times. This approach effectively includes both the unperturbed bootstrapped error from the original sample and the uncertainty in the measured SFR$_{\rm sSFR}$ of each massive galaxy.

In essence, this calculation corresponds to assuming that the massive galaxies in the field which have not been observed by ALMA have a similar sSFR to the ALMA-detected galaxies at each redshift, and that the resulting inferred SFR {\it for these massive galaxies} is overwhelmingly dust enshrouded. The latter assumption is relatively safe, given the properties of the ALMA-detected galaxies studied here, and indeed the properties of all known high-mass star-forming galaxies. The former assumption is likely also reasonable at high-redshift (where quiescent galaxies are extremely rare) but becomes dubious at lower redshift and obviously is completely incorrect by the present day (when the vast majority of massive galaxies are passive). We therefore restricted this calculation to $z >2$. 

These resulting `corrected' values of $\rho_{\rm SFR\,IR}$, derived by {\it adding} the directly-measured contributions of those massive galaxies detected by ALMA to the estimated additional contributions of the other high-mass galaxies in the field, are tabulated in the third row of Table \ref{tab:sfrd}, again in integer redshift bins but this time only for $z > 2$. 
These values, and the associated errors are plotted as the bright {\it filled} red band in 
Fig.\,\ref{fig:CSFH}. At $z > 3$ this still lies reassuringly below previous estimates of total $\rho_{\rm SFR}$ shown in Fig.\,\ref{fig:CSFH} as derived by \citet{2014ARA&A..52..415M} (converted to a Chabrier IMF; dashed black curve) or \citet{2016Msngr.166...48D} (solid black curve) where the latter relation is given by:

\vspace*{-0.04in}

\begin{equation}
\label{eq:rho_total}
\rho_{\rm SFR\,total} = \frac{0.175}{10^{-0.9 (z -1.4)}+10^{0.2(z-1.4)}}
\end{equation}

\vspace*{0.07in}
\noindent
Interestingly, however, our `corrected' evolution of $\rho_{\rm SFR\,IR}$ (as indicated by the bright filled red band) in fact follows rather well the evolution of dust-enshrouded SFR density inferred by \citet{2016Msngr.166...48D} by simply subtracting Equation (\ref{eq:rho_uv}) from Equation (\ref{eq:rho_total}), {\it i.e.},:

\vspace*{-0.04in}

\begin{equation}
\label{eq:rho_ir}
\rho_{\rm SFR\,IR} = \rho_{\rm SFR\,total} - \rho_{\rm SFR\,UV}
\end{equation}

\vspace*{0.07in}
\noindent
which is plotted as the solid brown curve in Fig.\,\ref{fig:CSFH}. This new measurement of $\rho_{\rm SFR\,IR}$ at $z > 2$ presented here thus appears to confirm the conclusions of several previous studies that $z\simeq 4$ represents the transition redshift between $\rho_{\rm SFR\,IR}$ and $\rho_{\rm SFR\,UV}$ with UV visible SFR becoming increasingly dominant at all higher redshifts \citep{2017MNRAS.466..861D, 2024A&A...688A..55M}. We note that this high-redshift decline of the relative contribution of dust-enshrouded star formation to $\rho_{\rm SFR\,total}$ can be explained simply by the relative rarity of high-mass galaxies ($M_* > 10^{10}\,{\rm M_{\odot}}$) at early times, as is evident from Fig. \ref{fig:M*_z}. 

Whilst massive quiescent galaxies are known to be rare at $z>2$, an unexpectedly large number of such objects have been found in early JWST data. We therefore perform a check on our assumption that all massive $z>2$ galaxies in the field have SFRs comparable to our ALMA sample by excluding the robust $z>2$ quiescent galaxy sample of \citet{2025arXiv250906913S} from our estimate. This sample was selected via a {\sc Bagpipes} fitting approach described in \cite{Carnall2023}, which has now been extensively spectroscopically validated \citep{Carnall2024}. This robust quiescent sample has 37 objects in common with our high-mass PRIMER COSMOS galaxy sample (though none in our ALMA-detected sample), the removal of which reduces our `corrected' measure of $\rho_{\rm SFR\,IR}$ by $\simeq9$\% in our $2 < z \leq 3$ bin and $\simeq4$\% in our $3 < z \leq 4$ bin. We therefore conclude that the presence of massive quiescent galaxies at $z>2$ in PRIMER COSMOS does not strongly affect our $\rho_{\rm SFR\,IR}$ estimates.

Finally, in Fig. \ref{fig:CSFH2} we again show our new raw and corrected estimates of the evolution of dust-enshrouded star-formation rate density, $\rho_{\rm SFR\,IR}$, but this time compared against a broad range of recent results from the literature. In general it can be seen that our results now categorically exclude most of the very high values of  $\rho_{\rm SFR\,IR}$ at $z > 3$ proposed by \citet{2016MNRAS.461.1100R}, \citet{2018ApJ...862...77C}, \citet{2020A&A...643A...8G}, and \citet{2021A&A...649A.152K} before the advent of {\it JWST}. They also exclude almost all of the (highly-uncertain) region suggested by \citet{2024A&A...690A..84T} from their recent analysis of the wider A$^3$COSMOS sample (where the redshift information is in general much poorer than achieved here), and the upper regions of the uncertain high-redshift extrapolation proposed by \citet{2018ApJ...869...71Z}. By contrast, our results are consistent with the upper end of the range derived by \citet{2021ApJ...909..165Z}, align well with the results of \citet{2017MNRAS.466..861D} out to $z \simeq 4$, and are completely consistent with the result derived at $z \simeq 4.5$ by \citet{2024A&A...688A..55M} from A3COSMOS. At the highest redshifts, our results are in excellent agreement with the evolution of dust-enshrouded 
$\rho_{\rm SFR\,IR}$ inferred by \citet{2019A&A...624A..98W} and our measurement is consistent with the values derived at $z \simeq 7$ from the REBELS survey \citep{2023MNRAS.518.6142A, 2023MNRAS.522.3926B} and from the ACLS \citep{2024ApJS..275...36F}, albeit some of these previous $z \simeq 7$ measurements are highly uncertain.

In summary, while a complete  ALMA survey of the PRIMER COSMOS field remains a highly desirable (albeit demanding) observational goal, through the analysis of the existing fairly dense ALMA observations of the field, in tandem with the PRIMER COSMOS {\it JWST}+{\it HST}+VISTA near/mid-infrared imaging, we have been able to derive a new estimate of the high-redshift evolution of dust-enshrouded star-formation rate density which is significantly more complete, robust and well-constrained than previously achieved. Solutions in which dust-enshrouded star-formation rate density ($\rho_{\rm SFR\,IR}$) exceeds UV-visible star-formation rate density ($\rho_{\rm SFR\,UV}$) at $z > 5$ are now effectively excluded by our analysis, even when we make the assumption that all massive galaxies in the field at $z > 2$ are as highly star-forming as the 128 ALMA-detected sources studied in detail here. That said, dust-enshrouded star formation clearly persists to higher redshifts than inferred from some earlier extrapolations from lower-redshift ALMA studies, but $\rho_{\rm SFR\,IR}$ is certainly lower at high redshift than inferred from many earlier investigations, especially those based on (relatively) poor-resolution {\it Herschel} surveys, which in many cases lacked secure galaxy identifications with robust redshift information. 
Our final result for the high-redshift evolution of dust-enshrouded star-formation rate density, $\rho_{\rm SFR\,IR}$, transpires to be very well described by the relation given by \citet{2016Msngr.166...48D} (and provided explicitly here in Equation (\ref{eq:rho_ir}) from the difference between Equations (\ref{eq:rho_total}) and (\ref{eq:rho_uv})), as shown in Fig. \ref{fig:CSFH}. Extrapolation of our results (according to this relation) predicts that, by $z \simeq 8$, dust-enshrouded star-formation activity is expected to contribute only 20\% of the total cosmic star-formation rate density, $\rho_{\rm SFR}$, with this contribution dropping to $\simeq 5$\% by $z = 10$. 

Of course, these are simply extrapolations, and the results presented here do not provide direct constraints beyond $z \simeq 7$. Nonetheless, this serves to demonstrate that even given the evidence that UV-visible star formation dominates at $z > 4$, dust may still be present at $z > 10$. This possibility is not yet ruled out by recent reports that galaxies discovered with {\it JWST} at $z \simeq 11$ are largely dust-free \citep[e.g.,][]{2024MNRAS.531..997C, 2025arXiv250111099C, 2025A&A...694A.215F} because the galaxy samples uncovered to date at such early times have yet to reveal galaxies with stellar masses $M_* > 10^{10}\,{\rm M_{\odot}}$ \citep{2024MNRAS.533.3222D, 2025arXiv250103217D}, should they exist.

\section{Summary \& Conclusions}
\label{sec:sum}
We have used the wealth of archival ALMA data in the COSMOS field to assemble a robust sample of 128 unique (sub)mm sources that lie within the central $\simeq 175$\,arcmin$^2$ region covered by the new {\it JWST} NIRCam and MIRI imaging from the PRIMER survey. This ALMA sample, being assembled from numerous pointed follow-up observations, does not represent a complete (sub)mm survey of the field. However, it spans two orders-of-magnitude in terms of sub-mm flux density, $S_{\rm 870\,\mu m}$, enabling us to explore the properties of dust-enshrouded star-forming galaxies from the brightest (sub)mm emitters uncovered by wide-area single-dish surveys (AS2COSMOS detected 17 of the 128 sources in our sample) down to the much fainter (sub)mm emitters recently uncovered by, for example, the REBELS programme.

The depth, red sensitivity, and high angular resolution (and hence high positional accuracy) 
of the {\it JWST} observations have enabled us to identify a secure and unique galaxy counterpart for {\it every one} of the 128 ALMA sources in our sample (aided, of course, by the high positional accuracy provided by ALMA). This high level of 100\% identification completeness is accompanied by a pleasingly high level of redshift completeness: an extensive search of the optical-infrared-millimetre spectroscopic literature revealed that 52\% of the sources in our sample have secure spectroscopic redshifts. This spectroscopic redshift information (extending to $z \simeq 7$), coupled with the extensive multi-frequency photometry provided by {\it JWST} PRIMER (and, where necessary/helpful, UltraVISTA), has enabled us to produce robust photometric redshifts for the galaxies in the remaining half of the ALMA-detected sample (checked, where possible, by the plausibility of multi-band ALMA photometry). We provide the positions and basic photometric properties of all 128 sources in Appendix \ref{sec:append_a}, Table \ref{tab:CAT1}, with colour postage-stamp images of all the galaxies presented in Appendix \ref{sec:append_b}, Figure \ref{fig:stamp1}.

Armed with this complete and robust information, we undertook a detailed exploration of the basic physical properties of the ALMA-detected galaxies. We made 3 different estimates of SFR for each source: (1) the SFR inferred from fitting the optical-infrared SED allowing for dust attenuation (using {\sc bagpipes}), (2) the `raw' UV-visible SFR, calculated from the emission at $\lambda_{\rm rest} = 1500$\,\AA, and (3) the `raw' dust-obscured SFR inferred from the (sub)mm photometry. As expected for (sub)mm detected sources, SFR measure (3) (i.e., the SFR inferred from the (sub)mm) was found to be much larger than SFR measure (2), with the raw UV-visible SFR being essentially negligible in comparison for most sources. Nonetheless, for completeness, we added SFR measures (2) and (3) to obtain a best estimate of total `raw' SFR, independent of the SED fitting. As has been shown in previous studies, we found that SFR estimate (1) works well for moderately obscured sources, but falls short of the total raw (sub)mm+UV SFR for the more highly-obscured objects. Thus, for all subsequent analysis, we adopted the total `raw' SFR (although we provide all SFR measures for all the sources in Appendix \ref{sec:append_a}, Table \ref{tab:SED1}). The star-formation rates of the ALMA-detected sources range from SFR $\simeq 1000\,{\rm M{\odot}\,yr^{-1}}$ down to SFR $\simeq 20\,{\rm M{\odot}\,yr^{-1}}$, reflecting the high dynamic range of our sample.

We used the {\sc bagpipes} SED fitting to determine the stellar masses, $M_*$ of all the ALMA-detected galaxies, as well as the masses of all other galaxies in the field imaged with {\it JWST}. Consistent with previous studies, we found that virtually all (126/128) of the ALMA-detected galaxies have high stellar masses, $M_* > 10^{10}\, {\rm M_{\odot}}$, but we are now able to show that this appears to be {\it independent of redshift}, at least out to $z \simeq 7$ (the effective redshift limit of the present study). Plotting the completely independent measures of `raw' total SFR against $M_*$ we find that, with very few exceptions, the ALMA-detected sources lie on the evolving `main sequence' (MS) of star-forming galaxies, and we are able to use our data to establish the sSFR of the ALMA-detected galaxies as a function of redshift (i.e., the redshift evolution of the MS). 

The unusually high quality of our sample (in terms of robust and complete redshift and SFR information)  has enabled us to make a robust estimate of the contribution of the ALMA-detected galaxies to cosmic star-formation rate density, $\rho_{\rm SFR}$. However, this represents only a lower limit (albeit a robust one, which can already exclude some previous estimates), to the true contribution of dust-enshrouded activity to cosmic star-formation rate density. This is because, as mentioned above, despite delivering 128 sources, the ALMA imaging is highly incomplete, with the individual pointings covering $< 20$\% of the PRIMER COSMOS area. We therefore used our knowledge of all other comparably massive galaxies in the field (yet to be observed with ALMA), to produce a new, completeness-corrected estimate of dust-enshrouded $\rho_{\rm SFR}$ from $z \simeq 2$ out to $z \simeq 7$. 

Our new relatively well-constrained estimate of the high-redshift evolution of dust-enshrouded $\rho_{\rm SFR\,IR}$ is accurate enough to now rule out several previous claims that dust-enshrouded SFR continues unabated beyond cosmic noon. Instead, it confirms the conclusions of several other recent studies that UV-visible star formation dominates $\rho_{\rm SFR}$ beyond $z \simeq 4-5$, a result that can be explained simply by the inevitable decline in the numbers of massive galaxies at high redshift (rather than requiring a change in the nature of dust, e.g., dust temperature, grain size, etc.). We note that our final results at $z > 2$ align very well by the parametric fit to the evolution of dust-enshrouded $\rho_{\rm SFR\,IR}$ derived by \citet{2016Msngr.166...48D}. Further extrapolation of this relation indicates that, while dust-enshrouded star formation inevitably continues to decline with increasing redshift, it could still have made a significant contribution at higher redshifts, amounting to most likely $\simeq 20$\% of total $\rho_{\rm SFR}$ at $z \simeq 8$, and potentially still as much as $\simeq 5$\% at $z \simeq 10$. 

Finally, we note that to improve on this study really requires the provision of deep, complete and contiguous ALMA imaging of the PRIMER COSMOS field, to remove the need for the incompleteness correction to $\rho_{\rm SFR\,IR}$ for high-mass galaxies applied here (which could in fact have led to an over-estimate) and to enable stacking to assess the contribution of galaxies with $M_* < 10^{10}\,{\rm M_{\odot}}$ (as previously undertaken in much smaller fields such as the HUDF). That said, the quality of the available {\it JWST} data means that our current 128-source sample still merits further investigation, and we will present the results of a more detailed analysis of the morphologies and SED properties of the ALMA-detected PRIMER COSMOS galaxies in a subsequent paper.

\section*{Acknowledgements}

JSD, DJM and LB acknowledge the support of the Royal Society through the award of a Royal Society University Research Professorship to JSD. DJM, CD, RJM, and JSD acknowledge the support of the UK Science and Technology Facilities Council (STFC). ACC and SDS acknowledge support from a UKRI Frontier Research Guarantee Grant (PI Carnall; grant reference EP/Y037065/1). 
RSE acknowledges financial support from the Peter and Patricia Gruber Foundation.
FC acknowledges support from a UKRI Frontier Research Guarantee Grant (PI Cullen; grant reference EP/X021025/1).
RB acknowledges support from an STFC Ernest Rutherford Fellowship [grant number ST/T003596/1].

This work is based in part on observations made with the NASA/ESA/CSA {\it James Webb Space Telescope}. The data were obtained from the Mikulski Archive for Space Telescopes at the Space Telescope Science Institute, which is operated by the Association of Universities for Research in Astronomy, Inc., under NASA contract NAS 5-03127 for JWST. This work is based in part on observations obtained with the NASA/ESA Hubble Space Telescope, retrieved from the Mikulski Archive for Space Telescopes at the Space Telescope Science Institute (STScI). STScI is operated by the Association of Universities for Research in Astronomy, Inc. under NASA contract NAS 5-26555. This work is also based in part on observations made with the Spitzer Space Telescope, which is operated by the Jet Propulsion Laboratory, California Institute of Technology under NASA contract 1407.
ALMA is a partnership of ESO (representing its member states), NSF (USA), and NINS (Japan), together with NRC (Canada), MOST and ASIAA (Taiwan), and KASI (Republic of Korea), in cooperation with the Republic of Chile. The Joint ALMA Observatory is operated by ESO, AUI/NRAO, and NAOJ. This work is based in part on VISTA observations collected at the European Southern Observatory under ESO programme ID 179.A-2005 and 198.A-2003 and on data products produced by CALET and the Cambridge Astronomy Survey Unit on behalf of the UltraVISTA consortium. 
 This work is based in part on observations obtained with MegaPrime/MegaCam, a joint project of CFHT and CEA/IRFU, at the Canada–France–Hawaii Telescope (CFHT) which is operated by the National Research Council (NRC) of Canada, the Institut National des Science de l'Univers of the Centre National de la Recherche Scientifique (CNRS) of France, and the University of Hawaii. This work is based in part on data products produced at Terapix available at the Canadian Astronomy Data Centre as part of the Canada–France–Hawaii Telescope Legacy Survey, a collaborative project of NRC and CNRS. 
Some of the NIRSpec data products utilised here to provide spectroscopic redshifts were retrieved from the Dawn JWST Archive (DJA). DJA is an initiative of the Cosmic Dawn Center (DAWN), which is funded by the Danish National Research Foundation under grant DNRF140. 
This research used the facilities of the Canadian Astronomy Data Centre operated by the National Research Council of Canada with the support of the Canadian Space Agency. 
This research has made use of NASA’s Astrophysics Data System Bibliographic Services. 

For the purpose of open access, the author has applied a Creative Commons Attribution (CC BY) licence to any Author Accepted Manuscript version arising from this submission. 

\section*{Data Availability}
All {\it JWST} and {\it HST} data products are available via the Mikulski Archive for Space Telescopes (https://mast.stsci.edu). Additional data products are available from the authors upon reasonable request.\\




\bibliographystyle{mnras}
\bibliography{Liu+2025} 

@ARTICLE{Carnall2024,
       author = {{Carnall}, A.~C. and {Cullen}, F. and {McLure}, R.~J. and {McLeod}, D.~J. and {Begley}, R. and {Donnan}, C.~T. and {Dunlop}, J.~S. and {Shapley}, A.~E. and {Rowlands}, K. and {Almaini}, O. and {Arellano-C{\'o}rdova}, K.~Z. and {Barrufet}, L. and {Cimatti}, A. and {Ellis}, R.~S. and {Grogin}, N.~A. and {Hamadouche}, M.~L. and {Illingworth}, G.~D. and {Koekemoer}, A.~M. and {Leung}, H. -H. and {Lovell}, C.~C. and {P{\'e}rez-Gonz{\'a}lez}, P.~G. and {Santini}, P. and {Stanton}, T.~M. and {Wild}, V.},
        title = "{The JWST EXCELS survey: too much, too young, too fast? Ultra-massive quiescent galaxies at 3 < z < 5}",
      journal = {\mnras},
         year = 2024,
        month = oct,
       volume = {534},
       number = {1},
        pages = {325-348},
          doi = {10.1093/mnras/stae2092},
archivePrefix = {arXiv},
       eprint = {2405.02242},
 primaryClass = {astro-ph.GA},
       adsurl = {https://ui.adsabs.harvard.edu/abs/2024MNRAS.534..325C}
}

@ARTICLE{Carnall2023,
       author = {{Carnall}, A.~C. and {McLeod}, D.~J. and {McLure}, R.~J. and {Dunlop}, J.~S. and {Begley}, R. and {Cullen}, F. and {Donnan}, C.~T. and {Hamadouche}, M.~L. and {Jewell}, S.~M. and {Jones}, E.~W. and {Pollock}, C.~L. and {Wild}, V.},
        title = "{A surprising abundance of massive quiescent galaxies at 3 < z < 5 in the first data from JWST CEERS}",
      journal = {\mnras},
         year = 2023,
        month = apr,
       volume = {520},
       number = {3},
        pages = {3974-3985},
          doi = {10.1093/mnras/stad369},
archivePrefix = {arXiv},
       eprint = {2208.00986},
 primaryClass = {astro-ph.GA},
       adsurl = {https://ui.adsabs.harvard.edu/abs/2023MNRAS.520.3974C}
}

@ARTICLE{2019ApJS..244...40L,
       author = {{Liu}, Daizhong and {Lang}, P. and {Magnelli}, B. and {Schinnerer}, E. and {Leslie}, S. and {Fudamoto}, Y. and {Bondi}, M. and {Groves}, B. and {Jim{\'e}nez-Andrade}, E. and {Harrington}, K. and {Karim}, A. and {Oesch}, P.~A. and {Sargent}, M. and {Vardoulaki}, E. and {B{\v{a}}descu}, T. and {Moser}, L. and {Bertoldi}, F. and {Battisti}, A. and {da Cunha}, E. and {Zavala}, J. and {Vaccari}, M. and {Davidzon}, I. and {Riechers}, D. and {Aravena}, M.},
        title = "{Automated Mining of the ALMA Archive in the COSMOS Field (A$^{3}$COSMOS). I. Robust ALMA Continuum Photometry Catalogs and Stellar Mass and Star Formation Properties for {\ensuremath{\sim}}700 Galaxies at z = 0.5-6}",
      journal = {\apjs},
         year = 2019,
        month = oct,
       volume = {244},
       number = {2},
          eid = {40},
        pages = {40},
          doi = {10.3847/1538-4365/ab42da},
archivePrefix = {arXiv},
       eprint = {1910.12872},
 primaryClass = {astro-ph.GA},
       adsurl = {https://ui.adsabs.harvard.edu/abs/2019ApJS..244...40L}
}

@ARTICLE{2012A&A...544A.156M,
       author = {{McCracken}, H.~J. and {Milvang-Jensen}, B. and {Dunlop}, J. and {Franx}, M. and {Fynbo}, J.~P.~U. and {Le F{\`e}vre}, O. and {Holt}, J. and {Caputi}, K.~I. and {Goranova}, Y. and {Buitrago}, F. and {Emerson}, J.~P. and {Freudling}, W. and {Hudelot}, P. and {L{\'o}pez-Sanjuan}, C. and {Magnard}, F. and {Mellier}, Y. and {M{\o}ller}, P. and {Nilsson}, K.~K. and {Sutherland}, W. and {Tasca}, L. and {Zabl}, J.},
        title = "{UltraVISTA: a new ultra-deep near-infrared survey in COSMOS}",
      journal = {\aap},
         year = 2012,
        month = aug,
       volume = {544},
          eid = {A156},
        pages = {A156},
          doi = {10.1051/0004-6361/201219507},
archivePrefix = {arXiv},
       eprint = {1204.6586},
 primaryClass = {astro-ph.CO},
       adsurl = {https://ui.adsabs.harvard.edu/abs/2012A&A...544A.156M}
}

@MISC{2021jwst.prop.1837D,
       author = {{Dunlop}, James S. and {Abraham}, Roberto G. and {Ashby}, Matthew L.~N. and {Bagley}, Micaela and {Best}, Philip N. and {Bongiorno}, Angela and {Bouwens}, Rychard and {Bowler}, Rebecca A.~A. and {Brammer}, Gabriel and {Bremer}, Malcolm and {Calabro'}, Antonello and {Carnall}, Adam and {Castellano}, Marco and {Cirasuolo}, Michele and {Conselice}, Christopher and {Cullen}, Fergus and {Dave}, Romeel and {Dayal}, Pratika and {Dekel}, Avishai and {Dickinson}, Mark and {Duncan}, Kenneth James and {Elbaz}, David and {Ellis}, Richard S. and {Ferguson}, Harry C. and {Ferrara}, Andrea and {Finkelstein}, Steven L. and {Fontana}, Adriano and {Furlanetto}, Steven and {Fynbo}, Johan P.~U. and {Gallerani}, Simona and {Gardner}, Jonathan P. and {Giavalisco}, Mauro and {Grazian}, Andrea and {Grogin}, Norman and {Harikane}, Yuichi and {Hopkins}, Philip F. and {Ilbert}, Olivier and {Illingworth}, Garth D. and {Juneau}, Stephanie and {Jung}, Intae and {Kartaltepe}, Jeyhan and {Kassin}, Susan and {Kauffmann}, Olivier Benjamin and {Khochfar}, Sadegh and {Kirkpatrick}, Allison and {Kocevski}, Dale D. and {Koekemoer}, Anton M. and {Labbe}, Ivo and {Laporte}, Nicolas and {Larson}, Rebecca L. and {Lucas}, Ray A. and {Magee}, Daniel K. and {Mason}, Charlotte and {McCracken}, Henry Joy and {McLeod}, Derek and {McLure}, Ross and {Merlin}, Emiliano and {Mesinger}, Andrei and {Milvang-Jensen}, Bo and {Newman}, Jeffrey Allen and {Oesch}, Pascal and {Ouchi}, Masami and {Pacifici}, Camilla and {Papovich}, Casey and {Peacock}, John and {Peeples}, Molly and {Pentericci}, Laura and {Perez-Gonzalez}, Pablo G. and {Pirzkal}, Norbert and {Pope}, Alexandra and {Pye}, John P. and {Reddy}, Naveen A. and {Robertson}, Brant and {Salvato}, Mara and {Santini}, Paola and {Schaerer}, Daniel and {Shapley}, Alice E. and {Simons}, Raymond and {Smit}, Renske and {Smith}, Britton D. and {Snyder}, Greg and {Somerville}, Rachel S. and {Stanway}, Elizabeth R. and {Stefanon}, Mauro and {Tasca}, Lidia and {Tikkanen}, Tuomo and {Tresse}, Laurence and {Trump}, Jonathan R. and {Whitaker}, Katherine E. and {Wilkins}, Stephen Matthew and {Wright}, Gillian and {Wyithe}, J. Stuart B. and {van Dokkum}, Pieter and {van der Werf}, Paul},
        title = "{PRIMER: Public Release IMaging for Extragalactic Research}",
 howpublished = {JWST Proposal. Cycle 1, ID. \#1837},
         year = 2021,
        month = mar,
        pages = {1837},
       adsurl = {https://ui.adsabs.harvard.edu/abs/2021jwst.prop.1837D}
}

@ARTICLE{2019ApJ...880...43S,
       author = {{Simpson}, J.~M. and {Smail}, Ian and {Swinbank}, A.~M. and {Chapman}, S.~C. and {Chen}, Chian-Chou and {Geach}, J.~E. and {Matsuda}, Y. and {Wang}, R. and {Wang}, Wei-Hao and {Yang}, Y. and {Ao}, Y. and {Asquith}, R. and {Bourne}, N. and {Coogan}, R.~T. and {Coppin}, K. and {Gullberg}, B. and {Hine}, N.~K. and {Ho}, L.~C. and {Hwang}, H.~S. and {Ivison}, R.~J. and {Kato}, Y. and {Lacaille}, K. and {Lewis}, A.~J.~R. and {Liu}, D. and {Micha{\l}owski}, M.~J. and {Oteo}, I. and {Sawicki}, M. and {Scholtz}, J. and {Smith}, D. and {Thomson}, A.~P. and {Wardlow}, J.~L.},
        title = "{The East Asian Observatory SCUBA-2 Survey of the COSMOS Field: Unveiling 1147 Bright Sub-millimeter Sources across 2.6 Square Degrees}",
      journal = {\apj},
         year = 2019,
        month = jul,
       volume = {880},
       number = {1},
          eid = {43},
        pages = {43},
          doi = {10.3847/1538-4357/ab23ff},
archivePrefix = {arXiv},
       eprint = {1912.02229},
 primaryClass = {astro-ph.GA},
       adsurl = {https://ui.adsabs.harvard.edu/abs/2019ApJ...880...43S}
}

@ARTICLE{2024arXiv240814546L,
       author = {{Long}, Arianna S. and {Casey}, Caitlin M. and {McKinney}, Jed and {Zavala}, Jorge A. and {Akins}, Hollis B. and {Cooper}, Olivia R. and {Lambrides}, Matthieu Bethermin Erini L. and {Franco}, Maximilien and {Caputi}, Karina and {Champagne}, Jaclyn B. and {Man}, Allison W.~S. and {Treister}, Ezequiel and {Manning}, Sinclaire M. and {Sanders}, David B. and {Talia}, Margherita and {Aravena}, Manuel and {Clements}, D.~L. and {da Cunha}, Elisabete and {Faisst}, Andreas L. and {Gentile}, Fabrizio and {Hodge}, Jacqueline and {Brammer}, Gabriel and {Brusa}, Marcella and {Finkelstein}, Steven L. and {Fujimoto}, Seiji and {Hayward}, Christopher C. and {Ilbert}, Olivier and {Jolly}, Jean-Baptiste and {Kartaltepe}, Jeyhan S. and {Knudsen}, Kirsten and {Koekemoer}, Anton M. and {Liu}, Daizhong and {Magdis}, Georgios and {McCracken}, Henry Joy and {Rhodes}, Jason and {Robertson}, Brant E. and {Scoville}, Nick and {Sheth}, Kartik and {Smolcic}, Vernesa and {Spilker}, Justin and {Taniguchi}, Yoshiaki and {Toft}, Sune and {Urry}, C. Megan and {Yun}, Min},
        title = "{The Extended Mapping Obscuration to Reionization with ALMA (Ex-MORA) Survey: 5$\sigma$ Source Catalog and Redshift Distribution}",
      journal = {arXiv e-prints},
         year = 2024,
        month = aug,
          eid = {arXiv:2408.14546},
        pages = {arXiv:2408.14546},
          doi = {10.48550/arXiv.2408.14546},
archivePrefix = {arXiv},
       eprint = {2408.14546},
 primaryClass = {astro-ph.GA},
       adsurl = {https://ui.adsabs.harvard.edu/abs/2024arXiv240814546L}
}

@software{2011ascl.soft08009A,
       author = {{Arnouts}, S. and {Ilbert}, O.},
        title = "{LePHARE: Photometric Analysis for Redshift Estimate}",
 howpublished = {Astrophysics Source Code Library, record ascl:1108.009},
         year = 2011,
        month = aug,
          eid = {ascl:1108.009},
       adsurl = {https://ui.adsabs.harvard.edu/abs/2011ascl.soft08009A}
}

@ARTICLE{2008ApJ...686.1503B,
       author = {{Brammer}, Gabriel B. and {van Dokkum}, Pieter G. and {Coppi}, Paolo},
        title = "{EAZY: A Fast, Public Photometric Redshift Code}",
      journal = {\apj},
         year = 2008,
        month = oct,
       volume = {686},
       number = {2},
        pages = {1503-1513},
          doi = {10.1086/591786},
archivePrefix = {arXiv},
       eprint = {0807.1533},
 primaryClass = {astro-ph},
       adsurl = {https://ui.adsabs.harvard.edu/abs/2008ApJ...686.1503B}
}

@ARTICLE{2007ApJS..172....1S,
       author = {{Scoville}, N. and {Aussel}, H. and {Brusa}, M. and {Capak}, P. and {Carollo}, C.~M. and {Elvis}, M. and {Giavalisco}, M. and {Guzzo}, L. and {Hasinger}, G. and {Impey}, C. and {Kneib}, J. -P. and {LeFevre}, O. and {Lilly}, S.~J. and {Mobasher}, B. and {Renzini}, A. and {Rich}, R.~M. and {Sanders}, D.~B. and {Schinnerer}, E. and {Schminovich}, D. and {Shopbell}, P. and {Taniguchi}, Y. and {Tyson}, N.~D.},
        title = "{The Cosmic Evolution Survey (COSMOS): Overview}",
      journal = {\apjs},
         year = 2007,
        month = sep,
       volume = {172},
       number = {1},
        pages = {1-8},
          doi = {10.1086/516585},
archivePrefix = {arXiv},
       eprint = {astro-ph/0612305},
 primaryClass = {astro-ph},
       adsurl = {https://ui.adsabs.harvard.edu/abs/2007ApJS..172....1S}
}

@ARTICLE{2007MNRAS.379.1599L,
       author = {{Lawrence}, A. and {Warren}, S.~J. and {Almaini}, O. and {Edge}, A.~C. and {Hambly}, N.~C. and {Jameson}, R.~F. and {Lucas}, P. and {Casali}, M. and {Adamson}, A. and {Dye}, S. and {Emerson}, J.~P. and {Foucaud}, S. and {Hewett}, P. and {Hirst}, P. and {Hodgkin}, S.~T. and {Irwin}, M.~J. and {Lodieu}, N. and {McMahon}, R.~G. and {Simpson}, C. and {Smail}, I. and {Mortlock}, D. and {Folger}, M.},
        title = "{The UKIRT Infrared Deep Sky Survey (UKIDSS)}",
      journal = {\mnras},
         year = 2007,
        month = aug,
       volume = {379},
       number = {4},
        pages = {1599-1617},
          doi = {10.1111/j.1365-2966.2007.12040.x},
archivePrefix = {arXiv},
       eprint = {astro-ph/0604426},
 primaryClass = {astro-ph},
       adsurl = {https://ui.adsabs.harvard.edu/abs/2007MNRAS.379.1599L}
}

@ARTICLE{2011ApJS..197...35G,
       author = {{Grogin}, Norman A. and {Kocevski}, Dale D. and {Faber}, S.~M. and {Ferguson}, Henry C. and {Koekemoer}, Anton M. and {Riess}, Adam G. and {Acquaviva}, Viviana and {Alexander}, David M. and {Almaini}, Omar and {Ashby}, Matthew L.~N. and {Barden}, Marco and {Bell}, Eric F. and {Bournaud}, Fr{\'e}d{\'e}ric and {Brown}, Thomas M. and {Caputi}, Karina I. and {Casertano}, Stefano and {Cassata}, Paolo and {Castellano}, Marco and {Challis}, Peter and {Chary}, Ranga-Ram and {Cheung}, Edmond and {Cirasuolo}, Michele and {Conselice}, Christopher J. and {Roshan Cooray}, Asantha and {Croton}, Darren J. and {Daddi}, Emanuele and {Dahlen}, Tomas and {Dav{\'e}}, Romeel and {de Mello}, Du{\'\i}lia F. and {Dekel}, Avishai and {Dickinson}, Mark and {Dolch}, Timothy and {Donley}, Jennifer L. and {Dunlop}, James S. and {Dutton}, Aaron A. and {Elbaz}, David and {Fazio}, Giovanni G. and {Filippenko}, Alexei V. and {Finkelstein}, Steven L. and {Fontana}, Adriano and {Gardner}, Jonathan P. and {Garnavich}, Peter M. and {Gawiser}, Eric and {Giavalisco}, Mauro and {Grazian}, Andrea and {Guo}, Yicheng and {Hathi}, Nimish P. and {H{\"a}ussler}, Boris and {Hopkins}, Philip F. and {Huang}, Jia-Sheng and {Huang}, Kuang-Han and {Jha}, Saurabh W. and {Kartaltepe}, Jeyhan S. and {Kirshner}, Robert P. and {Koo}, David C. and {Lai}, Kamson and {Lee}, Kyoung-Soo and {Li}, Weidong and {Lotz}, Jennifer M. and {Lucas}, Ray A. and {Madau}, Piero and {McCarthy}, Patrick J. and {McGrath}, Elizabeth J. and {McIntosh}, Daniel H. and {McLure}, Ross J. and {Mobasher}, Bahram and {Moustakas}, Leonidas A. and {Mozena}, Mark and {Nandra}, Kirpal and {Newman}, Jeffrey A. and {Niemi}, Sami-Matias and {Noeske}, Kai G. and {Papovich}, Casey J. and {Pentericci}, Laura and {Pope}, Alexandra and {Primack}, Joel R. and {Rajan}, Abhijith and {Ravindranath}, Swara and {Reddy}, Naveen A. and {Renzini}, Alvio and {Rix}, Hans-Walter and {Robaina}, Aday R. and {Rodney}, Steven A. and {Rosario}, David J. and {Rosati}, Piero and {Salimbeni}, Sara and {Scarlata}, Claudia and {Siana}, Brian and {Simard}, Luc and {Smidt}, Joseph and {Somerville}, Rachel S. and {Spinrad}, Hyron and {Straughn}, Amber N. and {Strolger}, Louis-Gregory and {Telford}, Olivia and {Teplitz}, Harry I. and {Trump}, Jonathan R. and {van der Wel}, Arjen and {Villforth}, Carolin and {Wechsler}, Risa H. and {Weiner}, Benjamin J. and {Wiklind}, Tommy and {Wild}, Vivienne and {Wilson}, Grant and {Wuyts}, Stijn and {Yan}, Hao-Jing and {Yun}, Min S.},
        title = "{CANDELS: The Cosmic Assembly Near-infrared Deep Extragalactic Legacy Survey}",
      journal = {\apjs},
         year = 2011,
        month = dec,
       volume = {197},
       number = {2},
          eid = {35},
        pages = {35},
          doi = {10.1088/0067-0049/197/2/35},
archivePrefix = {arXiv},
       eprint = {1105.3753},
 primaryClass = {astro-ph.CO},
       adsurl = {https://ui.adsabs.harvard.edu/abs/2011ApJS..197...35G}
}

@ARTICLE{2021ApJ...909..165Z,
       author = {{Zavala}, J.~A. and {Casey}, C.~M. and {Manning}, S.~M. and {Aravena}, M. and {Bethermin}, M. and {Caputi}, K.~I. and {Clements}, D.~L. and {Cunha}, E. da and {Drew}, P. and {Finkelstein}, S.~L. and {Fujimoto}, S. and {Hayward}, C. and {Hodge}, J. and {Kartaltepe}, J.~S. and {Knudsen}, K. and {Koekemoer}, A.~M. and {Long}, A.~S. and {Magdis}, G.~E. and {Man}, A.~W.~S. and {Popping}, G. and {Sanders}, D. and {Scoville}, N. and {Sheth}, K. and {Staguhn}, J. and {Toft}, S. and {Treister}, E. and {Vieira}, J.~D. and {Yun}, M.~S.},
        title = "{The Evolution of the IR Luminosity Function and Dust-obscured Star Formation over the Past 13 Billion Years}",
      journal = {\apj},
         year = 2021,
        month = mar,
       volume = {909},
       number = {2},
          eid = {165},
        pages = {165},
          doi = {10.3847/1538-4357/abdb27},
archivePrefix = {arXiv},
       eprint = {2101.04734},
 primaryClass = {astro-ph.GA},
       adsurl = {https://ui.adsabs.harvard.edu/abs/2021ApJ...909..165Z}
}

@ARTICLE{2021ApJ...923..215C,
       author = {{Casey}, Caitlin M. and {Zavala}, Jorge A. and {Manning}, Sinclaire M. and {Aravena}, Manuel and {B{\'e}thermin}, Matthieu and {Caputi}, Karina I. and {Champagne}, Jaclyn B. and {Clements}, David L. and {Drew}, Patrick and {Finkelstein}, Steven L. and {Fujimoto}, Seiji and {Hayward}, Christopher C. and {Dekel}, Anton M. and {Kokorev}, Vasily and {Lagos}, Claudia del P. and {Long}, Arianna S. and {Magdis}, Georgios E. and {Man}, Allison W.~S. and {Mitsuhashi}, Ikki and {Popping}, Gerg{\"o} and {Spilker}, Justin and {Staguhn}, Johannes and {Talia}, Margherita and {Toft}, Sune and {Treister}, Ezequiel and {Weaver}, John R. and {Yun}, Min},
        title = "{Mapping Obscuration to Reionization with ALMA (MORA): 2 mm Efficiently Selects the Highest-redshift Obscured Galaxies}",
      journal = {\apj},
         year = 2021,
        month = dec,
       volume = {923},
       number = {2},
          eid = {215},
        pages = {215},
          doi = {10.3847/1538-4357/ac2eb4},
archivePrefix = {arXiv},
       eprint = {2110.06930},
 primaryClass = {astro-ph.GA},
       adsurl = {https://ui.adsabs.harvard.edu/abs/2021ApJ...923..215C}
}

@dataset{2012yCat.2317....0H,
       author = {{Hudelot}, P. and {Cuillandre}, J. -Ch. and {Withington}, K. and {Goranova}, Y. and {McCracken}, H. and {Magnard}, F. and {Mellier}, Y. and {Regnault}, N. and {Betoule}, M. and {Aussel}, H. and {Kavelaars}, J.~J. and {Fernique}, P. and {Bonnarel}, F. and {Ochsenbein}, F. and {Ilbert}, O.},
        title = "{VizieR Online Data Catalog: The CFHTLS Survey (T0007 release) (Hudelot+ 2012)}",
 howpublished = {VizieR On-line Data Catalog: II/317.  Originally published in: SPIE Conf. 2012},
         year = 2012,
        month = sep,
          eid = {II/317},
       adsurl = {https://ui.adsabs.harvard.edu/abs/2012yCat.2317....0H}
}

@ARTICLE{2016ApJ...822...46F,
       author = {{Furusawa}, Hisanori and {Kashikawa}, Nobunari and {Kobayashi}, Masakazu A.~R. and {Dunlop}, James S. and {Shimasaku}, Kazuhiro and {Takata}, Tadafumi and {Sekiguchi}, Kazuhiro and {Naito}, Yoshiaki and {Furusawa}, Junko and {Ouchi}, Masami and {Nakata}, Fumiaki and {Yasuda}, Naoki and {Okura}, Yuki and {Taniguchi}, Yoshiaki and {Yamada}, Toru and {Kajisawa}, Masaru and {Fynbo}, Johan P.~U. and {Le F{\`e}vre}, Olivier},
        title = "{A New Constraint on the Ly{\ensuremath{\alpha}} Fraction of UV Very Bright Galaxies at Redshift 7}",
      journal = {\apj},
         year = 2016,
        month = may,
       volume = {822},
       number = {1},
          eid = {46},
        pages = {46},
          doi = {10.3847/0004-637X/822/1/46},
archivePrefix = {arXiv},
       eprint = {1604.02214},
 primaryClass = {astro-ph.GA},
       adsurl = {https://ui.adsabs.harvard.edu/abs/2016ApJ...822...46F}
}

@ARTICLE{2021MNRAS.503.4413M,
       author = {{McLeod}, D.~J. and {McLure}, R.~J. and {Dunlop}, J.~S. and {Cullen}, F. and {Carnall}, A.~C. and {Duncan}, K.},
        title = "{The evolution of the galaxy stellar-mass function over the last 12 billion years from a combination of ground-based and HST surveys}",
      journal = {\mnras},
         year = 2021,
        month = may,
       volume = {503},
       number = {3},
        pages = {4413-4435},
          doi = {10.1093/mnras/stab731},
archivePrefix = {arXiv},
       eprint = {2009.03176},
 primaryClass = {astro-ph.GA},
       adsurl = {https://ui.adsabs.harvard.edu/abs/2021MNRAS.503.4413M}
}

@ARTICLE{2014MNRAS.440.2810B,
       author = {{Bowler}, R.~A.~A. and {Dunlop}, J.~S. and {McLure}, R.~J. and {Rogers}, A.~B. and {McCracken}, H.~J. and {Milvang-Jensen}, B. and {Furusawa}, H. and {Fynbo}, J.~P.~U. and {Taniguchi}, Y. and {Afonso}, J. and {Bremer}, M.~N. and {Le F{\`e}vre}, O.},
        title = "{The bright end of the galaxy luminosity function at z≃7: before the onset of mass quenching?}",
      journal = {\mnras},
         year = 2014,
        month = may,
       volume = {440},
       number = {3},
        pages = {2810-2842},
          doi = {10.1093/mnras/stu449},
archivePrefix = {arXiv},
       eprint = {1312.5643},
 primaryClass = {astro-ph.CO},
       adsurl = {https://ui.adsabs.harvard.edu/abs/2014MNRAS.440.2810B}
}

@ARTICLE{2020A&A...641A...6P,
       author = {{Planck Collaboration} and {Aghanim}, N. and {Akrami}, Y. and {Ashdown}, M. and {Aumont}, J. and {Baccigalupi}, C. and {Ballardini}, M. and {Banday}, A.~J. and {Barreiro}, R.~B. and {Bartolo}, N. and {Basak}, S. and {Battye}, R. and {Benabed}, K. and {Bernard}, J. -P. and {Bersanelli}, M. and {Bielewicz}, P. and {Bock}, J.~J. and {Bond}, J.~R. and {Borrill}, J. and {Bouchet}, F.~R. and {Boulanger}, F. and {Bucher}, M. and {Burigana}, C. and {Butler}, R.~C. and {Calabrese}, E. and {Cardoso}, J. -F. and {Carron}, J. and {Challinor}, A. and {Chiang}, H.~C. and {Chluba}, J. and {Colombo}, L.~P.~L. and {Combet}, C. and {Contreras}, D. and {Crill}, B.~P. and {Cuttaia}, F. and {de Bernardis}, P. and {de Zotti}, G. and {Delabrouille}, J. and {Delouis}, J. -M. and {Di Valentino}, E. and {Diego}, J.~M. and {Dor{\'e}}, O. and {Douspis}, M. and {Ducout}, A. and {Dupac}, X. and {Dusini}, S. and {Efstathiou}, G. and {Elsner}, F. and {En{\ss}lin}, T.~A. and {Eriksen}, H.~K. and {Fantaye}, Y. and {Farhang}, M. and {Fergusson}, J. and {Fernandez-Cobos}, R. and {Finelli}, F. and {Forastieri}, F. and {Frailis}, M. and {Fraisse}, A.~A. and {Franceschi}, E. and {Frolov}, A. and {Galeotta}, S. and {Galli}, S. and {Ganga}, K. and {G{\'e}nova-Santos}, R.~T. and {Gerbino}, M. and {Ghosh}, T. and {Gonz{\'a}lez-Nuevo}, J. and {G{\'o}rski}, K.~M. and {Gratton}, S. and {Gruppuso}, A. and {Gudmundsson}, J.~E. and {Hamann}, J. and {Handley}, W. and {Hansen}, F.~K. and {Herranz}, D. and {Hildebrandt}, S.~R. and {Hivon}, E. and {Huang}, Z. and {Jaffe}, A.~H. and {Jones}, W.~C. and {Karakci}, A. and {Keih{\"a}nen}, E. and {Keskitalo}, R. and {Kiiveri}, K. and {Kim}, J. and {Kisner}, T.~S. and {Knox}, L. and {Krachmalnicoff}, N. and {Kunz}, M. and {Kurki-Suonio}, H. and {Lagache}, G. and {Lamarre}, J. -M. and {Lasenby}, A. and {Lattanzi}, M. and {Lawrence}, C.~R. and {Le Jeune}, M. and {Lemos}, P. and {Lesgourgues}, J. and {Levrier}, F. and {Lewis}, A. and {Liguori}, M. and {Lilje}, P.~B. and {Lilley}, M. and {Lindholm}, V. and {L{\'o}pez-Caniego}, M. and {Lubin}, P.~M. and {Ma}, Y. -Z. and {Mac{\'\i}as-P{\'e}rez}, J.~F. and {Maggio}, G. and {Maino}, D. and {Mandolesi}, N. and {Mangilli}, A. and {Marcos-Caballero}, A. and {Maris}, M. and {Martin}, P.~G. and {Martinelli}, M. and {Mart{\'\i}nez-Gonz{\'a}lez}, E. and {Matarrese}, S. and {Mauri}, N. and {McEwen}, J.~D. and {Meinhold}, P.~R. and {Melchiorri}, A. and {Mennella}, A. and {Migliaccio}, M. and {Millea}, M. and {Mitra}, S. and {Miville-Desch{\^e}nes}, M. -A. and {Molinari}, D. and {Montier}, L. and {Morgante}, G. and {Moss}, A. and {Natoli}, P. and {N{\o}rgaard-Nielsen}, H.~U. and {Pagano}, L. and {Paoletti}, D. and {Partridge}, B. and {Patanchon}, G. and {Peiris}, H.~V. and {Perrotta}, F. and {Pettorino}, V. and {Piacentini}, F. and {Polastri}, L. and {Polenta}, G. and {Puget}, J. -L. and {Rachen}, J.~P. and {Reinecke}, M. and {Remazeilles}, M. and {Renzi}, A. and {Rocha}, G. and {Rosset}, C. and {Roudier}, G. and {Rubi{\~n}o-Mart{\'\i}n}, J.~A. and {Ruiz-Granados}, B. and {Salvati}, L. and {Sandri}, M. and {Savelainen}, M. and {Scott}, D. and {Shellard}, E.~P.~S. and {Sirignano}, C. and {Sirri}, G. and {Spencer}, L.~D. and {Sunyaev}, R. and {Suur-Uski}, A. -S. and {Tauber}, J.~A. and {Tavagnacco}, D. and {Tenti}, M. and {Toffolatti}, L. and {Tomasi}, M. and {Trombetti}, T. and {Valenziano}, L. and {Valiviita}, J. and {Van Tent}, B. and {Vibert}, L. and {Vielva}, P. and {Villa}, F. and {Vittorio}, N. and {Wandelt}, B.~D. and {Wehus}, I.~K. and {White}, M. and {White}, S.~D.~M. and {Zacchei}, A. and {Zonca}, A.},
        title = "{Planck 2018 results. VI. Cosmological parameters}",
      journal = {\aap},
         year = 2020,
        month = sep,
       volume = {641},
          eid = {A6},
        pages = {A6},
          doi = {10.1051/0004-6361/201833910},
archivePrefix = {arXiv},
       eprint = {1807.06209},
 primaryClass = {astro-ph.CO},
       adsurl = {https://ui.adsabs.harvard.edu/abs/2020A&A...641A...6P}
}

@ARTICLE{2018A&A...611A..22S,
       author = {{Schreiber}, C. and {Labb{\'e}}, I. and {Glazebrook}, K. and {Bekiaris}, G. and {Papovich}, C. and {Costa}, T. and {Elbaz}, D. and {Kacprzak}, G.~G. and {Nanayakkara}, T. and {Oesch}, P. and {Pannella}, M. and {Spitler}, L. and {Straatman}, C. and {Tran}, K. -V. and {Wang}, T.},
        title = "{Jekyll \& Hyde: quiescence and extreme obscuration in a pair of massive galaxies 1.5 Gyr after the Big Bang}",
      journal = {\aap},
         year = 2018,
        month = mar,
       volume = {611},
          eid = {A22},
        pages = {A22},
          doi = {10.1051/0004-6361/201731917},
archivePrefix = {arXiv},
       eprint = {1709.03505},
 primaryClass = {astro-ph.GA},
       adsurl = {https://ui.adsabs.harvard.edu/abs/2018A&A...611A..22S}
}

@ARTICLE{2023ApJ...951..111C,
       author = {{Cataldi}, Gianni and {Aikawa}, Yuri and {Iwasaki}, Kazunari and {Marino}, Sebastian and {Brandeker}, Alexis and {Hales}, Antonio and {Henning}, Thomas and {Higuchi}, Aya E. and {Hughes}, A. Meredith and {Janson}, Markus and {Kral}, Quentin and {Matr{\`a}}, Luca and {Mo{\'o}r}, Attila and {Olofsson}, G{\"o}ran and {Redfield}, Seth and {Roberge}, Aki},
        title = "{Primordial or Secondary? Testing Models of Debris Disk Gas with ALMA}",
      journal = {\apj},
         year = 2023,
        month = jul,
       volume = {951},
       number = {2},
          eid = {111},
        pages = {111},
          doi = {10.3847/1538-4357/acd6f3},
archivePrefix = {arXiv},
       eprint = {2305.12093},
 primaryClass = {astro-ph.EP},
       adsurl = {https://ui.adsabs.harvard.edu/abs/2023ApJ...951..111C}
}

@ARTICLE{2018MNRAS.480.4379C,
       author = {{Carnall}, A.~C. and {McLure}, R.~J. and {Dunlop}, J.~S. and {Dav{\'e}}, R.},
        title = "{Inferring the star formation histories of massive quiescent galaxies with BAGPIPES: evidence for multiple quenching mechanisms}",
      journal = {\mnras},
         year = 2018,
        month = nov,
       volume = {480},
       number = {4},
        pages = {4379-4401},
          doi = {10.1093/mnras/sty2169},
archivePrefix = {arXiv},
       eprint = {1712.04452},
 primaryClass = {astro-ph.GA},
       adsurl = {https://ui.adsabs.harvard.edu/abs/2018MNRAS.480.4379C}
}

@ARTICLE{2024A&A...690L..16J,
       author = {{Jin}, Shuowen and {Sillassen}, Nikolaj B. and {Hodge}, Jacqueline and {Magdis}, Georgios E. and {Rizzo}, Francesca and {Casey}, Caitlin and {Koekemoer}, Anton M. and {Valentino}, Francesco and {Kokorev}, Vasily and {Magnelli}, Benjamin and {Gobat}, Raphael and {Gillman}, Steven and {Franco}, Maximilien and {Faisst}, Andreas and {Kartaltepe}, Jeyhan and {Schinnerer}, Eva and {Toft}, Sune and {Algera}, Hiddo S.~B. and {Harish}, Santosh and {Lee}, Minju and {Liu}, Daizhong and {Shuntov}, Marko and {Talia}, Margherita and {Vijayan}, Aswin},
        title = "{A photo-z cautionary tale: Redshift confirmation of COSBO-7 at z = 2.625}",
      journal = {\aap},
         year = 2024,
        month = oct,
       volume = {690},
          eid = {L16},
        pages = {L16},
          doi = {10.1051/0004-6361/202451445},
archivePrefix = {arXiv},
       eprint = {2407.07585},
 primaryClass = {astro-ph.GA},
       adsurl = {https://ui.adsabs.harvard.edu/abs/2024A&A...690L..16J}
}

@ARTICLE{2014ApJS..214...15S,
       author = {{Speagle}, J.~S. and {Steinhardt}, C.~L. and {Capak}, P.~L. and {Silverman}, J.~D.},
        title = "{A Highly Consistent Framework for the Evolution of the Star-Forming ``Main Sequence'' from z \raisebox{-0.5ex}\textasciitilde 0-6}",
      journal = {\apjs},
         year = 2014,
        month = oct,
       volume = {214},
       number = {2},
          eid = {15},
        pages = {15},
          doi = {10.1088/0067-0049/214/2/15},
archivePrefix = {arXiv},
       eprint = {1405.2041},
 primaryClass = {astro-ph.GA},
       adsurl = {https://ui.adsabs.harvard.edu/abs/2014ApJS..214...15S}
}

@ARTICLE{2015ApJS..220...12S,
       author = {{Silverman}, J.~D. and {Kashino}, D. and {Sanders}, D. and {Kartaltepe}, J.~S. and {Arimoto}, N. and {Renzini}, A. and {Rodighiero}, G. and {Daddi}, E. and {Zahid}, J. and {Nagao}, T. and {Kewley}, L.~J. and {Lilly}, S.~J. and {Sugiyama}, N. and {Baronchelli}, I. and {Capak}, P. and {Carollo}, C.~M. and {Chu}, J. and {Hasinger}, G. and {Ilbert}, O. and {Juneau}, S. and {Kajisawa}, M. and {Koekemoer}, A.~M. and {Kovac}, K. and {Le F{\`e}vre}, O. and {Masters}, D. and {McCracken}, H.~J. and {Onodera}, M. and {Schulze}, A. and {Scoville}, N. and {Strazzullo}, V. and {Taniguchi}, Y.},
        title = "{The FMOS-COSMOS Survey of Star-forming Galaxies at z\raisebox{-0.5ex}\textasciitilde1.6. III. Survey Design, Performance, and Sample Characteristics}",
      journal = {\apjs},
         year = 2015,
        month = sep,
       volume = {220},
       number = {1},
          eid = {12},
        pages = {12},
          doi = {10.1088/0067-0049/220/1/12},
archivePrefix = {arXiv},
       eprint = {1409.0447},
 primaryClass = {astro-ph.GA},
       adsurl = {https://ui.adsabs.harvard.edu/abs/2015ApJS..220...12S}
}

@ARTICLE{2018A&A...618A..85S,
       author = {{Schreiber}, C. and {Glazebrook}, K. and {Nanayakkara}, T. and {Kacprzak}, G.~G. and {Labb{\'e}}, I. and {Oesch}, P. and {Yuan}, T. and {Tran}, K. -V. and {Papovich}, C. and {Spitler}, L. and {Straatman}, C.},
        title = "{Near infrared spectroscopy and star-formation histories of 3 {\ensuremath{\leq}} z {\ensuremath{\leq}} 4 quiescent galaxies}",
      journal = {\aap},
         year = 2018,
        month = oct,
       volume = {618},
          eid = {A85},
        pages = {A85},
          doi = {10.1051/0004-6361/201833070},
archivePrefix = {arXiv},
       eprint = {1807.02523},
 primaryClass = {astro-ph.GA},
       adsurl = {https://ui.adsabs.harvard.edu/abs/2018A&A...618A..85S}
}

@ARTICLE{2021A&A...646A..96C,
       author = {{Circosta}, C. and {Mainieri}, V. and {Lamperti}, I. and {Padovani}, P. and {Bischetti}, M. and {Harrison}, C.~M. and {Kakkad}, D. and {Zanella}, A. and {Vietri}, G. and {Lanzuisi}, G. and {Salvato}, M. and {Brusa}, M. and {Carniani}, S. and {Cicone}, C. and {Cresci}, G. and {Feruglio}, C. and {Husemann}, B. and {Mannucci}, F. and {Marconi}, A. and {Perna}, M. and {Piconcelli}, E. and {Puglisi}, A. and {Saintonge}, A. and {Schramm}, M. and {Vignali}, C. and {Zappacosta}, L.},
        title = "{SUPER. IV. CO(J = 3-2) properties of active galactic nucleus hosts at cosmic noon revealed by ALMA}",
      journal = {\aap},
         year = 2021,
        month = feb,
       volume = {646},
          eid = {A96},
        pages = {A96},
          doi = {10.1051/0004-6361/202039270},
archivePrefix = {arXiv},
       eprint = {2012.07965},
 primaryClass = {astro-ph.GA},
       adsurl = {https://ui.adsabs.harvard.edu/abs/2021A&A...646A..96C}
}

@ARTICLE{2023A&A...679A.129R,
       author = {{Rizzo}, F. and {Roman-Oliveira}, F. and {Fraternali}, F. and {Frickmann}, D. and {Valentino}, F.~M. and {Brammer}, G. and {Zanella}, A. and {Kokorev}, V. and {Popping}, G. and {Whitaker}, K.~E. and {Kohandel}, M. and {Magdis}, G.~E. and {Di Mascolo}, L. and {Ikeda}, R. and {Jin}, S. and {Toft}, S.},
        title = "{The ALMA-ALPAKA survey. I. High-resolution CO and [CI] kinematics of star-forming galaxies at z = 0.5-3.5}",
      journal = {\aap},
         year = 2023,
        month = nov,
       volume = {679},
          eid = {A129},
        pages = {A129},
          doi = {10.1051/0004-6361/202346444},
archivePrefix = {arXiv},
       eprint = {2303.16227},
 primaryClass = {astro-ph.GA},
       adsurl = {https://ui.adsabs.harvard.edu/abs/2023A&A...679A.129R}
}

@ARTICLE{2017MNRAS.466..861D,
       author = {{Dunlop}, J.~S. and {McLure}, R.~J. and {Biggs}, A.~D. and {Geach}, J.~E. and {Micha{\l}owski}, M.~J. and {Ivison}, R.~J. and {Rujopakarn}, W. and {van Kampen}, E. and {Kirkpatrick}, A. and {Pope}, A. and {Scott}, D. and {Swinbank}, A.~M. and {Targett}, T.~A. and {Aretxaga}, I. and {Austermann}, J.~E. and {Best}, P.~N. and {Bruce}, V.~A. and {Chapin}, E.~L. and {Charlot}, S. and {Cirasuolo}, M. and {Coppin}, K. and {Ellis}, R.~S. and {Finkelstein}, S.~L. and {Hayward}, C.~C. and {Hughes}, D.~H. and {Ibar}, E. and {Jagannathan}, P. and {Khochfar}, S. and {Koprowski}, M.~P. and {Narayanan}, D. and {Nyland}, K. and {Papovich}, C. and {Peacock}, J.~A. and {Rieke}, G.~H. and {Robertson}, B. and {Vernstrom}, T. and {Werf}, P.~P. van der and {Wilson}, G.~W. and {Yun}, M.},
        title = "{A deep ALMA image of the Hubble Ultra Deep Field}",
      journal = {\mnras},
         year = 2017,
        month = apr,
       volume = {466},
       number = {1},
        pages = {861-883},
          doi = {10.1093/mnras/stw3088},
archivePrefix = {arXiv},
       eprint = {1606.00227},
 primaryClass = {astro-ph.GA},
       adsurl = {https://ui.adsabs.harvard.edu/abs/2017MNRAS.466..861D}
}

@ARTICLE{2003PASP..115..763C,
       author = {{Chabrier}, Gilles},
        title = "{Galactic Stellar and Substellar Initial Mass Function}",
      journal = {\pasp},
         year = 2003,
        month = jul,
       volume = {115},
       number = {809},
        pages = {763-795},
          doi = {10.1086/376392},
archivePrefix = {arXiv},
       eprint = {astro-ph/0304382},
 primaryClass = {astro-ph},
       adsurl = {https://ui.adsabs.harvard.edu/abs/2003PASP..115..763C}
}

@ARTICLE{1998ARA&A..36..189K,
       author = {{Kennicutt}, Jr., Robert C.},
        title = "{Star Formation in Galaxies Along the Hubble Sequence}",
      journal = {\araa},
         year = 1998,
        month = jan,
       volume = {36},
        pages = {189-232},
          doi = {10.1146/annurev.astro.36.1.189},
archivePrefix = {arXiv},
       eprint = {astro-ph/9807187},
 primaryClass = {astro-ph},
       adsurl = {https://ui.adsabs.harvard.edu/abs/1998ARA&A..36..189K}
}

@ARTICLE{2014ARA&A..52..415M,
       author = {{Madau}, Piero and {Dickinson}, Mark},
        title = "{Cosmic Star-Formation History}",
      journal = {\araa},
         year = 2014,
        month = aug,
       volume = {52},
        pages = {415-486},
          doi = {10.1146/annurev-astro-081811-125615},
archivePrefix = {arXiv},
       eprint = {1403.0007},
 primaryClass = {astro-ph.CO},
       adsurl = {https://ui.adsabs.harvard.edu/abs/2014ARA&A..52..415M}
}

@ARTICLE{2021ApJ...907..122M,
       author = {{Mitsuhashi}, I. and {Matsuda}, Y. and {Smail}, Ian and {Hayatsu}, N.~H. and {Simpson}, J.~M. and {Swinbank}, A.~M. and {Umehata}, H. and {Dudzevi{\v{c}}i{\={u}}t{\.{e}}}, U. and {Birkin}, J.~E. and {Ikarashi}, S. and {Chen}, Chian-Chou and {Tadaki}, K. and {Yajima}, H. and {Harikane}, Y. and {Inami}, H. and {Chapman}, S.~C. and {Hatsukade}, B. and {Iono}, D. and {Bunker}, A. and {Ao}, Y. and {Saito}, T. and {Ueda}, J. and {Sakamoto}, S.},
        title = "{FIR-luminous [C II] Emitters in the ALMA-SCUBA-2 COSMOS Survey (AS2COSMOS): The Nature of Submillimeter Galaxies in a 10 Comoving Megaparsec-scale Structure at z {\ensuremath{\sim}} 4.6}",
      journal = {\apj},
         year = 2021,
        month = feb,
       volume = {907},
       number = {2},
          eid = {122},
        pages = {122},
          doi = {10.3847/1538-4357/abcc72},
archivePrefix = {arXiv},
       eprint = {2011.09917},
 primaryClass = {astro-ph.GA},
       adsurl = {https://ui.adsabs.harvard.edu/abs/2021ApJ...907..122M}
}

@ARTICLE{2019ApJ...887...55C,
       author = {{Casey}, Caitlin M. and {Zavala}, Jorge A. and {Aravena}, Manuel and {B{\'e}thermin}, Matthieu and {Caputi}, Karina I. and {Champagne}, Jaclyn B. and {Clements}, David L. and {da Cunha}, Elisabete and {Drew}, Patrick and {Finkelstein}, Steven L. and {Hayward}, Christopher C. and {Kartaltepe}, Jeyhan S. and {Knudsen}, Kirsten and {Koekemoer}, Anton M. and {Magdis}, Georgios E. and {Man}, Allison and {Manning}, Sinclaire M. and {Scoville}, Nick Z. and {Sheth}, Kartik and {Spilker}, Justin and {Staguhn}, Johannes and {Talia}, Margherita and {Taniguchi}, Yoshiaki and {Toft}, Sune and {Treister}, Ezequiel and {Yun}, Min},
        title = "{Physical Characterization of an Unlensed, Dusty Star-forming Galaxy at z = 5.85}",
      journal = {\apj},
         year = 2019,
        month = dec,
       volume = {887},
       number = {1},
          eid = {55},
        pages = {55},
          doi = {10.3847/1538-4357/ab52ff},
archivePrefix = {arXiv},
       eprint = {1910.13331},
 primaryClass = {astro-ph.GA},
       adsurl = {https://ui.adsabs.harvard.edu/abs/2019ApJ...887...55C}
}

@ARTICLE{2018Natur.553..178S,
       author = {{Smit}, Renske and {Bouwens}, Rychard J. and {Carniani}, Stefano and {Oesch}, Pascal A. and {Labb{\'e}}, Ivo and {Illingworth}, Garth D. and {van der Werf}, Paul and {Bradley}, Larry D. and {Gonzalez}, Valentino and {Hodge}, Jacqueline A. and {Holwerda}, Benne W. and {Maiolino}, Roberto and {Zheng}, Wei},
        title = "{Rotation in [C II]-emitting gas in two galaxies at a redshift of 6.8}",
      journal = {\nat},
         year = 2018,
        month = jan,
       volume = {553},
       number = {7687},
        pages = {178-181},
          doi = {10.1038/nature24631},
archivePrefix = {arXiv},
       eprint = {1706.04614},
 primaryClass = {astro-ph.GA},
       adsurl = {https://ui.adsabs.harvard.edu/abs/2018Natur.553..178S}
}

@ARTICLE{2019ApJ...880...15K,
       author = {{Kaasinen}, M. and {Scoville}, N. and {Walter}, F. and {Da Cunha}, E. and {Popping}, G. and {Pavesi}, R. and {Darvish}, B. and {Casey}, C.~M. and {Riechers}, D.~A. and {Glover}, S.},
        title = "{The Molecular Gas Reservoirs of z {\ensuremath{\sim}} 2 Galaxies: A Comparison of CO(1-0) and Dust-based Molecular Gas Masses}",
      journal = {\apj},
         year = 2019,
        month = jul,
       volume = {880},
       number = {1},
          eid = {15},
        pages = {15},
          doi = {10.3847/1538-4357/ab253b},
archivePrefix = {arXiv},
       eprint = {1905.11417},
 primaryClass = {astro-ph.GA},
       adsurl = {https://ui.adsabs.harvard.edu/abs/2019ApJ...880...15K}
}

@ARTICLE{2019ApJ...887..183Z,
       author = {{Zavala}, J.~A. and {Casey}, C.~M. and {Scoville}, N. and {Champagne}, J.~B. and {Chiang}, Y. and {Dannerbauer}, H. and {Drew}, P. and {Fu}, H. and {Spilker}, J. and {Spitler}, L. and {Tran}, K.~V. and {Treister}, E. and {Toft}, S.},
        title = "{On the Gas Content, Star Formation Efficiency, and Environmental Quenching of Massive Galaxies in Protoclusters at z {\ensuremath{\approx}} 2.0-2.5}",
      journal = {\apj},
         year = 2019,
        month = dec,
       volume = {887},
       number = {2},
          eid = {183},
        pages = {183},
          doi = {10.3847/1538-4357/ab5302},
archivePrefix = {arXiv},
       eprint = {1910.13457},
 primaryClass = {astro-ph.GA},
       adsurl = {https://ui.adsabs.harvard.edu/abs/2019ApJ...887..183Z}
}

@ARTICLE{2017MNRAS.471.4155K,
       author = {{Koprowski}, M.~P. and {Dunlop}, J.~S. and {Micha{\l}owski}, M.~J. and {Coppin}, K.~E.~K. and {Geach}, J.~E. and {McLure}, R.~J. and {Scott}, D. and {van der Werf}, P.~P.},
        title = "{The evolving far-IR galaxy luminosity function and dust-obscured star formation rate density out to z≃5.}",
      journal = {\mnras},
         year = 2017,
        month = nov,
       volume = {471},
       number = {4},
        pages = {4155-4169},
          doi = {10.1093/mnras/stx1843},
archivePrefix = {arXiv},
       eprint = {1706.00426},
 primaryClass = {astro-ph.GA},
       adsurl = {https://ui.adsabs.harvard.edu/abs/2017MNRAS.471.4155K}
}

@ARTICLE{2016Msngr.166...48D,
       author = {{Dunlop}, J.~S.},
        title = "{A Deep ALMA Image of the Hubble Ultra Deep Field}",
      journal = {The Messenger},
         year = 2016,
        month = dec,
       volume = {166},
        pages = {48-52},
       adsurl = {https://ui.adsabs.harvard.edu/abs/2016Msngr.166...48D}
}

@ARTICLE{2015ApJS..219...29M,
       author = {{Momcheva}, Ivelina G. and {Williams}, Kurtis A. and {Cool}, Richard J. and {Keeton}, Charles R. and {Zabludoff}, Ann I.},
        title = "{A Spectroscopic Survey of the Fields of 28 Strong Gravitational Lenses}",
      journal = {\apjs},
         year = 2015,
        month = aug,
       volume = {219},
       number = {2},
          eid = {29},
        pages = {29},
          doi = {10.1088/0067-0049/219/2/29},
archivePrefix = {arXiv},
       eprint = {1503.02074},
 primaryClass = {astro-ph.CO},
       adsurl = {https://ui.adsabs.harvard.edu/abs/2015ApJS..219...29M}
}

@ARTICLE{2012ApJS..200...13B,
       author = {{Brammer}, Gabriel B. and {van Dokkum}, Pieter G. and {Franx}, Marijn and {Fumagalli}, Mattia and {Patel}, Shannon and {Rix}, Hans-Walter and {Skelton}, Rosalind E. and {Kriek}, Mariska and {Nelson}, Erica and {Schmidt}, Kasper B. and {Bezanson}, Rachel and {da Cunha}, Elisabete and {Erb}, Dawn K. and {Fan}, Xiaohui and {F{\"o}rster Schreiber}, Natascha and {Illingworth}, Garth D. and {Labb{\'e}}, Ivo and {Leja}, Joel and {Lundgren}, Britt and {Magee}, Dan and {Marchesini}, Danilo and {McCarthy}, Patrick and {Momcheva}, Ivelina and {Muzzin}, Adam and {Quadri}, Ryan and {Steidel}, Charles C. and {Tal}, Tomer and {Wake}, David and {Whitaker}, Katherine E. and {Williams}, Anna},
        title = "{3D-HST: A Wide-field Grism Spectroscopic Survey with the Hubble Space Telescope}",
      journal = {\apjs},
         year = 2012,
        month = jun,
       volume = {200},
       number = {2},
          eid = {13},
        pages = {13},
          doi = {10.1088/0067-0049/200/2/13},
archivePrefix = {arXiv},
       eprint = {1204.2829},
 primaryClass = {astro-ph.CO},
       adsurl = {https://ui.adsabs.harvard.edu/abs/2012ApJS..200...13B}
}

@ARTICLE{2015ApJS..218...15K,
       author = {{Kriek}, Mariska and {Shapley}, Alice E. and {Reddy}, Naveen A. and {Siana}, Brian and {Coil}, Alison L. and {Mobasher}, Bahram and {Freeman}, William R. and {de Groot}, Laura and {Price}, Sedona H. and {Sanders}, Ryan and {Shivaei}, Irene and {Brammer}, Gabriel B. and {Momcheva}, Ivelina G. and {Skelton}, Rosalind E. and {van Dokkum}, Pieter G. and {Whitaker}, Katherine E. and {Aird}, James and {Azadi}, Mojegan and {Kassis}, Marc and {Bullock}, James S. and {Conroy}, Charlie and {Dav{\'e}}, Romeel and {Kere{\v{s}}}, Du{\v{s}}an and {Krumholz}, Mark},
        title = "{The MOSFIRE Deep Evolution Field (MOSDEF) Survey: Rest-frame Optical Spectroscopy for \raisebox{-0.5ex}\textasciitilde1500 H-selected Galaxies at 1.37 < z < 3.8}",
      journal = {\apjs},
         year = 2015,
        month = jun,
       volume = {218},
       number = {2},
          eid = {15},
        pages = {15},
          doi = {10.1088/0067-0049/218/2/15},
archivePrefix = {arXiv},
       eprint = {1412.1835},
 primaryClass = {astro-ph.GA},
       adsurl = {https://ui.adsabs.harvard.edu/abs/2015ApJS..218...15K}
}

@ARTICLE{2018ApJ...858...77H,
       author = {{Hasinger}, G. and {Capak}, P. and {Salvato}, M. and {Barger}, A.~J. and {Cowie}, L.~L. and {Faisst}, A. and {Hemmati}, S. and {Kakazu}, Y. and {Kartaltepe}, J. and {Masters}, D. and {Mobasher}, B. and {Nayyeri}, H. and {Sanders}, D. and {Scoville}, N.~Z. and {Suh}, H. and {Steinhardt}, C. and {Yang}, Fengwei},
        title = "{The DEIMOS 10K Spectroscopic Survey Catalog of the COSMOS Field}",
      journal = {\apj},
         year = 2018,
        month = may,
       volume = {858},
       number = {2},
          eid = {77},
        pages = {77},
          doi = {10.3847/1538-4357/aabacf},
archivePrefix = {arXiv},
       eprint = {1803.09251},
 primaryClass = {astro-ph.GA},
       adsurl = {https://ui.adsabs.harvard.edu/abs/2018ApJ...858...77H}
}

@ARTICLE{2021ApJS..256...44V,
       author = {{van der Wel}, Arjen and {Bezanson}, Rachel and {D'Eugenio}, Francesco and {Straatman}, Caroline and {Franx}, Marijn and {van Houdt}, Josha and {Maseda}, Michael V. and {Gallazzi}, Anna and {Wu}, Po-Feng and {Pacifici}, Camilla and {Barisic}, Ivana and {Brammer}, Gabriel B. and {Munoz-Mateos}, Juan Carlos and {Vervalcke}, Sarah and {Zibetti}, Stefano and {Sobral}, David and {de Graaff}, Anna and {Calhau}, Joao and {Kaushal}, Yasha and {Muzzin}, Adam and {Bell}, Eric F. and {van Dokkum}, Pieter G.},
        title = "{The Large Early Galaxy Astrophysics Census (LEGA-C) Data Release 3: 3000 High-quality Spectra of K$_{s}$-selected Galaxies at z > 0.6}",
      journal = {\apjs},
         year = 2021,
        month = oct,
       volume = {256},
       number = {2},
          eid = {44},
        pages = {44},
          doi = {10.3847/1538-4365/ac1356},
archivePrefix = {arXiv},
       eprint = {2108.00744},
 primaryClass = {astro-ph.GA},
       adsurl = {https://ui.adsabs.harvard.edu/abs/2021ApJS..256...44V}
}

@ARTICLE{2007ApJS..172...70L,
       author = {{Lilly}, S.~J. and {Le F{\`e}vre}, O. and {Renzini}, A. and {Zamorani}, G. and {Scodeggio}, M. and {Contini}, T. and {Carollo}, C.~M. and {Hasinger}, G. and {Kneib}, J. -P. and {Iovino}, A. and {Le Brun}, V. and {Maier}, C. and {Mainieri}, V. and {Mignoli}, M. and {Silverman}, J. and {Tasca}, L.~A.~M. and {Bolzonella}, M. and {Bongiorno}, A. and {Bottini}, D. and {Capak}, P. and {Caputi}, K. and {Cimatti}, A. and {Cucciati}, O. and {Daddi}, E. and {Feldmann}, R. and {Franzetti}, P. and {Garilli}, B. and {Guzzo}, L. and {Ilbert}, O. and {Kampczyk}, P. and {Kovac},, K. and {Lamareille}, F. and {Leauthaud}, A. and {Le Borgne}, J. -F. and {McCracken}, H.~J. and {Marinoni}, C. and {Pello}, R. and {Ricciardelli}, E. and {Scarlata}, C. and {Vergani}, D. and {Sanders}, D.~B. and {Schinnerer}, E. and {Scoville}, N. and {Taniguchi}, Y. and {Arnouts}, S. and {Aussel}, H. and {Bardelli}, S. and {Brusa}, M. and {Cappi}, A. and {Ciliegi}, P. and {Finoguenov}, A. and {Foucaud}, S. and {Franceschini}, A. and {Halliday}, C. and {Impey}, C. and {Knobel}, C. and {Koekemoer}, A. and {Kurk}, J. and {Maccagni}, D. and {Maddox}, S. and {Marano}, B. and {Marconi}, G. and {Meneux}, B. and {Mobasher}, B. and {Moreau}, C. and {Peacock}, J.~A. and {Porciani}, C. and {Pozzetti}, L. and {Scaramella}, R. and {Schiminovich}, D. and {Shopbell}, P. and {Smail}, I. and {Thompson}, D. and {Tresse}, L. and {Vettolani}, G. and {Zanichelli}, A. and {Zucca}, E.},
        title = "{zCOSMOS: A Large VLT/VIMOS Redshift Survey Covering 0 < z < 3 in the COSMOS Field}",
      journal = {\apjs},
         year = 2007,
        month = sep,
       volume = {172},
       number = {1},
        pages = {70-85},
          doi = {10.1086/516589},
archivePrefix = {arXiv},
       eprint = {astro-ph/0612291},
 primaryClass = {astro-ph},
       adsurl = {https://ui.adsabs.harvard.edu/abs/2007ApJS..172...70L}
}

@software{2022zndo...7299500B,
       author = {{Brammer}, Gabriel},
        title = "{msaexp: NIRSpec analyis tools}",
         year = 2023,
        month = sep,
          eid = {10.5281/zenodo.7299500},
          doi = {10.5281/zenodo.7299500},
      version = {0.6.17},
    publisher = {Zenodo},
       adsurl = {https://ui.adsabs.harvard.edu/abs/2022zndo...7299500B}
}

@ARTICLE{2024arXiv240905948D,
       author = {{de Graaff}, Anna and {Brammer}, Gabriel and {Weibel}, Andrea and {Lewis}, Zach and {Maseda}, Michael V. and {Oesch}, Pascal A. and {Bezanson}, Rachel and {Boogaard}, Leindert A. and {Cleri}, Nikko J. and {Cooper}, Olivia R. and {Gottumukkala}, Rashmi and {Greene}, Jenny E. and {Hirschmann}, Michaela and {Hviding}, Raphael E. and {Katz}, Harley and {Labb{\'e}}, Ivo and {Leja}, Joel and {Matthee}, Jorryt and {McConachie}, Ian and {Miller}, Tim B. and {Naidu}, Rohan P. and {Price}, Sedona H. and {Rix}, Hans-Walter and {Setton}, David J. and {Suess}, Katherine A. and {Wang}, Bingjie and {Whitaker}, Katherine E. and {Williams}, Christina C.},
        title = "{RUBIES: a complete census of the bright and red distant Universe with JWST/NIRSpec}",
      journal = {arXiv e-prints},
         year = 2024,
        month = sep,
          eid = {arXiv:2409.05948},
        pages = {arXiv:2409.05948},
          doi = {10.48550/arXiv.2409.05948},
archivePrefix = {arXiv},
       eprint = {2409.05948},
 primaryClass = {astro-ph.GA},
       adsurl = {https://ui.adsabs.harvard.edu/abs/2024arXiv240905948D}
}

@ARTICLE{2024Sci...384..890H,
       author = {{Heintz}, Kasper E. and {Watson}, Darach and {Brammer}, Gabriel and {Vejlgaard}, Simone and {Hutter}, Anne and {Strait}, Victoria B. and {Matthee}, Jorryt and {Oesch}, Pascal A. and {Jakobsson}, P{\'a}ll and {Tanvir}, Nial R. and {Laursen}, Peter and {Naidu}, Rohan P. and {Mason}, Charlotte A. and {Killi}, Meghana and {Jung}, Intae and {Hsiao}, Tiger Yu-Yang and {Abdurro'uf} and {Coe}, Dan and {Arrabal Haro}, Pablo and {Finkelstein}, Steven L. and {Toft}, Sune},
        title = "{Strong damped Lyman-{\ensuremath{\alpha}} absorption in young star-forming galaxies at redshifts 9 to 11}",
      journal = {Science},
         year = 2024,
        month = may,
       volume = {384},
       number = {6698},
        pages = {890-894},
          doi = {10.1126/science.adj0343},
archivePrefix = {arXiv},
       eprint = {2306.00647},
 primaryClass = {astro-ph.GA},
       adsurl = {https://ui.adsabs.harvard.edu/abs/2024Sci...384..890H}
}

@ARTICLE{2009ApJS..184..218L,
       author = {{Lilly}, Simon J. and {Le Brun}, Vincent and {Maier}, Christian and {Mainieri}, Vincenzo and {Mignoli}, Marco and {Scodeggio}, Marco and {Zamorani}, Gianni and {Carollo}, Marcella and {Contini}, Thierry and {Kneib}, Jean-Paul and {Le F{\`e}vre}, Olivier and {Renzini}, Alvio and {Bardelli}, Sandro and {Bolzonella}, Micol and {Bongiorno}, Angela and {Caputi}, Karina and {Coppa}, Graziano and {Cucciati}, Olga and {de la Torre}, Sylvain and {de Ravel}, Loic and {Franzetti}, Paolo and {Garilli}, Bianca and {Iovino}, Angela and {Kampczyk}, Pawel and {Kovac}, Katarina and {Knobel}, Christian and {Lamareille}, Fabrice and {Le Borgne}, Jean-Francois and {Pello}, Roser and {Peng}, Yingjie and {P{\'e}rez-Montero}, Enrique and {Ricciardelli}, Elena and {Silverman}, John D. and {Tanaka}, Masayuki and {Tasca}, Lidia and {Tresse}, Laurence and {Vergani}, Daniela and {Zucca}, Elena and {Ilbert}, Olivier and {Salvato}, Mara and {Oesch}, Pascal and {Abbas}, Umi and {Bottini}, Dario and {Capak}, Peter and {Cappi}, Alberto and {Cassata}, Paolo and {Cimatti}, Andrea and {Elvis}, Martin and {Fumana}, Marco and {Guzzo}, Luigi and {Hasinger}, Gunther and {Koekemoer}, Anton and {Leauthaud}, Alexei and {Maccagni}, Dario and {Marinoni}, Christian and {McCracken}, Henry and {Memeo}, Pierdomenico and {Meneux}, Baptiste and {Porciani}, Cristiano and {Pozzetti}, Lucia and {Sanders}, David and {Scaramella}, Roberto and {Scarlata}, Claudia and {Scoville}, Nick and {Shopbell}, Patrick and {Taniguchi}, Yoshiaki},
        title = "{The zCOSMOS 10k-Bright Spectroscopic Sample}",
      journal = {\apjs},
         year = 2009,
        month = oct,
       volume = {184},
       number = {2},
        pages = {218-229},
          doi = {10.1088/0067-0049/184/2/218},
       adsurl = {https://ui.adsabs.harvard.edu/abs/2009ApJS..184..218L}
}

@ARTICLE{2019MNRAS.482.3135B,
       author = {{Bourne}, N. and {Dunlop}, J.~S. and {Simpson}, J.~M. and {Rowlands}, K.~E. and {Geach}, J.~E. and {McLeod}, D.~J.},
        title = "{The relationship between dust and [C I] at z = 1 and beyond}",
      journal = {\mnras},
         year = 2019,
        month = jan,
       volume = {482},
       number = {3},
        pages = {3135-3161},
          doi = {10.1093/mnras/sty2773},
archivePrefix = {arXiv},
       eprint = {1810.01640},
 primaryClass = {astro-ph.GA},
       adsurl = {https://ui.adsabs.harvard.edu/abs/2019MNRAS.482.3135B}
}

@ARTICLE{2021ApJ...913..110C,
       author = {{Champagne}, Jaclyn B. and {Casey}, Caitlin M. and {Zavala}, Jorge A. and {Cooray}, Asantha and {Dannerbauer}, Helmut and {Fabian}, Andrew and {Hayward}, Christopher C. and {Long}, Arianna S. and {Spilker}, Justin S.},
        title = "{Comprehensive Gas Characterization of a z = 2.5 Protocluster: A Cluster Core Caught in the Beginning of Virialization?}",
      journal = {\apj},
         year = 2021,
        month = jun,
       volume = {913},
       number = {2},
          eid = {110},
        pages = {110},
          doi = {10.3847/1538-4357/abf4e6},
archivePrefix = {arXiv},
       eprint = {2104.02098},
 primaryClass = {astro-ph.GA},
       adsurl = {https://ui.adsabs.harvard.edu/abs/2021ApJ...913..110C}
}

@ARTICLE{2020ApJ...900....1F,
       author = {{Fujimoto}, Seiji and {Silverman}, John D. and {Bethermin}, Matthieu and {Ginolfi}, Michele and {Jones}, Gareth C. and {Le F{\`e}vre}, Olivier and {Dessauges-Zavadsky}, Miroslava and {Rujopakarn}, Wiphu and {Faisst}, Andreas L. and {Fudamoto}, Yoshinobu and {Cassata}, Paolo and {Morselli}, Laura and {Maiolino}, Roberto and {Schaerer}, Daniel and {Capak}, Peter and {Yan}, Lin and {Vallini}, Livia and {Toft}, Sune and {Loiacono}, Federica and {Zamorani}, Gianni and {Talia}, Margherita and {Narayanan}, Desika and {Hathi}, Nimish P. and {Lemaux}, Brian C. and {Boquien}, M{\'e}d{\'e}ric and {Amorin}, Ricardo and {Ibar}, Edo and {Koekemoer}, Anton M. and {M{\'e}ndez-Hern{\'a}ndez}, Hugo and {Bardelli}, Sandro and {Vergani}, Daniela and {Zucca}, Elena and {Romano}, Michael and {Cimatti}, Andrea},
        title = "{The ALPINE-ALMA [C II] Survey: Size of Individual Star-forming Galaxies at z = 4-6 and Their Extended Halo Structure}",
      journal = {\apj},
         year = 2020,
        month = sep,
       volume = {900},
       number = {1},
          eid = {1},
        pages = {1},
          doi = {10.3847/1538-4357/ab94b3},
archivePrefix = {arXiv},
       eprint = {2003.00013},
 primaryClass = {astro-ph.GA},
       adsurl = {https://ui.adsabs.harvard.edu/abs/2020ApJ...900....1F}
}

@ARTICLE{2023ApJ...942...24S,
       author = {{Sanders}, Ryan L. and {Shapley}, Alice E. and {Jones}, Tucker and {Shivaei}, Irene and {Popping}, Gerg{\"o} and {Reddy}, Naveen A. and {Dav{\'e}}, Romeel and {Price}, Sedona H. and {Mobasher}, Bahram and {Kriek}, Mariska and {Coil}, Alison L. and {Siana}, Brian},
        title = "{CO Emission, Molecular Gas, and Metallicity in Main-sequence Star-forming Galaxies at z {\ensuremath{\sim}} 2.3}",
      journal = {\apj},
         year = 2023,
        month = jan,
       volume = {942},
       number = {1},
          eid = {24},
        pages = {24},
          doi = {10.3847/1538-4357/aca46f},
archivePrefix = {arXiv},
       eprint = {2204.06937},
 primaryClass = {astro-ph.GA},
       adsurl = {https://ui.adsabs.harvard.edu/abs/2023ApJ...942...24S}
}

@ARTICLE{2018ApJ...869...27V,
       author = {{Valentino}, Francesco and {Magdis}, Georgios E. and {Daddi}, Emanuele and {Liu}, Daizhong and {Aravena}, Manuel and {Bournaud}, Fr{\'e}d{\'e}ric and {Cibinel}, Anna and {Cormier}, Diane and {Dickinson}, Mark E. and {Gao}, Yu and {Jin}, Shuowen and {Juneau}, St{\'e}phanie and {Kartaltepe}, Jeyhan and {Lee}, Min-Young and {Madden}, Suzanne C. and {Puglisi}, Annagrazia and {Sanders}, David and {Silverman}, John},
        title = "{A Survey of Atomic Carbon [C I] in High-redshift Main-sequence Galaxies}",
      journal = {\apj},
         year = 2018,
        month = dec,
       volume = {869},
       number = {1},
          eid = {27},
        pages = {27},
          doi = {10.3847/1538-4357/aaeb88},
archivePrefix = {arXiv},
       eprint = {1810.11029},
 primaryClass = {astro-ph.GA},
       adsurl = {https://ui.adsabs.harvard.edu/abs/2018ApJ...869...27V}
}

@ARTICLE{2020ApJ...890...24V,
       author = {{Valentino}, Francesco and {Magdis}, Georgios E. and {Daddi}, Emanuele and {Liu}, Daizhong and {Aravena}, Manuel and {Bournaud}, Fr{\'e}d{\'e}ric and {Cortzen}, Isabella and {Gao}, Yu and {Jin}, Shuowen and {Juneau}, St{\'e}phanie and {Kartaltepe}, Jeyhan S. and {Kokorev}, Vasily and {Lee}, Min-Young and {Madden}, Suzanne C. and {Narayanan}, Desika and {Popping}, Gerg{\"o} and {Puglisi}, Annagrazia},
        title = "{The Properties of the Interstellar Medium of Galaxies across Time as Traced by the Neutral Atomic Carbon [C I]}",
      journal = {\apj},
         year = 2020,
        month = feb,
       volume = {890},
       number = {1},
          eid = {24},
        pages = {24},
          doi = {10.3847/1538-4357/ab6603},
archivePrefix = {arXiv},
       eprint = {2001.01734},
 primaryClass = {astro-ph.GA},
       adsurl = {https://ui.adsabs.harvard.edu/abs/2020ApJ...890...24V}
}

@ARTICLE{2020A&A...643A...1L,
       author = {{Le F{\`e}vre}, O. and {B{\'e}thermin}, M. and {Faisst}, A. and {Jones}, G.~C. and {Capak}, P. and {Cassata}, P. and {Silverman}, J.~D. and {Schaerer}, D. and {Yan}, L. and {Amorin}, R. and {Bardelli}, S. and {Boquien}, M. and {Cimatti}, A. and {Dessauges-Zavadsky}, M. and {Giavalisco}, M. and {Hathi}, N.~P. and {Fudamoto}, Y. and {Fujimoto}, S. and {Ginolfi}, M. and {Gruppioni}, C. and {Hemmati}, S. and {Ibar}, E. and {Koekemoer}, A. and {Khusanova}, Y. and {Lagache}, G. and {Lemaux}, B.~C. and {Loiacono}, F. and {Maiolino}, R. and {Mancini}, C. and {Narayanan}, D. and {Morselli}, L. and {M{\'e}ndez-Hern{\`a}ndez}, Hugo and {Oesch}, P.~A. and {Pozzi}, F. and {Romano}, M. and {Riechers}, D. and {Scoville}, N. and {Talia}, M. and {Tasca}, L.~A.~M. and {Thomas}, R. and {Toft}, S. and {Vallini}, L. and {Vergani}, D. and {Walter}, F. and {Zamorani}, G. and {Zucca}, E.},
        title = "{The ALPINE-ALMA [CII] survey. Survey strategy, observations, and sample properties of 118 star-forming galaxies at 4 < z < 6}",
      journal = {\aap},
         year = 2020,
        month = nov,
       volume = {643},
          eid = {A1},
        pages = {A1},
          doi = {10.1051/0004-6361/201936965},
archivePrefix = {arXiv},
       eprint = {1910.09517},
 primaryClass = {astro-ph.CO},
       adsurl = {https://ui.adsabs.harvard.edu/abs/2020A&A...643A...1L}
}

@ARTICLE{2022ApJ...929..159C,
       author = {{Chen}, Chian-Chou and {Liao}, Cheng-Lin and {Smail}, Ian and {Swinbank}, A.~M. and {Ao}, Y. and {Bunker}, A.~J. and {Chapman}, S.~C. and {Hatsukade}, B. and {Ivison}, R.~J. and {Lee}, Minju M. and {Serjeant}, Stephen and {Umehata}, Hideki and {Wang}, Wei-Hao and {Zhao}, Y.},
        title = "{An ALMA Spectroscopic Survey of the Brightest Submillimeter Galaxies in the SCUBA-2-COSMOS Field (AS2COSPEC): Survey Description and First Results}",
      journal = {\apj},
         year = 2022,
        month = apr,
       volume = {929},
       number = {2},
          eid = {159},
        pages = {159},
          doi = {10.3847/1538-4357/ac61df},
archivePrefix = {arXiv},
       eprint = {2112.07430},
 primaryClass = {astro-ph.GA},
       adsurl = {https://ui.adsabs.harvard.edu/abs/2022ApJ...929..159C}
}

@ARTICLE{BC03,
       author = {{Bruzual}, G. and {Charlot}, S.},
        title = "{Stellar population synthesis at the resolution of 2003}",
      journal = {\mnras},
         year = 2003,
        month = oct,
       volume = {344},
       number = {4},
        pages = {1000-1028},
          doi = {10.1046/j.1365-8711.2003.06897.x},
archivePrefix = {arXiv},
       eprint = {astro-ph/0309134},
 primaryClass = {astro-ph},
       adsurl = {https://ui.adsabs.harvard.edu/abs/2003MNRAS.344.1000B}
}

@ARTICLE{Fioc1997,
       author = {{Fioc}, M. and {Rocca-Volmerange}, B.},
        title = "{PEGASE: a UV to NIR spectral evolution model of galaxies. Application to the calibration of bright galaxy counts.}",
      journal = {\aap},
         year = 1997,
        month = oct,
       volume = {326},
        pages = {950-962},
          doi = {10.48550/arXiv.astro-ph/9707017},
archivePrefix = {arXiv},
       eprint = {astro-ph/9707017},
 primaryClass = {astro-ph},
       adsurl = {https://ui.adsabs.harvard.edu/abs/1997A&A...326..950F}
}

@ARTICLE{Fioc1999,
       author = {{Fioc}, Michel and {Rocca-Volmerange}, Brigitte},
        title = "{PEGASE.2, a metallicity-consistent spectral evolution model of galaxies: the documentation and the code}",
      journal = {arXiv e-prints},
         year = 1999,
        month = dec,
          eid = {astro-ph/9912179},
        pages = {astro-ph/9912179},
archivePrefix = {arXiv},
       eprint = {astro-ph/9912179},
 primaryClass = {astro-ph},
       adsurl = {https://ui.adsabs.harvard.edu/abs/1999astro.ph.12179F}
}

@ARTICLE{Fioc2019,
       author = {{Fioc}, Michel and {Rocca-Volmerange}, Brigitte},
        title = "{P{\'E}GASE.3: A code for modeling the UV-to-IR/submm spectral and chemical evolution of galaxies with dust}",
      journal = {\aap},
         year = 2019,
        month = mar,
       volume = {623},
          eid = {A143},
        pages = {A143},
          doi = {10.1051/0004-6361/201833556},
archivePrefix = {arXiv},
       eprint = {1902.07929},
 primaryClass = {astro-ph.GA},
       adsurl = {https://ui.adsabs.harvard.edu/abs/2019A&A...623A.143F}
}

@ARTICLE{Larson2023,
       author = {{Larson}, Rebecca L. and {Hutchison}, Taylor A. and {Bagley}, Micaela and {Finkelstein}, Steven L. and {Yung}, L.~Y. Aaron and {Somerville}, Rachel S. and {Hirschmann}, Michaela and {Brammer}, Gabriel and {Holwerda}, Benne W. and {Papovich}, Casey and {Morales}, Alexa M. and {Wilkins}, Stephen M.},
        title = "{Spectral Templates Optimal for Selecting Galaxies at z > 8 with the JWST}",
      journal = {\apj},
         year = 2023,
        month = dec,
       volume = {958},
       number = {2},
          eid = {141},
        pages = {141},
          doi = {10.3847/1538-4357/acfed4},
archivePrefix = {arXiv},
       eprint = {2211.10035},
 primaryClass = {astro-ph.GA},
       adsurl = {https://ui.adsabs.harvard.edu/abs/2023ApJ...958..141L}
}

@ARTICLE{Hainline2024,
       author = {{Hainline}, Kevin N. and {Johnson}, Benjamin D. and {Robertson}, Brant and {Tacchella}, Sandro and {Helton}, Jakob M. and {Sun}, Fengwu and {Eisenstein}, Daniel J. and {Simmonds}, Charlotte and {Topping}, Michael W. and {Whitler}, Lily and {Willmer}, Christopher N.~A. and {Rieke}, Marcia and {Suess}, Katherine A. and {Hviding}, Raphael E. and {Cameron}, Alex J. and {Alberts}, Stacey and {Baker}, William M. and {Baum}, Stefi and {Bhatawdekar}, Rachana and {Bonaventura}, Nina and {Boyett}, Kristan and {Bunker}, Andrew J. and {Carniani}, Stefano and {Charlot}, Stephane and {Chevallard}, Jacopo and {Chen}, Zuyi and {Curti}, Mirko and {Curtis-Lake}, Emma and {D'Eugenio}, Francesco and {Egami}, Eiichi and {Endsley}, Ryan and {Hausen}, Ryan and {Ji}, Zhiyuan and {Looser}, Tobias J. and {Lyu}, Jianwei and {Maiolino}, Roberto and {Nelson}, Erica and {Pusk{\'a}s}, D{\'a}vid and {Rawle}, Tim and {Sandles}, Lester and {Saxena}, Aayush and {Smit}, Renske and {Stark}, Daniel P. and {Williams}, Christina C. and {Willott}, Chris and {Witstok}, Joris},
        title = "{The Cosmos in Its Infancy: JADES Galaxy Candidates at z > 8 in GOODS-S and GOODS-N}",
      journal = {\apj},
         year = 2024,
        month = mar,
       volume = {964},
       number = {1},
          eid = {71},
        pages = {71},
          doi = {10.3847/1538-4357/ad1ee4},
archivePrefix = {arXiv},
       eprint = {2306.02468},
 primaryClass = {astro-ph.GA},
       adsurl = {https://ui.adsabs.harvard.edu/abs/2024ApJ...964...71H}
}

@ARTICLE{2024ApJ...961..226L,
       author = {{Liao}, Cheng-Lin and {Chen}, Chian-Chou and {Wang}, Wei-Hao and {Smail}, Ian and {Ao}, Y. and {Chapman}, S.~C. and {Dudzevi{\v{c}}i{\={u}}t{\.{e}}}, U. and {Frias Castillo}, M. and {Lee}, Minju M. and {Serjeant}, Stephen and {Swinbank}, A.~M. and {Taylor}, Dominic J. and {Umehata}, Hideki and {Zhao}, Y.},
        title = "{An ALMA Spectroscopic Survey of the Brightest Submillimeter Galaxies in the SCUBA-2-COSMOS Field (AS2COSPEC): Physical Properties of z = 2{\textendash}5 Ultra- and Hyperluminous Infrared Galaxies}",
      journal = {\apj},
         year = 2024,
        month = feb,
       volume = {961},
       number = {2},
          eid = {226},
        pages = {226},
          doi = {10.3847/1538-4357/ad148c},
archivePrefix = {arXiv},
       eprint = {2311.17417},
 primaryClass = {astro-ph.GA},
       adsurl = {https://ui.adsabs.harvard.edu/abs/2024ApJ...961..226L}
}

@ARTICLE{2022ApJ...931..160B,
       author = {{Bouwens}, R.~J. and {Smit}, R. and {Schouws}, S. and {Stefanon}, M. and {Bowler}, R. and {Endsley}, R. and {Gonzalez}, V. and {Inami}, H. and {Stark}, D. and {Oesch}, P. and {Hodge}, J. and {Aravena}, M. and {da Cunha}, E. and {Dayal}, P. and {de Looze}, I. and {Ferrara}, A. and {Fudamoto}, Y. and {Graziani}, L. and {Li}, C. and {Nanayakkara}, T. and {Pallottini}, A. and {Schneider}, R. and {Sommovigo}, L. and {Topping}, M. and {van der Werf}, P. and {Algera}, H. and {Barrufet}, L. and {Hygate}, A. and {Labb{\'e}}, I. and {Riechers}, D. and {Witstok}, J.},
        title = "{Reionization Era Bright Emission Line Survey: Selection and Characterization of Luminous Interstellar Medium Reservoirs in the z > 6.5 Universe}",
      journal = {\apj},
         year = 2022,
        month = jun,
       volume = {931},
       number = {2},
          eid = {160},
        pages = {160},
          doi = {10.3847/1538-4357/ac5a4a},
archivePrefix = {arXiv},
       eprint = {2106.13719},
 primaryClass = {astro-ph.GA},
       adsurl = {https://ui.adsabs.harvard.edu/abs/2022ApJ...931..160B}
}

@ARTICLE{2022MNRAS.515.3126I,
       author = {{Inami}, Hanae and {Algera}, Hiddo S.~B. and {Schouws}, Sander and {Sommovigo}, Laura and {Bouwens}, Rychard and {Smit}, Renske and {Stefanon}, Mauro and {Bowler}, Rebecca A.~A. and {Endsley}, Ryan and {Ferrara}, Andrea and {Oesch}, Pascal and {Stark}, Daniel and {Aravena}, Manuel and {Barrufet}, Laia and {da Cunha}, Elisabete and {Dayal}, Pratika and {De Looze}, Ilse and {Fudamoto}, Yoshinobu and {Gonzalez}, Valentino and {Graziani}, Luca and {Hodge}, Jacqueline A. and {Hygate}, Alexander P.~S. and {Nanayakkara}, Themiya and {Pallottini}, Andrea and {Riechers}, Dominik A. and {Schneider}, Raffaella and {Topping}, Michael and {van der Werf}, Paul},
        title = "{The ALMA REBELS Survey: dust continuum detections at z > 6.5}",
      journal = {\mnras},
         year = 2022,
        month = sep,
       volume = {515},
       number = {3},
        pages = {3126-3143},
          doi = {10.1093/mnras/stac1779},
archivePrefix = {arXiv},
       eprint = {2203.15136},
 primaryClass = {astro-ph.GA},
       adsurl = {https://ui.adsabs.harvard.edu/abs/2022MNRAS.515.3126I}
}

@ARTICLE{2023MNRAS.522..449B,
       author = {{Barrufet}, L. and {Oesch}, P.~A. and {Weibel}, A. and {Brammer}, G. and {Bezanson}, R. and {Bouwens}, R. and {Fudamoto}, Y. and {Gonzalez}, V. and {Gottumukkala}, R. and {Illingworth}, G. and {Heintz}, K.~E. and {Holden}, B. and {Labbe}, I. and {Magee}, D. and {Naidu}, R.~P. and {Nelson}, E. and {Stefanon}, M. and {Smit}, R. and {van Dokkum}, P. and {Weaver}, J.~R. and {Williams}, C.~C.},
        title = "{Unveiling the nature of infrared bright, optically dark galaxies with early JWST data}",
      journal = {\mnras},
         year = 2023,
        month = jun,
       volume = {522},
       number = {1},
        pages = {449-456},
          doi = {10.1093/mnras/stad947},
archivePrefix = {arXiv},
       eprint = {2207.14733},
 primaryClass = {astro-ph.GA},
       adsurl = {https://ui.adsabs.harvard.edu/abs/2023MNRAS.522..449B}
}

@ARTICLE{2024MNRAS.527.5808B,
       author = {{Bowler}, R.~A.~A. and {Inami}, H. and {Sommovigo}, L. and {Smit}, R. and {Algera}, H.~S.~B. and {Aravena}, M. and {Barrufet}, L. and {Bouwens}, R. and {da Cunha}, E. and {Cullen}, F. and {Dayal}, P. and {De Looze}, I. and {Dunlop}, J.~S. and {Fudamoto}, Y. and {Mauerhofer}, V. and {McLure}, R.~J. and {Stefanon}, M. and {Schneider}, R. and {Ferrara}, A. and {Graziani}, L. and {Hodge}, J.~A. and {Nanayakkara}, T. and {Palla}, M. and {Schouws}, S. and {Stark}, D.~P. and {van der Werf}, P.~P.},
        title = "{The ALMA REBELS survey: obscured star formation in massive Lyman-break galaxies at z= 4-8 revealed by the IRX-{\ensuremath{\beta}} and M$_{{\ensuremath{\star}}}$ relations}",
      journal = {\mnras},
         year = 2024,
        month = jan,
       volume = {527},
       number = {3},
        pages = {5808-5828},
          doi = {10.1093/mnras/stad3578},
archivePrefix = {arXiv},
       eprint = {2309.17386},
 primaryClass = {astro-ph.GA},
       adsurl = {https://ui.adsabs.harvard.edu/abs/2024MNRAS.527.5808B}
}

@ARTICLE{2023ApJ...954...31C,
       author = {{Casey}, Caitlin M. and {Kartaltepe}, Jeyhan S. and {Drakos}, Nicole E. and {Franco}, Maximilien and {Harish}, Santosh and {Paquereau}, Louise and {Ilbert}, Olivier and {Rose}, Caitlin and {Cox}, Isabella G. and {Nightingale}, James W. and {Robertson}, Brant E. and {Silverman}, John D. and {Koekemoer}, Anton M. and {Massey}, Richard and {McCracken}, Henry Joy and {Rhodes}, Jason and {Akins}, Hollis B. and {Allen}, Natalie and {Amvrosiadis}, Aristeidis and {Arango-Toro}, Rafael C. and {Bagley}, Micaela B. and {Bongiorno}, Angela and {Capak}, Peter L. and {Champagne}, Jaclyn B. and {Chartab}, Nima and {Ch{\'a}vez Ortiz}, {\'O}scar A. and {Chworowsky}, Katherine and {Cooke}, Kevin C. and {Cooper}, Olivia R. and {Darvish}, Behnam and {Ding}, Xuheng and {Faisst}, Andreas L. and {Finkelstein}, Steven L. and {Fujimoto}, Seiji and {Gentile}, Fabrizio and {Gillman}, Steven and {Gould}, Katriona M.~L. and {Gozaliasl}, Ghassem and {Hayward}, Christopher C. and {He}, Qiuhan and {Hemmati}, Shoubaneh and {Hirschmann}, Michaela and {Jahnke}, Knud and {Jin}, Shuowen and {Khostovan}, Ali Ahmad and {Kokorev}, Vasily and {Lambrides}, Erini and {Laigle}, Clotilde and {Larson}, Rebecca L. and {Leung}, Gene C.~K. and {Liu}, Daizhong and {Liaudat}, Tobias and {Long}, Arianna S. and {Magdis}, Georgios and {Mahler}, Guillaume and {Mainieri}, Vincenzo and {Manning}, Sinclaire M. and {Maraston}, Claudia and {Martin}, Crystal L. and {McCleary}, Jacqueline E. and {McKinney}, Jed and {McPartland}, Conor J.~R. and {Mobasher}, Bahram and {Pattnaik}, Rohan and {Renzini}, Alvio and {Rich}, R. Michael and {Sanders}, David B. and {Sattari}, Zahra and {Scognamiglio}, Diana and {Scoville}, Nick and {Sheth}, Kartik and {Shuntov}, Marko and {Sparre}, Martin and {Suzuki}, Tomoko L. and {Talia}, Margherita and {Toft}, Sune and {Trakhtenbrot}, Benny and {Urry}, C. Megan and {Valentino}, Francesco and {Vanderhoof}, Brittany N. and {Vardoulaki}, Eleni and {Weaver}, John R. and {Whitaker}, Katherine E. and {Wilkins}, Stephen M. and {Yang}, Lilan and {Zavala}, Jorge A.},
        title = "{COSMOS-Web: An Overview of the JWST Cosmic Origins Survey}",
      journal = {\apj},
         year = 2023,
        month = sep,
       volume = {954},
       number = {1},
          eid = {31},
        pages = {31},
          doi = {10.3847/1538-4357/acc2bc},
archivePrefix = {arXiv},
       eprint = {2211.07865},
 primaryClass = {astro-ph.GA},
       adsurl = {https://ui.adsabs.harvard.edu/abs/2023ApJ...954...31C}
}

@ARTICLE{merlin2015,
       author = {{Merlin}, E. and {Fontana}, A. and {Ferguson}, H.~C. and {Dunlop}, J.~S. and {Elbaz}, D. and {Bourne}, N. and {Bruce}, V.~A. and {Buitrago}, F. and {Castellano}, M. and {Schreiber}, C. and {Grazian}, A. and {McLure}, R.~J. and {Okumura}, K. and {Shu}, X. and {Wang}, T. and {Amor{\'\i}n}, R. and {Boutsia}, K. and {Cappelluti}, N. and {Comastri}, A. and {Derriere}, S. and {Faber}, S.~M. and {Santini}, P.},
        title = "{T-PHOT: A new code for PSF-matched, prior-based, multiwavelength extragalactic deconfusion photometry}",
      journal = {\aap},
         year = 2015,
        month = oct,
       volume = {582},
          eid = {A15},
        pages = {A15},
          doi = {10.1051/0004-6361/201526471},
archivePrefix = {arXiv},
       eprint = {1505.02516},
 primaryClass = {astro-ph.IM},
       adsurl = {https://ui.adsabs.harvard.edu/abs/2015A&A...582A..15M}
}

@ARTICLE{2020MNRAS.495.3409S,
       author = {{Simpson}, J.~M. and {Smail}, Ian and {Dudzevi{\v{c}}i{\={u}}t{\.{e}}}, U. and {Matsuda}, Y. and {Hsieh}, B. -C. and {Wang}, W. -H. and {Swinbank}, A.~M. and {Stach}, S.~M. and {An}, Fang Xia and {Birkin}, J.~E. and {Ao}, Y. and {Bunker}, A.~J. and {Chapman}, S.~C. and {Chen}, Chian-Chou and {Coppin}, K.~E.~K. and {Ikarashi}, S. and {Ivison}, R.~J. and {Mitsuhashi}, I. and {Saito}, T. and {Umehata}, H. and {Wang}, R. and {Zhao}, Y.},
        title = "{An ALMA survey of the brightest sub-millimetre sources in the SCUBA-2-COSMOS field}",
      journal = {\mnras},
         year = 2020,
        month = jul,
       volume = {495},
       number = {3},
        pages = {3409-3430},
          doi = {10.1093/mnras/staa1345},
archivePrefix = {arXiv},
       eprint = {2003.05484},
 primaryClass = {astro-ph.GA},
       adsurl = {https://ui.adsabs.harvard.edu/abs/2020MNRAS.495.3409S}
}

@ARTICLE{2024A&A...691A.299G,
       author = {{Gillman}, Steven and {Smail}, Ian and {Gullberg}, Bitten and {Swinbank}, A.~M. and {Vijayan}, Aswin P. and {Lee}, Minju and {Brammer}, Gabe and {Dudzevi{\v{c}}i{\={u}}t{\.{e}}}, Ugn{\.{e}} and {Greve}, Thomas R. and {Almaini}, Omar and {Brinch}, Malte and {Chapman}, Scott C. and {Chen}, Chian-Chou and {Ikarashi}, Soh and {Matsuda}, Yuichi and {Wang}, Wei-Hao and {Walter}, Fabian and {van der Werf}, Paul P.},
        title = "{The structure of massive star-forming galaxies from JWST and ALMA: Dusty, high-redshift disc galaxies}",
      journal = {\aap},
         year = 2024,
        month = nov,
       volume = {691},
          eid = {A299},
        pages = {A299},
          doi = {10.1051/0004-6361/202451006},
archivePrefix = {arXiv},
       eprint = {2406.03544},
 primaryClass = {astro-ph.GA},
       adsurl = {https://ui.adsabs.harvard.edu/abs/2024A&A...691A.299G}
}

@ARTICLE{2024A&A...685A...1A,
       author = {{Adscheid}, Sylvia and {Magnelli}, Benjamin and {Liu}, Daizhong and {Bertoldi}, Frank and {Delvecchio}, Ivan and {Gruppioni}, Carlotta and {Schinnerer}, Eva and {Traina}, Alberto and {B{\'e}thermin}, Matthieu and {Gkogkou}, Athanasia},
        title = "{A$^{3}$COSMOS and A$^{3}$GOODSS: Continuum source catalogues and multi-band number counts}",
      journal = {\aap},
         year = 2024,
        month = may,
       volume = {685},
          eid = {A1},
        pages = {A1},
          doi = {10.1051/0004-6361/202348407},
archivePrefix = {arXiv},
       eprint = {2403.03125},
 primaryClass = {astro-ph.GA},
       adsurl = {https://ui.adsabs.harvard.edu/abs/2024A&A...685A...1A}
}

@MISC{2024jwst.prop.6585C,
       author = {{Coulter}, David and {Engesser}, Mike and {Pierel}, Justin and {Akins}, Hollis and {Casey}, Caitlin M. and {DeCoursey}, Christa Noel and {Egami}, Eiichi and {Fox}, Ori Dosovitz and {Franco}, Maximilien and {Fujimoto}, Seiji and {Gezari}, Suvi and {Gomez}, Sebastian and {Guolo}, Muryel and {Harish}, Santosh and {Jencson}, Jacob and {Karmen}, Mitchell and {Kartaltepe}, Jeyhan and {Koekemoer}, Anton M. and {McCracken}, Henry Joy and {Moriya}, Takashi and {Quimby}, Robert M. and {Rest}, Armin and {Rhodes}, Jason D. and {Robertson}, Brant and {Shahbandeh}, Melissa and {Shuntov}, Marko and {Siebert}, Matthew Ryan and {Strolger}, Louis-Gregory and {Suzuki}, Nao and {Wang}, Qinan and {Yang}, Lilan and {Zenati}, Yossef},
        title = "{The High-z Menagerie: A Rare Chance to Study the Early and Exotic Transient Universe}",
 howpublished = {JWST Proposal. Cycle 2, ID. \#6585},
         year = 2024,
        month = mar,
        pages = {6585},
       adsurl = {https://ui.adsabs.harvard.edu/abs/2024jwst.prop.6585C}
}

@ARTICLE{Kron1980,
       author = {{Kron}, R.~G.},
        title = "{Photometry of a complete sample of faint galaxies.}",
      journal = {\apjs},
         year = 1980,
        month = jun,
       volume = {43},
        pages = {305-325},
          doi = {10.1086/190669},
       adsurl = {https://ui.adsabs.harvard.edu/abs/1980ApJS...43..305K}
}

@ARTICLE{mcleod2024,
       author = {{McLeod}, D.~J. and {Donnan}, C.~T. and {McLure}, R.~J. and {Dunlop}, J.~S. and {Magee}, D. and {Begley}, R. and {Carnall}, A.~C. and {Cullen}, F. and {Ellis}, R.~S. and {Hamadouche}, M.~L. and {Stanton}, T.~M.},
        title = "{The galaxy UV luminosity function at z ≃ 11 from a suite of public JWST ERS, ERO, and Cycle-1 programs}",
      journal = {\mnras},
         year = 2024,
        month = jan,
       volume = {527},
       number = {3},
        pages = {5004-5022},
          doi = {10.1093/mnras/stad3471},
archivePrefix = {arXiv},
       eprint = {2304.14469},
 primaryClass = {astro-ph.GA},
       adsurl = {https://ui.adsabs.harvard.edu/abs/2024MNRAS.527.5004M}
}

@ARTICLE{2007MNRAS.380..199I,
       author = {{Ivison}, R.~J. and {Greve}, T.~R. and {Dunlop}, J.~S. and {Peacock}, J.~A. and {Egami}, E. and {Smail}, Ian and {Ibar}, E. and {van Kampen}, E. and {Aretxaga}, I. and {Babbedge}, T. and {Biggs}, A.~D. and {Blain}, A.~W. and {Chapman}, S.~C. and {Clements}, D.~L. and {Coppin}, K. and {Farrah}, D. and {Halpern}, M. and {Hughes}, D.~H. and {Jarvis}, M.~J. and {Jenness}, T. and {Jones}, J.~R. and {Mortier}, A.~M.~J. and {Oliver}, S. and {Papovich}, C. and {P{\'e}rez-Gonz{\'a}lez}, P.~G. and {Pope}, A. and {Rawlings}, S. and {Rieke}, G.~H. and {Rowan-Robinson}, M. and {Savage}, R.~S. and {Scott}, D. and {Seigar}, M. and {Serjeant}, S. and {Simpson}, C. and {Stevens}, J.~A. and {Vaccari}, M. and {Wagg}, J. and {Willott}, C.~J.},
        title = "{The SCUBA HAlf Degree Extragalactic Survey - III. Identification of radio and mid-infrared counterparts to submillimetre galaxies}",
      journal = {\mnras},
         year = 2007,
        month = sep,
       volume = {380},
       number = {1},
        pages = {199-228},
          doi = {10.1111/j.1365-2966.2007.12044.x},
archivePrefix = {arXiv},
       eprint = {astro-ph/0702544},
 primaryClass = {astro-ph},
       adsurl = {https://ui.adsabs.harvard.edu/abs/2007MNRAS.380..199I}
}

@ARTICLE{2020A&A...643A...8G,
       author = {{Gruppioni}, C. and {B{\'e}thermin}, M. and {Loiacono}, F. and {Le F{\`e}vre}, O. and {Capak}, P. and {Cassata}, P. and {Faisst}, A.~L. and {Schaerer}, D. and {Silverman}, J. and {Yan}, L. and {Bardelli}, S. and {Boquien}, M. and {Carraro}, R. and {Cimatti}, A. and {Dessauges-Zavadsky}, M. and {Ginolfi}, M. and {Fujimoto}, S. and {Hathi}, N.~P. and {Jones}, G.~C. and {Khusanova}, Y. and {Koekemoer}, A.~M. and {Lagache}, G. and {Lemaux}, B.~C. and {Oesch}, P.~A. and {Pozzi}, F. and {Riechers}, D.~A. and {Rodighiero}, G. and {Romano}, M. and {Talia}, M. and {Vallini}, L. and {Vergani}, D. and {Zamorani}, G. and {Zucca}, E.},
        title = "{The ALPINE-ALMA [CII] survey. The nature, luminosity function, and star formation history of dusty galaxies up to z ≃ 6}",
      journal = {\aap},
         year = 2020,
        month = nov,
       volume = {643},
          eid = {A8},
        pages = {A8},
          doi = {10.1051/0004-6361/202038487},
archivePrefix = {arXiv},
       eprint = {2006.04974},
 primaryClass = {astro-ph.GA},
       adsurl = {https://ui.adsabs.harvard.edu/abs/2020A&A...643A...8G}
}

@ARTICLE{2023MNRAS.522.3926B,
       author = {{Barrufet}, L. and {Oesch}, P.~A. and {Bouwens}, R. and {Inami}, H. and {Sommovigo}, L. and {Algera}, H. and {da Cunha}, E. and {Aravena}, M. and {Dayal}, P. and {Ferrara}, A. and {Fudamoto}, Y. and {Gonzalez}, V. and {Graziani}, L. and {Hygate}, A.~P.~S. and {de Looze}, I. and {Nanayakkara}, T. and {Pallottini}, A. and {Schneider}, R. and {Stefanon}, M. and {Topping}, M. and {van der Werf}, P.},
        title = "{The ALMA REBELS Survey: the first infrared luminosity function measurement at z {\ensuremath{\sim}} 7}",
      journal = {\mnras},
         year = 2023,
        month = jul,
       volume = {522},
       number = {3},
        pages = {3926-3934},
          doi = {10.1093/mnras/stad1259},
archivePrefix = {arXiv},
       eprint = {2303.11321},
 primaryClass = {astro-ph.GA},
       adsurl = {https://ui.adsabs.harvard.edu/abs/2023MNRAS.522.3926B}
}

@ARTICLE{2015MNRAS.451.3419G,
       author = {{Gruppioni}, C. and {Calura}, F. and {Pozzi}, F. and {Delvecchio}, I. and {Berta}, S. and {De Lucia}, G. and {Fontanot}, F. and {Franceschini}, A. and {Marchetti}, L. and {Menci}, N. and {Monaco}, P. and {Vaccari}, M.},
        title = "{Star formation in Herschel's Monsters versus semi-analytic models}",
      journal = {\mnras},
         year = 2015,
        month = aug,
       volume = {451},
       number = {4},
        pages = {3419-3426},
          doi = {10.1093/mnras/stv1204},
archivePrefix = {arXiv},
       eprint = {1506.01518},
 primaryClass = {astro-ph.GA},
       adsurl = {https://ui.adsabs.harvard.edu/abs/2015MNRAS.451.3419G}
}

@ARTICLE{2024ApJS..275...36F,
       author = {{Fujimoto}, Seiji and {Kohno}, Kotaro and {Ouchi}, Masami and {Oguri}, Masamune and {Kokorev}, Vasily and {Brammer}, Gabriel and {Sun}, Fengwu and {Gonz{\'a}lez-L{\'o}pez}, Jorge and {Bauer}, Franz E. and {Caminha}, Gabriel B. and {Hatsukade}, Bunyo and {Richard}, Johan and {Smail}, Ian and {Tsujita}, Akiyoshi and {Ueda}, Yoshihiro and {Uematsu}, Ryosuke and {Zitrin}, Adi and {Coe}, Dan and {Kneib}, Jean-Paul and {Postman}, Marc and {Umetsu}, Keiichi and {Lagos}, Claudia del P. and {Popping}, Gerg{\"o} and {Ao}, Yiping and {Bradley}, Larry and {Caputi}, Karina and {Dessauges-Zavadsky}, Miroslava and {Egami}, Eiichi and {Espada}, Daniel and {Ivison}, R.~J. and {Jauzac}, Mathilde and {Knudsen}, Kirsten K. and {Koekemoer}, Anton M. and {Magdis}, Georgios E. and {Mahler}, Guillaume and {Mu{\~n}oz Arancibia}, A.~M. and {Rawle}, Timothy and {Shimasaku}, Kazuhiro and {Toft}, Sune and {Umehata}, Hideki and {Valentino}, Francesco and {Wang}, Tao and {Wang}, Wei-Hao},
        title = "{ALMA Lensing Cluster Survey: Deep 1.2 mm Number Counts and Infrared Luminosity Functions at z ≃ 1{\textendash}8}",
      journal = {\apjs},
         year = 2024,
        month = dec,
       volume = {275},
       number = {2},
          eid = {36},
        pages = {36},
          doi = {10.3847/1538-4365/ad5ae2},
archivePrefix = {arXiv},
       eprint = {2303.01658},
 primaryClass = {astro-ph.GA},
       adsurl = {https://ui.adsabs.harvard.edu/abs/2024ApJS..275...36F}
}

@ARTICLE{2024A&A...688A..55M,
       author = {{Magnelli}, Benjamin and {Adscheid}, Sylvia and {Wang}, Tsan-Ming and {Ciesla}, Laure and {Daddi}, Emanuele and {Delvecchio}, Ivan and {Elbaz}, David and {Fudamoto}, Yoshinobu and {Fukushima}, Shuma and {Franco}, Maximilien and {G{\'o}mez-Guijarro}, Carlos and {Gruppioni}, Carlotta and {Jim{\'e}nez-Andrade}, Eric F. and {Liu}, Daizhong and {Oesch}, Pascal and {Schinnerer}, Eva and {Traina}, Alberto},
        title = "{A$^{3}$COSMOS: Measuring the cosmic dust-attenuated star formation rate density at 4 < z < 5}",
      journal = {\aap},
         year = 2024,
        month = aug,
       volume = {688},
          eid = {A55},
        pages = {A55},
          doi = {10.1051/0004-6361/202450081},
archivePrefix = {arXiv},
       eprint = {2405.18086},
 primaryClass = {astro-ph.GA},
       adsurl = {https://ui.adsabs.harvard.edu/abs/2024A&A...688A..55M}
}

@ARTICLE{2013ApJ...770...57B,
       author = {{Behroozi}, Peter S. and {Wechsler}, Risa H. and {Conroy}, Charlie},
        title = "{The Average Star Formation Histories of Galaxies in Dark Matter Halos from z = 0-8}",
      journal = {\apj},
         year = 2013,
        month = jun,
       volume = {770},
       number = {1},
          eid = {57},
        pages = {57},
          doi = {10.1088/0004-637X/770/1/57},
archivePrefix = {arXiv},
       eprint = {1207.6105},
 primaryClass = {astro-ph.CO},
       adsurl = {https://ui.adsabs.harvard.edu/abs/2013ApJ...770...57B}
}

@ARTICLE{2019A&A...624A..98W,
       author = {{Wang}, L. and {Pearson}, W.~J. and {Cowley}, W. and {Trayford}, J.~W. and {B{\'e}thermin}, M. and {Gruppioni}, C. and {Hurley}, P. and {Micha{\l}owski}, M.~J.},
        title = "{Multi-wavelength de-blended Herschel view of the statistical properties of dusty star-forming galaxies across cosmic time}",
      journal = {\aap},
         year = 2019,
        month = apr,
       volume = {624},
          eid = {A98},
        pages = {A98},
          doi = {10.1051/0004-6361/201834093},
archivePrefix = {arXiv},
       eprint = {1902.09172},
 primaryClass = {astro-ph.GA},
       adsurl = {https://ui.adsabs.harvard.edu/abs/2019A&A...624A..98W}
}

@ARTICLE{2013A&A...553A.132M,
       author = {{Magnelli}, B. and {Popesso}, P. and {Berta}, S. and {Pozzi}, F. and {Elbaz}, D. and {Lutz}, D. and {Dickinson}, M. and {Altieri}, B. and {Andreani}, P. and {Aussel}, H. and {B{\'e}thermin}, M. and {Bongiovanni}, A. and {Cepa}, J. and {Charmandaris}, V. and {Chary}, R. -R. and {Cimatti}, A. and {Daddi}, E. and {F{\"o}rster Schreiber}, N.~M. and {Genzel}, R. and {Gruppioni}, C. and {Harwit}, M. and {Hwang}, H.~S. and {Ivison}, R.~J. and {Magdis}, G. and {Maiolino}, R. and {Murphy}, E. and {Nordon}, R. and {Pannella}, M. and {P{\'e}rez Garc{\'\i}a}, A. and {Poglitsch}, A. and {Rosario}, D. and {Sanchez-Portal}, M. and {Santini}, P. and {Scott}, D. and {Sturm}, E. and {Tacconi}, L.~J. and {Valtchanov}, I.},
        title = "{The deepest Herschel-PACS far-infrared survey: number counts and infrared luminosity functions from combined PEP/GOODS-H observations}",
      journal = {\aap},
         year = 2013,
        month = may,
       volume = {553},
          eid = {A132},
        pages = {A132},
          doi = {10.1051/0004-6361/201321371},
archivePrefix = {arXiv},
       eprint = {1303.4436},
 primaryClass = {astro-ph.CO},
       adsurl = {https://ui.adsabs.harvard.edu/abs/2013A&A...553A.132M}
}

@ARTICLE{2011A&A...528A..35M,
       author = {{Magnelli}, B. and {Elbaz}, D. and {Chary}, R.~R. and {Dickinson}, M. and {Le Borgne}, D. and {Frayer}, D.~T. and {Willmer}, C.~N.~A.},
        title = "{Evolution of the dusty infrared luminosity function from z = 0 to z = 2.3 using observations from Spitzer}",
      journal = {\aap},
         year = 2011,
        month = apr,
       volume = {528},
          eid = {A35},
        pages = {A35},
          doi = {10.1051/0004-6361/200913941},
archivePrefix = {arXiv},
       eprint = {1101.2467},
 primaryClass = {astro-ph.CO},
       adsurl = {https://ui.adsabs.harvard.edu/abs/2011A&A...528A..35M}
}

@ARTICLE{2016MNRAS.461.1100R,
       author = {{Rowan-Robinson}, Michael and {Oliver}, Seb and {Wang}, Lingyu and {Farrah}, Duncan and {Clements}, David L. and {Gruppioni}, Carlotta and {Marchetti}, Lucia and {Rigopoulou}, Dimitra and {Vaccari}, Mattia},
        title = "{The star formation rate density from z = 1 to 6}",
      journal = {\mnras},
         year = 2016,
        month = sep,
       volume = {461},
       number = {1},
        pages = {1100-1111},
          doi = {10.1093/mnras/stw1169},
archivePrefix = {arXiv},
       eprint = {1605.03937},
 primaryClass = {astro-ph.GA},
       adsurl = {https://ui.adsabs.harvard.edu/abs/2016MNRAS.461.1100R}
}

@ARTICLE{2018ApJ...869...71Z,
       author = {{Zavala}, J.~A. and {Casey}, C.~M. and {da Cunha}, E. and {Spilker}, J. and {Staguhn}, J. and {Hodge}, J. and {Drew}, P.~M.},
        title = "{Constraining the Volume Density of Dusty Star-forming Galaxies through the First 3 mm Number Counts from ALMA}",
      journal = {\apj},
         year = 2018,
        month = dec,
       volume = {869},
       number = {1},
          eid = {71},
        pages = {71},
          doi = {10.3847/1538-4357/aaecd2},
archivePrefix = {arXiv},
       eprint = {1810.12300},
 primaryClass = {astro-ph.GA},
       adsurl = {https://ui.adsabs.harvard.edu/abs/2018ApJ...869...71Z}
}

@ARTICLE{2018ApJ...862...77C,
       author = {{Casey}, Caitlin M. and {Zavala}, Jorge A. and {Spilker}, Justin and {da Cunha}, Elisabete and {Hodge}, Jacqueline and {Hung}, Chao-Ling and {Staguhn}, Johannes and {Finkelstein}, Steven L. and {Drew}, Patrick},
        title = "{The Brightest Galaxies in the Dark Ages: Galaxies{\textquoteright} Dust Continuum Emission during the Reionization Era}",
      journal = {\apj},
         year = 2018,
        month = jul,
       volume = {862},
       number = {1},
          eid = {77},
        pages = {77},
          doi = {10.3847/1538-4357/aac82d},
archivePrefix = {arXiv},
       eprint = {1805.10301},
 primaryClass = {astro-ph.GA},
       adsurl = {https://ui.adsabs.harvard.edu/abs/2018ApJ...862...77C}
}

@ARTICLE{2024A&A...690A..84T,
       author = {{Traina}, A. and {Magnelli}, B. and {Gruppioni}, C. and {Delvecchio}, I. and {Parente}, M. and {Calura}, F. and {Bisigello}, L. and {Feltre}, A. and {Pozzi}, F. and {Vallini}, L.},
        title = "{A$^{3}$COSMOS: Dust mass function and dust mass density at 0.5 < z < 6}",
      journal = {\aap},
         year = 2024,
        month = oct,
       volume = {690},
          eid = {A84},
        pages = {A84},
          doi = {10.1051/0004-6361/202451113},
archivePrefix = {arXiv},
       eprint = {2407.09607},
 primaryClass = {astro-ph.GA},
       adsurl = {https://ui.adsabs.harvard.edu/abs/2024A&A...690A..84T}
}

@ARTICLE{2023MNRAS.518.6142A,
       author = {{Algera}, Hiddo S.~B. and {Inami}, Hanae and {Oesch}, Pascal A. and {Sommovigo}, Laura and {Bouwens}, Rychard J. and {Topping}, Michael W. and {Schouws}, Sander and {Stefanon}, Mauro and {Stark}, Daniel P. and {Aravena}, Manuel and {Barrufet}, Laia and {da Cunha}, Elisabete and {Dayal}, Pratika and {Endsley}, Ryan and {Ferrara}, Andrea and {Fudamoto}, Yoshinobu and {Gonzalez}, Valentino and {Graziani}, Luca and {Hodge}, Jacqueline A. and {Hygate}, Alexander P.~S. and {de Looze}, Ilse and {Nanayakkara}, Themiya and {Schneider}, Raffaella and {van der Werf}, Paul P.},
        title = "{The ALMA REBELS survey: the dust-obscured cosmic star formation rate density at redshift 7}",
      journal = {\mnras},
         year = 2023,
        month = feb,
       volume = {518},
       number = {4},
        pages = {6142-6157},
          doi = {10.1093/mnras/stac3195},
archivePrefix = {arXiv},
       eprint = {2208.08243},
 primaryClass = {astro-ph.GA},
       adsurl = {https://ui.adsabs.harvard.edu/abs/2023MNRAS.518.6142A}
}

@ARTICLE{2025ApJ...980...12S,
       author = {{Sun}, Fengwu and {Wang}, Feige and {Yang}, Jinyi and {Champagne}, Jaclyn B. and {Decarli}, Roberto and {Fan}, Xiaohui and {Ba{\~n}ados}, Eduardo and {Cai}, Zheng and {Colina}, Luis and {Egami}, Eiichi and {Hennawi}, Joseph F. and {Jin}, Xiangyu and {Jun}, Hyunsung D. and {Khusanova}, Yana and {Li}, Mingyu and {Li}, Zihao and {Lin}, Xiaojing and {Liu}, Weizhe and {Meyer}, Romain A. and {Pudoka}, Maria A. and {Rieke}, George H. and {Shen}, Yue and {Tee}, Wei Leong and {Venemans}, Bram and {Walter}, Fabian and {Wu}, Yunjing and {Zhang}, Huanian and {Zou}, Siwei},
        title = "{A SPectroscopic Survey of Biased Halos in the Reionization Era (ASPIRE): Spectroscopically Complete Census of Obscured Cosmic Star Formation Rate Density at z = 4{\textendash}6}",
      journal = {\apj},
         year = 2025,
        month = feb,
       volume = {980},
       number = {1},
          eid = {12},
        pages = {12},
          doi = {10.3847/1538-4357/ad9d0e},
archivePrefix = {arXiv},
       eprint = {2412.06894},
 primaryClass = {astro-ph.GA},
       adsurl = {https://ui.adsabs.harvard.edu/abs/2025ApJ...980...12S}
}

@ARTICLE{2021A&A...649A.152K,
       author = {{Khusanova}, Y. and {Bethermin}, M. and {Le F{\`e}vre}, O. and {Capak}, P. and {Faisst}, A.~L. and {Schaerer}, D. and {Silverman}, J.~D. and {Cassata}, P. and {Yan}, L. and {Ginolfi}, M. and {Fudamoto}, Y. and {Loiacono}, F. and {Amorin}, R. and {Bardelli}, S. and {Boquien}, M. and {Cimatti}, A. and {Dessauges-Zavadsky}, M. and {Gruppioni}, C. and {Hathi}, N.~P. and {Jones}, G.~C. and {Koekemoer}, A.~M. and {Lagache}, G. and {Maiolino}, R. and {Lemaux}, B.~C. and {Oesch}, P. and {Pozzi}, F. and {Riechers}, D.~A. and {Romano}, M. and {Talia}, M. and {Toft}, S. and {Vergani}, D. and {Zamorani}, G. and {Zucca}, E.},
        title = "{The ALPINE-ALMA [CII] survey. Obscured star formation rate density and main sequence of star-forming galaxies at z > 4}",
      journal = {\aap},
         year = 2021,
        month = may,
       volume = {649},
          eid = {A152},
        pages = {A152},
          doi = {10.1051/0004-6361/202038944},
archivePrefix = {arXiv},
       eprint = {2007.08384},
 primaryClass = {astro-ph.GA},
       adsurl = {https://ui.adsabs.harvard.edu/abs/2021A&A...649A.152K}
}

@ARTICLE{2018ApJ...862...78C,
       author = {{Casey}, Caitlin M. and {Hodge}, Jacqueline and {Zavala}, Jorge A. and {Spilker}, Justin and {da Cunha}, Elisabete and {Staguhn}, Johannes and {Finkelstein}, Steven L. and {Drew}, Patrick},
        title = "{An Analysis of ALMA Deep Fields and the Perceived Dearth of High-z Galaxies}",
      journal = {\apj},
         year = 2018,
        month = jul,
       volume = {862},
       number = {1},
          eid = {78},
        pages = {78},
          doi = {10.3847/1538-4357/aacd11},
archivePrefix = {arXiv},
       eprint = {1806.05603},
 primaryClass = {astro-ph.GA},
       adsurl = {https://ui.adsabs.harvard.edu/abs/2018ApJ...862...78C}
}

@ARTICLE{2017A&A...602A...5N,
       author = {{Novak}, M. and {Smol{\v{c}}i{\'c}}, V. and {Delhaize}, J. and {Delvecchio}, I. and {Zamorani}, G. and {Baran}, N. and {Bondi}, M. and {Capak}, P. and {Carilli}, C.~L. and {Ciliegi}, P. and {Civano}, F. and {Ilbert}, O. and {Karim}, A. and {Laigle}, C. and {Le F{\`e}vre}, O. and {Marchesi}, S. and {McCracken}, H. and {Miettinen}, O. and {Salvato}, M. and {Sargent}, M. and {Schinnerer}, E. and {Tasca}, L.},
        title = "{The VLA-COSMOS 3 GHz Large Project: Cosmic star formation history since z   5}",
      journal = {\aap},
         year = 2017,
        month = jun,
       volume = {602},
          eid = {A5},
        pages = {A5},
          doi = {10.1051/0004-6361/201629436},
archivePrefix = {arXiv},
       eprint = {1703.09724},
 primaryClass = {astro-ph.GA},
       adsurl = {https://ui.adsabs.harvard.edu/abs/2017A&A...602A...5N}
}

@ARTICLE{2018ApJ...853..172L,
       author = {{Liu}, Daizhong and {Daddi}, Emanuele and {Dickinson}, Mark and {Owen}, Frazer and {Pannella}, Maurilio and {Sargent}, Mark and {B{\'e}thermin}, Matthieu and {Magdis}, Georgios and {Gao}, Yu and {Shu}, Xinwen and {Wang}, Tao and {Jin}, Shuowen and {Inami}, Hanae},
        title = "{{\textquotedblleft}Super-deblended{\textquotedblright} Dust Emission in Galaxies. I. The GOODS-North Catalog and the Cosmic Star Formation Rate Density out to Redshift 6}",
      journal = {\apj},
         year = 2018,
        month = feb,
       volume = {853},
       number = {2},
          eid = {172},
        pages = {172},
          doi = {10.3847/1538-4357/aaa600},
archivePrefix = {arXiv},
       eprint = {1703.05281},
 primaryClass = {astro-ph.GA},
       adsurl = {https://ui.adsabs.harvard.edu/abs/2018ApJ...853..172L}
}

@ARTICLE{2022ApJ...941...10V,
       author = {{van der Vlugt}, D. and {Hodge}, J.~A. and {Algera}, H.~S.~B. and {Smail}, I. and {Leslie}, S.~K. and {Radcliffe}, J.~F. and {Riechers}, D.~A. and {R{\"o}ttgering}, H.},
        title = "{An Ultra-deep Multiband Very Large Array (VLA) Survey of the Faint Radio Sky (COSMOS-XS): New Constraints on the Cosmic Star Formation History}",
      journal = {\apj},
         year = 2022,
        month = dec,
       volume = {941},
       number = {1},
          eid = {10},
        pages = {10},
          doi = {10.3847/1538-4357/ac99db},
archivePrefix = {arXiv},
       eprint = {2204.04167},
 primaryClass = {astro-ph.GA},
       adsurl = {https://ui.adsabs.harvard.edu/abs/2022ApJ...941...10V}
}

@ARTICLE{2020ApJ...902..112B,
       author = {{Bouwens}, Rychard and {Gonz{\'a}lez-L{\'o}pez}, Jorge and {Aravena}, Manuel and {Decarli}, Roberto and {Novak}, Mladen and {Stefanon}, Mauro and {Walter}, Fabian and {Boogaard}, Leindert and {Carilli}, Chris and {Dudzevi{\v{c}}i{\={u}}t{\.{e}}}, Ugn{\.{e}} and {Smail}, Ian and {Daddi}, Emanuele and {da Cunha}, Elisabete and {Ivison}, Rob and {Nanayakkara}, Themiya and {Cortes}, Paulo and {Cox}, Pierre and {Inami}, Hanae and {Oesch}, Pascal and {Popping}, Gerg{\"o} and {Riechers}, Dominik and {van der Werf}, Paul and {Weiss}, Axel and {Fudamoto}, Yoshi and {Wagg}, Jeff},
        title = "{The ALMA Spectroscopic Survey Large Program: The Infrared Excess of z = 1.5-10 UV-selected Galaxies and the Implied High-redshift Star Formation History}",
      journal = {\apj},
         year = 2020,
        month = oct,
       volume = {902},
       number = {2},
          eid = {112},
        pages = {112},
          doi = {10.3847/1538-4357/abb830},
archivePrefix = {arXiv},
       eprint = {2009.10727},
 primaryClass = {astro-ph.GA},
       adsurl = {https://ui.adsabs.harvard.edu/abs/2020ApJ...902..112B}
}

@ARTICLE{2023A&A...672A..18X,
       author = {{Xiao}, M. -Y. and {Elbaz}, D. and {G{\'o}mez-Guijarro}, C. and {Leroy}, L. and {Bing}, L. -J. and {Daddi}, E. and {Magnelli}, B. and {Franco}, M. and {Zhou}, L. and {Dickinson}, M. and {Wang}, T. and {Rujopakarn}, W. and {Magdis}, G.~E. and {Treister}, E. and {Inami}, H. and {Demarco}, R. and {Sargent}, M.~T. and {Shu}, X. and {Kartaltepe}, J.~S. and {Alexander}, D.~M. and {B{\'e}thermin}, M. and {Bournaud}, F. and {Ciesla}, L. and {Ferguson}, H.~C. and {Finkelstein}, S.~L. and {Giavalisco}, M. and {Gu}, Q. -S. and {Iono}, D. and {Juneau}, S. and {Lagache}, G. and {Leiton}, R. and {Messias}, H. and {Motohara}, K. and {Mullaney}, J. and {Nagar}, N. and {Pannella}, M. and {Papovich}, C. and {Pope}, A. and {Schreiber}, C. and {Silverman}, J.},
        title = "{The hidden side of cosmic star formation at z > 3. Bridging optically dark and Lyman-break galaxies with GOODS-ALMA}",
      journal = {\aap},
         year = 2023,
        month = apr,
       volume = {672},
          eid = {A18},
        pages = {A18},
          doi = {10.1051/0004-6361/202245100},
archivePrefix = {arXiv},
       eprint = {2210.03135},
 primaryClass = {astro-ph.GA},
       adsurl = {https://ui.adsabs.harvard.edu/abs/2023A&A...672A..18X}
}

@ARTICLE{2023ApJ...946L..16P,
       author = {{P{\'e}rez-Gonz{\'a}lez}, Pablo G. and {Barro}, Guillermo and {Annunziatella}, Marianna and {Costantin}, Luca and {Garc{\'\i}a-Argum{\'a}nez}, {\'A}ngela and {McGrath}, Elizabeth J. and {M{\'e}rida}, Rosa M. and {Zavala}, Jorge A. and {Arrabal Haro}, Pablo and {Bagley}, Micaela B. and {Backhaus}, Bren E. and {Behroozi}, Peter and {Bell}, Eric F. and {Bisigello}, Laura and {Buat}, V{\'e}ronique and {Calabr{\`o}}, Antonello and {Casey}, Caitlin M. and {Cleri}, Nikko J. and {Coogan}, Rosemary T. and {Cooper}, M.~C. and {Cooray}, Asantha R. and {Dekel}, Avishai and {Dickinson}, Mark and {Elbaz}, David and {Ferguson}, Henry C. and {Finkelstein}, Steven L. and {Fontana}, Adriano and {Franco}, Maximilien and {Gardner}, Jonathan P. and {Giavalisco}, Mauro and {G{\'o}mez-Guijarro}, Carlos and {Grazian}, Andrea and {Grogin}, Norman A. and {Guo}, Yuchen and {Huertas-Company}, Marc and {Jogee}, Shardha and {Kartaltepe}, Jeyhan S. and {Kewley}, Lisa J. and {Kirkpatrick}, Allison and {Kocevski}, Dale D. and {Koekemoer}, Anton M. and {Long}, Arianna S. and {Lotz}, Jennifer M. and {Lucas}, Ray A. and {Papovich}, Casey and {Pirzkal}, Nor and {Ravindranath}, Swara and {Somerville}, Rachel S. and {Tacchella}, Sandro and {Trump}, Jonathan R. and {Wang}, Weichen and {Wilkins}, Stephen M. and {Wuyts}, Stijn and {Yang}, Guang and {Yung}, L.~Y. Aaron},
        title = "{CEERS Key Paper. IV. A Triality in the Nature of HST-dark Galaxies}",
      journal = {\apjl},
         year = 2023,
        month = mar,
       volume = {946},
       number = {1},
          eid = {L16},
        pages = {L16},
          doi = {10.3847/2041-8213/acb3a5},
archivePrefix = {arXiv},
       eprint = {2211.00045},
 primaryClass = {astro-ph.GA},
       adsurl = {https://ui.adsabs.harvard.edu/abs/2023ApJ...946L..16P}
}

@ARTICLE{2019ApJ...884..154W,
       author = {{Williams}, Christina C. and {Labbe}, Ivo and {Spilker}, Justin and {Stefanon}, Mauro and {Leja}, Joel and {Whitaker}, Katherine and {Bezanson}, Rachel and {Narayanan}, Desika and {Oesch}, Pascal and {Weiner}, Benjamin},
        title = "{Discovery of a Dark, Massive, ALMA-only Galaxy at z {\ensuremath{\sim}} 5-6 in a Tiny 3 mm Survey}",
      journal = {\apj},
         year = 2019,
        month = oct,
       volume = {884},
       number = {2},
          eid = {154},
        pages = {154},
          doi = {10.3847/1538-4357/ab44aa},
archivePrefix = {arXiv},
       eprint = {1905.11996},
 primaryClass = {astro-ph.GA},
       adsurl = {https://ui.adsabs.harvard.edu/abs/2019ApJ...884..154W}
}

@ARTICLE{2021A&A...646A..76L,
       author = {{Loiacono}, Federica and {Decarli}, Roberto and {Gruppioni}, Carlotta and {Talia}, Margherita and {Cimatti}, Andrea and {Zamorani}, Gianni and {Pozzi}, Francesca and {Yan}, Lin and {Lemaux}, Brian C. and {Riechers}, Dominik A. and {Le F{\`e}vre}, Olivier and {B{\`e}thermin}, Matthieu and {Capak}, Peter and {Cassata}, Paolo and {Faisst}, Andreas and {Schaerer}, Daniel and {Silverman}, John D. and {Bardelli}, Sandro and {Boquien}, M{\'e}d{\'e}ric and {Burkutean}, Sandra and {Dessauges-Zavadsky}, Miroslava and {Fudamoto}, Yoshinobu and {Fujimoto}, Seiji and {Ginolfi}, Michele and {Hathi}, Nimish P. and {Jones}, Gareth C. and {Khusanova}, Yana and {Koekemoer}, Anton M. and {Lagache}, Guilaine and {Lubin}, Lori M. and {Massardi}, Marcella and {Oesch}, Pascal and {Romano}, Michael and {Vallini}, Livia and {Vergani}, Daniela and {Zucca}, Elena},
        title = "{The ALPINE-ALMA [C II] survey. Luminosity function of serendipitous [C II] line emitters at z {\ensuremath{\sim}} 5}",
      journal = {\aap},
         year = 2021,
        month = feb,
       volume = {646},
          eid = {A76},
        pages = {A76},
          doi = {10.1051/0004-6361/202038607},
archivePrefix = {arXiv},
       eprint = {2006.04837},
 primaryClass = {astro-ph.GA},
       adsurl = {https://ui.adsabs.harvard.edu/abs/2021A&A...646A..76L}
}

@ARTICLE{2017MNRAS.469..492M,
       author = {{Micha{\l}owski}, Micha{\l} J. and {Dunlop}, J.~S. and {Koprowski}, M.~P. and {Cirasuolo}, M. and {Geach}, J.~E. and {Bowler}, R.~A.~A. and {Mortlock}, A. and {Caputi}, K.~I. and {Aretxaga}, I. and {Arumugam}, V. and {Chen}, Chian-Chou and {McLure}, R.~J. and {Birkinshaw}, M. and {Bourne}, N. and {Farrah}, D. and {Ibar}, E. and {van der Werf}, P. and {Zemcov}, M.},
        title = "{The SCUBA-2 Cosmology Legacy Survey: the nature of bright submm galaxies from 2 deg$^{2}$ of 850-{\ensuremath{\mu}}m imaging}",
      journal = {\mnras},
         year = 2017,
        month = jul,
       volume = {469},
       number = {1},
        pages = {492-515},
          doi = {10.1093/mnras/stx861},
archivePrefix = {arXiv},
       eprint = {1610.02409},
 primaryClass = {astro-ph.GA},
       adsurl = {https://ui.adsabs.harvard.edu/abs/2017MNRAS.469..492M}
}

@ARTICLE{2020MNRAS.494.3828D,
       author = {{Dudzevi{\v{c}}i{\={u}}t{\.{e}}}, U. and {Smail}, Ian and {Swinbank}, A.~M. and {Stach}, S.~M. and {Almaini}, O. and {da Cunha}, E. and {An}, Fang Xia and {Arumugam}, V. and {Birkin}, J. and {Blain}, A.~W. and {Chapman}, S.~C. and {Chen}, C. -C. and {Conselice}, C.~J. and {Coppin}, K.~E.~K. and {Dunlop}, J.~S. and {Farrah}, D. and {Geach}, J.~E. and {Gullberg}, B. and {Hartley}, W.~G. and {Hodge}, J.~A. and {Ivison}, R.~J. and {Maltby}, D.~T. and {Scott}, D. and {Simpson}, C.~J. and {Simpson}, J.~M. and {Thomson}, A.~P. and {Walter}, F. and {Wardlow}, J.~L. and {Weiss}, A. and {van der Werf}, P.},
        title = "{An ALMA survey of the SCUBA-2 CLS UDS field: physical properties of 707 sub-millimetre galaxies}",
      journal = {\mnras},
         year = 2020,
        month = may,
       volume = {494},
       number = {3},
        pages = {3828-3860},
          doi = {10.1093/mnras/staa769},
archivePrefix = {arXiv},
       eprint = {1910.07524},
 primaryClass = {astro-ph.GA},
       adsurl = {https://ui.adsabs.harvard.edu/abs/2020MNRAS.494.3828D}
}

@ARTICLE{2023MNRAS.523.6082C,
       author = {{Cochrane}, R.~K. and {Kondapally}, R. and {Best}, P.~N. and {Sabater}, J. and {Duncan}, K.~J. and {Smith}, D.~J.~B. and {Hardcastle}, M.~J. and {R{\"o}ttgering}, H.~J.~A. and {Prandoni}, I. and {Haskell}, P. and {G{\"u}rkan}, G. and {Miley}, G.~K.},
        title = "{The LOFAR Two-metre Sky Survey: the radio view of the cosmic star formation history}",
      journal = {\mnras},
         year = 2023,
        month = aug,
       volume = {523},
       number = {4},
        pages = {6082-6102},
          doi = {10.1093/mnras/stad1602},
archivePrefix = {arXiv},
       eprint = {2305.15510},
 primaryClass = {astro-ph.GA},
       adsurl = {https://ui.adsabs.harvard.edu/abs/2023MNRAS.523.6082C}
}

@ARTICLE{2025arXiv250103217D,
       author = {{Donnan}, C.~T. and {Dunlop}, J.~S. and {McLure}, R.~J. and {McLeod}, D.~J. and {Cullen}, F.},
        title = "{No evidence (yet) for increased star-formation efficiency at early times}",
      journal = {arXiv e-prints},
         year = 2025,
        month = jan,
          eid = {arXiv:2501.03217},
        pages = {arXiv:2501.03217},
          doi = {10.48550/arXiv.2501.03217},
archivePrefix = {arXiv},
       eprint = {2501.03217},
 primaryClass = {astro-ph.GA},
       adsurl = {https://ui.adsabs.harvard.edu/abs/2025arXiv250103217D}
}

@ARTICLE{2024MNRAS.533.3222D,
       author = {{Donnan}, C.~T. and {McLure}, R.~J. and {Dunlop}, J.~S. and {McLeod}, D.~J. and {Magee}, D. and {Arellano-C{\'o}rdova}, K.~Z. and {Barrufet}, L. and {Begley}, R. and {Bowler}, R.~A.~A. and {Carnall}, A.~C. and {Cullen}, F. and {Ellis}, R.~S. and {Fontana}, A. and {Illingworth}, G.~D. and {Grogin}, N.~A. and {Hamadouche}, M.~L. and {Koekemoer}, A.~M. and {Liu}, F. -Y. and {Mason}, C. and {Santini}, P. and {Stanton}, T.~M.},
        title = "{JWST PRIMER: a new multifield determination of the evolving galaxy UV luminosity function at redshifts z ≃ 9 - 15}",
      journal = {\mnras},
         year = 2024,
        month = sep,
       volume = {533},
       number = {3},
        pages = {3222-3237},
          doi = {10.1093/mnras/stae2037},
archivePrefix = {arXiv},
       eprint = {2403.03171},
 primaryClass = {astro-ph.GA},
       adsurl = {https://ui.adsabs.harvard.edu/abs/2024MNRAS.533.3222D}
}

@ARTICLE{2024MNRAS.531..997C,
       author = {{Cullen}, F. and {McLeod}, D.~J. and {McLure}, R.~J. and {Dunlop}, J.~S. and {Donnan}, C.~T. and {Carnall}, A.~C. and {Keating}, L.~C. and {Magee}, D. and {Arellano-Cordova}, K.~Z. and {Bowler}, R.~A.~A. and {Begley}, R. and {Flury}, S.~R. and {Hamadouche}, M.~L. and {Stanton}, T.~M.},
        title = "{The ultraviolet continuum slopes of high-redshift galaxies: evidence for the emergence of dust-free stellar populations at z > 10}",
      journal = {\mnras},
         year = 2024,
        month = jun,
       volume = {531},
       number = {1},
        pages = {997-1020},
          doi = {10.1093/mnras/stae1211},
archivePrefix = {arXiv},
       eprint = {2311.06209},
 primaryClass = {astro-ph.GA},
       adsurl = {https://ui.adsabs.harvard.edu/abs/2024MNRAS.531..997C}
}

@ARTICLE{2025arXiv250111099C,
       author = {{Cullen}, F. and {Carnall}, A.~C. and {Scholte}, D. and {McLeod}, D.~J. and {McLure}, R.~J. and {Arellano-C{\'o}rdova}, K.~Z. and {Stanton}, T. M and {Donnan}, C.~T. and {Dunlop}, J.~S. and {Shapley}, A.~E. and {Barrufet}, L. and {Begley}, R. and {Bondestam}, C. and {Cirasuolo}, M. and {Leung}, H. -H. and {Pollock}, C.~L. and {Stevenson}, S.},
        title = "{The JWST EXCELS survey: an extremely metal-poor galaxy at $z=8.271$ hosting an unusual population of massive stars}",
      journal = {arXiv e-prints},
         year = 2025,
        month = jan,
          eid = {arXiv:2501.11099},
        pages = {arXiv:2501.11099},
          doi = {10.48550/arXiv.2501.11099},
archivePrefix = {arXiv},
       eprint = {2501.11099},
 primaryClass = {astro-ph.GA},
       adsurl = {https://ui.adsabs.harvard.edu/abs/2025arXiv250111099C}
}

@ARTICLE{2025A&A...694A.215F,
       author = {{Ferrara}, A. and {Carniani}, S. and {di Mascia}, F. and {Bouwens}, R.~J. and {Oesch}, P. and {Schouws}, S.},
        title = "{ALMA observations of super-early galaxies: Attenuation-free model predictions}",
      journal = {\aap},
         year = 2025,
        month = feb,
       volume = {694},
          eid = {A215},
        pages = {A215},
          doi = {10.1051/0004-6361/202452368},
archivePrefix = {arXiv},
       eprint = {2409.17223},
 primaryClass = {astro-ph.GA},
       adsurl = {https://ui.adsabs.harvard.edu/abs/2025A&A...694A.215F}
}

@ARTICLE{2018MNRAS.476.3991M,
       author = {{McLure}, R.~J. and {Dunlop}, J.~S. and {Cullen}, F. and {Bourne}, N. and {Best}, P.~N. and {Khochfar}, S. and {Bowler}, R.~A.~A. and {Biggs}, A.~D. and {Geach}, J.~E. and {Scott}, D. and {Micha{\l}owski}, M.~J. and {Rujopakarn}, W. and {van Kampen}, E. and {Kirkpatrick}, A. and {Pope}, A.},
        title = "{Dust attenuation in 2 < z < 3 star-forming galaxies from deep ALMA observations of the Hubble Ultra Deep Field}",
      journal = {\mnras},
         year = 2018,
        month = may,
       volume = {476},
       number = {3},
        pages = {3991-4006},
          doi = {10.1093/mnras/sty522},
archivePrefix = {arXiv},
       eprint = {1709.06102},
 primaryClass = {astro-ph.GA},
       adsurl = {https://ui.adsabs.harvard.edu/abs/2018MNRAS.476.3991M}
}

@ARTICLE{2017MNRAS.465.1789G,
       author = {{Geach}, J.~E. and {Dunlop}, J.~S. and {Halpern}, M. and {Smail}, Ian and {van der Werf}, P. and {Alexander}, D.~M. and {Almaini}, O. and {Aretxaga}, I. and {Arumugam}, V. and {Asboth}, V. and {Banerji}, M. and {Beanlands}, J. and {Best}, P.~N. and {Blain}, A.~W. and {Birkinshaw}, M. and {Chapin}, E.~L. and {Chapman}, S.~C. and {Chen}, C. -C. and {Chrysostomou}, A. and {Clarke}, C. and {Clements}, D.~L. and {Conselice}, C. and {Coppin}, K.~E.~K. and {Cowley}, W.~I. and {Danielson}, A.~L.~R. and {Eales}, S. and {Edge}, A.~C. and {Farrah}, D. and {Gibb}, A. and {Harrison}, C.~M. and {Hine}, N.~K. and {Hughes}, D. and {Ivison}, R.~J. and {Jarvis}, M. and {Jenness}, T. and {Jones}, S.~F. and {Karim}, A. and {Koprowski}, M. and {Knudsen}, K.~K. and {Lacey}, C.~G. and {Mackenzie}, T. and {Marsden}, G. and {McAlpine}, K. and {McMahon}, R. and {Meijerink}, R. and {Micha{\l}owski}, M.~J. and {Oliver}, S.~J. and {Page}, M.~J. and {Peacock}, J.~A. and {Rigopoulou}, D. and {Robson}, E.~I. and {Roseboom}, I. and {Rotermund}, K. and {Scott}, Douglas and {Serjeant}, S. and {Simpson}, C. and {Simpson}, J.~M. and {Smith}, D.~J.~B. and {Spaans}, M. and {Stanley}, F. and {Stevens}, J.~A. and {Swinbank}, A.~M. and {Targett}, T. and {Thomson}, A.~P. and {Valiante}, E. and {Wake}, D.~A. and {Webb}, T.~M.~A. and {Willott}, C. and {Zavala}, J.~A. and {Zemcov}, M.},
        title = "{The SCUBA-2 Cosmology Legacy Survey: 850 {\ensuremath{\mu}}m maps, catalogues and number counts}",
      journal = {\mnras},
         year = 2017,
        month = feb,
       volume = {465},
       number = {2},
        pages = {1789-1806},
          doi = {10.1093/mnras/stw2721},
archivePrefix = {arXiv},
       eprint = {1607.03904},
 primaryClass = {astro-ph.GA},
       adsurl = {https://ui.adsabs.harvard.edu/abs/2017MNRAS.465.1789G}
}

@ARTICLE{1974ApJS...27...21O,
       author = {{Oke}, J.~B.},
        title = "{Absolute Spectral Energy Distributions for White Dwarfs}",
      journal = {\apjs},
         year = 1974,
        month = feb,
       volume = {27},
        pages = {21},
          doi = {10.1086/190287},
       adsurl = {https://ui.adsabs.harvard.edu/abs/1974ApJS...27...21O}
}

@ARTICLE{1983ApJ...266..713O,
       author = {{Oke}, J.~B. and {Gunn}, J.~E.},
        title = "{Secondary standard stars for absolute spectrophotometry.}",
      journal = {\apj},
         year = 1983,
        month = mar,
       volume = {266},
        pages = {713-717},
          doi = {10.1086/160817},
       adsurl = {https://ui.adsabs.harvard.edu/abs/1983ApJ...266..713O}
}

@ARTICLE{2011ApJS..197...36K,
       author = {{Koekemoer}, Anton M. and {Faber}, S.~M. and {Ferguson}, Henry C. and {Grogin}, Norman A. and {Kocevski}, Dale D. and {Koo}, David C. and {Lai}, Kamson and {Lotz}, Jennifer M. and {Lucas}, Ray A. and {McGrath}, Elizabeth J. and {Ogaz}, Sara and {Rajan}, Abhijith and {Riess}, Adam G. and {Rodney}, Steve A. and {Strolger}, Louis and {Casertano}, Stefano and {Castellano}, Marco and {Dahlen}, Tomas and {Dickinson}, Mark and {Dolch}, Timothy and {Fontana}, Adriano and {Giavalisco}, Mauro and {Grazian}, Andrea and {Guo}, Yicheng and {Hathi}, Nimish P. and {Huang}, Kuang-Han and {van der Wel}, Arjen and {Yan}, Hao-Jing and {Acquaviva}, Viviana and {Alexander}, David M. and {Almaini}, Omar and {Ashby}, Matthew L.~N. and {Barden}, Marco and {Bell}, Eric F. and {Bournaud}, Fr{\'e}d{\'e}ric and {Brown}, Thomas M. and {Caputi}, Karina I. and {Cassata}, Paolo and {Challis}, Peter J. and {Chary}, Ranga-Ram and {Cheung}, Edmond and {Cirasuolo}, Michele and {Conselice}, Christopher J. and {Roshan Cooray}, Asantha and {Croton}, Darren J. and {Daddi}, Emanuele and {Dav{\'e}}, Romeel and {de Mello}, Duilia F. and {de Ravel}, Loic and {Dekel}, Avishai and {Donley}, Jennifer L. and {Dunlop}, James S. and {Dutton}, Aaron A. and {Elbaz}, David and {Fazio}, Giovanni G. and {Filippenko}, Alexei V. and {Finkelstein}, Steven L. and {Frazer}, Chris and {Gardner}, Jonathan P. and {Garnavich}, Peter M. and {Gawiser}, Eric and {Gruetzbauch}, Ruth and {Hartley}, Will G. and {H{\"a}ussler}, Boris and {Herrington}, Jessica and {Hopkins}, Philip F. and {Huang}, Jia-Sheng and {Jha}, Saurabh W. and {Johnson}, Andrew and {Kartaltepe}, Jeyhan S. and {Khostovan}, Ali A. and {Kirshner}, Robert P. and {Lani}, Caterina and {Lee}, Kyoung-Soo and {Li}, Weidong and {Madau}, Piero and {McCarthy}, Patrick J. and {McIntosh}, Daniel H. and {McLure}, Ross J. and {McPartland}, Conor and {Mobasher}, Bahram and {Moreira}, Heidi and {Mortlock}, Alice and {Moustakas}, Leonidas A. and {Mozena}, Mark and {Nandra}, Kirpal and {Newman}, Jeffrey A. and {Nielsen}, Jennifer L. and {Niemi}, Sami and {Noeske}, Kai G. and {Papovich}, Casey J. and {Pentericci}, Laura and {Pope}, Alexandra and {Primack}, Joel R. and {Ravindranath}, Swara and {Reddy}, Naveen A. and {Renzini}, Alvio and {Rix}, Hans-Walter and {Robaina}, Aday R. and {Rosario}, David J. and {Rosati}, Piero and {Salimbeni}, Sara and {Scarlata}, Claudia and {Siana}, Brian and {Simard}, Luc and {Smidt}, Joseph and {Snyder}, Diana and {Somerville}, Rachel S. and {Spinrad}, Hyron and {Straughn}, Amber N. and {Telford}, Olivia and {Teplitz}, Harry I. and {Trump}, Jonathan R. and {Vargas}, Carlos and {Villforth}, Carolin and {Wagner}, Cory R. and {Wandro}, Pat and {Wechsler}, Risa H. and {Weiner}, Benjamin J. and {Wiklind}, Tommy and {Wild}, Vivienne and {Wilson}, Grant and {Wuyts}, Stijn and {Yun}, Min S.},
        title = "{CANDELS: The Cosmic Assembly Near-infrared Deep Extragalactic Legacy Survey{\textemdash}The Hubble Space Telescope Observations, Imaging Data Products, and Mosaics}",
      journal = {\apjs},
         year = 2011,
        month = dec,
       volume = {197},
       number = {2},
          eid = {36},
        pages = {36},
          doi = {10.1088/0067-0049/197/2/36},
archivePrefix = {arXiv},
       eprint = {1105.3754},
 primaryClass = {astro-ph.CO},
       adsurl = {https://ui.adsabs.harvard.edu/abs/2011ApJS..197...36K}
}

@misc{khostovan2025cosmosspectroscopicredshiftcompilation,
      title={COSMOS Spectroscopic Redshift Compilation (First Data Release): 165k Redshifts Encompassing Two Decades of Spectroscopy}, 
      author={Ali Ahmad Khostovan and Jeyhan S. Kartaltepe and Mara Salvato and Olivier Ilbert and Caitlin M. Casey and Hiddo Algera and Jacqueline Antwi-Danso and Andrew Battisti and Malte Brinch and Marcella Brusa and Antonello Calabro and Peter L. Capak and Nima Chartab and Olivia R. Cooper and Isa G. Cox and Behnam Darvish and Nicole E. Drakos and Andreas L. Faisst and Matthew R. George and Ghassem Gozaliasl and Santosh Harish and Gunther Hasinger and Hossein Hatamnia and Angela Iovino and Shuowen Jin and Daichi Kashino and Anton M. Koekemoer and Ronaldo Laishram and Khee-Gan Lee and Jitrapon Lertprasertpong and Simon J. Lilly and Daniel C. Masters and Bahram Mobasher and Tohru Nagao and Masato Onodera and Yingjie Peng and David B. Sanders and Ryan L. Sanders and Zahra Sattari and Nick Scoville and Ekta A. Shah and John D. Silverman and Nao Suzuki and Masayuki Tanaka and Sune Toft and Benny Trakhtenbrot and Jonathan R. Trump and Mattia Vaccari and Francesco Valentino and Brittany N. Vanderhoof and John R. Weaver and Min S. Yun and Jorge A. Zavala},
      year={2025},
      eprint={2503.00120},
      archivePrefix={arXiv},
      primaryClass={astro-ph.GA},
      url={https://arxiv.org/abs/2503.00120}, 
}

@ARTICLE{2016ApJ...828...21N,
       author = {{Nanayakkara}, Themiya and {Glazebrook}, Karl and {Kacprzak}, Glenn G. and {Yuan}, Tiantian and {Tran}, Kim-Vy and {Spitler}, Lee and {Kewley}, Lisa and {Straatman}, Caroline and {Cowley}, Michael and {Fisher}, David and {Labbe}, Ivo and {Tomczak}, Adam and {Allen}, Rebecca and {Alcorn}, Leo},
        title = "{ZFIRE: A KECK/MOSFIRE Spectroscopic Survey of Galaxies in Rich Environments at z {\ensuremath{\sim}} 2}",
      journal = {\apj},
         year = 2016,
        month = sep,
       volume = {828},
       number = {1},
          eid = {21},
        pages = {21},
          doi = {10.3847/0004-637X/828/1/21},
archivePrefix = {arXiv},
       eprint = {1607.00013},
 primaryClass = {astro-ph.GA},
       adsurl = {https://ui.adsabs.harvard.edu/abs/2016ApJ...828...21N}
}

@ARTICLE{2015A&A...576A..79L,
       author = {{Le F{\`e}vre}, O. and {Tasca}, L.~A.~M. and {Cassata}, P. and {Garilli}, B. and {Le Brun}, V. and {Maccagni}, D. and {Pentericci}, L. and {Thomas}, R. and {Vanzella}, E. and {Zamorani}, G. and {Zucca}, E. and {Amorin}, R. and {Bardelli}, S. and {Capak}, P. and {Cassar{\`a}}, L. and {Castellano}, M. and {Cimatti}, A. and {Cuby}, J.~G. and {Cucciati}, O. and {de la Torre}, S. and {Durkalec}, A. and {Fontana}, A. and {Giavalisco}, M. and {Grazian}, A. and {Hathi}, N.~P. and {Ilbert}, O. and {Lemaux}, B.~C. and {Moreau}, C. and {Paltani}, S. and {Ribeiro}, B. and {Salvato}, M. and {Schaerer}, D. and {Scodeggio}, M. and {Sommariva}, V. and {Talia}, M. and {Taniguchi}, Y. and {Tresse}, L. and {Vergani}, D. and {Wang}, P.~W. and {Charlot}, S. and {Contini}, T. and {Fotopoulou}, S. and {L{\'o}pez-Sanjuan}, C. and {Mellier}, Y. and {Scoville}, N.},
        title = "{The VIMOS Ultra-Deep Survey: \raisebox{-0.5ex}\textasciitilde10 000 galaxies with spectroscopic redshifts to study galaxy assembly at early epochs 2 < z ≃ 6}",
      journal = {\aap},
         year = 2015,
        month = apr,
       volume = {576},
          eid = {A79},
        pages = {A79},
          doi = {10.1051/0004-6361/201423829},
archivePrefix = {arXiv},
       eprint = {1403.3938},
 primaryClass = {astro-ph.CO},
       adsurl = {https://ui.adsabs.harvard.edu/abs/2015A&A...576A..79L}
}

@ARTICLE{2017A&A...600A.110T,
       author = {{Tasca}, L.~A.~M. and {Le F{\`e}vre}, O. and {Ribeiro}, B. and {Thomas}, R. and {Moreau}, C. and {Cassata}, P. and {Garilli}, B. and {Le Brun}, V. and {Lemaux}, B.~C. and {Maccagni}, D. and {Pentericci}, L. and {Schaerer}, D. and {Vanzella}, E. and {Zamorani}, G. and {Zucca}, E. and {Amorin}, R. and {Bardelli}, S. and {Cassar{\`a}}, L.~P. and {Castellano}, M. and {Cimatti}, A. and {Cucciati}, O. and {Durkalec}, A. and {Fontana}, A. and {Giavalisco}, M. and {Grazian}, A. and {Hathi}, N.~P. and {Ilbert}, O. and {Paltani}, S. and {Pforr}, J. and {Scodeggio}, M. and {Sommariva}, V. and {Talia}, M. and {Tresse}, L. and {Vergani}, D. and {Capak}, P. and {Charlot}, S. and {Contini}, T. and {de la Torre}, S. and {Dunlop}, J. and {Fotopoulou}, S. and {Guaita}, L. and {Koekemoer}, A. and {L{\'o}pez-Sanjuan}, C. and {Mellier}, Y. and {Salvato}, M. and {Scoville}, N. and {Taniguchi}, Y. and {Wang}, P.~W.},
        title = "{The VIMOS Ultra Deep Survey first data release: Spectra and spectroscopic redshifts of 698 objects up to z$_{spec}$   6 in CANDELS}",
      journal = {\aap},
         year = 2017,
        month = apr,
       volume = {600},
          eid = {A110},
        pages = {A110},
          doi = {10.1051/0004-6361/201527963},
archivePrefix = {arXiv},
       eprint = {1602.01842},
 primaryClass = {astro-ph.GA},
       adsurl = {https://ui.adsabs.harvard.edu/abs/2017A&A...600A.110T}
}

@ARTICLE{2017ApJ...841..111M,
       author = {{Masters}, Daniel C. and {Stern}, Daniel K. and {Cohen}, Judith G. and {Capak}, Peter L. and {Rhodes}, Jason D. and {Castander}, Francisco J. and {Paltani}, St{\'e}phane},
        title = "{The Complete Calibration of the Color-Redshift Relation (C3R2) Survey: Survey Overview and Data Release 1}",
      journal = {\apj},
         year = 2017,
        month = jun,
       volume = {841},
       number = {2},
          eid = {111},
        pages = {111},
          doi = {10.3847/1538-4357/aa6f08},
archivePrefix = {arXiv},
       eprint = {1704.06665},
 primaryClass = {astro-ph.CO},
       adsurl = {https://ui.adsabs.harvard.edu/abs/2017ApJ...841..111M}
}

@ARTICLE{2019ApJ...877...81M,
       author = {{Masters}, Daniel C. and {Stern}, Daniel K. and {Cohen}, Judith G. and {Capak}, Peter L. and {Stanford}, S. Adam and {Hernitschek}, Nina and {Galametz}, Audrey and {Davidzon}, Iary and {Rhodes}, Jason D. and {Sanders}, Dave and {Mobasher}, Bahram and {Castander}, Francisco and {Pruett}, Kerianne and {Fotopoulou}, Sotiria},
        title = "{The Complete Calibration of the Color-Redshift Relation (C3R2) Survey: Analysis and Data Release 2}",
      journal = {\apj},
         year = 2019,
        month = jun,
       volume = {877},
       number = {2},
          eid = {81},
        pages = {81},
          doi = {10.3847/1538-4357/ab184d},
archivePrefix = {arXiv},
       eprint = {1904.06394},
 primaryClass = {astro-ph.GA},
       adsurl = {https://ui.adsabs.harvard.edu/abs/2019ApJ...877...81M}
}

@ARTICLE{2023A&A...674A...1G,
       author = {{Gaia Collaboration} and {Vallenari}, A. and {Brown}, A.~G.~A. and {Prusti}, T. and {de Bruijne}, J.~H.~J. and {Arenou}, F. and {Babusiaux}, C. and {Biermann}, M. and {Creevey}, O.~L. and {Ducourant}, C. and {Evans}, D.~W. and {Eyer}, L. and {Guerra}, R. and {Hutton}, A. and {Jordi}, C. and {Klioner}, S.~A. and {Lammers}, U.~L. and {Lindegren}, L. and {Luri}, X. and {Mignard}, F. and {Panem}, C. and {Pourbaix}, D. and {Randich}, S. and {Sartoretti}, P. and {Soubiran}, C. and {Tanga}, P. and {Walton}, N.~A. and {Bailer-Jones}, C.~A.~L. and {Bastian}, U. and {Drimmel}, R. and {Jansen}, F. and {Katz}, D. and {Lattanzi}, M.~G. and {van Leeuwen}, F. and {Bakker}, J. and {Cacciari}, C. and {Casta{\~n}eda}, J. and {De Angeli}, F. and {Fabricius}, C. and {Fouesneau}, M. and {Fr{\'e}mat}, Y. and {Galluccio}, L. and {Guerrier}, A. and {Heiter}, U. and {Masana}, E. and {Messineo}, R. and {Mowlavi}, N. and {Nicolas}, C. and {Nienartowicz}, K. and {Pailler}, F. and {Panuzzo}, P. and {Riclet}, F. and {Roux}, W. and {Seabroke}, G.~M. and {Sordo}, R. and {Th{\'e}venin}, F. and {Gracia-Abril}, G. and {Portell}, J. and {Teyssier}, D. and {Altmann}, M. and {Andrae}, R. and {Audard}, M. and {Bellas-Velidis}, I. and {Benson}, K. and {Berthier}, J. and {Blomme}, R. and {Burgess}, P.~W. and {Busonero}, D. and {Busso}, G. and {C{\'a}novas}, H. and {Carry}, B. and {Cellino}, A. and {Cheek}, N. and {Clementini}, G. and {Damerdji}, Y. and {Davidson}, M. and {de Teodoro}, P. and {Nu{\~n}ez Campos}, M. and {Delchambre}, L. and {Dell'Oro}, A. and {Esquej}, P. and {Fern{\'a}ndez-Hern{\'a}ndez}, J. and {Fraile}, E. and {Garabato}, D. and {Garc{\'\i}a-Lario}, P. and {Gosset}, E. and {Haigron}, R. and {Halbwachs}, J. -L. and {Hambly}, N.~C. and {Harrison}, D.~L. and {Hern{\'a}ndez}, J. and {Hestroffer}, D. and {Hodgkin}, S.~T. and {Holl}, B. and {Jan{\ss}en}, K. and {Jevardat de Fombelle}, G. and {Jordan}, S. and {Krone-Martins}, A. and {Lanzafame}, A.~C. and {L{\"o}ffler}, W. and {Marchal}, O. and {Marrese}, P.~M. and {Moitinho}, A. and {Muinonen}, K. and {Osborne}, P. and {Pancino}, E. and {Pauwels}, T. and {Recio-Blanco}, A. and {Reyl{\'e}}, C. and {Riello}, M. and {Rimoldini}, L. and {Roegiers}, T. and {Rybizki}, J. and {Sarro}, L.~M. and {Siopis}, C. and {Smith}, M. and {Sozzetti}, A. and {Utrilla}, E. and {van Leeuwen}, M. and {Abbas}, U. and {{\'A}brah{\'a}m}, P. and {Abreu Aramburu}, A. and {Aerts}, C. and {Aguado}, J.~J. and {Ajaj}, M. and {Aldea-Montero}, F. and {Altavilla}, G. and {{\'A}lvarez}, M.~A. and {Alves}, J. and {Anders}, F. and {Anderson}, R.~I. and {Anglada Varela}, E. and {Antoja}, T. and {Baines}, D. and {Baker}, S.~G. and {Balaguer-N{\'u}{\~n}ez}, L. and {Balbinot}, E. and {Balog}, Z. and {Barache}, C. and {Barbato}, D. and {Barros}, M. and {Barstow}, M.~A. and {Bartolom{\'e}}, S. and {Bassilana}, J. -L. and {Bauchet}, N. and {Becciani}, U. and {Bellazzini}, M. and {Berihuete}, A. and {Bernet}, M. and {Bertone}, S. and {Bianchi}, L. and {Binnenfeld}, A. and {Blanco-Cuaresma}, S. and {Blazere}, A. and {Boch}, T. and {Bombrun}, A. and {Bossini}, D. and {Bouquillon}, S. and {Bragaglia}, A. and {Bramante}, L. and {Breedt}, E. and {Bressan}, A. and {Brouillet}, N. and {Brugaletta}, E. and {Bucciarelli}, B. and {Burlacu}, A. and {Butkevich}, A.~G. and {Buzzi}, R. and {Caffau}, E. and {Cancelliere}, R. and {Cantat-Gaudin}, T. and {Carballo}, R. and {Carlucci}, T. and {Carnerero}, M.~I. and {Carrasco}, J.~M. and {Casamiquela}, L. and {Castellani}, M. and {Castro-Ginard}, A. and {Chaoul}, L. and {Charlot}, P. and {Chemin}, L. and {Chiaramida}, V. and {Chiavassa}, A. and {Chornay}, N. and {Comoretto}, G. and {Contursi}, G. and {Cooper}, W.~J. and {Cornez}, T. and {Cowell}, S. and {Crifo}, F. and {Cropper}, M. and {Crosta}, M. and {Crowley}, C. and {Dafonte}, C. and {Dapergolas}, A. and {David}, M. and {David}, P. and {de Laverny}, P. and {De Luise}, F. and {De March}, R.},
        title = "{Gaia Data Release 3. Summary of the content and survey properties}",
      journal = {\aap},
         year = 2023,
        month = jun,
       volume = {674},
          eid = {A1},
        pages = {A1},
          doi = {10.1051/0004-6361/202243940},
archivePrefix = {arXiv},
       eprint = {2208.00211},
 primaryClass = {astro-ph.GA},
       adsurl = {https://ui.adsabs.harvard.edu/abs/2023A&A...674A...1G}
}

@ARTICLE{2024ApJ...968....4P,
       author = {{P{\'e}rez-Gonz{\'a}lez}, Pablo G. and {Barro}, Guillermo and {Rieke}, George H. and {Lyu}, Jianwei and {Rieke}, Marcia and {Alberts}, Stacey and {Williams}, Christina C. and {Hainline}, Kevin and {Sun}, Fengwu and {Pusk{\'a}s}, D{\'a}vid and {Annunziatella}, Marianna and {Baker}, William M. and {Bunker}, Andrew J. and {Egami}, Eiichi and {Ji}, Zhiyuan and {Johnson}, Benjamin D. and {Robertson}, Brant and {Rodr{\'\i}guez Del Pino}, Bruno and {Rujopakarn}, Wiphu and {Shivaei}, Irene and {Tacchella}, Sandro and {Willmer}, Christopher N.~A. and {Willott}, Chris},
        title = "{What Is the Nature of Little Red Dots and what Is Not, MIRI SMILES Edition}",
      journal = {\apj},
         year = 2024,
        month = jun,
       volume = {968},
       number = {1},
          eid = {4},
        pages = {4},
          doi = {10.3847/1538-4357/ad38bb},
archivePrefix = {arXiv},
       eprint = {2401.08782},
 primaryClass = {astro-ph.GA},
       adsurl = {https://ui.adsabs.harvard.edu/abs/2024ApJ...968....4P}
}

@ARTICLE{2024arXiv241119686O,
       author = {{{\"O}stlin}, G{\"o}ran and {P{\'e}rez-Gonz{\'a}lez}, Pablo G. and {Melinder}, Jens and {Gillman}, Steven and {Iani}, Edoardo and {Costantin}, Luca and {Boogaard}, Leindert A. and {Rinaldi}, Pierluigi and {Colina}, Luis and {N{\o}rgaard-Nielsen}, Hans Ulrik and {Dicken}, Daniel and {Greve}, Thomas R. and {Wright}, Gillian and {Alonso-Herrero}, Almudena and {Alvarez-Marquez}, Javier and {Annunziatella}, Marianna and {Bik}, Arjan and {Bosman}, Sarah E.~I. and {Caputi}, Karina I. and {Crespo Gomez}, Alejandro and {Eckart}, Andreas and {Garcia-Marin}, Macarena and {Hjorth}, Jens and {Ilbert}, Olivier and {Jermann}, Iris and {Kendrew}, Sarah and {Labiano}, Alvaro and {Langeroodi}, Danial and {Le Fevre}, Olivier and {Libralato}, Mattia and {Meyer}, Romain A. and {Moutard}, Thibaud and {Peissker}, Florian and {Pye}, John P. and {Tikkanen}, Tuomo V. and {Topinka}, Martin and {Walter}, Fabian and {Ward}, Martin and {van der Werf}, Paul and {van Dishoeck}, Ewine F. and {Henning}, Manuel G{\"u}del Thomas and {Lagage}, Pierre-Olivier and {Ray}, Tom P. and {Vandenbussche}, Bart},
        title = "{MIRI Deep Imaging Survey (MIDIS) of the Hubble Ultra Deep Field}",
      journal = {arXiv e-prints},
         year = 2024,
        month = nov,
          eid = {arXiv:2411.19686},
        pages = {arXiv:2411.19686},
          doi = {10.48550/arXiv.2411.19686},
archivePrefix = {arXiv},
       eprint = {2411.19686},
 primaryClass = {astro-ph.GA},
       adsurl = {https://ui.adsabs.harvard.edu/abs/2024arXiv241119686O}
}

@ARTICLE{2023ApJ...946L..12B,
       author = {{Bagley}, Micaela B. and {Finkelstein}, Steven L. and {Koekemoer}, Anton M. and {Ferguson}, Henry C. and {Arrabal Haro}, Pablo and {Dickinson}, Mark and {Kartaltepe}, Jeyhan S. and {Papovich}, Casey and {P{\'e}rez-Gonz{\'a}lez}, Pablo G. and {Pirzkal}, Nor and {Somerville}, Rachel S. and {Willmer}, Christopher N.~A. and {Yang}, Guang and {Yung}, L.~Y. Aaron and {Fontana}, Adriano and {Grazian}, Andrea and {Grogin}, Norman A. and {Hirschmann}, Michaela and {Kewley}, Lisa J. and {Kirkpatrick}, Allison and {Kocevski}, Dale D. and {Lotz}, Jennifer M. and {Medrano}, Aubrey and {Morales}, Alexa M. and {Pentericci}, Laura and {Ravindranath}, Swara and {Trump}, Jonathan R. and {Wilkins}, Stephen M. and {Calabr{\`o}}, Antonello and {Cooper}, M.~C. and {Costantin}, Luca and {de la Vega}, Alexander and {Hilbert}, Bryan and {Hutchison}, Taylor A. and {Larson}, Rebecca L. and {Lucas}, Ray A. and {McGrath}, Elizabeth J. and {Ryan}, Russell and {Wang}, Xin and {Wuyts}, Stijn},
        title = "{CEERS Epoch 1 NIRCam Imaging: Reduction Methods and Simulations Enabling Early JWST Science Results}",
      journal = {\apjl},
         year = 2023,
        month = mar,
       volume = {946},
       number = {1},
          eid = {L12},
        pages = {L12},
          doi = {10.3847/2041-8213/acbb08},
archivePrefix = {arXiv},
       eprint = {2211.02495},
 primaryClass = {astro-ph.IM},
       adsurl = {https://ui.adsabs.harvard.edu/abs/2023ApJ...946L..12B}
}

@ARTICLE{2016ApJ...833...72B,
       author = {{Bouwens}, Rychard J. and {Aravena}, Manuel and {Decarli}, Roberto and {Walter}, Fabian and {da Cunha}, Elisabete and {Labb{\'e}}, Ivo and {Bauer}, Franz E. and {Bertoldi}, Frank and {Carilli}, Chris and {Chapman}, Scott and {Daddi}, Emanuele and {Hodge}, Jacqueline and {Ivison}, Rob J. and {Karim}, Alex and {Le Fevre}, Olivier and {Magnelli}, Benjamin and {Ota}, Kazuaki and {Riechers}, Dominik and {Smail}, Ian R. and {van der Werf}, Paul and {Weiss}, Axel and {Cox}, Pierre and {Elbaz}, David and {Gonzalez-Lopez}, Jorge and {Infante}, Leopoldo and {Oesch}, Pascal and {Wagg}, Jeff and {Wilkins}, Steve},
        title = "{ALMA Spectroscopic Survey in the Hubble Ultra Deep Field: The Infrared Excess of UV-Selected z = 2-10 Galaxies as a Function of UV-Continuum Slope and Stellar Mass}",
      journal = {\apj},
         year = 2016,
        month = dec,
       volume = {833},
       number = {1},
          eid = {72},
        pages = {72},
          doi = {10.3847/1538-4357/833/1/72},
archivePrefix = {arXiv},
       eprint = {1606.05280},
 primaryClass = {astro-ph.GA},
       adsurl = {https://ui.adsabs.harvard.edu/abs/2016ApJ...833...72B}
}

@MISC{2021jwst.prop.1810B,
       author = {{Belli}, Sirio and {Conroy}, Charlie and {Davies}, Rebecca and {Emami}, Razieh and {Johnson}, Benjamin D. and {Leja}, Joel and {Mendel}, Jon Trevor and {Naidu}, Rohan and {Nelson}, Erica and {Tacchella}, Sandro and {Terrazas}, Bryan and {Weinberger}, Rainer},
        title = "{The Stellar and Gas Content of Galaxies at Cosmic Noon}",
 howpublished = {JWST Proposal. Cycle 1, ID. \#1810},
         year = 2021,
        month = mar,
        pages = {1810},
       adsurl = {https://ui.adsabs.harvard.edu/abs/2021jwst.prop.1810B}
}

@MISC{2021jwst.prop.2565G,
       author = {{Glazebrook}, Karl and {Nanayakkara}, Themiya and {Esdaile}, James and {Espejo}, Juan Manuel and {Jacobs}, Colin and {Kacprzak}, Glenn G. and {Labbe}, Ivo and {Marchesini}, Danilo and {Marsan}, Cemile and {Oesch}, Pascal and {Papovich}, Casey and {Remus}, Rhea-Silvia and {Schreiber}, Corentin and {Straatman}, Caroline and {Tran}, Kim-Vy and {Urbina}, Claudia Lagos},
        title = "{How Many Quiescent Galaxies are There at 3}",
 howpublished = {JWST Proposal. Cycle 1, ID. \#2565},
         year = 2021,
        month = mar,
        pages = {2565},
       adsurl = {https://ui.adsabs.harvard.edu/abs/2021jwst.prop.2565G}
}

@ARTICLE{2025MNRAS.537.3245B,
       author = {{Begley}, R. and {McLure}, R.~J. and {Cullen}, F. and {McLeod}, D.~J. and {Dunlop}, J.~S. and {Carnall}, A.~C. and {Stanton}, T.~M. and {Shapley}, A.~E. and {Cochrane}, R. and {Donnan}, C.~T. and {Ellis}, R.~S. and {Fontana}, A. and {Grogin}, N.~A. and {Koekemoer}, A.~M.},
        title = "{The evolution of [O III] + H{\ensuremath{\beta}} equivalent width from z ≃ 3-8: implications for the production and escape of ionizing photons during reionization}",
      journal = {\mnras},
         year = 2025,
        month = mar,
       volume = {537},
       number = {4},
        pages = {3245-3264},
          doi = {10.1093/mnras/staf211},
archivePrefix = {arXiv},
       eprint = {2410.10988},
 primaryClass = {astro-ph.GA},
       adsurl = {https://ui.adsabs.harvard.edu/abs/2025MNRAS.537.3245B}
}

@ARTICLE{2018A&A...609A.130L,
       author = {{Lagache}, G. and {Cousin}, M. and {Chatzikos}, M.},
        title = "{The [CII] 158 {\ensuremath{\mu}}m line emission in high-redshift galaxies}",
      journal = {\aap},
         year = 2018,
        month = jan,
       volume = {609},
          eid = {A130},
        pages = {A130},
          doi = {10.1051/0004-6361/201732019},
archivePrefix = {arXiv},
       eprint = {1711.00798},
 primaryClass = {astro-ph.GA},
       adsurl = {https://ui.adsabs.harvard.edu/abs/2018A&A...609A.130L}
}

@ARTICLE{2025arXiv250805740B,
       author = {{Barrufet}, L. and {Dunlop}, J.~S. and {Begley}, R. and {Flury}, S. and {McLeod}, D.~J. and {Arellano-Cordova}, K. and {Carnall}, A. and {Cullen}, F. and {Donnan}, C.~T. and {Liu}, F. and {McLure}, R. and {Scholte}, D. and {Stanton}, T.~M. and {Cochrane}, R. and {Conselice}, C. and {Ellis}, R. and {P{\'e}rez-Gonz{\'a}lez}, P.~G. and {Gottumukkala}, R. and {Grogin}, N.~A. and {Illingworth}, G.~D. and {Koekemoer}, A.~M. and {Magee}, D. and {Michalowski}, M.},
        title = "{Strength in Numbers: Red Galaxies Bolster the Cosmic Star Formation Rate Density at z > 3}",
      journal = {arXiv e-prints},
         year = 2025,
        month = aug,
          eid = {arXiv:2508.05740},
        pages = {arXiv:2508.05740},
          doi = {10.48550/arXiv.2508.05740},
archivePrefix = {arXiv},
       eprint = {2508.05740},
 primaryClass = {astro-ph.GA},
       adsurl = {https://ui.adsabs.harvard.edu/abs/2025arXiv250805740B}
}

@ARTICLE{2025arXiv250906913S,
       author = {{Stevenson}, Struan D. and {Carnall}, Adam C. and {Leung}, Ho-Hin and {Taylor}, Elizabeth and {Cullen}, Fergus and {Dunlop}, James S. and {McLeod}, Derek J. and {McLure}, Ross J. and {Begley}, Ryan and {Arellano-C{\'o}rdova}, Karla Z. and {Barrufet}, Laia and {Bondestam}, Cecilia and {Donnan}, Callum T. and {Ellis}, Richard S. and {Grogin}, Norman A. and {Liu}, Feng-Yuan and {Koekemoer}, Anton M. and {P{\'e}rez-Gonz{\'a}lez}, Pablo G. and {Rowlands}, Kate and {Sanders}, Ryan L. and {Scholte}, Dirk and {Shapley}, Alice E. and {Skarbinski}, Maya and {Stanton}, Thomas M. and {Wild}, Vivienne},
        title = "{PRIMER \& JADES reveal an abundance of massive quiescent galaxies at 2 < z < 5}",
      journal = {arXiv e-prints},
         year = 2025,
        month = sep,
          eid = {arXiv:2509.06913},
        pages = {arXiv:2509.06913},
          doi = {10.48550/arXiv.2509.06913},
archivePrefix = {arXiv},
       eprint = {2509.06913},
 primaryClass = {astro-ph.GA},
       adsurl = {https://ui.adsabs.harvard.edu/abs/2025arXiv250906913S}
}

@ARTICLE{2014ApJ...781...34G,
       author = {{Gonz{\'a}lez}, Valentino and {Bouwens}, Rychard and {Illingworth}, Garth and {Labb{\'e}}, Ivo and {Oesch}, Pascal and {Franx}, Marijn and {Magee}, Dan},
        title = "{Slow Evolution of the Specific Star Formation Rate at z > 2: The Impact of Dust, Emission Lines, and a Rising Star Formation History}",
      journal = {\apj},
         year = 2014,
        month = jan,
       volume = {781},
       number = {1},
          eid = {34},
        pages = {34},
          doi = {10.1088/0004-637X/781/1/34},
archivePrefix = {arXiv},
       eprint = {1208.4362},
 primaryClass = {astro-ph.CO},
       adsurl = {https://ui.adsabs.harvard.edu/abs/2014ApJ...781...34G}
}

@ARTICLE{2015A&A...581A..54T,
       author = {{Tasca}, L.~A.~M. and {Le F{\`e}vre}, O. and {Hathi}, N.~P. and {Schaerer}, D. and {Ilbert}, O. and {Zamorani}, G. and {Lemaux}, B.~C. and {Cassata}, P. and {Garilli}, B. and {Le Brun}, V. and {Maccagni}, D. and {Pentericci}, L. and {Thomas}, R. and {Vanzella}, E. and {Zucca}, E. and {Amorin}, R. and {Bardelli}, S. and {Cassar{\`a}}, L.~P. and {Castellano}, M. and {Cimatti}, A. and {Cucciati}, O. and {Durkalec}, A. and {Fontana}, A. and {Giavalisco}, M. and {Grazian}, A. and {Paltani}, S. and {Ribeiro}, B. and {Scodeggio}, M. and {Sommariva}, V. and {Talia}, M. and {Tresse}, L. and {Vergani}, D. and {Capak}, P. and {Charlot}, S. and {Contini}, T. and {de la Torre}, S. and {Dunlop}, J. and {Fotopoulou}, S. and {Koekemoer}, A. and {L{\'o}pez-Sanjuan}, C. and {Mellier}, Y. and {Pforr}, J. and {Salvato}, M. and {Scoville}, N. and {Taniguchi}, Y. and {Wang}, P.~W.},
        title = "{The evolving star formation rate: M$_{{\ensuremath{\star}}}$ relation and sSFR since z ≃ 5 from the VUDS spectroscopic survey}",
      journal = {\aap},
         year = 2015,
        month = sep,
       volume = {581},
          eid = {A54},
        pages = {A54},
          doi = {10.1051/0004-6361/201425379},
archivePrefix = {arXiv},
       eprint = {1411.5687},
 primaryClass = {astro-ph.GA},
       adsurl = {https://ui.adsabs.harvard.edu/abs/2015A&A...581A..54T}
}

@ARTICLE{2023MNRAS.519.1526P,
       author = {{Popesso}, P. and {Concas}, A. and {Cresci}, G. and {Belli}, S. and {Rodighiero}, G. and {Inami}, H. and {Dickinson}, M. and {Ilbert}, O. and {Pannella}, M. and {Elbaz}, D.},
        title = "{The main sequence of star-forming galaxies across cosmic times}",
      journal = {\mnras},
         year = 2023,
        month = feb,
       volume = {519},
       number = {1},
        pages = {1526-1544},
          doi = {10.1093/mnras/stac3214},
archivePrefix = {arXiv},
       eprint = {2203.10487},
 primaryClass = {astro-ph.GA},
       adsurl = {https://ui.adsabs.harvard.edu/abs/2023MNRAS.519.1526P}
}




\appendix
\section{Source catalogues}
\label{sec:append_a}
Here we present the final sample of 128 {\it JWST}-detected ALMA sources in the PRIMER COSMOS footprint. Basic observational properties such as positions and flux densities are tabulated in Table \ref{tab:CAT1}, while the redshifts (spectroscopic and photometric) and derived physical properties of the galaxies are given in Table \ref{tab:SED1}.


\begin{table*}
\centering
\caption{The complete catalogue of ALMA sources within the PRIMER COSMOS footprint. Only the highest-S/N ALMA detection for each source is listed here. For the {\it JWST} PRIMER detections, we list two representative fluxes from NIRCam (F444W) and MIRI (F770W), respectively, while the coordinates are based on F356W imaging. The $f_{\rm F770W}$ has a typical uncertainty of 10\%. The flag indicates if the source is also detected by AS2COSMOS \citep[`S',][]{2020MNRAS.495.3409S} or Ex-MORA \citep[`E',][]{2024arXiv240814546L} or both. $^{\dag}$ALMA properties of ID 024 are taken from Ex-MORA \citep{2024arXiv240814546L}.}
\label{tab:CAT1}
\setlength{\tabcolsep}{5.0pt} 
\begin{tabularx}{\textwidth}{c|cccr|ccrr|ccr|c}
\hline
ID &\multicolumn{4}{c}{ALMA}&\multicolumn{4}{c}{JWST NIRCam \& MIRI}&\multicolumn{3}{c}{UltraVISTA}&Flag\\
\hline
&R.A.&Dec.&$S_{\nu}$ &$\lambda$\phantom{0,}&R.A.&Dec.&$S_{\rm{F444W}}\phantom{00}$ & $S_{\rm{F770W}}$& R.A.&Dec.&$S_{\rm{Ks}}$\phantom{000}\\
& (deg)&(deg)&(mJy)&($\rm{\mu}$m)&(deg)&(deg)&($\rm{\mu}$Jy)\phantom{00}&($\rm{\mu}$Jy)&(deg)&(deg)&($\rm{\mu}$Jy)\phantom{00}&\\
\hline
001&150.06505&2.26361&5.13 (0.10)&1287&   150.06509&2.26363&0.55 (0.03)& 2.94 &   150.06519&2.26365&0.55 (0.07)& E, S    \\
002 & 150.09991&2.29722&10.45 (0.60)\phantom{'}&873&    150.10004&2.29739&1.67 (0.08)& 4.46 &  150.10011&2.29714&66.65 (3.33)& S   \\
003 & 150.13264&2.21187&2.82 (0.10)&1287&    150.13265&2.21184&12.15 (0.61)& 18.80 &  150.13266&2.21193&5.84 (0.29) & S  \\
004 & 150.09856&2.36537&8.63 (0.39)&873&    150.09855&2.36537&7.22 (0.36)& - &   150.09857&2.36536&2.46 (0.12) & S    \\
005 & 150.10541&2.31282&8.15 (0.37)&873&    150.10539&2.31283&6.35 (0.32)&  10.92 &  150.10543&2.31284&2.66 (0.13) & E, S   \\
006 & 150.20803&2.38314&8.01 (0.37)&873&    -&-&-& 8.66 &  150.20803&2.38310&1.09 (0.05)  & S \\
007 & 150.16352&2.37252&7.20 (0.36)&873&    150.16352&2.37248&6.93 (0.35)& 15.02 &  150.16357&2.37246&5.95 (0.30) & S  \\
008 & 150.04823&2.25145&6.90 (0.38)&873&    -&-&-& 11.08&    150.04826&2.25148&1.99 (0.10) & E, S   \\
009 & 150.10063&2.33483&6.76 (0.32)&873&    150.10063&2.33483&0.31 (0.02)& 1.46&   -&-&-  & E, S  \\
010 & 150.13894&2.43376&2.17 (0.09)&1249&    150.13896&2.43377&7.40 (0.37)& 13.37 &  150.13900&2.43376&2.22 (0.11) & E, S  \\
011 & 150.10394&2.18642&6.15 (0.31)&873&    150.10393&2.18642&15.13 (0.76)& 15.54 &  150.10396&2.18642&5.13 (0.26)  &  \\
012 & 150.05243&2.24559&5.77 (0.28)&873&    -&-&-& 15.88&    150.05247&2.24554&9.72 (0.49)  & S  \\
013 & 150.14320&2.35603&5.72 (0.41)&870&    150.14323&2.35602&2.06 (0.10)& 5.01 &  150.14313&2.35591&1.22 (0.07)  & S \\
014 & 150.14728&2.47409&3.31 (0.17)&1035&    150.14725&2.47407&0.81 (0.04)& 2.87 &  -&-&-  &  \\
015 & 150.03142&2.19637&1.85 (0.09)&1249&    -&-&-&6.27 &    150.03147&2.19644&1.69 (0.08)& S   \\
016 & 150.10445&2.43537&5.23 (0.46)&873&    150.10445&2.43540&0.12 (0.01)&  0.86 &  -&-&- & S   \\
017 & 150.12926&2.46431&5.16 (0.41)&873&    150.12928&2.46431&3.22 (0.16)&  - & 150.12931&2.46431&0.76 (0.06)  &  \\
018 & 150.15026&2.36414&0.12 (0.04)&3010&    150.15024&2.36413&7.16 (0.36)& 11.77 & 150.15023&2.36412&3.87 (0.19) &   \\
019 & 150.15373&2.32798&1.90 (0.12)&1200&    150.15374&2.32798&4.67 (0.23)& 2.33 &  150.15377&2.32797&1.00 (0.05)  &  \\
020 & 150.09852&2.32085&4.83 (0.39)&873&    150.09853&2.32087&2.67 (0.13)&  6.53&  150.09863&2.32082&1.12 (0.06)    \\
021 & 150.08063&2.39027&0.11 (0.02)&3088&    150.08061&2.39027&66.30 (3.31)& - & 150.08065&2.39026&222.43 (9.12) &   \\
022 & 150.17326&2.46435&4.36 (0.35)&873&    150.17324&2.46432&12.71 (0.64)& 16.77 &  150.17327&2.46431&7.91 (0.40)&   \\
023 & 150.07593&2.21182&1.36 (0.11)&1287&   150.07594&2.21180&10.99 (0.55)& 11.42 &   150.07598&2.21176&8.88 (0.44)&    \\
024$^{\dag}$ & 150.04096 &2.28048& 0.33 (0.06) & 2040 &  - & - & - & 6.27 & 150.04104 & 2.28044 & 0.52 (0.05) & E  \\
025 & 150.10981&2.25775&3.90 (0.12)&889&    150.10982&2.25776&0.09 (0.01)& 0.41 &  -&-&- & E  \\
026 & 150.10588&2.42880&4.09 (0.25)&873&    150.10587&2.42881&2.49 (0.12)&  6.12 &  -&-&- &   \\
027 & 150.10631&2.25158&3.68 (0.47)&873&    150.10631&2.25157&11.36 (0.57)&  10.59 &  150.10642&2.25159&3.88 (0.19) &   \\
028 & 150.03650&2.19838&1.23 (0.08)&1249&    -&-&-& 8.75&    150.03657&2.19837&1.89 (0.09)&    \\
029 & 150.03738&2.34073&2.32 (0.12)&1009&    -&-&-& 4.63 &    150.03729&2.34060&1.79 (0.09)&  E  \\
030 & 150.17488&2.35278&3.62 (0.34)&870&    150.17487&2.35274&1.54 (0.08)& 4.07 &  150.17490&2.35269&0.52 (0.04)  &  \\
031 & 150.10572&2.43481&3.57 (0.43)&873&    150.10574&2.43483&0.38 (0.02)&  1.32 &  -&-&-  & S \\
032 & 150.17184&2.24070&1.07 (0.07)&1287&    150.17185&2.24071&11.14 (0.56)& - &  150.17186&2.24067&3.54 (0.18)  &  \\
033 & 150.17425&2.42973&3.26 (0.29)&870&    150.17428&2.42973&0.67 (0.03)& 2.78 &  150.17432&2.42975&0.17 (0.04)&    \\
034 & 150.07192&2.19020&3.22 (0.53)&873&    150.07194&2.19017&4.40 (0.22)& 8.84 &  150.07198&2.19017&1.66 (0.08)  &  \\
035 & 150.13655&2.23243&0.99 (0.13)&1287&    150.13653&2.23242&3.59 (0.18)& 8.23 &  -&-&-  &  \\
036 & 150.13914&2.43197&1.05 (0.10)&1249&    150.13916&2.43198&4.43 (0.22)& 8.13 &  150.13922&2.43198&1.42 (0.07) & S  \\
037 & 150.03954&2.37208&3.04 (0.52)&873&    -&-&-& 11.72 &  150.03960&2.37205&6.27 (0.31) &   \\
038 & 150.11126&2.40319&3.04 (0.59)&873&    150.11129&2.40318&5.65 (0.28)& - &  150.11132&2.40320&1.50 (0.08)  &  \\
039 & 150.03719&2.24460&0.94 (0.20)&1287&    -&-&-& 54.91 &   150.03722&2.24463&73.06 (3.65) &   \\
040 & 150.13639&2.22533&0.91 (0.09)&1287&    150.13638&2.22531&8.41 (0.42)& 11.80 & 150.13641&2.22530&4.91 (0.25) &   \\
041 & 150.03641&2.31719&0.79 (0.06)&1333&    -&-&-& 10.04 &    150.03651&2.31718&20.84 (1.04)&    \\
042 & 150.14415&2.37632&2.36 (0.43)&873&    150.14417&2.37632&2.18 (0.11)& 4.39 &  150.14421&2.37633&0.60 (0.05)  &  \\
043 & 150.05312&2.19807&2.80 (0.58)&873&    -&-&-& 15.38  & 150.05322&2.19809&22.36 (1.12)  &  \\
044 & 150.14283&2.41751&2.62 (0.44)&873&    150.14285&2.41747&1.91 (0.10)& 4.10 &  150.14289&2.41745&0.79 (0.05) &   \\
045 & 150.03652&2.32122&0.72 (0.06)&1333&    -&-&-& 27.45&    150.03654&2.32125&1.01 (0.05)&    \\
046 & 150.12059&2.41810&2.55 (0.28)&873&    150.12058&2.41809&1.47 (0.07)& 4.13 &  150.12063&2.41811&0.33 (0.04)  &  \\
047 & 150.09940&2.40492&0.77 (0.13)&1287&    150.0994&2.40492&3.58 (0.18)& -  & 150.09945&2.40490&1.31 (0.07) &   \\
048 & 150.21011&2.31165&1.44 (0.08)&1042&    -&-&-& 37.58  & 150.21019&2.31167&52.54 (2.63)  &  \\
049 & 150.11211&2.31401&1.08 (0.21)&1131&    150.11213&2.31400&4.81 (0.24)& - &  150.11219&2.31399&2.70 (0.14) &   \\
050 & 150.14174&2.42569&0.91 (0.05)&1195&    150.14176&2.42568&3.90 (0.20)& 7.78  & 150.14181&2.42574&1.01 (0.05) &   \\
051 & 150.07469&2.21648&0.71 (0.04)&1287&  150.07469&2.21648&5.28 (0.26)& 6.01 &   150.07473&2.21650&2.00 (0.10)&    \\
052 & 150.09458&2.33633&0.84 (0.08)&1215&    150.09459&2.33635&17.28 (0.86)& 11.73  & 150.09463&2.33635&29.09 (1.45) &   \\
053 & 150.18759&2.32248&2.21 (0.25)&873&    150.18759&2.32249&22.64 (1.13)& 17.19 &  150.18764&2.32249&16.51 (0.83) &   \\
054 & 150.16768&2.29873&0.81 (0.23)&1209&    150.16767&2.29875&10.21 (0.51)&  - & 150.16771&2.29875&3.81 (0.19) &   \\
055 & 150.10355&2.34608&0.66 (0.05)&1279&    150.10351&2.34608&9.98 (0.50)&  10.40 & 150.10353&2.34611&23.27 (1.16) &   \\
056 & 150.15365&2.24773&0.71 (0.12)&1249&    150.15365&2.24777&1.84 (0.09)& - &  150.15366&2.24773&0.55 (0.05)  &  \\
057 & 150.09824&2.16625&0.62 (0.07)&1287&    -&-&-& 7.01&   150.09826 &2.16623 & 0.58 (0.05)    \\
058 & 150.18712&2.38012&0.13 (0.02)&2160&    150.18715&2.38014&9.73 (0.49)& 13.29 &  150.18720&2.38014&4.65 (0.23)  &  \\
059 & 150.09251&2.39816&0.60 (0.08)&1287&  150.09251&2.39816&0.69 (0.03)& - &   -&-&-    \\
060 & 150.22026&2.31933&0.04 (0.01)&3187&    -&-&-& 4.62 &  150.22035&2.31925&2.70 (0.13) &   \\

		\hline
\end{tabularx}
\end{table*}

\begin{table*}
\contcaption{}
\label{tab:CAT2}
\centering
\setlength{\tabcolsep}{5.5pt} 
\begin{tabularx}{\textwidth}{c|cccr|ccrr|ccr|c}
\hline
ID &\multicolumn{4}{c}{ALMA}&\multicolumn{4}{c}{JWST NIRCam \& MIRI}&\multicolumn{3}{c}{UltraVISTA}&Flag\\
\hline
&R.A.&Dec.&$S_{\nu}$ &$\lambda$\phantom{0,}&R.A.&Dec.&$S_{\rm{F444W}}\phantom{00}$ &$S_{\rm{F770W}}$ & R.A.&Dec.&$S_{\rm{Ks}}$\phantom{000}\\
&(deg)&(deg)&(mJy)&($\rm{\mu}$m)&(deg)&(deg)&($\rm{\mu}$Jy)\phantom{00}&($\rm{\mu}$Jy)&(deg)&(deg)&($\rm{\mu}$Jy)\phantom{00}&\\
\hline

061 & 150.12996&2.25271&1.92 (0.43)&873&    150.12997&2.25268&5.84 (0.29)& - &  150.13001&2.25268&1.59 (0.08) &   \\
062 & 150.09596&2.36527&1.90 (0.26)&870&   150.09596&2.36526&1.39 (0.07)& - &   150.09599&2.36525&0.25 (0.06) & S   \\
063 & 150.12913&2.28456&1.14 (0.17)&1031&    150.12914&2.28456&0.71 (0.04)& 2.04 &  150.12918&2.28458&0.32 (0.05) &   \\
064 & 150.06143&2.3788&1.29 (0.03)&989&    150.06145&2.37871&5.05 (0.25)& - &   150.06149&2.37872&2.63 (0.13)&    \\
065 & 150.10520&2.32581&0.69 (0.11)&1213&    150.10520&2.32582&3.46 (0.17)&  5.32  & 150.10528&2.32589&1.46 (0.07)  &  \\
066 & 150.18425&2.25212&0.55 (0.12)&1287&    150.18423&2.25210&0.51 (0.03)&  - & -&-&-  &  \\
067 & 150.10520&2.15661&0.55 (0.11)&1249&    -&-&-&  5.91  &150.10522&2.15664&2.24 (0.11)  &  \\
068 & 150.09808&2.16577&0.48 (0.09)&1287&   -&-&-& 24.92 &   150.09813&2.16581&20.62 (1.03)    \\
069 & 150.11574&2.43112&1.52 (0.23)&873&    150.11575&2.43112&0.52 (0.03)& 1.58 &  150.11570&2.43110&0.12 (0.06)  &  \\
070 & 150.06451&2.32914&1.51 (0.23)&873&    150.06446&2.32916&6.17 (0.31)& 8.84&   150.06456&2.32904&14.11 (0.71)&    \\
071 & 150.12287&2.36095&1.51 (0.28)&873&    150.12289&2.36096&4.90 (0.25)& 6.98 &  150.12294&2.36097&2.72 (0.14)  &  \\
072 & 150.07740&2.18536&1.48 (0.36)&873&    150.07740&2.18536&4.30 (0.22)& 4.88 &  150.07745&2.18537&1.47 (0.07)  &  \\
073 & 150.04297&2.34090&0.57 (0.05)&1193&    -&-&-& 12.18 &   150.04304&2.34091&21.46 (1.07)  &  \\
074 & 150.09885&2.16626&0.45 (0.09)&1287&    -&-&-& 20.42 &  150.09887&2.16623&14.09 (0.70) &   \\
075 & 150.14299&2.35590&1.44 (0.38)&873&    150.14300&2.35589&2.39 (0.12)& 5.87  & 150.14313&2.35591&1.22 (0.07) &   \\
076 & 150.11256&2.37658&1.45 (0.34)&870&    150.11252&2.37654&3.15 (0.16)& - &  150.11256&2.37653&0.84 (0.05)  &  \\
077 & 150.13085&2.20895&0.48 (0.11)&1249&    150.13087&2.20894&3.29 (0.16)& 5.51 &  150.13088&2.20903&1.31 (0.07)  &  \\
078 & 150.07716&2.38036&0.44 (0.03)&1287&   150.07716&2.38037&3.87 (0.19)& - &   150.07717&2.38040&2.12 (0.11)&    \\
079 & 150.03003&2.31411&0.40 (0.08)&1326&    -&-&-& 10.15 &   150.03003&2.31415&3.75 (0.19)  &  \\
080 & 150.10981&2.25740&1.31 (0.09)&889&    150.10979&2.25739&0.10 (0.01)& 0.24  & -&-&-  &  \\
081 & 150.15865&2.46835&1.38 (0.22)&873&    150.15858&2.46835&0.82 (0.04)& 0.06 & 150.15862&2.46833&0.40 (0.05) &   \\
082 & 150.16687&2.23582&0.40 (0.09)&1287&    150.16687&2.23582&12.98 (0.65)& -  & 150.16691&2.23580&5.36 (0.27)  &  \\
083 & 150.10698&2.42895&1.30 (0.31)&873&    150.10697&2.42894&2.07 (0.10)& 3.63  & 150.10701&2.42912&10.31 (0.52)  &  \\
084 & 150.05911&2.21987&0.35 (0.08)&1344&    150.05922&2.21979&17.05 (0.85)&  14.30  &150.05915&2.21994&27.93 (1.40)&    \\
085 & 150.09400&2.24583&0.40 (0.12)&1287&    150.09401&2.24585&6.93 (0.35)& 6.45 &  150.09406&2.24588&3.46 (0.17) &   \\
086 & 150.07066&2.30517&0.43 (0.10)&1249&    150.07066&2.30512&1.70 (0.09)& 3.84 & 150.07065&2.30518&1.58 (0.08) &   \\
087 & 150.15175&2.33459&0.37 (0.03)&1310&    150.15178&2.33459&22.33 (1.12)&  15.11 & 150.15184&2.33459&16.05 (0.80) &   \\
088 & 150.11262&2.40668&1.25 (0.23)&873&    150.11260&2.40668&1.03 (0.05)& - &  -&-&-  &  \\
089 & 150.08802&2.39511&1.24 (0.15)&873&    150.08801&2.39510&2.49 (0.12)& -  & 150.08804&2.39509&1.28 (0.08) &   \\
090 & 150.13653&2.26058&1.33 (0.23)&847&    150.13650&2.26058&2.36 (0.12)& 2.94 &  150.13653&2.26055&0.28 (0.04)  &  \\
091 & 150.10283&2.38454&1.22 (0.29)&870&    150.10283&2.38451&3.39 (0.17)& - &  150.10287&2.38452&0.64 (0.05)  &  \\
092 & 150.15025&2.47514&1.19 (0.26)&873&    150.15022&2.47515&-& 368.11 &   150.15027&2.47515&87.40 (4.37)  &  \\
093 & 150.12637&2.43288&1.24 (0.10)&857&    150.12635&2.43287&15.48 (0.77)& 12.64 &  150.12639&2.43287&13.08 (0.65)  &  \\
094 & 150.07725&2.21141&0.46 (0.06)&1187&    150.07724&2.21141&5.18 (0.26)& 5.23  & 150.07725&2.21141&1.85 (0.09)  &  \\
095 & 150.14722&2.33718&1.12 (0.31)&885&    150.14718&2.33718&25.94 (1.30)& 20.27 &  150.14722&2.33717&60.72 (3.04)  &  \\
096 & 150.17078&2.36860&1.14 (0.33)&870&    150.17079&2.36862&1.85 (0.09)& 3.41 &  150.17079&2.36857&0.55 (0.04)  &  \\
097 & 150.03737&2.27182&1.10 (0.24)&873&    -&-&-& 9.05  & 150.03745&2.27185&6.47 (0.32)   & \\
098 & 150.14479&2.21969&1.09 (0.33)&873&    150.14481&2.21971&7.25 (0.36)& 5.47 &  150.14476&2.21972&3.74 (0.19) &   \\
099 & 150.19617&2.48181&0.38 (0.08)&1241&    -&-&-&   9.56 &150.19624&2.48179&4.93 (0.25) &   \\
100 & 150.07933&2.34058&1.05 (0.29)&873&    150.07932&2.34056&6.60 (0.33)& 7.35 &  150.07938&2.34057&3.84 (0.19)  &  \\
101 & 150.11431&2.36983&0.32 (0.09)&1287&    150.11431&2.36985&1.31 (0.07)& - &  150.11439&2.36989&0.82 (0.05)  &  \\
102 & 150.09545&2.24667&0.99 (0.28)&873&   150.09546&2.24667&6.05 (0.30)& 8.08 &   150.09551&2.24668&1.52 (0.08)    \\
103 & 150.15710&2.33230&0.37 (0.06)&1200    &150.15711&2.33231&0.84 (0.04)& 1.51 &  -&-&-  &  \\
104 & 150.11237&2.37526&0.29 (0.05)&1287&    150.11235&2.37524&2.51 (0.13)& - &  150.11240&2.37522&2.14 (0.11) &   \\
105 & 150.11510&2.32127&0.28 (0.05)&1287&    150.11510&2.32127&4.51 (0.23)& 3.05 &  150.11514&2.32128&1.37 (0.07)  &  \\
106 & 150.10750&2.34435&0.28 (0.07)&1279&    150.10753&2.34433&2.20 (0.11)&  - & 150.10758&2.34430&0.89 (0.06) &   \\
107 & 150.03846&2.33944&0.56 (0.06)&1009&    -&-&-& 3.41&    150.03848&2.33947&0.64 (0.05)&    \\
108 & 150.06669&2.25772&0.26 (0.08)&1287&    150.06669&2.25772&3.07 (0.15)& 4.12 &  150.06674&2.25772&1.70 (0.08) &   \\
109 & 150.09532&2.33895&0.31 (0.07)&1215&    150.09532&2.33897&2.14 (0.11)& 3.66  & 150.09537&2.33898&1.36 (0.07) &   \\
110 & 150.10875&2.20871&0.79 (0.20)&873&    150.10874&2.20870&0.93 (0.05)& 2.25 & 150.10875&2.20873&0.21 (0.05) &   \\
111 & 150.08425&2.21847&0.26 (0.08)&1213&    150.08432&2.21848&1.19 (0.06)& 2.25 &  150.08433&2.21854&1.64 (0.08)  &  \\
112 & 150.07528&2.37943&0.21 (0.03)&1287&    150.07524&2.37941&5.61 (0.28)&  - & 150.07527&2.37941&4.91 (0.25)  &  \\
113 & 150.14127&2.27816&0.21 (0.05)&1287&    150.14127&2.27819&6.22 (0.31)&  10.74 & 150.14130&2.27818&2.04 (0.10)   & \\
114 & 150.04203&2.33898&0.26 (0.05)&1193&    -&-&-& 2.91&    150.04209&2.33900&1.40 (0.07) &    \\
115 & 150.10282&2.37940&0.19 (0.04)&1287&    150.10284&2.37940&4.24 (0.21)& - &  150.10289&2.37942&1.03 (0.05) &   \\
116 & 150.07574&2.38072&0.18 (0.04)&1287&    150.07584&2.38073&3.08 (0.15)& - &  150.07585&2.38067&3.56 (0.18) &   \\
117 & 150.14728&2.28236&0.18 (0.04)&1287&    150.14727&2.28232&7.41 (0.37)& 7.30 &  150.14732&2.28229&3.03 (0.15) &   \\
118 & 150.03921&2.33710&0.36 (0.07)&1014&    -&-&-& 0.70 &  150.03932&2.33720&0.42 (0.04)  &  \\
119 & 150.19437&2.27329&0.17 (0.05)&1287&    150.19435&2.27331&5.06 (0.25)& 6.26 &  150.19438&2.27334&3.36 (0.17) &   \\
120 & 150.19728&2.48045&0.18 (0.04)&1241&    -&-&-& 0.41  & 150.19735&2.48043&0.18 (0.04)  &  \\
121 & 150.12358&2.22071&0.17 (0.03)&1203&    150.12357&2.22070&0.95 (0.05)& 1.70 &  150.12362&2.22072&0.26 (0.04) &   \\
122 & 150.18430&2.26436&0.45 (0.07)&873&    150.18428&2.26437&3.27 (0.16)& 3.08 &  150.18431&2.26436&1.90 (0.09) &   \\
123 & 150.06978&2.38156&0.20 (0.06)&1131&    150.06982&2.38151&1.62 (0.08)& - &  150.06985&2.38155&1.01 (0.06) &   \\
        \hline
\end{tabularx}
\end{table*}

\begin{table*}
\centering
\contcaption{}
\label{tab:CAT3}
\setlength{\tabcolsep}{5.7pt} 
\begin{tabularx}{\textwidth}{c|cccr|ccrr|ccr|c}
\hline
ID &\multicolumn{4}{c}{ALMA}&\multicolumn{4}{c}{JWST NIRCam \& MIRI}&\multicolumn{3}{c}{UltraVISTA}&Flag\\
\hline
&R.A.&Dec.&$S_{\nu}$ &$\lambda$\phantom{0,}&R.A.&Dec.&$S_{\rm{F444W}}$ & $S_{\rm{F770W}}$ &R.A.&Dec.&$S_{\rm{Ks}}$\phantom{000}\\
&(deg)&(deg)&(mJy)&($\rm{\mu}$m)&(deg)&(deg)&($\rm{\mu}$Jy)&($\rm{\mu}$Jy)&(deg)&(deg)&($\rm{\mu}$Jy)\phantom{00}&\\
\hline
124 & 150.13584&2.25791&0.33 (0.08)&847&    150.13582&2.25789&0.53 (0.03)&  0.66  & 150.13587&2.25789&0.50 (0.04) &   \\
125 & 150.17157&2.28730&0.30 (0.06)&866&    150.17157&2.28729&0.24 (0.01)& 0.27  & 150.17149&2.28731&0.49 (0.05) &   \\
126 & 150.12578&2.26662&0.08 (0.01)&1284&    150.12574&2.26660&0.41 (0.02)& - &  150.12577&2.26656&0.27 (0.04) &   \\
127 & 150.12487&2.26946&0.06 (0.01)&1284&    150.12487&2.26947&5.59 (0.28)&  -  & 150.12492&2.26947&5.41 (0.27) &   \\
128 & 150.03687&2.25772&0.16 (0.03)&873&    -&-&-&  10.42  &150.03690&2.25781&2.53 (0.13) &   \\
		\hline
\end{tabularx}
\end{table*}

\onecolumn

\begin{table*}
\renewcommand{\arraystretch}{1.3}
\caption{The derived physical properties of the 128 sources in our sample. References for the best quality redshifts (`Ref.') are provided for the spectroscopic redshifts, which are rounded to three decimal places. For those from DJA, we refer to the observation programmes. The photometric redshifts are rounded to two decimal places, and flagged as `J' or `U' depending on whether they are derived form the {\it JWST} PRIMER or UltraVISTA data respectively. In the final column (`Flag'), we also use the `J' and `U' flags to indicate the origin of the photometry used to derive the physical properties of each galaxy. $^{\dag}$The M$_*$ and A$_{\rm V}$ of the lensed galaxy ID 002 are taken from \citet{2024A&A...690L..16J}. All its properties are corrected for lensing with the magnification factor provided there.}
\label{tab:SED1}
\begin{tabularx}{\textwidth}{clXrlcllllc}
\hline
ID& $z$   & Ref. & \multicolumn{1}{c}{f$_{\rm{870\mu m}}$} & \multicolumn{1}{c}{SFR$_{\rm{FIR}}$}       & \multicolumn{1}{c}{SFR$_{\rm{UV}}$}        & \multicolumn{1}{c}{SFR$_{\rm{raw}}$}       & \multicolumn{1}{c}{SFR$_{\rm{corr}}$}          & \multicolumn{1}{c}{M$_*$}                        & \multicolumn{1}{c}{A$_{\rm V}$}                       & Flag \\
&       &            & \multicolumn{1}{c}{(mJy)}               & \multicolumn{1}{c}{(M$_{\odot}$yr$^{-1}$)} & \multicolumn{1}{c}{(M$_{\odot}$yr$^{-1}$)} & \multicolumn{1}{c}{(M$_{\odot}$yr$^{-1}$)} & \multicolumn{1}{c}{(M$_{\odot}$yr$^{-1}$)}     & \multicolumn{1}{c}{(logM$_{\odot}$)}             &                             &                     \\ \hline
001 & 4.596 & {\citet{2022ApJ...929..159C}}          & 16.58 $\pm$ 0.32    & 1490 $\pm$ 28          & 0.90  &   1491 $\pm$ 28      & 1762 $_{- 694 }^{+ 1557 }$ & 10.82 $_{- 0.07 }^{+ 0.07 }$ & 2.88 $_{- 0.26 }^{+ 0.29 }$ & J                   \\
002$^{\dag}$ &  2.625                   & {\citet{2024A&A...690L..16J}}                              & 10.54 $\pm$ 0.61                        & 263 $^{+147}_{-68}$                               & 0.38                                       & 264 $^{+147}_{-68}$                              & 9 $_{- 5 }^{+ 4 }$                       & 10.84$^{+0.19}_{-0.14}$          & 1.95 $\pm$ 0.01 & J                                       \\
003 & 2.104 & {\citet{2019ApJ...880...15K}}          & 9.13 $\pm$ 0.32     & 821 $\pm$ 29           & 0.21  &   821 $\pm$ 29     & 271 $_{- 51 }^{+ 51 }$     & 11.69 $_{- 0.08 }^{+ 0.06 }$ & 3.42 $_{- 0.43 }^{+ 0.23 }$ & J                   \\
004 & 2.364 & {\citet{{2012ApJS..200...13B, 2015ApJS..219...29M}}}         & 8.71 $\pm$ 0.40      & 783 $\pm$ 35           & 0.15  &   783 $\pm$ 35    & 176 $_{- 28 }^{+ 38 }$     & 11.51 $_{- 0.06 }^{+ 0.07 }$ & 3.03 $_{- 0.19 }^{+ 0.31 }$ & J                   \\
005 & 2.283  & \citet{khostovan2025cosmosspectroscopicredshiftcompilation}          & 8.22 $\pm$ 0.38     & 739 $\pm$ 33           & 0.29  &   739 $\pm$ 33     & 324 $_{- 152 }^{+ 244 }$   & 11.34 $_{- 0.19 }^{+ 0.11 }$ & 2.71 $_{- 0.36 }^{+ 0.26 }$ & J                   \\
006 & 2.59                  & U                              & 8.08 $\pm$ 0.37                         & 726 $\pm$ 33                               & 0.39  &   727 $\pm$ 33 & 55 $_{- 12 }^{+ 12 }$  &   10.95 $_{- 0.06 }^{+ 0.06 }$  &   1.88 $_{- 0.24 }^{+ 0.16 }$  &   U  \\
007 & 2.086                   & {\citet{{2012ApJS..200...13B, 2015ApJS..219...29M}}}                              & 7.27 $\pm$ 0.36                         & 653 $\pm$ 32                               & 4.12  &   657 $\pm$ 32  & 417 $_{- 29 }^{+ 28 }$  &   10.95 $_{- 0.01 }^{+ 0.02 }$  &   2.06 $_{- 0.03 }^{+ 0.04 }$  &   J  \\
008 & 2.32  & U          & 6.96 $\pm$ 0.39     & 625 $\pm$ 34           & 0.07  &   625 $\pm$ 34    & 107 $_{- 26 }^{+ 25 }$     & 11.32 $_{- 0.06 }^{+ 0.07 }$ & 2.68 $_{- 0.31 }^{+ 0.30 }$  & U                   \\
009 & 4.27  & J          & 6.82 $\pm$ 0.32     & 613 $\pm$ 29           & 0.03  &   613 $\pm$ 29   & 73 $_{- 18 }^{+ 20 }$      & 10.86 $_{- 0.08 }^{+ 0.10 }$  & 3.25 $_{- 0.29 }^{+ 0.28 }$ & J                   \\
010 & 2.510 & {\citet{2019ApJ...887..183Z}}          & 6.41 $\pm$ 0.25     & 576 $\pm$ 22           & 0.25  &   576 $\pm$ 22    & 145 $_{- 28 }^{+ 52 }$     & 11.34 $_{- 0.09 }^{+ 0.09 }$ & 3.22 $_{- 0.23 }^{+ 0.31 }$ & J                   \\
011 & 1.914                   & {\citet{{2012ApJS..200...13B, 2015ApJS..219...29M}}}                              & 6.20 $\pm$ 0.31                          & 557 $\pm$ 28                               & 0.06  &   557 $\pm$ 28     & 118 $_{- 30 }^{+ 21 }$   &  11.45 $_{- 0.05 }^{+ 0.05 }$   &  3.03 $_{- 0.28 }^{+ 0.19 }$  &   J                                        \\
012 & 2.438  & \citet{khostovan2025cosmosspectroscopicredshiftcompilation}          & 5.82 $\pm$ 0.28     & 523 $\pm$ 25           & 1.31  &   525 $\pm$ 25         & 513 $_{- 167 }^{+ 219 }$   & 11.39 $_{- 0.08 }^{+ 0.08 }$ & 1.55 $_{- 0.26 }^{+ 0.22 }$ & U                   \\
013 & 2.988 & {\citet{2019ApJ...887..183Z}}          & 5.71 $\pm$ 0.41     & 513 $\pm$ 36           & 0.12  &   513 $\pm$ 36   & 80 $_{- 14 }^{+ 22 }$      & 11.02 $_{- 0.08 }^{+ 0.07 }$ & 2.79 $_{- 0.21 }^{+ 0.21 }$ & J                   \\
014 & 5.62 & J          & 5.57 $\pm$ 0.29     & 500 $\pm$ 26           & 0.48  &   500 $\pm$ 26   & 967 $_{- 230 }^{+ 258 }$   & 10.80 $_{- 0.11 }^{+ 0.16 }$  & 2.60 $_{- 0.17 }^{+ 0.19 }$  & J                   \\
015 & 3.40  & U          & 5.48 $\pm$ 0.28     & 492 $\pm$ 24           & 2.86  &   495 $\pm$ 24           & 216 $_{- 57 }^{+ 56 }$     & 10.69 $_{- 0.07 }^{+ 0.08 }$ & 1.08 $_{- 0.19 }^{+ 0.14 }$ & U                   \\
016 & 4.615 & {\citet{2021ApJ...907..122M}}          & 5.28 $\pm$ 0.46     & 474 $\pm$ 41           & 0.00  &   474 $\pm$ 41      & 35 $_{- 35 }^{+ 20 }$      & 10.63 $_{- 0.12 }^{+ 0.13 }$ & 4.13 $_{- 0.41 }^{+ 0.49 }$ & J                   \\
017 & 3.62       & J    & 5.21 $\pm$ 0.42                         & 468 $\pm$ 37                               & 0.64  &   469 $\pm$ 37   & 170 $_{- 37 }^{+ 35 }$  &   11.32 $_{- 0.06 }^{+ 0.06 }$  &   2.14 $_{- 0.17 }^{+ 0.22 }$  &   J                              \\
018 & 2.464                   & {\citet{2015ApJS..218...15K}}                              & 5.05 $\pm$ 1.54                         & 453 $\pm$ 138                              & 1.77  &   455 $\pm$ 138 & 192 $_{- 30 }^{+ 42 }$  &   11.50 $_{- 0.05 }^{+ 0.06 }$  &   2.51 $_{- 0.16 }^{+ 0.24 }$  &   J                                  \\
019 & 2.68 & J          & 4.97 $\pm$ 0.30      & 446 $\pm$ 27           & 0.12  &   446 $\pm$ 27    & 140 $_{- 21 }^{+ 32 }$     & 11.33 $_{- 0.06 }^{+ 0.06 }$ & 3.18 $_{- 0.15 }^{+ 0.22 }$ & J                   \\
020 & 2.86 & J          & 4.88 $\pm$ 0.39     & 438 $\pm$ 35           & 0.17  &   438 $\pm$ 35   & 380 $_{- 63 }^{+ 61 }$     & 10.88 $_{- 0.08 }^{+ 0.09 }$ & 3.79 $_{- 0.10 }^{+ 0.09 }$  & J                   \\
021 & 0.353                   & {\citet{khostovan2025cosmosspectroscopicredshiftcompilation}}                              & 4.79 $\pm$ 0.91                         & 430 $\pm$ 82                               & 0.01  &   430 $\pm$ 82  & 0 $_{- 0 }^{+ 0 }$  &   11.40 $_{- 0.01 }^{+ 0.01 }$  &   0.79 $_{- 0.10 }^{+ 0.09 }$  &   J        \\
022 & 2.25 & J          & 4.40 $\pm$ 0.36      & 395 $\pm$ 32           & 4.45  &   399 $\pm$ 32   & 183 $_{- 53 }^{+ 97 }$     & 11.30 $_{- 0.08 }^{+ 0.08 }$  & 1.47 $_{- 0.19 }^{+ 0.25 }$ & J                   \\
023 & 2.103 & {\citet{2023A&A...679A.129R}}          & 4.39 $\pm$ 0.37     & 394 $\pm$ 33           & 0.50  &   394 $\pm$ 33     & 184 $_{- 31 }^{+ 40 }$     & 11.53 $_{- 0.07 }^{+ 0.07 }$ & 2.20 $_{- 0.23 }^{+ 0.27 }$  & J                   \\
024 & 3.83                  & U                              & 4.19 $\pm$ 0.77                         & 376 $\pm$ 69                               & 0.04  &   376 $\pm$ 69 & 272 $_{- 105 }^{+ 116 }$  &   11.56 $_{- 0.11 }^{+ 0.09 }$  &   3.24 $_{- 0.39 }^{+ 0.29 }$  &   U  \\
025 &  5.850  & {\citet{2019ApJ...887...55C}}          & 4.16 $\pm$ 0.12     & 373 $\pm$ 11           & 0.06  &   373 $\pm$ 11    & 68 $_{- 20 }^{+ 32 }$      & 10.61 $_{- 0.12 }^{+ 0.12 }$ & 2.94 $_{- 0.31 }^{+ 0.33 }$ & J                   \\
026 & 3.47  & J          & 4.13 $\pm$ 0.25     & 371 $\pm$ 22           & 0.33  &   372 $\pm$ 22     & 157 $_{- 46 }^{+ 218 }$    & 11.09 $_{- 0.25 }^{+ 0.10 }$  & 2.51 $_{- 0.27 }^{+ 0.27 }$ & J                   \\
027 & 2.163 & {\citet{{2012ApJS..200...13B, 2015ApJS..219...29M}}}          & 3.72 $\pm$ 0.48     & 334 $\pm$ 42           & 0.04  &   334 $\pm$ 42      & 90 $_{- 26 }^{+ 18 }$      & 11.34 $_{- 0.05 }^{+ 0.06 }$ & 2.64 $_{- 0.19 }^{+ 0.20 }$  & J                   \\
028 & 2.17  & U          & 3.64 $\pm$ 0.23     & 327 $\pm$ 20           & 0.23  &   327 $\pm$ 20   & 52 $_{- 8 }^{+ 11 }$       & 11.00 $_{- 0.07 }^{+ 0.06 }$  & 1.95 $_{- 0.18 }^{+ 0.20 }$  & U                   \\
029 & 2.81  & U          & 3.62 $\pm$ 0.19     & 325 $\pm$ 17           & 0.57  &   325 $\pm$ 17   & 66 $_{- 12 }^{+ 19 }$      & 10.97 $_{- 0.06 }^{+ 0.05 }$ & 1.21 $_{- 0.14 }^{+ 0.19 }$ & U                   \\
030 & 3.21  & J          & 3.62 $\pm$ 0.34     & 325 $\pm$ 30           & 0.05  &   325 $\pm$ 30   & 59 $_{- 14 }^{+ 15 }$      & 10.93 $_{- 0.07 }^{+ 0.08 }$ & 2.77 $_{- 0.23 }^{+ 0.33 }$ & J                   \\
031 & 4.621 & {\citet{2021ApJ...907..122M}}          & 3.61 $\pm$ 0.43     & 324 $\pm$ 38           & 0.21  &   324 $\pm$ 38    & 91 $_{- 23 }^{+ 26 }$      & 10.90 $_{- 0.08 }^{+ 0.08 }$  & 2.20 $_{- 0.25 }^{+ 0.26 }$  & J                   \\
032 & 2.012 & {\citet{{2012ApJS..200...13B, 2015ApJS..219...29M}}}          & 3.45 $\pm$ 0.23     & 310 $\pm$ 20           & 0.08  &   310 $\pm$ 20    & 81 $_{- 15 }^{+ 16 }$      & 11.24 $_{- 0.06 }^{+ 0.06 }$ & 2.67 $_{- 0.29 }^{+ 0.34 }$ & J                   \\
033 & 4.30  & J          & 3.27 $\pm$ 0.29     & 293 $\pm$ 26           & 0.01  &   293 $\pm$ 26   & 220 $_{- 52 }^{+ 71 }$     & 11.32 $_{- 0.08 }^{+ 0.09 }$ & 3.96 $_{- 0.30 }^{+ 0.31 }$  & J                   \\
034 & 3.55      & J                              & 3.25 $\pm$ 0.54                         & 292 $\pm$ 48                               & 0.00  &   292 $\pm$ 48 & 0 $_{- 0 }^{+ 339 }$  &   11.79 $_{- 0.11 }^{+ 0.07 }$  &   2.10 $_{- 0.37 }^{+ 0.59 }$  &   J                                       \\
035 & 2.53  & J          & 3.20 $\pm$ 0.41      & 287 $\pm$ 37           & 1.36  &   289 $\pm$ 37   & 48 $_{- 5 }^{+ 3 }$        & 10.20 $_{- 0.02 }^{+ 0.04 }$  & 0.27 $_{- 0.04 }^{+ 0.02 }$ & J                   \\
036 & 2.69 & J          & 3.11 $\pm$ 0.30      & 279 $\pm$ 26           & 0.01  &   279 $\pm$ 26  & 147 $_{- 147 }^{+ 59 }$    & 11.54 $_{- 0.07 }^{+ 0.14 }$ & 3.42 $_{- 0.64 }^{+ 0.24 }$ & J                   \\
037 & 1.84                  & U                              & 3.07 $\pm$ 0.52                         & 275 $\pm$ 46                               & 0.45  &   276 $\pm$ 46  & 152 $_{- 52 }^{+ 62 }$  &   11.14 $_{- 0.05 }^{+ 0.10 }$  &   1.58 $_{- 0.26 }^{+ 0.22 }$  &   U \\
038 & 2.60      & J                              & 3.07 $\pm$ 0.60                          & 276 $\pm$ 53                               & 0.08  &   276 $\pm$ 53   & 97 $_{- 17 }^{+ 25 }$  &   11.20 $_{- 0.05 }^{+ 0.07 }$  &   2.64 $_{- 0.20 }^{+ 0.34 }$  &   J                                        \\
039 & 2.536                   & {\citet{khostovan2025cosmosspectroscopicredshiftcompilation}}                     & 3.04 $\pm$ 0.66           & 273 $\pm$ 59                               & 127.55  &   400 $\pm$ 64    & 8386 $_{- 337 }^{+ 304 }$  &   11.02 $_{- 0.02 }^{+ 0.03 }$  &   0.61 $_{- 0.01 }^{+ 0.02 }$  &   U        \\
040 & 1.780                    & {\citet{{2012ApJS..200...13B, 2015ApJS..219...29M}}}                              & 2.94 $\pm$ 0.28                         & 263 $\pm$ 24                               & 0.89  &   264 $\pm$ 24    & 490 $_{- 71 }^{+ 43 }$  &   10.87 $_{- 0.06 }^{+ 0.05 }$  &   3.12 $_{- 0.08 }^{+ 0.07 }$  &   J \\
041 & 1.463 & {\citet{2015ApJS..220...12S}}          & 2.83 $\pm$ 0.23     & 254 $\pm$ 20           & 2.06  &   256 $\pm$ 20     & 637 $_{- 92 }^{+ 92 }$     & 10.86 $_{- 0.06 }^{+ 0.07 }$ & 1.24 $_{- 0.07 }^{+ 0.06 }$ & U                   \\
042 & 3.46       & J                              & 2.38 $\pm$ 0.43                         & 214 $\pm$ 38                               & 0.51  &   214 $\pm$ 38 & 73 $_{- 14 }^{+ 15 }$  &   10.97 $_{- 0.04 }^{+ 0.05 }$  &   1.89 $_{- 0.17 }^{+ 0.18 }$  &   J                                       \\
\hline
\end{tabularx}
\end{table*}

\begin{table*}
\contcaption{}
\label{tab:SED2}
\renewcommand{\arraystretch}{1.3}
\begin{tabularx}{\textwidth}{clXrlcllllc}
\hline
ID & \multicolumn{1}{l}{$z$} & \multicolumn{1}{l}{Ref.} & \multicolumn{1}{c}{f$_{\rm{870\mu m}}$} & \multicolumn{1}{c}{SFR$_{\rm{FIR}}$}       & \multicolumn{1}{c}{SFR$_{\rm{UV}}$}        & \multicolumn{1}{c}{SFR$_{\rm{raw}}$}       & \multicolumn{1}{c}{SFR$_{\rm{corr}}$}      & \multicolumn{1}{c}{M$_*$}            & \multicolumn{1}{c}{A$_{\rm V}$}   & \multicolumn{1}{c}{Flag} \\
\multicolumn{1}{c}{}   & \multicolumn{1}{c}{}    & \multicolumn{1}{c}{}           & \multicolumn{1}{c}{(mJy)}               & \multicolumn{1}{c}{(M$_{\odot}$yr$^{-1}$)} & \multicolumn{1}{c}{(M$_{\odot}$yr$^{-1}$)} & \multicolumn{1}{c}{(M$_{\odot}$yr$^{-1}$)} & \multicolumn{1}{c}{(M$_{\odot}$yr$^{-1}$)} & \multicolumn{1}{c}{(logM$_{\odot}$)} &                             &                                         \\
\hline
043 & 1.324                   & {\citet{2015ApJS..218...15K}}                              & 2.83 $\pm$ 0.58                         & 254 $\pm$ 52                               & 0.46  &   254 $\pm$ 52  & 244 $_{- 57 }^{+ 78 }$  &   10.94 $_{- 0.06 }^{+ 0.06 }$  &   1.57 $_{- 0.18 }^{+ 0.18 }$  &   U                                \\
044 & 3.53                  & J                              & 2.64 $\pm$ 0.44                         & 237 $\pm$ 39                               & 0.20  &   237 $\pm$ 39  & 115 $_{- 29 }^{+ 27 }$  &   11.18 $_{- 0.06 }^{+ 0.07 }$  &   2.34 $_{- 0.22 }^{+ 0.29 }$  &   J                                       \\
045 & 2.89  & U          & 2.60 $\pm$ 0.22      & 233 $\pm$ 20           & 0.33  &   233 $\pm$ 20    & 93 $_{- 20 }^{+ 22 }$      & 11.12 $_{- 0.07 }^{+ 0.08 }$ & 1.79 $_{- 0.21 }^{+ 0.22 }$ & U                   \\
046 & 4.200 & J          & 2.57 $\pm$ 0.29     & 231 $\pm$ 25           & 0.14  &   231 $\pm$ 25   & 159 $_{- 38 }^{+ 39 }$     & 11.21 $_{- 0.07 }^{+ 0.07 }$ & 2.85 $_{- 0.25 }^{+ 0.30 }$  & J       \\
047 & 2.62       & J                              & 2.50 $\pm$ 0.43                          & 225 $\pm$ 38                               & 0.35  &   225 $\pm$ 38 & 101 $_{- 27 }^{+ 58 }$  &   11.02 $_{- 0.09 }^{+ 0.08 }$  &   2.47 $_{- 0.23 }^{+ 0.23 }$  &   J  \\
048 & 0.748 & {\citet{2021ApJS..256...44V}}          & 2.47 $\pm$ 0.14     & 222 $\pm$ 12           & 0.47  &   222 $\pm$ 12    & 38 $_{- 5 }^{+ 8 }$        & 11.16 $_{- 0.08 }^{+ 0.04 }$ & 1.37 $_{- 0.14 }^{+ 0.13 }$ & U                   \\
049 & 1.736 & {\citet{{2012ApJS..200...13B, 2015ApJS..219...29M}}}          & 2.37 $\pm$ 0.45     & 213 $\pm$ 40           & 0.02  &   213 $\pm$ 40    & 43 $_{- 25 }^{+ 14 }$      & 11.14 $_{- 0.07 }^{+ 0.12 }$ & 3.47 $_{- 0.32 }^{+ 0.20 }$  & J                   \\
050 & 2.59      & J                              & 2.35 $\pm$ 0.13                         & 211 $\pm$ 11                               & 0.03  &   211 $\pm$ 11  & 89 $_{- 16 }^{+ 24 }$  &   11.19 $_{- 0.07 }^{+ 0.07 }$  &   3.41 $_{- 0.24 }^{+ 0.36 }$  &   J \\
051 & 2.73 & J          & 2.31 $\pm$ 0.13     & 207 $\pm$ 12           & 0.06  &   207 $\pm$ 12    & 101 $_{- 31 }^{+ 22 }$     & 11.28 $_{- 0.05 }^{+ 0.05 }$ & 2.58 $_{- 0.36 }^{+ 0.25 }$ & J                   \\
052 & 0.929                   & {\citet{2019MNRAS.482.3135B}}                              & 2.29 $\pm$ 0.22                         & 206 $\pm$ 19                               & 0.25  &   206 $\pm$ 19  & 37 $_{- 5 }^{+ 5 }$  &   11.16 $_{- 0.05 }^{+ 0.05 }$  &   1.75 $_{- 0.18 }^{+ 0.26 }$  &   J    \\
053 & 1.522                   & {\citet{{2012ApJS..200...13B, 2015ApJS..219...29M}}}                              & 2.23 $\pm$ 0.26                         & 200 $\pm$ 22                               & 0.33  &   200 $\pm$ 22  & 107 $_{- 14 }^{+ 16 }$  &   11.45 $_{- 0.05 }^{+ 0.05 }$  &   1.72 $_{- 0.16 }^{+ 0.21 }$  &   J                                        \\
054 & 1.831                   & {\citet{{2012ApJS..200...13B, 2015ApJS..219...29M}}}                              & 2.17 $\pm$ 0.62                         & 195 $\pm$ 55                               & 0.00  &   195 $\pm$ 55  & 0 $_{- 0 }^{+ 3 }$  &   11.53 $_{- 0.10 }^{+ 0.05 }$  &   3.20 $_{- 0.34 }^{+ 0.20 }$  &   J  \\
055 & 1.030                    & {\citet{2019MNRAS.482.3135B}}                              & 2.10 $\pm$ 0.17                          & 188 $\pm$ 14                               & 0.64  &   189 $\pm$ 14 & 42 $_{- 5 }^{+ 7 }$  &   11.15 $_{- 0.06 }^{+ 0.04 }$  &   1.65 $_{- 0.19 }^{+ 0.25 }$  &   J   \\
056 & 2.94       & J                              & 2.10 $\pm$ 0.35                          & 189 $\pm$ 31                               & 0.13  &   189 $\pm$ 31  & 91 $_{- 19 }^{+ 67 }$  &   11.00 $_{- 0.10 }^{+ 0.09 }$  &   2.92 $_{- 0.22 }^{+ 0.25 }$  &   J                                        \\
057 & 5.36  & U          & 2.01 $\pm$ 0.24     & 181 $\pm$ 21           & 0.04  &   181 $\pm$ 21   & 0 $_{- 0 }^{+ 327 }$       & 11.74 $_{- 0.20 }^{+ 0.20 }$   & 1.30 $_{- 0.65 }^{+ 0.69 }$  & U                   \\
058 & 2.380                    & {\citet{{2012ApJS..200...13B, 2015ApJS..219...29M}}}                              & 1.97 $\pm$ 0.37                         & 177 $\pm$ 33                               & 0.77  &   177 $\pm$ 33  & 104 $_{- 37 }^{+ 41 }$  &   11.17 $_{- 0.06 }^{+ 0.07 }$  &   1.56 $_{- 0.21 }^{+ 0.20 }$  &   J  \\
059 & 3.50  & J          & 1.94 $\pm$ 0.24     & 174 $\pm$ 21           & 0.04  &   174 $\pm$ 21    & 33 $_{- 18 }^{+ 11 }$      & 10.66 $_{- 0.08 }^{+ 0.08 }$ & 2.33 $_{- 0.32 }^{+ 0.32 }$ & J                   \\
060 & 2.55                 & U                              & 1.94 $\pm$ 0.43                         & 174 $\pm$ 38                               & 2.06  &   176 $\pm$ 38 & 260 $_{- 58 }^{+ 59 }$  &   10.75 $_{- 0.05 }^{+ 0.06 }$  &   1.30 $_{- 0.16 }^{+ 0.14 }$  &   U  \\
061 & 2.33                  & J                              & 1.94 $\pm$ 0.43                         & 174 $\pm$ 38                               & 0.19  &   174 $\pm$ 38  & 68 $_{- 10 }^{+ 11 }$  &   11.10 $_{- 0.05 }^{+ 0.05 }$  &   2.69 $_{- 0.16 }^{+ 0.19 }$  &   J                               \\
062 & 3.422 &   {\citet{2024jwst.prop.6585C}}       & 1.91 $\pm$ 0.26     & 171 $\pm$ 23           & 0.07  &   171 $\pm$ 23     & 52 $_{- 11 }^{+ 15 }$      & 10.83 $_{- 0.08 }^{+ 0.07 }$ & 2.40 $_{- 0.20 }^{+ 0.21 }$   & J                   \\
063 & 4.19  & J          & 1.90 $\pm$ 0.28      & 170 $\pm$ 25           & 0.20  &   170 $\pm$ 25   & 110 $_{- 25 }^{+ 24 }$     & 11.03 $_{- 0.07 }^{+ 0.08 }$ & 2.70 $_{- 0.21 }^{+ 0.26 }$  & J                   \\
064 & 3.709 & {\citet{2018A&A...618A..85S}} & 1.89 $\pm$ 0.04     & 170 $\pm$ 3            & 0.22  &   170 $\pm$ 3      & 0 $_{- 0 }^{+ 0 }$         & 10.92 $_{- 0.03 }^{+ 0.03 }$ & 0.60 $_{- 0.11 }^{+ 0.10 }$   & J                   \\
065 & 2.60       & J                              & 1.87 $\pm$ 0.30                          & 168 $\pm$ 27                               & 0.34  &   168 $\pm$ 27 & 60 $_{- 9 }^{+ 33 }$  &   10.92 $_{- 0.11 }^{+ 0.05 }$  &   2.25 $_{- 0.15 }^{+ 0.22 }$  &   J \\
066 & 4.14      & J                              & 1.79 $\pm$ 0.39                         & 160 $\pm$ 35                               & 0.59  &   161 $\pm$ 35  & 173 $_{- 107 }^{+ 109 }$  &   10.66 $_{- 0.11 }^{+ 0.12 }$  &   2.70 $_{- 0.39 }^{+ 0.22 }$  &   J \\
067 & 1.63                 & U                              & 1.62 $\pm$ 0.32                         & 145 $\pm$ 29                               & 0.02  &   145 $\pm$ 29  & 21 $_{- 3 }^{+ 5 }$  &   10.71 $_{- 0.05 }^{+ 0.06 }$  &   1.94 $_{- 0.21 }^{+ 0.27 }$  &   U                                       \\
068 & 1.11  & U          & 1.54 $\pm$ 0.29     & 138 $\pm$ 25           & 0.02  &   138 $\pm$ 25   & 61 $_{- 8 }^{+ 13 }$       & 11.26 $_{- 0.08 }^{+ 0.05 }$ & 2.48 $_{- 0.17 }^{+ 0.16 }$ & U                   \\
069 & 3.07  & J          & 1.53 $\pm$ 0.23     & 137 $\pm$ 21           & 0.02  &   138 $\pm$ 21   & 30 $_{- 11 }^{+ 9 }$       & 10.67 $_{- 0.09 }^{+ 0.09 }$ & 3.28 $_{- 0.34 }^{+ 0.29 }$ & J                   \\
070 & 2.450 &  {\citet{2021A&A...646A..96C}}      & 1.53 $\pm$ 0.24     & 137 $\pm$ 21           & 0.85  &   138 $\pm$ 21     & 202 $_{- 34 }^{+ 23 }$     & 10.64 $_{- 0.05 }^{+ 0.10 }$  & 1.80 $_{- 0.09 }^{+ 0.09 }$  & J                   \\
071 & 1.793                   & {\citet{2021ApJ...913..110C}}                              & 1.52 $\pm$ 0.28                         & 136 $\pm$ 25                               & 0.18  &   137 $\pm$ 25  & 96 $_{- 32 }^{+ 35 }$  &   10.99 $_{- 0.07 }^{+ 0.08 }$  &   3.17 $_{- 0.25 }^{+ 0.25 }$  &   J                                       \\
072 & 2.70      & J                              & 1.49 $\pm$ 0.36                         & 134 $\pm$ 32                               & 0.37  &   134 $\pm$ 32   & 61 $_{- 19 }^{+ 13 }$  &   11.06 $_{- 0.06 }^{+ 0.05 }$  &   1.78 $_{- 0.21 }^{+ 0.26 }$  &   J                                        \\
073 & 0.919                   & {\citet{2018ApJ...858...77H}}                              & 1.48 $\pm$ 0.12                         & 133 $\pm$ 10                               & 0.13  &   133 $\pm$ 10 & 43 $_{- 6 }^{+ 7 }$  &   11.20 $_{- 0.04 }^{+ 0.05 }$  &   1.80 $_{- 0.18 }^{+ 0.24 }$  &   U \\
074 & 1.12                  & U                              & 1.46 $\pm$ 0.30                          & 131 $\pm$ 26                               & 0.03  &   131 $\pm$ 26  & 39 $_{- 8 }^{+ 18 }$  &   10.96 $_{- 0.10 }^{+ 0.06 }$  &   1.96 $_{- 0.19 }^{+ 0.25 }$  &   U \\
075 & 2.989 & {\citet{2024jwst.prop.6585C}}        & 1.46 $\pm$ 0.39                         & 131 $\pm$ 34                               & 0.94  &   131 $\pm$ 34  & 190 $_{- 20 }^{+ 17 }$  &   10.39 $_{- 0.05 }^{+ 0.06 }$  &   1.73 $_{- 0.07 }^{+ 0.08 }$  &   J  \\
076 & 2.474 & {\citet{2018A&A...618A..85S}}  & 1.45 $\pm$ 0.34                         & 130 $\pm$ 30                               & 0.01  &   130 $\pm$ 30       & 88 $_{- 25 }^{+ 23 }$                      & 11.22 $_{- 0.07 }^{+ 0.09 }$         & 3.51 $_{- 0.31 }^{+ 0.33 }$ & J                                       \\
077 & 2.98  & J          & 1.42 $\pm$ 0.32     & 127 $\pm$ 28           & 0.03  &   127 $\pm$ 28   & 99 $_{- 28 }^{+ 23 }$      & 11.19 $_{- 0.06 }^{+ 0.07 }$ & 2.96 $_{- 0.31 }^{+ 0.32 }$ & J                   \\
078 & 2.467  & \citet{khostovan2025cosmosspectroscopicredshiftcompilation}          & 1.41 $\pm$ 0.11     & 126 $\pm$ 9            & 0.79  &   127 $\pm$ 9    & 65 $_{- 17 }^{+ 57 }$      & 10.90 $_{- 0.15 }^{+ 0.07 }$  & 1.86 $_{- 0.22 }^{+ 0.28 }$ & J                   \\
079 & 1.11                  & U                              & 1.41 $\pm$ 0.27                         & 126 $\pm$ 24                               & 0.02  &   126 $\pm$ 24 & 20 $_{- 3 }^{+ 7 }$  &   10.73 $_{- 0.08 }^{+ 0.06 }$  &   2.11 $_{- 0.21 }^{+ 0.28 }$  &   U \\
080 & 5.852                   & {\citet{2019ApJ...887...55C}}                              & 1.40 $\pm$ 0.09                          & 125 $\pm$ 8                                & 0.02  &   125 $\pm$ 8  & 27 $_{- 27 }^{+ 12 }$  &   10.36 $_{- 0.12 }^{+ 0.13 }$  &   2.41 $_{- 0.32 }^{+ 0.30 }$  &   J \\
081 & 3.69                  & J                              & 1.40 $\pm$ 0.23                          & 125 $\pm$ 20                               & 0.01  &   125 $\pm$ 20  & 0 $_{- 0 }^{+ 0 }$  &   10.40 $_{- 0.02 }^{+ 0.02 }$  &   0.50 $_{- 0.05 }^{+ 0.05 }$  &   J  \\
082 & 1.610                    & {\citet{{2012ApJS..200...13B, 2015ApJS..219...29M}}}                              & 1.31 $\pm$ 0.28                         & 117 $\pm$ 24                               & 0.05  &   117 $\pm$ 24  & 60 $_{- 13 }^{+ 11 }$  &   11.21 $_{- 0.07 }^{+ 0.06 }$  &   2.81 $_{- 0.27 }^{+ 0.31 }$  &   J  \\
083 & 1.68       & J                              & 1.31 $\pm$ 0.31                         & 118 $\pm$ 28                               & 0.82  &   118 $\pm$ 28  & 43 $_{- 4 }^{+ 5 }$  &   10.01 $_{- 0.04 }^{+ 0.03 }$  &   1.65 $_{- 0.05 }^{+ 0.04 }$  &   J  \\
084 & 1.139                   & {\citet{2018ApJ...869...27V}}                              & 1.29 $\pm$ 0.31                         & 115 $\pm$ 27                               & 0.16  &   115 $\pm$ 27  & 42 $_{- 6 }^{+ 5 }$  &   11.16 $_{- 0.05 }^{+ 0.05 }$  &   2.30 $_{- 0.17 }^{+ 0.15 }$  &   J      \\
085 & 1.746                   & {\citet{{2012ApJS..200...13B, 2015ApJS..219...29M}}}       & 1.29 $\pm$ 0.38                         & 115 $\pm$ 34                               & 1.94  &   117 $\pm$ 34  &  57 $_{- 3 }^{+ 4 }$  &   10.58 $_{- 0.02 }^{+ 0.01 }$  &   1.45 $_{- 0.04 }^{+ 0.05 }$  &   J             \\
086 & 3.256                   & {\citet{2015ApJS..218...15K}}                              & 1.28 $\pm$ 0.31                         & 114 $\pm$ 27                               & 1.44  &   116 $\pm$ 27     & 183 $_{- 37 }^{+ 34 }$                     & 10.05 $_{- 0.10 }^{+ 0.11 }$          & 1.15 $_{- 0.05 }^{+ 0.06 }$ & J                                       \\

\hline
\end{tabularx}
\end{table*}

\begin{table*}
\contcaption{}
\label{tab:SED3}
\renewcommand{\arraystretch}{1.3}
\begin{tabularx}{\textwidth}{clXrlcllllc}
\hline
ID & \multicolumn{1}{l}{$z$} & \multicolumn{1}{l}{Ref.} & \multicolumn{1}{c}{f$_{\rm{870\mu m}}$} & \multicolumn{1}{c}{SFR$_{\rm{FIR}}$}       & \multicolumn{1}{c}{SFR$_{\rm{UV}}$}        & \multicolumn{1}{c}{SFR$_{\rm{raw}}$}       & \multicolumn{1}{c}{SFR$_{\rm{corr}}$}      & \multicolumn{1}{c}{M$_*$}            & \multicolumn{1}{c}{A$_{\rm V}$}   & \multicolumn{1}{c}{Flag} \\
\multicolumn{1}{c}{}   & \multicolumn{1}{c}{}    & \multicolumn{1}{c}{}           & \multicolumn{1}{c}{(mJy)}               & \multicolumn{1}{c}{(M$_{\odot}$yr$^{-1}$)} & \multicolumn{1}{c}{(M$_{\odot}$yr$^{-1}$)} & \multicolumn{1}{c}{(M$_{\odot}$yr$^{-1}$)} & \multicolumn{1}{c}{(M$_{\odot}$yr$^{-1}$)} & \multicolumn{1}{c}{(logM$_{\odot}$)} &                             &                                         \\
\hline
087 & 1.214                   & {\citet{2019MNRAS.482.3135B}}                              & 1.27 $\pm$ 0.10                          & 113 $\pm$ 8                                & 0.00  &   113 $\pm$ 8 & 0 $_{- 0 }^{+ 0 }$  &   11.63 $_{- 0.04 }^{+ 0.03 }$  &   2.16 $_{- 0.29 }^{+ 0.23 }$  &   J     \\
088 & 5.17       & J                             & 1.26 $\pm$ 0.24                         & 113 $\pm$ 21                               & 0.59  &   113 $\pm$ 21  & 793 $_{- 321 }^{+ 461 }$  &   11.01 $_{- 0.15 }^{+ 0.15 }$  &   2.61 $_{- 0.31 }^{+ 0.38 }$  &   J \\
089 & 3.291                  & {\citet{2021jwst.prop.1810B}}                              & 1.25 $\pm$ 0.15                         & 112 $\pm$ 13                               & 2.04  &   114 $\pm$ 13  & 55 $_{- 7 }^{+ 16 }$  &   10.81 $_{- 0.05 }^{+ 0.04 }$  &   0.94 $_{- 0.10 }^{+ 0.15 }$  &   J  \\
090 & 2.71 & J          & 1.23 $\pm$ 0.21     & 110 $\pm$ 19           & 0.02  &   110 $\pm$ 19   & 27 $_{- 6 }^{+ 6 }$        & 10.65 $_{- 0.05 }^{+ 0.07 }$ & 2.41 $_{- 0.19 }^{+ 0.42 }$ & J                   \\
091 & 2.488  &  {\citet{2024jwst.prop.6585C}} & 1.22 $\pm$ 0.29                         & 109 $\pm$ 26                               & 0.01  &   109 $\pm$ 26  & 52 $_{- 14 }^{+ 13 }$                      & 10.98 $_{- 0.07 }^{+ 0.08 }$         & 3.23 $_{- 0.37 }^{+ 0.31 }$ & J                                       \\
092 & 0.688 & {\citet{2021ApJS..256...44V}}                              & 1.20 $\pm$ 0.26                          & 107 $\pm$ 23                               & 0.15  &   108 $\pm$ 23  & 283 $_{- 27 }^{+ 32 }$  &   11.13 $_{- 0.03 }^{+ 0.04 }$  &   3.96 $_{- 0.06 }^{+ 0.06 }$  &   J                \\
093 & 1.27                    & {\citet{2018ApJ...869...27V, 2020ApJ...890...24V}}                             & 1.18 $\pm$ 0.10          & 106 $\pm$ 8                   & 0.41  &   106 $\pm$ 8   & 40 $_{- 6 }^{+ 11 }$  &   10.97 $_{- 0.10 }^{+ 0.06 }$  &   1.50 $_{- 0.14 }^{+ 0.15 }$  &   J   \\
094 & 1.746                   & {\citet{{2012ApJS..200...13B, 2015ApJS..219...29M}}}                              & 1.18 $\pm$ 0.16                         & 105 $\pm$ 14                               & 0.03  &   105 $\pm$ 14   & 33 $_{- 5 }^{+ 5 }$  &   10.91 $_{- 0.07 }^{+ 0.05 }$  &   3.25 $_{- 0.25 }^{+ 0.24 }$  &   J \\
095 & 0.726                   & {\citet{2021ApJS..256...44V}}                              & 1.18 $\pm$ 0.33                         & 106 $\pm$ 29                               & 1.90  &   108 $\pm$ 29   & 29 $_{- 4 }^{+ 3 }$  &   11.12 $_{- 0.03 }^{+ 0.04 }$   &  0.85 $_{- 0.11 }^{+ 0.13 }$  &   J                                       \\
096 & 2.75       & J                              & 1.14 $\pm$ 0.33                         & 102 $\pm$ 29                               & 0.06  &   102 $\pm$ 29   & 49 $_{- 9 }^{+ 15 }$                       & 10.84 $_{- 0.07 }^{+ 0.08 }$         & 2.84 $_{- 0.19 }^{+ 0.27 }$ & J                                       \\
097 & 1.62                   & U                              & 1.11 $\pm$ 0.24                         & 99 $\pm$ 21                                & 0.88  &   100 $\pm$ 21   & 116 $_{- 27 }^{+ 29 }$  &   10.69 $_{- 0.05 }^{+ 0.05 }$  &   1.28 $_{- 0.17 }^{+ 0.14 }$  &   U                                         \\
098 & 1.82        & J                              & 1.10 $\pm$ 0.34                          & 99 $\pm$ 30                                & 0.15  &   99 $\pm$ 30  & 55 $_{- 8 }^{+ 8 }$  &   11.11 $_{- 0.04 }^{+ 0.05 }$  &   2.03 $_{- 0.15 }^{+ 0.22 }$  &   J                                       \\
099 & 1.51      & U                              & 1.09 $\pm$ 0.23                         & 97 $\pm$ 20                                & 0.04  &   97 $\pm$ 20  & 27 $_{- 4 }^{+ 3 }$  &   10.87 $_{- 0.04 }^{+ 0.03 }$  &   1.68 $_{- 0.14 }^{+ 0.14 }$  &   U \\
100 & 2.676                   & {\citet{khostovan2025cosmosspectroscopicredshiftcompilation}}                              & 1.06 $\pm$ 0.29                         & 95 $\pm$ 26                                & 9.29  &   104 $\pm$ 26      & 209 $_{- 23 }^{+ 22 }$    & 10.27 $_{- 0.05 }^{+ 0.05 }$   &  0.60 $_{- 0.02 }^{+ 0.02 }$  &   J                         \\
101 & 4.41       & J                              & 1.05 $\pm$ 0.29                         & 94 $\pm$ 26                                & 8.12  &   102 $\pm$ 26  & 158 $_{- 31 }^{+ 34 }$  &   10.22 $_{- 0.06 }^{+ 0.07 }$  &   0.64 $_{- 0.06 }^{+ 0.07 }$  &   J  \\
102 & 2.204 & {\citet{{2012ApJS..200...13B, 2015ApJS..219...29M}}}          & 1.00 $\pm$ 0.28      & 89 $\pm$ 25            & 0.19  &   89 $\pm$ 25    & 90 $_{- 23 }^{+ 27 }$      & 10.71 $_{- 0.08 }^{+ 0.08 }$ & 2.53 $_{- 0.16 }^{+ 0.12 }$ & J                   \\
103 & 3.20       & J                              & 0.98 $\pm$ 0.16                         & 88 $\pm$ 13                                & 0.58  &   88 $\pm$ 13 & 22 $_{- 3 }^{+ 4 }$  &   10.48 $_{- 0.05 }^{+ 0.04 }$  &   1.16 $_{- 0.12 }^{+ 0.15 }$  &   J \\
104 & 2.642 & {\citet{{2012ApJS..200...13B, 2015ApJS..219...29M}}}          & 0.94 $\pm$ 0.17     & 84 $\pm$ 15            & 0.59  &   85 $\pm$ 15    & 51 $_{- 9 }^{+ 20 }$       & 10.85 $_{- 0.06 }^{+ 0.06 }$ & 1.40 $_{- 0.22 }^{+ 0.26 }$  & J                   \\
105 & 1.54      & J                              & 0.90 $\pm$ 0.17                          & 80 $\pm$ 15                                & 0.09  &   80 $\pm$ 15 & 8 $_{- 1 }^{+ 1 }$  &   10.33 $_{- 0.03 }^{+ 0.03 }$  &   1.58 $_{- 0.09 }^{+ 0.09 }$  &   J  \\
106 & 2.19                  & J                              & 0.90 $\pm$ 0.24                          & 81 $\pm$ 21                                & 0.07  &   81 $\pm$ 21 & 16 $_{- 3 }^{+ 5 }$  &   10.46 $_{- 0.06 }^{+ 0.06 }$  &   1.91 $_{- 0.17 }^{+ 0.20 }$  &   J  \\
107 & 2.21  & U          & 0.88 $\pm$ 0.09     & 79 $\pm$ 7             & 0.16  &   79 $\pm$ 7   & 13 $_{- 3 }^{+ 3 }$        & 10.41 $_{- 0.06 }^{+ 0.06 }$ & 1.44 $_{- 0.20 }^{+ 0.22 }$  & U                   \\
108 & 1.968                   & {\citet{{2012ApJS..200...13B, 2015ApJS..219...29M}}}                              & 0.85 $\pm$ 0.27                         & 76 $\pm$ 23                                & 0.12  &   76 $\pm$ 23  & 47 $_{- 7 }^{+ 8 }$  &   10.99 $_{- 0.06 }^{+ 0.05 }$  &   2.91 $_{- 0.25 }^{+ 0.26 }$  &   J  \\
109 & 2.98       & J                              & 0.84 $\pm$ 0.19                         & 75 $\pm$ 17                                & 3.56  &   79 $\pm$ 17 & 136 $_{- 26 }^{+ 16 }$  &   10.32 $_{- 0.04 }^{+ 0.08 }$  &   1.10 $_{- 0.07 }^{+ 0.05 }$  &   J \\
110 & 3.46      & J                              & 0.80 $\pm$ 0.20                           & 71 $\pm$ 17                                & 0.23  &   71 $\pm$ 17  & 36 $_{- 6 }^{+ 13 }$  &   10.60 $_{- 0.06 }^{+ 0.06 }$  &   1.98 $_{- 0.18 }^{+ 0.21 }$  &   J  \\
111 & 3.529                   & {\citet{2015ApJS..218...15K}}                              & 0.70 $\pm$ 0.21                          & 63 $\pm$ 19                                & 2.67  &   66 $\pm$ 19 & 59 $_{- 10 }^{+ 21 }$  &   10.75 $_{- 0.07 }^{+ 0.05 }$  &   0.86 $_{- 0.11 }^{+ 0.15 }$  &   J     \\
112 & 2.195                   & {\citet{{2012ApJS..200...13B, 2015ApJS..219...29M}}}                              & 0.69 $\pm$ 0.11                         & 62 $\pm$ 10                                & 3.31  &   65 $\pm$ 10 & 120 $_{- 28 }^{+ 23 }$  &   10.64 $_{- 0.06 }^{+ 0.07 }$  &   0.99 $_{- 0.13 }^{+ 0.08 }$  &   J  \\
113 & 2.193                   & {\citet{{2012ApJS..200...13B, 2015ApJS..219...29M}}}                              & 0.69 $\pm$ 0.15                         & 61 $\pm$ 13                                & 0.38  &   62 $\pm$ 13  & 85 $_{- 24 }^{+ 40 }$  &   11.01 $_{- 0.08 }^{+ 0.06 }$  &   2.56 $_{- 0.23 }^{+ 0.17 }$  &   J  \\
114 & 1.88  & U          & 0.67 $\pm$ 0.13     & 60 $\pm$ 11            & 0.18  &   60 $\pm$ 11    & 15 $_{- 2 }^{+ 2 }$        & 10.49 $_{- 0.05 }^{+ 0.05 }$ & 1.38 $_{- 0.14 }^{+ 0.17 }$ & U                   \\
115 & 2.514                   & {\citet{2021jwst.prop.2565G}}                              & 0.63 $\pm$ 0.13                         & 56 $\pm$ 11                                & 0.06  &   56 $\pm$ 11 & 45 $_{- 9 }^{+ 15 }$  &   10.86 $_{- 0.07 }^{+ 0.07 }$  &   2.17 $_{- 0.26 }^{+ 0.34 }$  &   J    \\
116 & 2.469                   & {\citet{2023ApJ...942...24S}}                              & 0.59 $\pm$ 0.12                         & 53 $\pm$ 10                                & 0.87  &   53 $\pm$ 10 &  136 $_{- 44 }^{+ 25 }$  &   10.60 $_{- 0.08 }^{+ 0.10 }$  &   1.81 $_{- 0.17 }^{+ 0.08 }$  &   J        \\
117 & 1.599                   & {\citet{{2012ApJS..200...13B, 2015ApJS..219...29M}}}                              & 0.58 $\pm$ 0.14                         & 51 $\pm$ 12                                & 0.01  &   51 $\pm$ 12  & 27 $_{- 27 }^{+ 11 }$  &   11.08 $_{- 0.10 }^{+ 0.26 }$  &   3.19 $_{- 0.21 }^{+ 0.16 }$  &   J  \\
118 & 5.536                   & {\citet{2020ApJ...900....1F}}                              & 0.56 $\pm$ 0.11                         & 50 $\pm$ 10                                & 4.97  &   55 $\pm$ 10 & 46 $_{- 9 }^{+ 12 }$  &   10.50 $_{- 0.08 }^{+ 0.07 }$  &   0.29 $_{- 0.09 }^{+ 0.11 }$  &   U     \\
119 & 2.216                   & {\citet{{2012ApJS..200...13B, 2015ApJS..219...29M}}}                              & 0.54 $\pm$ 0.17                         & 48 $\pm$ 14                                & 0.48  &   48 $\pm$ 14    & 91 $_{- 32 }^{+ 36 }$  &   10.88 $_{- 0.09 }^{+ 0.12 }$  &   1.82 $_{- 0.22 }^{+ 0.23 }$  &   J  \\
120 & 3.82                  & U                              & 0.52 $\pm$ 0.13                         & 46 $\pm$ 11                                & 0.32  &   46 $\pm$ 11  & 20 $_{- 5 }^{+ 6 }$  &   10.35 $_{- 0.07 }^{+ 0.07 }$  &   1.05 $_{- 0.20 }^{+ 0.22 }$  &   U  \\
121 & 2.90  & J          & 0.46 $\pm$ 0.07     & 41 $\pm$ 6             & 0.23  &   41 $\pm$ 6   & 17 $_{- 3 }^{+ 4 }$        & 10.36 $_{- 0.05 }^{+ 0.04 }$ & 1.66 $_{- 0.14 }^{+ 0.18 }$ & J                   \\
122 & 1.646                   & {\citet{{2012ApJS..200...13B, 2015ApJS..219...29M}}}                              & 0.45 $\pm$ 0.08                         & 40 $\pm$ 6                                 & 1.07  &   41 $\pm$ 6   & 27 $_{- 4 }^{+ 4 }$  &   10.14 $_{- 0.04 }^{+ 0.04 }$  &   1.28 $_{- 0.10 }^{+ 0.08 }$  &   J \\
123 & 2.58              & J                              & 0.44 $\pm$ 0.13                         & 39 $\pm$ 11                                & 0.33  &   40 $\pm$ 11 & 33 $_{- 5 }^{+ 7 }$  &   10.72 $_{- 0.05 }^{+ 0.05 }$  &   1.64 $_{- 0.16 }^{+ 0.21 }$  &   J   \\
124 & 4.447                   & {\citet{2020A&A...643A...1L}}                              & 0.31 $\pm$ 0.08                         & 27 $\pm$ 6                                 & 11.67  &   39 $\pm$ 7  & 29 $_{- 3 }^{+ 4 }$  &   9.76 $_{- 0.04 }^{+ 0.05 }$  &   0.19 $_{- 0.03 }^{+ 0.04 }$  &   J    \\
125 & 4.578                   & {\citet{2020ApJ...900....1F}}                              & 0.30 $\pm$ 0.06                          & 26 $\pm$ 5                                 & 3.39  &   30 $\pm$ 5 & 13 $_{- 2 }^{+ 3 }$  &   9.41 $_{- 0.06 }^{+ 0.08 }$  &   0.34 $_{- 0.06 }^{+ 0.06 }$  &   J       \\
126 & 6.854 & {\citet{2018Natur.553..178S}}          & 0.27 $\pm$ 0.04     & 24 $\pm$ 3             & 14.61  &   38 $\pm$ 4     & 28 $_{- 4 }^{+ 4 }$        & 10.19 $_{- 0.05 }^{+ 0.05 }$ & 0.16 $_{- 0.05 }^{+ 0.05 }$ & J                   \\
127 & 0.945                   & {\citet{{2012ApJS..200...13B, 2015ApJS..219...29M}}}                              & 0.18 $\pm$ 0.03                         & 16 $\pm$ 3                                 & 0.02  &   16 $\pm$ 3   & 9 $_{- 2 }^{+ 1 }$  &   10.48 $_{- 0.08 }^{+ 0.05 }$  &   2.67 $_{- 0.43 }^{+ 0.15 }$  &   J  \\
128 & 2.33                 & U                              & 0.16 $\pm$ 0.03                         & 14 $\pm$ 2                                 & 0.48  &   15 $\pm$ 2  & 18 $_{- 7 }^{+ 13 }$  &   11.04 $_{- 0.04 }^{+ 0.04 }$  &   1.02 $_{- 0.29 }^{+ 0.30 }$  &   U \\

\hline
\end{tabularx}
\end{table*}

\clearpage

\section{Colour (RGB) postage-stamp images of the near-mid infrared ALMA-detected galaxies}
\label{sec:append_b}
In this Appendix we provide three-colour (RGB) postage-stamp images of all the near-mid infrared galaxy counterparts of the 128 ALMA sources in our final PRIMER COSMOS sample. Where NIRCam imaging was available these were created using the F444W (red), F277W (green) and F115W (blue) images, after convolving the imaging in the latter two filters to the same resolution as the F444W imaging. Where NIRCam imaging was not available (i.e., in the regions with only {\it JWST} MIRI coverage) we used the MIRI  $18\,\mu$m (red) and $7.7\,\mu$m (yellow) imaging to produce the pseudo-colour images (albeit these inevitably have poorer resolution than the NIRCam examples).

\newcommand{\imagegrid}[1]{
    \begin{figure}
        \centering
        \renewcommand{\thesubfigure}{\arabic{subfigure}}
        \setlength{\abovecaptionskip}{-5pt}  
        \setlength{\belowcaptionskip}{-10pt} 
        \setlength{\floatsep}{0pt}           
        \setlength{\textfloatsep}{0pt}       
        \setlength{\intextsep}{0pt}          
        \setlength{\parskip}{0pt}            
        #1
        \caption{RGB postage-stamp images of the near-mid infrared galaxy counterparts of all 128 ALMA sources in our final PRIMER COSMOS sample. For sources covered by NIRCam, the images are produced using F444W (red), F277W (green), and F115W (blue) after convolving the shorter wavelength images to the same resolution as the F444W imaging. Where NIRCam imaging was not available (i.e., in the regions with only {\it JWST} MIRI coverage) we used the MIRI  $18\,\mu$m (red) and $7.7\,\mu$m (yellow) imaging to produce the pseudo-colour images (albeit these inevitably have poorer resolution than the NIRCam examples). The ALMA coordinates from the highest-S/N detection (see Table \ref{tab:CAT1}) are marked as white crosses. The size of each image is $3.5''\times3.5''$, with an axis tick every arcsec.}
        \label{fig:stamp1}
    \end{figure}
}

\newcommand{\imagegridd}[1]{
    \begin{figure}
        \centering
        \renewcommand{\thesubfigure}{\arabic{subfigure}}
        #1

        \contcaption{}
        \label{fig:stamp2}
    \end{figure}
}

\newcommand{\imagegriddd}[1]{

    \begin{figure}
        \centering
        \renewcommand{\thesubfigure}{\arabic{subfigure}}
        #1
        \contcaption{}
        \label{fig:stamp3}
    \end{figure}
}

\newpage
\imagegrid{
\subfloat{\includegraphics[width=0.99\textwidth]{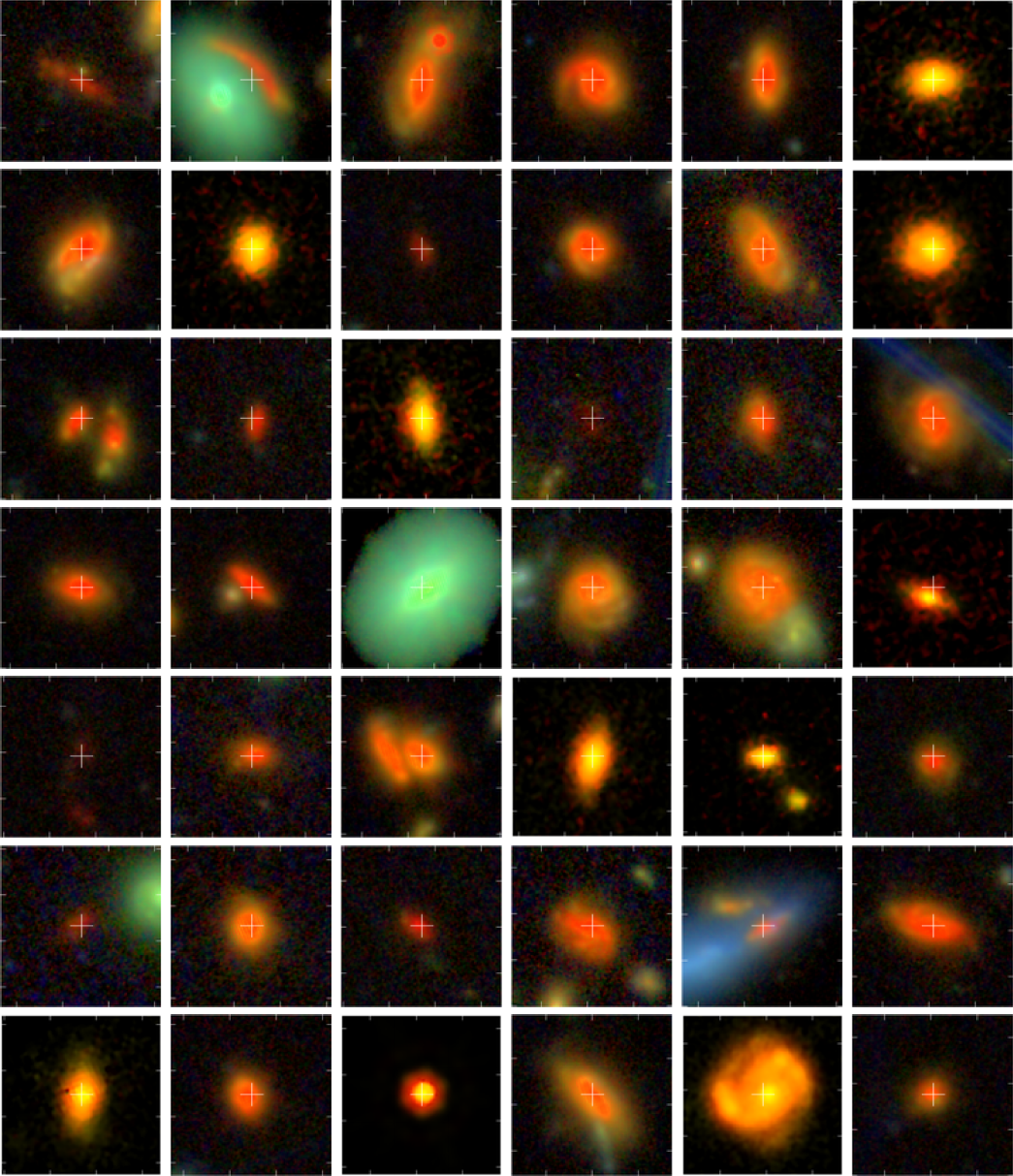}}
}

\newpage
\imagegridd{
\subfloat{\includegraphics[width=0.99\textwidth]{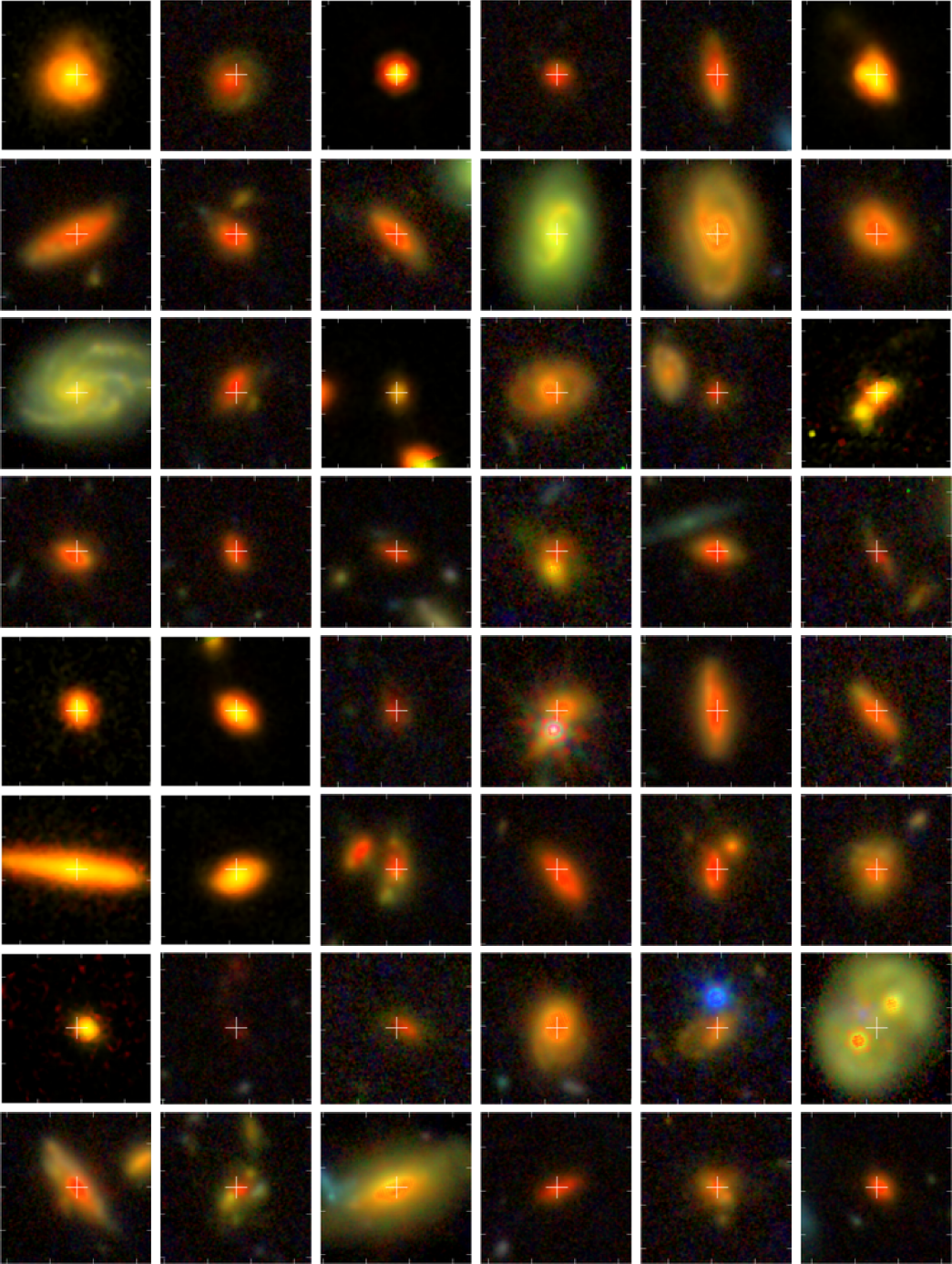}}
}

\newpage
\imagegriddd{
\subfloat{\includegraphics[width=0.99\textwidth]{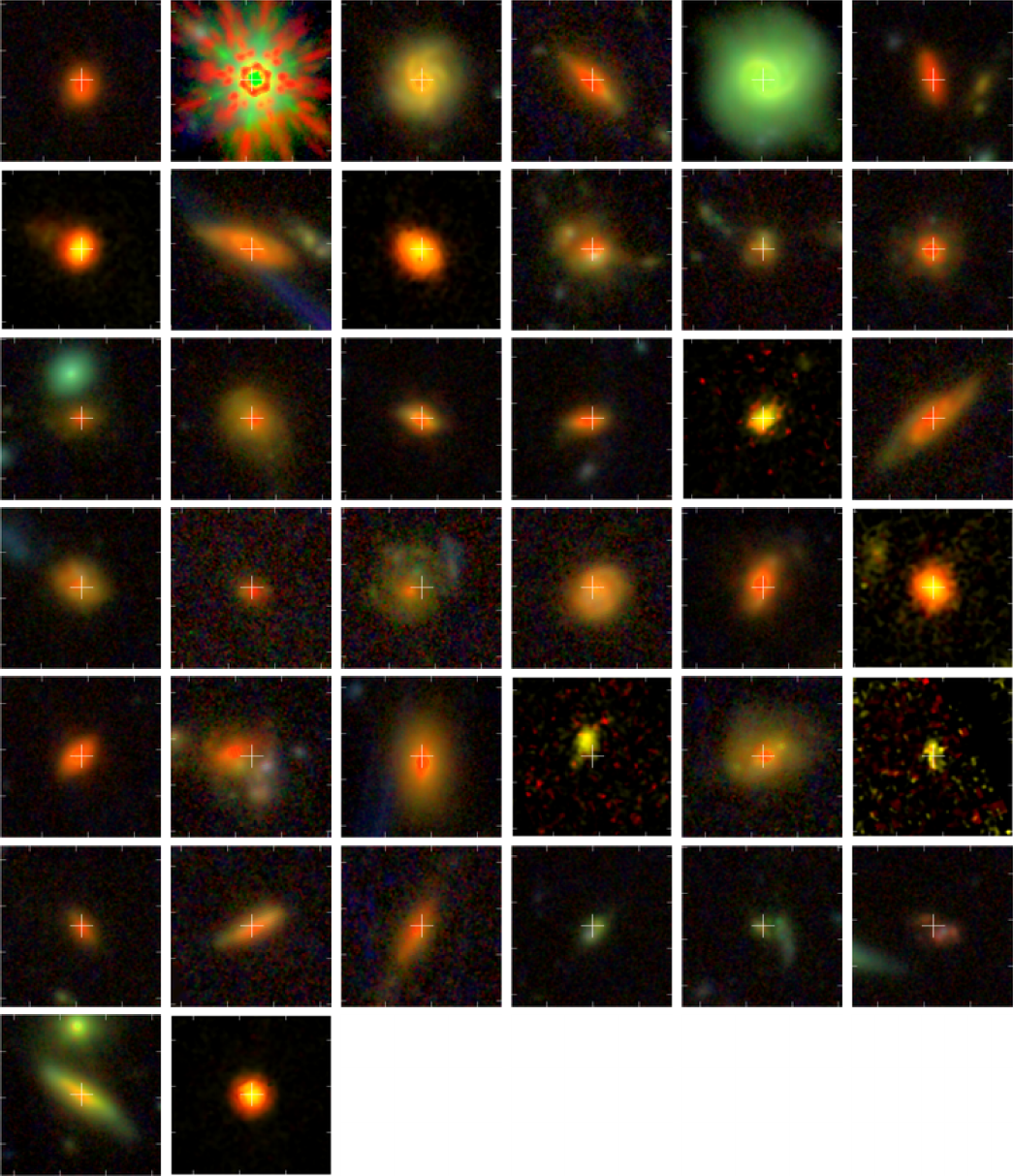}}
}

\clearpage
\bsp	
\label{lastpage}
\end{document}